\documentclass[a4paper,12pt]{book}
\usepackage{times}
\usepackage[T1]{fontenc}
\usepackage[utf8]{inputenc}
\usepackage[english]{babel}
\usepackage{verbatim} 
\usepackage{bbm} 

\usepackage{geometry}
\usepackage{setspace}
\RequirePackage{pstricks}
\newrgbcolor{Gray}{0.5 0.5 0.5}
\usepackage{graphicx}
\usepackage{floatflt} 
\usepackage[%
  font={small,rm,md,it}
]{caption,subfig} 

\usepackage[title]{appendix}
\renewcommand{\appendixtocname}{Appendices for this chapter}

\usepackage{fancyhdr}
\newcommand{\HeaderTextColor}{\Gray}
\newcommand{\MyHeadingChapterMark}{
    \sffamily \footnotesize \HeaderTextColor
    {\bfseries \small \thepage} \quad  
    {\scshape \leftmark}
}
\newcommand{\MyHeadingSectionMark}{
  \nouppercase{
    \sffamily \footnotesize \HeaderTextColor 
    \rightmark \quad {\bfseries \small \thepage}
  } 
}
\usepackage{bm} 
\usepackage{amsmath,amssymb,amsthm,amsxtra}
\usepackage{dsfont} 

\usepackage{hyperref}

\newcommand{\be}{\begin{equation}}
\newcommand{\ee}{\end{equation}}
\newcommand{\ben}{\begin{eqnarray}}
\newcommand{\een}{\end{eqnarray}}

\newcommand{\nn}{\nonumber}

\begin{document}


\thispagestyle{empty}

\vspace{11cm}

\centerline{\Large{\bf  Josinaldo Menezes da Silva}}

\vspace{3.5cm}
\centerline{\huge{\bf Cosmological Consequences of}}
\centerline{\huge{\bf Topological Defects:}}
\centerline{\Huge{\bf Dark Energy and }}
\centerline{\Huge{\bf Varying Fundamental Constants}}
\vspace{2.5cm}
\vspace{3.5cm}

\begin{figure}
\begin{center}
\includegraphics*[width=8cm]{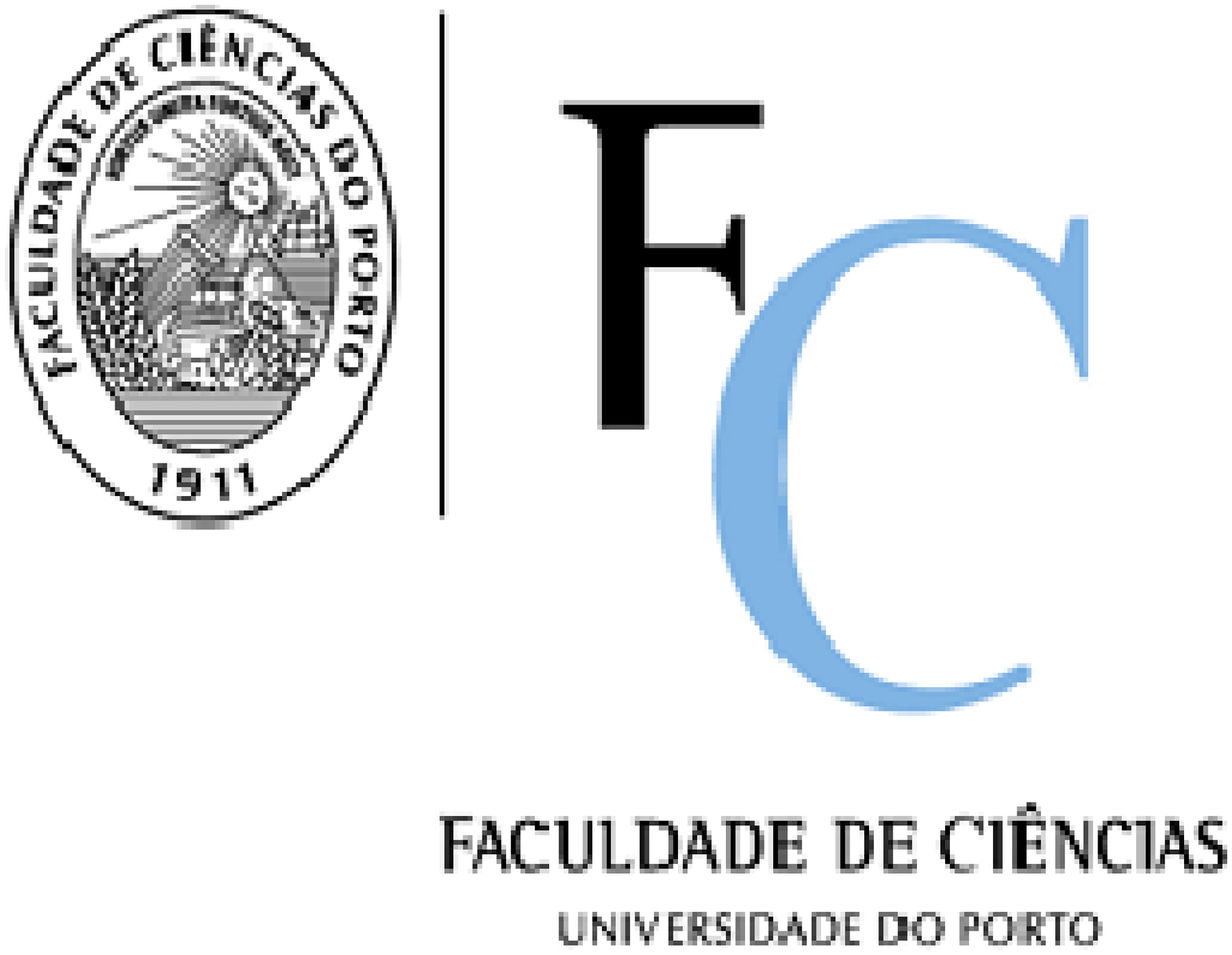}
\end{center}
\end{figure}

\centerline{{\bf Departamento de F\'\i sica }}
\centerline{{\bf Faculdade de Ci\^encias da Universidade do Porto}}
\centerline{{\bf Portugal}}
\centerline{{\bf August 2007}}

\columnsep 0.0cm

\newpage

\begin{center}
\LARGE{{\sf}}
\end{center}
\bigskip
\newpage
\thispagestyle{empty}
\setlength{\baselineskip}{25pt}
\centerline{\Large{\rm  Josinaldo Menezes da Silva}}
\bigskip
\vspace{1cm}
\centerline{\Huge{\rm Cosmological Consequences of}}
\centerline{\Huge{\rm Topological Defects:}}
\centerline{\Huge{\rm Dark Energy and }}
\centerline{\Huge{\rm Varying Fundamental Constants}}
\vspace{1.5cm}
\centerline{\textit{Tese submetida \`a Faculdade de Ci\^encias da Universidade do Porto}}
\centerline{\textit{para obten\c c\~ao do grau de Doutor em F\'\i sica}}
\vspace{1.5cm}
\centerline{{\rm Departamento de F\'\i sica }}
\centerline{{\rm Faculdade de Ci\^encias da Universidade do Porto}}
\begin{figure}
\begin{center}
\includegraphics*[width=8cm]{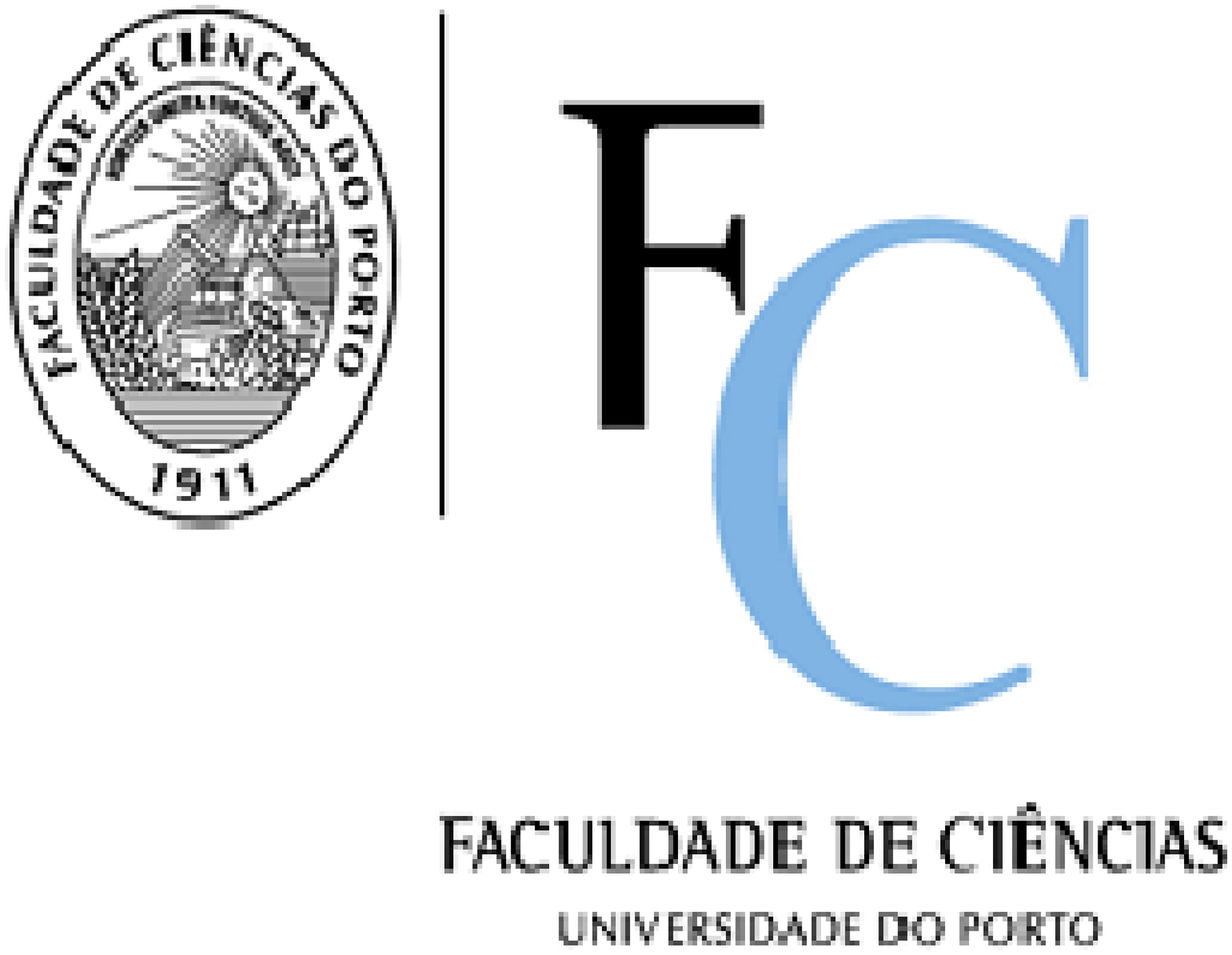}
\end{center}
\end{figure}
\vspace{1cm}
\centerline{{\rm Portugal}}
\centerline{{\rm August 2007}}
\columnsep 0.0cm

\newpage
\newpage

\begin{center}
\LARGE{{\sf}}
\end{center}
\bigskip
\newpage

\newpage
        \vspace*{7cm}
        \begin{flushright}

To my parents,\\
wife \&
daughters.

        \end{flushright}
\newpage

\begin{center}
\LARGE{{\sf}}
\end{center}
\bigskip
\newpage

\vspace{40.5cm}
\begin{center}
\textit{{\bf A Aeronave}}

\vspace*{1.5cm}
\textit{Cidindo a vastid\~ao do Azul Profundo,}\\
\textit{Sulcando o espa\c co, devassando a terra,}\\
\textit{A aeronave que um mist\'erio encerra}\\
\textit{Vai pelo espa\c co acompanhando o mundo.}

\vspace*{0.5cm}
\textit{E na esteira sem fim da az\'ulea esfera}\\
\textit{Ei-la embalada n'amplid\~ao dos mares,}\\
\textit{Fitando o abismo sepulcral dos mares,}\\
\textit{Vencendo o azul que ante si s'erguera.}

\vspace*{0.5cm}
\textit{Voa, se eleva em busca do infinito,}\\
\textit{\'E como um despertador de estranho mito,}\\
\textit{Auroreando a humana consci\^encia.}

\vspace*{0.5cm}
\textit{Cheia da luz do cintilar de um astro,}\\
\textit{Deixa ver na fulg\^encia do seu rastro}\\
\textit{A trajet\'oria augusta da Ci\^encia.}

\vspace*{1.0cm}
{\bf Augusto dos Anjos (1884-1914)}\\
{\rm Cruz do Esp\'\i rito Santo, Para\'\i ba.}
\end{center}
\newpage

\begin{center}
\LARGE{{\sf}}
\end{center}
\bigskip
\newpage

\begin{center}
\LARGE{{\sf Financial Support}}
\end{center}

\vspace{6cm}	

\begin{figure}[ht]
\begin{center}
\includegraphics*[width=12cm]{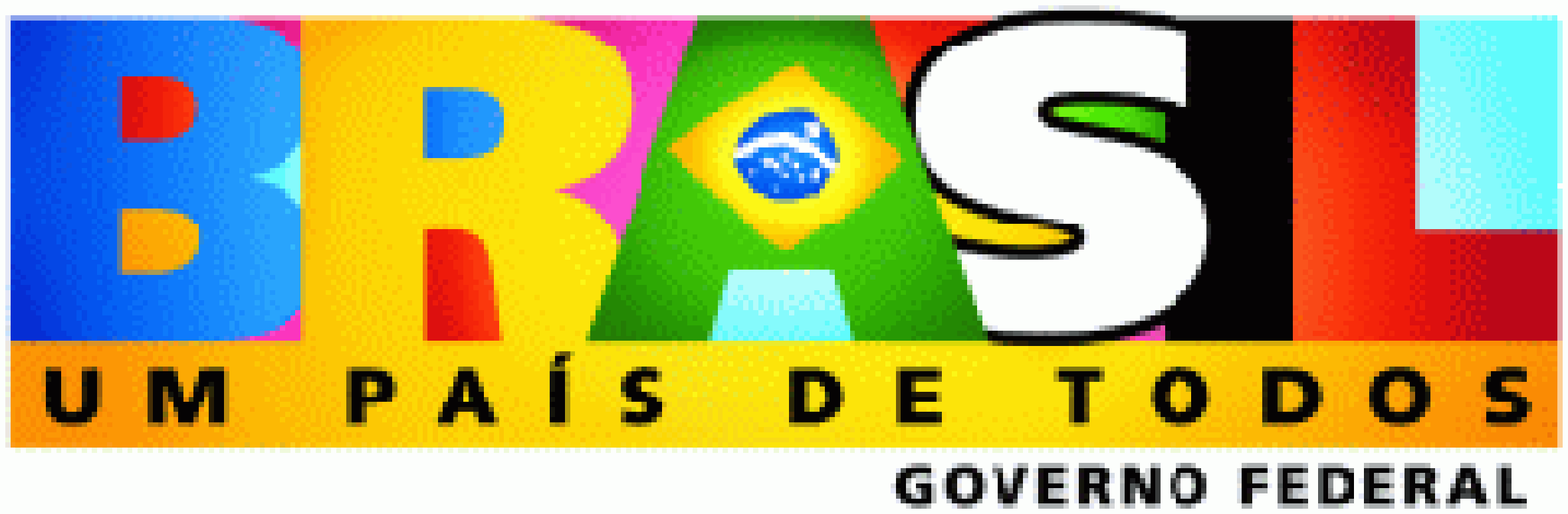}
\end{center}
\end{figure}

The work developed during the PhD program was completely supported by the Brazilian Government
through the Grant BEX 1970/02-0 from Capes - \textit{Coordena\c c\~ao de Aperfei\c coamento de Pessoal de N\'\i vel Superior}.

\makeatletter
\def\cleardoublepage{\clearpage\if@twoside \ifodd\c@page\else
  \hbox{}
  \vspace*{\fill}

  \vspace{\fill}
  \thispagestyle{empty}
  \newpage
  \if@twocolumn\hbox{}\newpage\fi\fi\fi}
\makeatother

\newpage

\begin{center}
\LARGE{{\sf}}
\end{center}
\bigskip
\newpage

\newpage

\begin{center}
\LARGE{{\sf Acknowledgments}}
\end{center}
\bigskip
The work developed in these doctoral years was possible due to help and encouragement of many people from here and far away.

First, I would like to thank my supervisors Caroline Santos and Pedro Avelino for their constant support and encouragement. 
I am also grateful to Carlos Martins for providing helpful comments and lessons related to the work carried out during these years.
Also, I would like to thank Roberto Menezes and Joana Oliveira for fruitfully collaboration.

I would like acknowledge to the Physics Department of University of Porto members for their hospitality and support. Particularly, I would like to thank my colleagues: Eduardo, Lu\'\i s, Miguel, Jo\~ao Penedones, Jo\~ao Viana, V\'\i tor, Joana, Carmen, Teresa, Aires, Pedro Gil, Welberth and Sanderson. I would also like to thank the Portuguese people who helped me and my family to settle during the first months. Thank you for teaching us about the culture and heritage of this beautiful land.  

I am specially grateful to my family and friends in Brazil, who gave me support for starting, developing and concluding my PhD program. I would like to thank all professors and students from UFPB and CEST, my friends and each member of my family. Specially, I would like to mention my mother and my father who thought me the correct way to win at life, and obviously 
my wife Fl\'avia and my daughters Laissa and Mel because they gave me energy and motivation to conclude this work. A special kiss for you three.  

Last but not least, my warmest thanks to  F\'abio, Fl\'avio, Zuleide, Heraldo, Viviane, Alexandre, Eli, Beto, Caio, Edilma, Elias, Salete, Luciana, Jos\'elia, Adalto, S\'\i lvia, Welberth, Suelen, Fernanda, Sanderson, Neto, Nayara, Pedro, Helena, In\^es, Lino, Rita and Jorge.
\newpage

\begin{center}
\LARGE{{\sf}}
\end{center}
\bigskip
\newpage
\begin{center}
\LARGE{{\sf Resumo}}
\end{center}
\bigskip

Defeitos topol\'ogicos ter\~ao sido formados no universo primitivo sendo as suas propriedades dependentes dos detalhes da quebra espont\^anea de simetria que os gerou. Neste trabalho n\'os consideramos consequ\^encias cosmol\'ogicas de paredes do dom\'\i nio, cordas c\'osmicas e monopolos magn\'eticos.

Paredes de dom\'\i nio, formadas quando uma simetria discreta \'e quebrada, s\~ao o exemplo mais simples de um defeito topol\'ogico.
Neste trabalho investigamos a possibilidade de redes de paredes de dom\'\i nio poderem
explicar 
a actual expans\~ao acelerada do universo.
Discutimos condi\c c\~oes necess\'arias para a obten\c c\~ao de uma rede de paredes 
de dom\'\i nio bidimensional frustrada
e propomos uma classe de modelos que, no limite de grande n\'umero $N$ de campos escalares acoplados, aproxima-se do ent\~ao chamado modelo `ideal' (em termos de seu potencial para produzir frustra\c c\~ao da rede). Utilizando os resultados das maiores e mais precisas simula\c c\~oes de teoria de campo tridimensionais de redes de paredes de dom\'\i nio com jun\c c\~oes, encontramos evid\^encias incisivas para uma aproxima\c c\~ao gradual 
a uma solu\c c\~ao invariante de escala cujos par\^ametros apresentam 
uma ligeira depend\^encia em $N$. Conjecturamos que, apesar de ser poss\'\i vel a constru\c c\~ao (\`a m\~ao) de redes est\'aveis, nenhuma destas redes seriam produtos directos de paredes de dom\'\i nio formadas em transi\c c\~oes de fase cosmol\'ogicas.

Cordas C\'osmicas e Monopolos Magn\'eticos podem ser formados quando simetrias cont\'\i nuas $U(1)$ e $SU(2)$ s\~ao espontaneamente quebradas, respectivamente.
Investigamos cordas c\'osmicas e monopolos magn\'eticos em modelos tipo-Bekenstein e mostramos que existe uma classe de modelos que ainda permitem as solu\c c\~oes cl\'assicas de v\'ortice de Nielsen-Olesen e de monopolo de 't Hooft-Polyakov.  
Contudo, em geral as solu\c c\~oes est\'aticas de cordas e monopolos em modelos tipo-Bekenstein s\~ao diferentes das cl\'assicas com a energia electromagn\'etica dentro de seus n\'ucleos gerando varia\c c\~oes espaciais da constante de estrutura fina na vizinhan\c ca dos defeitos.
Consideramos modelos com um fun\c c\~ao cin\'etica gen\'erica e mostramos que constrangimentos provenientes do Princ\'\i pio de Equival\^encia imp\~oem limites muito fortes \`as varia\c c\~oes de $\alpha$ induzidas pelas redes de cordas c\'osmicas em escalas cosmol\'ogicas. 

Finalmente, estudamos a evolu\c c\~ao das varia\c c\~oes espaciais da constante de estrutura fina induzidas por perturba\c c\~oes de densidade n\~ao-lineares.
Mostramos que os resultados obtidos utilizando o modelo de colapso esf\'erico para uma inomogeneidade de comprimento de onda infinito s\~ao inconsistentes com os resultados de um estudo local usando
gravidade linearizada e argumentamos em favor da segunda aproxima\c c\~ao.
Tamb\'em criticamos a sugest\~ao de 
que o valor de $\alpha$ de regi\~oes colapsadas
poderia ser significativamente
diferente do valor de $\alpha$ do universo de fundo, com base nos resultados obtidos.

\newpage
\begin{center}
\LARGE{{\sf Summary}}
\end{center}
\bigskip

Topological defects are expected to form in the early universe and their properties depend on the particular details of the spontaneous symmetry breaking that has generated them. In this work we consider cosmological consequences of domain walls, cosmic strings and magnetic monopoles.

Domain walls, formed when a discrete symmetry is broken, are the simplest example of a topological defect. 
We investigated the domain wall networks as a possible candidate to explain the present accelerated expansion of the universe. 
We discuss various requirements that any stable lattice of frustrated walls must obey and propose a class of models which, in the limit of large number $N$ of coupled scalar fields, approaches the so-called `ideal' model (in terms of its potential to lead to network frustration). 
By using the results of the largest and most accurate three-dimensional field theory simulations of domain wall networks with junctions,
we find compelling evidence for a gradual approach to scaling, with the quantitative scaling parameters having only a mild dependence on $N$. 
We conjecture that, even though one can build (by hand) lattices that would be stable, no such lattices will ever come out of realistic domain wall forming cosmological phase transitions. 

Cosmic strings and magnetic monopoles 
can arise with $U(1)$ and $SU(2)$ spontaneous symmetry breaking, respectively.
We consider cosmic strings and magnetic monopoles in Bekenstein-type models
and show that there is a class of models of this type for which 
the classical Nielsen-Olesen vortex and 't Hooft-Polyakov monopoles are still valid solutions. However, 
in general static string and monopole solutions in Bekenstein-type models strongly 
depart from the standard ones with the electromagnetic 
energy concentrated inside their cores seeding spatial variations of 
the fine structure constant, $\alpha$.
We consider models with a generic gauge kinetic function and show that
Equivalence Principle constraints impose tight limits on the allowed
variations of $\alpha$ induced by string networks on cosmological scales. 

Finally, we study the evolution of the spatial variation of the fine-structure constant induced by non-linear density perturbations.
We show that the results obtained using the spherical infall model for an infinite wavelength inhomogeneity are inconsistent with the results of a local linearized gravity study and we argue in favor of the second approach. We also criticize the claim that the value of $\alpha$ inside collapsed regions could be significantly different from the background one on the basis of these findings. 

\newpage
\begin{center}
\LARGE{{\sf R\'esum\'e}}
\end{center}
\bigskip

Des d\'efauts topologiques ont du \^etre à l’origine de l’univers primitif, ses propri\'et\'es \'etant d\'ependantes des d\'etails de la casse spontan\'ee de sym\'etrie qui en sont à l’origine. Dans ce travail, nous consid\'ererons des cons\'equences cosmologiques de murs du domaine, cordes cosmiques e monopoles magn\'etiques.  

Les murs du domaine, form\'es quand une sym\'etrie discrète est cass\'ee, sont l’exemple le plus simple d’un d\'efaut topologique. Dans ce travail, nous investiguons la possibilit\'e que des r\'eseaux de murs de domaine puissent expliquer l’actuelle expansion acc\'el\'er\'ee de l’univers. Nous discutons des conditions n\'ecessaires pour l’obtention d‘un r\'eseau de murs de domaine bidimensionnel frustr\'e e nous proposons une classe de modèles qui, à la limite de grand num\'ero N de champs scalaires coupl\'es, s’approchent alors du modèle appel\'e id\'eal (vue sa capacit\'e à provoquer la frustration du r\'eseau). En utilisant les plus grands r\'esultats et les simulations plus pr\'ecises de th\'eorie de champs tridimensionnels de r\'eseaux de murs de domaine avec des jonctions, nous avons trouv\'e des \'evidences incisives pour une approximation graduelle à une solution invariante d’\'echelle dont les paramètres pr\'esentent une l\'egère d\'ependance en N.  Nous avons suppos\'e que, malgr\'e le fait de la possibilit\'e de la construction (manuelle) de r\'eseaux stables, aucuns de ces dits r\'eseaux ne serait le produit direct de murs de domaine form\'es de transitions de phase cosmologiques.

Des Cordes Cosmiques et des Monopoles Magn\'etiques sont form\'es quand des sym\'etries continues $U(1)$ et $SU(2)$ sont bris\'ees. Nous avons investigu\'e des cordes cosmiques ainsi que des monopole magn\'etiques dans des modèles de type Bekenstein et nous avons montr\'e qu’il y a une classe de modèles qui permettent encore les solutions classiques de vortex de Nielsen Olsen et de monopole de 't Hooft Polyakov. Cependant, les solutions statiques des cordes et monopoles dans les modèles de type Bekenstein sont g\'en\'eralement diff\'erentes des classiques avec l’\'energie electro-magnetique à l’int\'erieur de ses noyaux, g\'en\'erant des variations spatiales de la constante de structure fine près des d\'efauts. Nous avons consid\'er\'e des modèles avec une fonction cin\'etique g\'en\'erique e nous avons montr\'e que des d\'esagr\'ements provenant du Principe d’Equivalence imposent de très petites limites aux variations de $\alpha$.

Finalement, nous \'etudions l’\'evolution des variations spatiales de la constante de structure fine induites par des perturbations de densit\'e non lin\'eaires. Nous montrons que les r\'esultats obtenus en utilisant le modèle de collapse sph\'erique, pour une inhomog\'en\'eit\'e de longueur d’onde, sont peu conformes avec les r\'esultats d’une \'etude locale qui utilise la gravit\'e linearis\'ee et argument\'es en faveur de la deuxième approximation. Nous critiquons aussi la suggestion selon laquelle la valeur de $\alpha$ de r\'egions collaps\'ees pourrait \^etre significativement diff\'erente de la valeur de $\alpha$ de l’univers de fond, sur la base des r\'esultats obtenus.

\newpage
\begin{center}
\LARGE{{\sf List of Publications}}
\end{center}
\bigskip
\begin{itemize}
\item
\textit{Dynamics of Domain Wall Networks with Junctions}\\
P. P.~Avelino,
C.~J. A.~P. Martins,
J. Menezes,
R. Menezes and
J.~C. R.~E. Oliveira,\\
Work in Progress.
\item
\textit{Scaling of Cosmological Domain Wall Networks with Junctions}\\
P. P.~Avelino,
C.~J. A.~P. Martins,
J. Menezes,
R. Menezes and
J.~C. R.~E. Oliveira,\\
Phys. Lett. B 647: 63 (2007).

\item
\textit{Defect Junctions and Domain Wall Dynamics }\\
P. P.~Avelino,
C.~J. A.~P. Martins,
J. Menezes,
R. Menezes and
J.~C. R.~E. Oliveira,\\
Phys. Rev. D 73: 123520 (2006).

\item
\textit{Frustrated Expectations: Defect Networks and Dark Energy}\\
P. P.~Avelino,
C.~J. A.~P. Martins,
J. Menezes,
R. Menezes and
J.~C. R.~E. Oliveira,\\
Phys. Rev. D 73: 123519 (2006).

\item
\textit{The Evolution of the Fine-Structure Constant in the Non-Linear Regime}\\
P. P.~Avelino,
C.~J. A.~P. Martins,
J. Menezes and
C. Santos,\\
JCAP 0612: 018, (2006).

\item
\textit{Gravitational Effects of Varying-$\alpha$ Strings}\\
J. Menezes,
C. Santos and
P. P.~Avelino,\\
Int. J. Mod. Phys. A 21: 3295, (2006).

\item
\textit{Varying-$\alpha$ Monopoles}\\
J. Menezes,
C. Santos and
P. P.~Avelino,\\
Phys. Rev. D 72: 103504, (2005).

\item
\textit{Global Defects in Field Theory with Applications to Condensed Matter}\\
D. Bazeia,
J. Menezes and
R. Menezes,\\
Mod. Phys. Lett. B19: 801, (2005).

\item
\textit{Cosmic Strings in Bekenstein-type Models}\\
J. Menezes,
C. Santos and
P. P.~Avelino,\\
JCAP 0502: 003, (2005).

\end{itemize}

\newpage
\begin{center}
\LARGE{{\sf Declaration}}
\end{center}
\vspace{4cm}	
\bigskip
The work presented in this thesis, except where otherwise indicated, is my own and to the best of my knowledge original. The work described in Chapter \ref{cap2} was carried out in collaboration with Pedro Avelino, Carlos Martins, Roberto Menezes and Joana Oliveira.
The work described in Chapters \ref{cap3} and \ref{cap4} was developed in collaboration with Pedro Avelino and Caroline Santos. The work described in Chapter \ref{cap5} was carried out in collaboration with Pedro Avelino, Carlos Martins and Caroline Santos.
The complete set of results of this thesis has not been submitted for any other degree, diploma or qualification except the work described in Sections \ref{intr11} and \ref{intr122} which appeared first in the thesis \textit{Domain walls in cosmology: network evolution and dark energy scenarios} by Joana Oliveira.
Parts of this thesis first appeared in various papers \cite{Nosso,Nosso2,Nossog,Nosso3,IDEAL1,IDEAL2,IDEAL3}. Other sections of this manuscript will appear soon in a paper currently in preparation \cite{IDEAL4}.
\newpage

\begin{center}
\LARGE{{\sf}}
\end{center}
\bigskip
\newpage

\newpage

\begin{center}
\LARGE{{\sf Units}}
\end{center}
\bigskip
Throughout this thesis, the space-time metric $g_{\mu\nu}$ will have a signature $+---$. 
Greek alphabet indices run over space and time while Latin indices
over space only. 
Except if stated otherwise we shall assume fundamental units in which
\be
\hbar = c = k_B = G\,m_{Pl}^2 = 1,
\ee
then all quantities can be expressed in terms of energy in ${\rm GeV}$ ($1 {\rm GeV} = 10^9 {\rm eV}$). The conversion factors are
\ben
&& 1\, GeV = 1.60 \times 10^{-3} erg = 1.16 \times 10^{13} K = 1.78 \times 10^{-24} g,\\
&& 1\, GeV^{-1} = 1.97 \times 10^{-14} cm = 6.58 \times 10^{-25} s.
\een
The Planck time and mass are approximately
\ben
&& t_{Pl} \sim 5.4 \times 10^{-44} s,\\
&& m_{Pl} \sim 1.2 \times 10^{19} GeV.
\een
Astrophysical distances will usually be expressed in parsecs, with 
\be
1\, pc \approx 3.1 \times 10^{18} cm,
\ee
or ${\rm Mpc}$ ($1\, {\rm Mpc} = 10^6 {\rm pc}$).

\newpage

\begin{center}
\LARGE{{\sf}}
\end{center}
\bigskip
\newpage


\renewcommand{\baselinestretch}{1.0} %
\small\normalsize %

\tableofcontents %
\newpage

\begin{center}
\LARGE{{\sf}}
\end{center}
\bigskip
\newpage
\addcontentsline{toc}{chapter}{Figures Content}\listoffigures%

\renewcommand{\baselinestretch}{1.5}
\small\normalsize %

\chapter{\label{intro}{\sf Introduction}}
\section{Overview}
In the last decades, there were important developments in the interface between cosmology and particle physics. In this scenario, topological defects have a wide range of cosmological implications. Among these, it has been proposed that domain wall networks might be applied to explain the accelerated expansion of the universe at the present time
\cite{WMAP3}.
In the context of the varying fundamental constants theories
\cite{Webb1},
it has been claimed that the cosmic strings, magnetic monopoles and other compact objects  
might generate space-time variations of the fine-structure constant, $\alpha$, in their vicinity.
In this thesis we will study these issues. 

In this  chapter we first review the foundations, successes and problems of the standard cosmological model in Sec. \ref{sccc}. In Sec. \ref{sfc} we introduce the scalar fields and the topological defects. We also study the static one-dimensional soliton solution and some models of scalar fields in dark energy scenario. Finally, in Sec. \ref{vfc} we review the varying fundamental constants observational results. 
  
\section{The Standard Cosmological Model}\label{sccc}
The Standard Cosmological Model is based on the assumption that the universe is
homogeneous and isotropic on large scales.
The most general form of a line element which is invariant under
spatial rotations and translations is
\ben
\label{11}
ds^2 = g_{\mu\nu} dx^\mu dx^\nu 
= dt^2 - a^2(t)\,\left[\frac{dr^2}{1-k\,r^2}+r^2\,(d\theta^2+\sin^2\theta\,d\varphi^2) \right],
\een
where $g_{\mu\nu}$ is the metric tensor of the Friedmann-Robertson-Walker (FRW) geometry, $a(t)$ is the scale factor and $t$ is the cosmic time. Note that the value of $k$ determines if the universe is flat ($k=0$), closed ($k=+1$) or open ($k=-1$).
The line element (\ref{11}) describes a homogeneous and isotropic space-time and can be parametrized in terms of the conformal time coordinate $\tau$ as
\ben
\label{12}
ds^2 = a^2(\tau)\, \Big\{ d\tau^2 - \left[\frac{dr^2}{1-k\,r^2}+r^2\,(d\theta^2+\sin^2\theta\,d\varphi^2) \right] \Big\}.
\een

In order to describe the components of the energy of the universe, we consider perfect barotropic fluids, i.e., perfect fluids with a definite relation between pressure $p$ and energy density $\rho$. 
These are
described by the tensor
\be
T_{\mu\nu} = (p + \rho)\,u_\mu\,u_\nu - p\,g_{\mu\nu},
\ee
where $u_\mu = dx_\mu/ds$ and 
for example a gas of photons has
$p=\rho/3$.	 

General Relativity connects the evolution of the universe to its energy content through the Einstein equations 
\be
\label{16}
R_{\mu\nu} - \frac12 g_{\mu\nu}\,R = 8\pi\,G\,T_{\mu\nu},
\ee
where $R_{\mu\nu}$ is the Ricci tensor, i.e.,
\be
R_{\mu\nu}=\partial_\sigma \Gamma_{\mu\nu}^\sigma - \partial_\nu \Gamma_{\mu\sigma}^\sigma + \Gamma_{\mu\nu}^\sigma\,\Gamma_{\sigma\beta}^\beta - \Gamma_{\mu\sigma}^\beta\,\Gamma_{\beta\nu}^\sigma
\ee
and $R=R_\mu^\mu$ is the Ricci scalar. Using the line element (\ref{11}) one has 
\ben
\label{13}
&&H^2 = \frac{8\,\pi\,G}3\,\rho - \frac{k}{a^2},\\
\label{14}
&&\dot H = -4\pi G\,(\rho + p) + \frac{k}{a^2},\\
\label{15}
&&\dot \rho + 3\,H\,(\rho + p) = 0,
\een
where $H=\dot a /a$ is the Hubble parameter and $\rho$ and $p$ are the total energy density and pressure of the universe. 
The dot represents the derivative with respect to $t$.
Note that while Eq. (\ref{13}) is obtained from the $00$ component of Eq. (\ref{16}), Eq. (\ref{14}) results from a linear combination of the $ij$ and $00$ components.
On the other hand, Eq. (\ref{15}) is obtained by considering the covariant conservation of the energy-momentum tensor, $\nabla_\mu \, T^{\mu\nu} = 0$. 
Defining the critical density at a given time as
\be
\rho_{c} = \frac{3\,H^2}{8\,\pi\,G},
\ee  
we can rewrite Eq. (\ref{13}) as
\be
\label{curv}
\Omega - 1 = \frac{k}{a^2\,H^2},
\ee
where $\Omega$ is the density parameter $\Omega \equiv \rho/\rho_c$. According to Eq. (\ref{curv}), for $\rho = \rho_{c}$, $\rho < \rho_{c}$ and $\rho > \rho_{c}$, the universe is flat ($\Omega=1$), open ($\Omega < 1$) and closed ($\Omega> 1$), respectively. 
\subsection{The Energy Content of the Universe}
Let us assume that the total energy density of the universe receives contribution from three components: matter, radiation and a cosmological constant component $\Lambda$, i.e., $\rho= \rho_m + \rho_r + \rho_\Lambda$. Since the more recent observational results indicate that the universe is almost flat \cite{WMAP3}, we can write
\be
\label{17}
\Omega = \Omega_m + \Omega_\Lambda + \Omega_r = 1,
\ee
where $\Omega_i = \rho_i/\rho_{c}$ is the density parameter for each species.

Firstly, the contribution of the non-relativistic species ($p=0$) is parameterized by $\rho_m$, that is composed by a cold dark matter (CDM) component plus a baryonic one, $\Omega_m = \Omega_{dm}+\Omega_b$. 
The term \textit{cold dark matter} means that this component is non-relativistic and does not emit or absorb light. The Wilkinson Microwave Anisotropy Probe (WMAP) three year data \cite{WMAP3} combined with the Supernovae Ia results \cite{Perlmutter}--\cite{Riess2} indicate that
\be
h^2\,\Omega_{dm}^0 \approx 0.111,\,\,\,\,\,\,\,\,\,
h^2\,\Omega_{b}^0 \approx 0.023,
\ee
where the subscript $0$ denotes the present value of the corresponding quantity, and $h$ parametrizes the 
uncertainty of the value of the
Hubble parameter (the Hubble parameter is given by $H_0 = 100 {\rm h km s^{-1} Mpc^{-1}}$ with $0.70 \leq h \leq 0.73$).
Another well known evidence for cold dark matter comes from the rotation curves of spiral galaxies. 
On the other hand, further indirect evidence stems from Big Bang Nucleosynthesis (BBN) \cite{Bernstein}.

Secondly, the contribution of the radiation given by $\rho_r$ may be due to photons ($\Omega_\gamma$), neutrinos
($\Omega_\nu$) and relic gravitons ($\Omega_g$), i.e., $\Omega_{r} = \Omega_{\gamma } +\Omega_{\nu} +\Omega_{g}$. One has 
\be
\label{22}
h^2\,\Omega_{\gamma}^0 \approx 2.47 \times 10^{-5},\,\,\,\,\,\,\,
h^2\,\Omega_{\nu}^0 \approx 1.68 \times 10^{-5},\,\,\,\,\,\,\,
h^2\,\Omega_{g}^0 \lesssim 10^{-11},
\ee
where the bound on the abundance relic gravitons is 
based on the analysis of the integrated Sachs-Wolfe contribution \cite{Giovannini} and the neutrino abundance given above is for three species of massless neutrinos.

Finally, another dark component is present in the universe: the dark energy. 
This component has an equation of state
\be
\omega = \frac{p}{\rho} < - \frac13
\ee
and dominates the energy of the universe today. It is responsible for the recent acceleration of the universe.
The main evidence
for the existence of dark energy comes from the Supernovae Ia observations and the Cosmic Microwave Background.
According to the WMAP three year results combined with the Supernovae Ia ones \cite{Perlmutter}--\cite{Riess2},
\be
h^2\,\Omega_{\Lambda}^0 \approx 0.357.
\ee

Assuming that $h \simeq 0.72$, the fractional contribution of the various components of
the energy density of the universe today are
\be
\label{20}
\Omega_{dm}^0 \sim 0.24,\,\,\,\,\,\,\,
\Omega_{b}^0 \sim 0.02,\,\,\,\,\,\,\,
\Omega_{\Lambda}^0 \sim 0.74,\,\,\,\,\,\,\,
\Omega_{r}^0 \sim 8.0 \times 10^{-5}.
\ee
\subsection{Distances in an Expanding Universe}
Let us consider an object emitting electromagnetic radiation with wavelength $\lambda_e$. In an expanding universe, the wavelength 
received by an observer $\lambda_0$ is not equal to $\lambda_e$. Since $\dot a >0$, the observed wavelength will be 
larger than $\lambda_e$, i.e.,
\be
\lambda_0 = \frac{a(t_0)}{a(t_e)}\,\lambda_e.
\ee
We define the \textit{redshift} $z$ as
\be
z \equiv \frac{\lambda_0-\lambda_e}{\lambda_e}.
\ee
If the luminous signal was emitted when the universe had a fraction of $a/a_0$ of its present size, an observer today should observe
a redshift given by
\be
\label{rshift}
1 + z = \frac{a_0}{a},
\ee
where $a_0$ is the value of the scale factor today. We will take $a_0=1$, except if stated otherwise. 

In a flat universe light traveling freely from $t=0$ to $dt$ moves a comoving distance equal to $dt/a$. Therefore the total comoving distance the light could have traveled can be written as
\be
\eta = \int_0^t \frac{dt'}{a(t')}.
\ee 
Accordingly, no information could have propagated further than $\eta$ since $t=0$ is the beginning of time. In other words, two regions separated by a distance greater than $\eta$ are not causally connected. $\eta$ is then named \textit{particle horizon}.
In some particular cases, it is possible to express $\eta$ analytically in terms of $a$. For instance, for matter-dominated and radiation dominated universes, we have that $\eta \propto a^{1/2}$ and $\eta \propto a$, respectively.

The \textit{comoving distance} between a distant emitter and a local observer is given by
\be
\label{chic}
\chi(a) = \int_{t(a)}^{t_0}\, \frac{dt'}{a(t')}. 
\ee
Eq. (\ref{chic}) can be 
rewritten as
\be
\label{chic2}
\chi(a) = \int_{a}^1 \frac{da'}{a'^2\,H(a')},
\ee
where we have used $H=\dot{a}/a$ in order to change the integration variable. For a matter-dominated flat universe, one has $H \propto a^{-3/2}$, which yields 
\be
\chi(z) = \frac2{H_0}\,\left[1 - \frac1{\sqrt{1+z}} \right].
\ee
Note that for small $z$ the comoving distance is given by $z/H_0$, whereas it asymptotes to $2/H_0$ for large $z$

Let us now consider an object with physical size $l$. The angle $\theta$ subtended by this object is related to its distance by
\be
d_A = \frac{l}{\theta},
\ee
which is called \textit{angular diameter distance} and is valid for small $\theta$.
As the angle subtended is 
\be
\theta = \frac{l}{a}\,\frac1{\chi_a},
\ee
the angular diameter distance for a flat universe is
\be
d_A = \frac{\chi}{1+z}.
\ee
The angular diameter distance is then equal to the comoving distance at low redshift but
decreases with redshift at high redshift. One direct consequence is that an object at large redshift appears larger than it would appear at an intermediate one.

Another way of determining distances is to consider the flux from an object of known luminosity.
Let $L_s$ be the absolute luminosity, i.e., the power emitted by the a source in its rest frame. 
If one neglects the expansion of the universe and considers that the source is at the comoving coordinate $r=r_s$ and the detector is at $r=0$, the received energy flux is
\be
\label{fru}
\mathcal{F} = \frac{L_s}{4\,\pi\,a_0^2\,r_s^2},
\ee
where it is assumed that the light is emitted at $t_s$ and observed at $t_0$.

For an expanding universe there are two corrections to Eq. (\ref{fru}):
\be
\label{fluxoex}
\mathcal{F} = \frac{L_s}{4\,\pi\,a_0^2\,r_s^2\,(1+z)^2}.
\ee
The first factor $(1+z)$ is due to the redshift of photons as they travel from the source to the detector while the second one computes the fraction\footnote{This correction is derived by considering that two photons which are emitted with an interval of time $\delta t$ arrives separated by an interval of time $\delta t\,(a_0/a_s)$.} between the number of photons that are emitted by the source $n_\gamma^e$ and received by the detector $n_\gamma^r$ per unit of time, i.e.,
\be
\frac{n_\gamma^e}{n_\gamma^r} = 1 + z.
\ee
Therefore the \textit{luminosity density} is
\be
\label{dl}
d_L^2 = a_0^2\,r_s^2\,(1+z)^2.
\ee
\subsection{The Evolution of the Universe}
The covariant conservation equations lead to the following relations
\be
\rho_m = \rho_{m0}\,\left( \frac{a_0}{a} \right)^3,\,\,\,\,\,\,
\rho_r = \rho_{r0}\,\left( \frac{a_0}{a} \right)^4,\,\,\,\,\,\,
\rho_\Lambda = \rho_{\Lambda\,0}.
\ee
Substituting these relations in
the Friedmann equation (\ref{13}) one gets
\be
\label{18}
H^2 = H_0^2 \left[\Omega_m^0 \left( \frac{a_0}{a} \right)^3 + \Omega_r^0 \left( \frac{a_0}{a} \right)^4 + \Omega_\Lambda^0  - \Omega_k^0 \left( \frac{a_0}{a} \right)^2 \right],
\ee
where $\Omega_k^0 \equiv - k/(a_0^2\,H_0^2)$. 
Of course,
the history of the universe then depends upon the relative weight of the various physical components of the universe. 

Let us first consider a matter dominated universe ($\Omega_\Lambda^0=\Omega_r^0=0$). 
For $\Omega_k^0=0$, it has a decelerating expansion forever with $a(t) \propto t^{2/3}$ whereas for $\Omega_k^0 > 0$ or $\Omega_k^0 < 0$ the universe will collapse in the future or expand forever in a decelerated way. Note that we have neglected $\Omega_r^0$ and have considered $\Omega_m^0 =1$.

On the other hand, for $\Omega_\Lambda^0 \neq 0$ one finds a different destiny for the universe. In particular, $\Omega_k^0 = 0$ and $\Omega_r^0 = 0$ yield the expression
\be
\frac{a}{a_0} = \left(\frac{\Omega_m^0}{\Omega_\Lambda^0} \right)^{1/3}\,\left[\sinh\left( \frac32\,\sqrt{\Omega_{\Lambda}^0\,H_0\,(t-t_0)} \right) \right]^{2/3},
\ee 
that interpolates between a matter dominated universe expanding in a decelerated way $a(t) \propto t^{2/3}$ and an accelerating expanding one, that is dominated by a cosmological constant term. 

Going back in time, it is possible to describe the different epochs until the big explosion - {\bf the Big Bang}. 
The epoch at which the energy density in matter equals to that in radiation is called \textit{matter-radiation equality} and has special significance for the growth of large-scale structure and for the CMB anisotropies. The reason is that perturbations in the dark matter component grow at different rates in the matter and radiation eras. 
Specifically, 
for $z > z_{eq}$ the universe is radiation-dominated with the scale factor $a(t) \propto t^{1/2}$, while for $z < z_{eq}$ is matter-dominated until dark energy starts dominating with $a(t) \propto t^{2/3}$.

\subsection{Cornerstones of the Big Bang Model}
\subsubsection{The Cosmic Microwave Background}
The Cosmic Microwave Background (CMB), a picture of the universe when it was $3 \times 10^5$ years old,\footnote{This corresponds to a redshift $z \sim 1100$.} is
one of the cornerstones of the standard cosmological model. 

Penzias and Wilson \cite{Penzias} measured the CMB spectrum at a frequency of $4.08\, {\rm GHz}$ and estimated a temperature 3.5 K.
A crucial fact about the history of the universe described by the CMB measurements is that the collisions with electrons before last scattering implied that the photons were in equilibrium leading to a blackbody spectrum. 
Indeed, the blackbody nature of CMB spectrum has been studied and confirmed for a wide range of frequencies ranging from 0.6 GHz \cite{Sironi,Howell} up to $300\, {\rm GHz}$ \cite{Patridge}.

\subsubsection{The Primordial Nucleosynthesis}
Another pillar of the standard cosmological model is the Big Bang Nucleosynthesis. The nuclear reactions took place when the universe was between $0.01\,{\rm s}$ and $100\,{\rm s}$ old. 
As a result, a substantial amount of  $^4{\rm He}$, ${\rm D}$, $^3{\rm He}$ and $^7{\rm Li}$ were produced. 
The predicted abundances of these four light elements agree with the observations, as long as the baryon to photon ratio is between $4\times 10^{-10}$ and $7 \times 10^{-10}$ (corresponding to $0.015 \leq \Omega_b\,h^2 \leq 0.026$) \cite{KOLB}. 
The recent WMAP results indicate that $\Omega_b\,h^2 = 0.0023 \pm 0.0008$ \cite{WMAP3}.
Note that 
since the present universe is almost flat (\ref{20}), the results from the primordial nucleosynthesis imply that the
energy density of the universe is mostly constituted by a non-baryonic component.

\subsubsection{The Hubble Diagram and Supernovae Observations}
The Hubble diagram is the most direct evidence that the universe is expanding. 
The spectral shifts of $41$ galaxies were known in 1923. Among them, $36$ presented a systematic redshift. In 1929, Hubble showed empirically that the universe was expanding. He found that the velocity of the galaxies increases linearly with distance \cite{Freedman}.
Using the same principle of measuring the distance and redshift for distant objects, current observations use other objects with
known intrinsic brightness. 
The results from observations of the luminosity distance of high redshift supernovae indicate that the universe is accelerating.  
In 1998 two research groups (Riess \textit{et al.} and Perlmutter \textit{et al.} \cite{Perlmutter}--\cite{Riess2}) published the first sets of evidence that the universe is accelerating. 
Type Ia Supernovae are thought to
have a common absolute magnitude $M$ 
and consequently they are considered good standard candles.

The simplest explanation for the present accelerated expansion of the universe is a component of the energy density that remains invariant during the cosmic evolution known as \textit{Cosmological Constant}, $\Lambda$. Although this explanation appears well defined, one problem arises when we try to understand it in
the context of particle physics. 
We would expect that the value of its energy density to be equal to the energy of the vacuum in quantum field theory. In other words, we would expect
it has the order of the typical scale of early universe phase transitions which even at the scale QCD is $\rho_{\Lambda} \sim
10^{-3}\,{\rm GeV^4}$. Instead, the value is of the order of the critical density at the present day, that is, $\rho_{\Lambda} \sim {\rm 10^{-3} eV^4}$.
This difference between the expected and observed values is called
\textit{Cosmological Constant Problem}.

The
apparent magnitude $m$ of a source in an expanding universe is related to the logarithm of $\mathcal{F}$ given in Eq. (\ref{fluxoex}) as
\be
\label{mm}
m - M = 5 \log\left(\frac{d_L}{{\rm Mpc}} + 25 \right),
\ee
where $M$ is the absolute magnitude. Taking the apparent magnitude for two supernovae (one at 
low redshift and another at high redshift) \cite{Perlmutter} one has
\ben
&&1992P \rightarrow m = 16.08\,\,\,\,\,\,\,z=0.026,\\
&&1997ap \rightarrow m = 24.32\,\,\,\,\,\,\,z=0.83.
\een
Since at
low redshift ($z << 1$), the luminosity distance can be written as
\be
\label{dllr}
d_L(z) \simeq \frac{z}{H_0},
\ee
we apply 		
Eqs. (\ref{mm}) and (\ref{dllr})
to the $1992P$ supernovae data and find that the absolute magnitude and the luminosity distance are, respectively,  $M=-19.09$ and $d_L \simeq 1.16 H_0^{-1}$.

Using Eq. (\ref{dl}) we find that the
luminosity distance for 
a flat FRW universe (\ref{11}) can be written as
\be
\label{iiii}
d_L = \frac{1+z}{H_0} \int_0^z d\tilde{z} \frac1{\sqrt{\sum_i \Omega_i^0 (1+\tilde{z})^{3(1+\omega_i)}}}.
\ee
Accordingly, assuming that the universe contains only matter ($\Omega_m^0 = 1$), the luminosity distance is $d_L \simeq 0.95 H_0^{-1}$.
On the other hand, for $\Omega_m^0 = 0.3$ and $\Omega_\Lambda = 0.7$ we have that $d_L \simeq 1.23 H_0^{-1}$.
In other words, an universe dominated by
a cosmological constant shows a better agreement with the observational data.
Note that since the radiation-dominated period is much smaller than the total age of the universe, the integral coming from the region $z < 1000$ does not affect the total integral (\ref{iiii}).

Further evidence for the existence of dark energy can be found by computing the age of the universe. It can be found by
rewriting Eq. (\ref{18}) as
\be
t_0 = \int_{0}^{\infty} \frac{dz}{H_0(1+z)[\Omega_m^0 (1+z)^3 + \Omega_r^0 (1+z)^4 + \Omega_\Lambda^0 - \Omega_k^0 (1+z)^2]^{1/2}},
\label{23}
\ee

The investigation of the Globular clusters in the Milk Way shows that the age of the oldest stellar populations is $t_s=13.5 \pm 2\, {\rm Gyr}$ \cite{Jimenez} while for the globular cluster M4 this value is $t_s = 12.7 \pm 0.7 \,{\rm Gyr}$ \cite{Richer,Hansen}.

However, assuming that $\Omega_m^0 = 1$ and $\Omega_\Lambda^0 =0$ in Eq. (\ref{23}) we get that $t_0 = 8 - 10\, {\rm Gyr}$ which is smaller than $t_s$. 
A possible explanation is that the universe today is dominated by a cosmological constant.

\begin{figure}
\begin{center}
\includegraphics*[width=9cm]{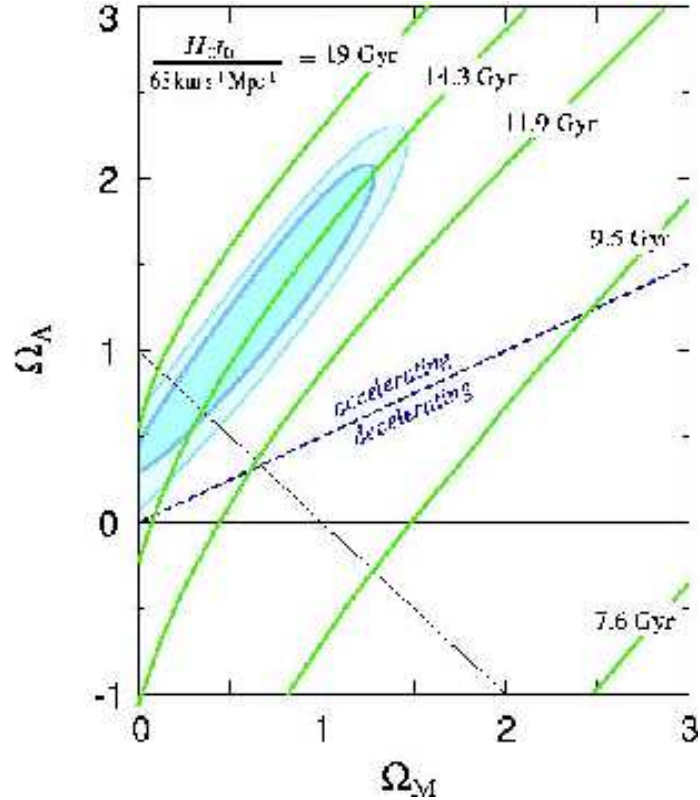}
\end{center}
\caption[Ages of the universe as a function of $\Omega_m^0$ and $\Omega_r^0$.]{Different ages of the universe for different values of $\Omega_m^0$ and $\Omega_\Lambda^0$.}  
  \label{pic4}
\end{figure}

This is illustrated in Fig. \ref{pic4} (taken from \cite{Sola}).

\subsection{Some Problems of the Standard Cosmological Model} 
Although the Big Bang theory successfully explains the Hubble expansion law, the CMB and the abundances of the light elements, the standard cosmological model does have its problems, and we now mention 
some of them.

First, the \textit{horizon problem} stems from the large-scale homogeneity and isotropy of the universe. The CMB comes
from regions which, according to the standard cosmological model, have never been causally in contact. The temperature of the radiation from two different regions is the same to within at least one part in $10^4$. Consequently, within 
the standard Big Bang model, the early universe was highly homogeneous and isotropic on scales much greater than the particle horizon.

Second, the question: \textit{Why is the present universe so close to flat?} summarizes the \textit{flatness problem}. 
Indeed, the energy density of the universe today is very close to
the critical density. On the other hand, the critical density is a point of unstable equilibrium, i.e., any deviations from $\Omega = 1$ grow in time. Accordingly, if today $\Omega \sim 1$ then this total density parameter was fine-tuned such that $|1-\Omega|\lesssim 10^{-58}$ at the Planck time.   
	
Finally, we emphasize that phase transitions in the early universe could have generated topological defects. For instance, the spontaneous symmetry breaking whose vacuum manifold contain non-contractible two-surfaces gives rise magnetic monopoles. This process is always present in
Grand Unified Theories (GUT), where the symmetry group $SU(3) \times U(1)$ is generated independent of the original initial group or the intermediate stages of the symmetry breaking \cite{Vilenki}. 
The existence of heavy monopoles is an inevitable prediction of GUT theories and the predicted abundance of this topological defect in the standard cosmological model is not consistent with the observations \cite{KOLB}.
This problem is named \textit{monopole problem}.

\section{\label{sfc}Scalar Fields and Topological Defects}
The water freezing or melting is an example of a \textit{first order phase transition}, where there is an associated discontinuous change of an order parameter. A \textit{second order phase transition} has a continuous order parameter which is not differentiable.
Like phase transitions in condensed matter systems, cosmological phase transitions are associated with \textit{spontaneous symmetry breaking} \cite{Vilenki}. 

Scalar fields can be used to describe the phenomenon of spontaneous symmetry breaking \cite{Manton}. 
Despite their simplicity, they have a key role in the understanding of particle physics, condensed matter and cosmology.

Let $\varphi$ be a real scalar field which depends on the space-time coordinates $x^\nu$, where $\nu = 0,1,2,3$, described by the action
\be
\label{ldw}
\mathcal{S} = \int d^4x \,\sqrt{-g}\,\left[ \frac12 \partial_\mu \varphi \partial^\mu \varphi - V(\varphi) \right],
\ee
where $V(\varphi)$ is the known $\lambda\varphi^4$ potential given by 
\be
\label{lldw}
V(\varphi) = \frac{\lambda}4\,(\varphi^2-\eta^2)^2
\ee
with $\lambda>0$ and $\eta$ real parameters.

The high temperature effective potential\footnote{We neglected the $\varphi$-independent terms.} for (\ref{lldw}) can be written as \cite{Vilenki}
\be
\label{temphigh}
V_{\rm eff}(\varphi,T) = m^2(T)\, \varphi^2 + \frac{\lambda}4\,\varphi^4,
\ee 
where $T$ is the temperature and $m$ is the effective mass of the field $\varphi$ in the symmetric state $\varphi=0$ given by
\be
m(T) = \frac{\lambda}{12}\,\sqrt{T^2 - 6\eta^2}. 
\ee

\begin{figure}
\begin{center}
\includegraphics*[width=9cm]{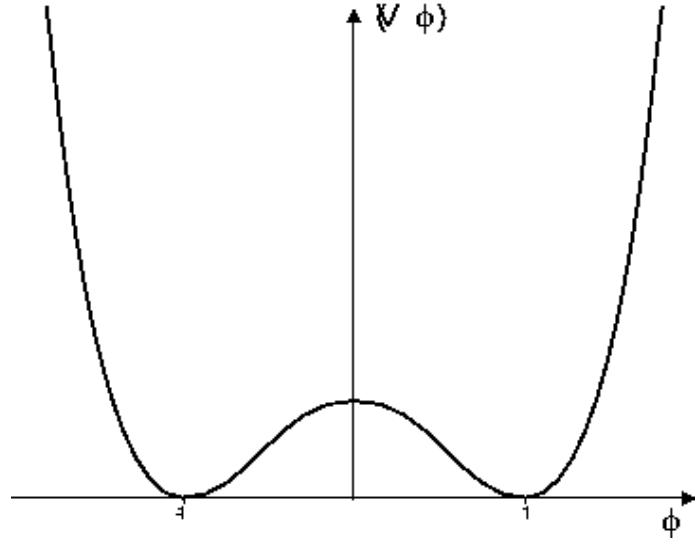}
\end{center}
\caption[The $\lambda \varphi^4$ potential.]{The $\lambda \varphi^4$ potential (\ref{lldw}) for $\lambda = \eta =1$.}
\label{lambdaphi4pot}
\end{figure}

We define the critical temperature $T_C=\sqrt{6}\,\eta$ such that 
\begin{itemize}
\item
For $T=T_C$, the mass vanishes.
\item
For $T>T_C$ one has $m^2(T)>0$ and the minimum of $V_{\rm eff}$ is at $\varphi=0$. Since the expectation value of $\varphi$ is null, the symmetry is restored.
\item
For $T<T_C$ one has $m^2(T)<0$. Since $\varphi=0$ becomes an unstable state, the field develops
a non-zero expectation value, say $\varphi = [(T_C^2-T^2)/6]^{1/2}$.
Given that $\varphi$ grows continuously from zero as the temperature decreases below $T_C$, this process is defined as a second-order phase transition.
\end{itemize}

The potential (\ref{lldw}) 
describes the most elementary type of topological defects (see Fig.  \ref{lambdaphi4pot}). The vacuum manifold, i.e., the set of minima of the potential forms a discrete set and intersection of these domains give rise to \textit{domain walls} that are solutions of the equation of motion
\be
\label{go}
\square \varphi + \lambda \varphi (\varphi^2 - \eta^2) = 0.
\ee

\subsection{Static One-Dimensional Soliton Solution}
Let us now focus on the static case. The equation of motion is now
\be
\label{stat}
\frac{d^2 \varphi}{dx^2} =  \frac{ d V(\varphi)}{d \varphi},
\ee
whose 
one-dimensional solutions must obey the boundary conditions 
\be
\lim_{x \to \pm \infty}\,\frac{d \varphi}{dx} \to 0,\,\,\,\,\,\,\,\lim_{x \to  \pm \infty}\,\varphi(x) \to \pm \eta,
\ee
in order that the energy 
\be
\label{enkink}
E = \int\, dx\, \left[ \frac12  \left( \frac{d \varphi}{dx} \right)^2 + \frac{\lambda}{4} (\varphi^2 - \eta)^2 \right]
\ee
be finite.

\begin{figure}
\begin{center}	
\includegraphics*[width=9cm]{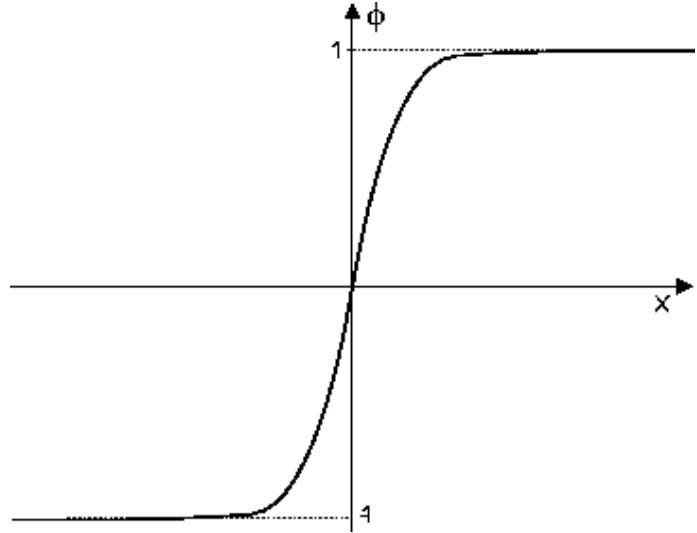}
\end{center}
\caption[The one-dimensional soliton solution for the $\lambda \varphi^4$ potential.]{The one-dimensional soliton solution for the $\lambda \varphi^4$ potential given in Fig. \ref{lambdaphi4pot} (we have assumed $\eta = \lambda =1$ and $x_0=0$).}
\label{lambdaphi4kink}
\end{figure}

The solutions are
\be
\label{kink}
\varphi = \pm \eta \tanh{\left( \sqrt{\frac{\lambda}{2}} \eta (x-x_0) \right)},
\ee
which are named \textit{kink}(+) and \textit{antikink}(-) and take $\varphi$ from $-\eta$ ($\eta$) at $x=-\infty$ to $\eta$ ($-\eta$) at $x=\infty$.  These
solutions are centered at $x_0$ and have width\footnote{The width of the topological defect determines the region in which the solution deviates from the vacua.} given by 
$\delta \sim (\sqrt{\lambda}\,\eta)^{-1}$.
Fig. \ref{lambdaphi4kink} shows the kink solution for $\lambda = \eta = 1$ and $x_0=0$.
Note that these solutions are Lorentz invariant and can be boosted up to arbitrary velocities.
We also stress that there is an important topological conservation law ensuring the stability of the soliton solution. 
The topological current is defined as $
j^\mu = \epsilon^{\mu\nu}\,\partial_\nu \varphi$, 
with the associated conserved charge
\be
\mathcal{Q} = \int_{-\infty}^{\infty} dx\,j^0 = \varphi(x \to \infty) - \varphi(x \to - \infty).
\ee
Therefore the kink (antikink) solution gives rise to $\mathcal{Q} \neq 0$, that is named \textit{topological charge}.
The one-dimensional solution given by (\ref{kink}) can be embedded in $(3,1)$ dimensions.  It is then a wall separating two regions with different 
domains (vacua) in the space. These topological defects are named \textit{domain walls} and carry surface tension which is identified with the energy of classical solutions in one-dimensional space.

Let us consider that $V(\varphi)$ can be written as 
\be
V(\varphi) = \frac12 \left[ \frac{dW(\varphi)}{d \varphi} \right]^2,
\ee
where $W(\varphi)$ is another function\footnote{In supersymmetric scenario, the function $W(\varphi)$ is named \textit{superpotential}.} of $\varphi$.
In this case the energy is
\be
\label{ienkink}
E = E_B + \int\, dx\, \frac12  \left( \frac{d \varphi}{dx} \mp \frac{dW(\varphi)}{d \varphi} \right)^2,
\ee
where
\be
E_B = | W[\varphi(x \to \infty)]-W[\varphi(x \to -\infty)] |
\ee
is the Bogomoln'yi energy bound \cite{bogomolnyi}. In this case, the solutions (\ref{kink}) can be found by solving the first order differential equations
\be
\label{fir}
\frac{d \varphi}{dx} = \pm \frac{dW(\varphi)}{d \varphi}
\ee  
and
are named BPS (Bogomoln'yi, Prasad, Sommerfield) states \cite{Bazeiaf1,Bazeiaf2}. It was shown in Refs. \cite{menezes0}--\cite{Menezes4} that Eqs. (\ref{fir}) factorize completely the equation of motion (\ref{stat}).

\subsection{Scalar Field Models and Dark Energy}
Let us consider a homogeneous universe described by
a scalar field $\varphi$ minimally coupled to gravity. 
The Lagrangian density is given by
\be
\label{lalala}
\mathcal{L} =  \frac12 \partial_\mu \varphi \partial^\mu \varphi - V(\varphi) ,
\ee
where 
$V(\varphi)$ is the potential. 

The energy momentum tensor of the field $\varphi$ is found by varying the action (\ref{ldw}) with respect to the metric $g_{\mu\nu}$, 
\be
T_{\mu\nu} = - \frac2{\sqrt{-g}}\,\frac{\delta S}{\delta g^{\mu\nu}},
\ee
where $\delta \sqrt{-g} = - (1/2)\,\sqrt{-g}\,g_{\mu\nu}\delta\,g^{\mu\nu}$. It gives
\be
T_{\mu\nu} = 
\partial_\mu \varphi \partial_\nu \varphi - g_{\mu\nu} \left[
\frac12 g^{\beta \kappa} \partial_{\beta} \varphi \partial_{\kappa} \varphi + V(\varphi)
\right].
\ee
Hence, the energy density in a flat Friedmann Robertson Walker universe is 
\be
\rho = \frac{\dot{\varphi}^2}2 + V(\varphi),
\ee
while the pressure is given by
\be
p = \frac{\dot{\varphi}^2}2 - V(\varphi),
\ee
where a dot indicates the derivative with respect to time.

The acceleration of the universe is obtained by combining Eqs. (\ref{13}) and (\ref{14}). It gives
\be
\frac{\ddot{a}}{a} = - \frac{8\pi G}3 \left[\dot{\varphi}^2 - V(\varphi) \right],
\ee
which
implies the existence of an accelerated universe is only possible if $\dot{\varphi}^2 < V(\varphi)$. 

Now we recall that
the equation of state for the field is given by 
\be
\omega_{\varphi} = \frac{\dot{\varphi}^2 - 2 V}{\dot{\varphi}^2 + 2 V},
\ee
while the continuity equation is
\be
\label{form}
\rho = \rho_0 e^{-\int 3(1+\omega_\varphi)\,\frac{da}{a}},
\ee
where $\rho_0$ is an integration constant. Note that the equation of state can vary between the values $[-1,1]$.
If $\omega_\varphi = -1$, one has the \textit{slow-roll limit}, $\dot{\varphi}^2 <<V(\varphi)$ and $\rho=$ const, while $\omega_\varphi =  1$ implies $\dot{\varphi}^2 >> V(\varphi)$, 
in which case the
energy density evolves as $a^{-6}$. 
On the other hand, $\omega_\varphi = -1/3$ is the border of acceleration and deceleration. Note that acceleration is realized if the energy density is $\rho  \propto a^{-m}$,
with
$0 \leq m < 2$.

\subsection{Inflation}
A period of very rapid expansion in the early universe may lead to a solution of
some of the problems of the standard cosmological model. This was named inflationary scenario and
was
first discussed by Guth \cite{Guth}. 
In the inflationary regime:  
\begin{itemize}
\item
Regions initially within the causal horizon are blown up to
sizes greater than the present Hubble radius. 
\item
The initial curvature decreases by a huge factor, yielding a
universe locally indistinguishable from a flat universe.
\item
All topological defects formed before inflation, for instance the magnetic monopoles, are diluted by an enormous factor. 
\end{itemize}

\section{Varying Fundamental Constants}\label{vfc}
 
\textit{Are there fundamental scalar fields in nature?} Several decades of accelerator experiments did not find them, even though they are a key ingredient in the standard model of particle physics. We recall that the Higgs particle is supposed to give mass to all other particles and make the theory gauge-invariant.
We emphasize that
there are a number of ways to change the standard model of particle physics in order to introduce a space-time variation on the fine-structure constant $\alpha$ \cite{maguei1,maguei2}. Since this constant measures the strength of the electromagnetic interaction, 
there are many different ways in which measurements of $\alpha$ can be made. 
To name just a few, locally one can use atomic clocks \cite{Marion} or the Oklo natural nuclear reactor \cite{Fujii,Lamoreaux}. On the other hand, on astrophysical and cosmological scales a lot of work has been done on measurements using quasar absorption systems \cite{Webb1}--\cite{Chand} and the Cosmic Microwave Background \cite{Avelino1}--\cite{Rocha}.
These different measurements probe very different environments, and therefore it is not trivial to compare and relate them. 
Simply comparing at face value numbers obtained at different redshifts, for example, it is at the very least too naive, and in most cases manifestly incorrect. Indeed, detailed comparisons can often only be made in a model-dependent way, meaning that one has to specify a cosmological model 
and/or a specific model for the evolution of $\alpha$ as a function of redshift. 
Simply assuming, for example, that $\alpha$ grows linearly with time (so that its time derivative is constant) is not satisfactory, as one can easily show that no sensible particle physics model will ever yield such a dependence for any significant redshift range. 

In order to explain the accelerated expansion of the universe, 
efforts have been made
to elaborate models in which
this acceleration is related to the variation of the fundamental constants. In particular, 
one can reconstruct the dark energy equation of state from variations in the fine-structure constant for a class of models where the quintessence field is  non-minimally coupled to the electromagnetic field. It has been claimed that variations on $\alpha$ would need to be measured to within an accuracy of at least $5 \times 10^{-7}$ to obtain a reconstructed equation of state with less than a twenty percent deviations from the true equation of state for $0 < z < 3$ \cite{nunes}--\cite{Avelinor}.

\subsection{Constraints on variation of $\alpha$}
\subsubsection{The Astrophysical Bounds to variation of $\alpha$}
The investigation of space-time variations of the 
fine-structure constant has dramatically increased due to the results of 
Webb \textit{et. al} \cite{Webb2}. The optical spectra of quasars are rich in absorption lines arising from gas clouds along 
our line of sight. 
Using the many-multiplet method, that exploits the information in many wavelength separations of absorption lines with different relativistic contributions to their fine structure, considerable gains in statistical significance were achieved. From a data set of $128$ objects at redshifts between $0.5$ and $3$, it is shown in Ref. \cite{Webb2} that the absorption spectra were consistent with 
\be
\frac{\Delta \alpha}{\alpha} \equiv \frac{\alpha-\alpha_0}{\alpha_0} = ( - 0.57 \pm 0.10 ) \times 10^{-5},
\ee
where $\alpha_0$ is the present value of the fine-structure constant.
However, 
the study of $23$ absorption systems from VLT-UVES quasars at $0.4 \leq z \leq 2.3$ developed by Chand 
\textit{et. al.} \cite{Chand} found a result consistent with no variation of $\alpha$, i.e.,
\be
\frac{\Delta \alpha}{\alpha}  = ( - 0.6 \pm 0.6 ) \times 10^{-6}.
\ee
This result was obtained by using a simplified version of the many-multiplet method and concerns remain about calibrations and the noisiness of the data fit. 
Also, Quast \textit{et. al.} and Lekshakov \textit{et. al.} \cite{Quast} observed single quasar absorption systems. They found results consistent with 
\be
\frac{\Delta \alpha}{\alpha} = ( - 0.1 \pm 1.7 ) \times 10^{-6}
\ee
and 
\be
\frac{\Delta \alpha}{\alpha} = ( 2.4 \pm 3.8 ) \times 10^{-6}
\ee
at $z = 1.15$ and $z=1.839$, respectively.

\subsubsection{Constraints on variation of $\alpha$ from CMB}
The value of the fine-structure constant at high redshift can be measured by investigating the CMB.
Since $\alpha$ is expected to be a monotonous function of time \cite{Damour2}, 
at this epoch any variations relative to the present-day value are expected to be larger compared to low redshift observations.
In fact, a variation of $\alpha$ would alter the ionization history of the universe and thus would have an impact on the CMB \cite{Avelino1,hannestad}. 
Since
the redshift of recombination would be changed due to a shift in the energy levels of hydrogen. Also, the Thomson scattering cross-section would also be modified.

According to Refs. \cite{Avelino1}--\cite{Martins2} and  \cite{sigurdson}, the overall limit on the variation of $\alpha$ is
\be
\frac{\Delta \alpha}{\alpha} \lesssim 10^{-2},
\ee 
for $z \sim 10^3$. This bound it is consistent with the limit found by Big Bang Nucleosynthesis ($z \sim 10^9 - 10^{10}$) \cite{Campbell,Cyburt}.
It is expected that the Planck mission will provide significant improvements in these bounds.
\subsubsection{The Oklo natural reactor bounds}
Oklo is the name of the place of a uranium mine in Gabon, West Africa, that supplied the uranium ore to the French government. In 1972, the abundance of $\rm{^{235}U}$
was found to be somewhat below the world standard. 
This indicated that a self-sustained fission reaction took place naturally in Oklo about $\rm{1.8\, Gyr}$ ago, during the period of Proterozoic, well before Fermi has invented the artificial reactor in 1942. Actually, natural reactors had been predicted by Paul K. Kuroda, 17 years before the Oklo phenomenon was discovered \cite{Kuroda}. 
Investigations of this phenomenon led to the conclusion that the resonant capture cross-section for thermal neutrons by $\rm{^{149}Sm}$ at $z \sim 0.15$ had created a $\rm{^{149}Sm/^{147}Sm}$ ratio at the reactor site that is depleted by the capture process 
\be
^{149}{\rm Sm} + n \to ^{150}{\rm Sm} + \gamma
\ee
to an observed value of only about $0.02$ compared to the value of about $0.9$ obtained in normal samples of Samarium. The need for this capture resonance to be in place two billion years ago at an energy level within about $90 {\rm MeV}$ of its current value leads to very strong bounds on all interaction coupling constants that contribute to the energy level. This result was noticed by Shlyakhter \cite{Shlyakter}. 

Results of recent investigations using realistic models of natural nuclear reactors
provide strong bounds on
the time evolution of $\alpha$. Gould \textit{et. al} use the epithermal spectral indices as a criteria for selecting such realistic reactor models \cite{Lamo}. Using numerical simulations, they calculated the change in the $\rm{^{149}Sm}$ effective neutron capture cross section as a function of a possible shift in the energy of the $93.7$ meV resonance. They found that a possible time variation in $\alpha$ over 2 billion years should be inside the range
\be
-0.24 \times 10^{-7} \leq \frac{\Delta \alpha}{\alpha} \leq 0.11 \times 10^{-7}.
\ee
On the other hand, Petrov \textit{et. al} used recent version of Monte Carlo codes (MCU REA and MNCP) for constructing a computer model of the Oklo reactor taking into account all details of design and composition \cite{Petrov}. They obtain
conservative
limits on the time variation of the fine-structure constant:
\be
-4 \times 10^{-8} \leq \frac{\Delta \alpha}{\alpha} \leq 3 \times 10^{-8}.
\ee
\subsubsection{Meteoritic limits on the variation of $\alpha$}
Peebles and Dicke studied the effects of varying $\alpha$ on the $\beta$-decay lifetime \cite{Peebles} by considering the ratio of Rhenium to Osmium in meteorites,
\be
_{187}^{75}Re \to _{187}^{76}Os + \bar{\nu_{e}} + e^{-},
\ee
which
is
very sensitive to the value of the fine-structure constant.
The analysis of the new meteoritic data together with laboratory measurements of the decay rates of long-lived beta isotopes indicated a time averaged limit of 
\be
\frac{\Delta \alpha}{\alpha} = ( 8 \pm 16 ) \times 10^{-7}
\ee
for a sample that spans the age of the solar system ($z \leq 0.45$) \cite{Oliveolive}.

\vspace{0.5cm}
\chapter{\label{cap2}{\sf Domain Wall Networks and Dark Energy}}
\section{\label{intr}Overview}
Topological defects are expected to form in the early universe \cite{KIBBLE}. Their properties depend on the particular details of the spontaneous symmetry breaking that generated them. 
The choice of the minimum after the phase transition
is uncorrelated on distances greater than the horizon, and defects are necessarily formed. 
Domain walls, formed when a discrete symmetry is broken, are the simplest example.
Their small scale properties, such as their physical width and tension are fixed and completely determined 
by the field theory model and are independent of the cosmology. 

The average energy density of a frozen domain wall network is given by
\be
\rho = \frac{\sigma\,A}{V} \propto \frac{a^2}{a^3} = a^{-1},
\ee
where $\sigma$ is the energy per area unit,
$A$ is the wall surface area and $V$ is the volume occupied by the domain wall network.

Hence the equation 
\be
\frac{d\rho}{dt}+3\,H\,\rho\,(1+w) = 0
\ee
implies that the equation of state of a frozen domain wall gas\footnote{For cosmic strings $\rho \propto a^{-2}$, which yields $w = -1/3$.} is $w = -2/3$. More generally, we have to take into account the dependence of the equation of state of domain walls on their velocity $v$. One finds \cite{SOLID}
\be
w \equiv \frac{p}{\rho} = -\frac23 + v^2.
\ee

In this chapter we will consider a domain wall network as a possible candidate to explain the accelerated expansion of the universe at the present time. 
We have shown in Chapter 1 that observational evidence from supernova observations and CMB indicates that the universe is accelerating today, i.e.,
\be
w < - \frac{(1+\Omega_m^0/\Omega_{\Lambda}^0)}3 \lesssim -1/2.
\ee
Therefore, if a domain wall network frustrates, i.e., is frozen in comoving coordinates ($v \sim 0$) it might explain the present accelerated expansion of the universe \cite{SOLID}.

The simplest domain wall networks  
described by a single scalar field were extensively studied and the results indicate that they reach a scaling regime \cite{SIMS1,SIMS2}. For this reason, 
in this thesis we consider
the possibility that the dynamics of more complex
domain wall networks produced by two or more coupled fields can 
lead to frustration.
We focus on the role of junctions, and we study several models where different kinds of junctions can appear.
Specifically, we consider a number of realizations of the two classes of models introduced in \cite{KUBOTANI,BAZEIA}.
The similarities and differences between them will allow us to point out the key mechanisms at play, and further characterize the differences between models with stable Y-type junctions, models with stable X-type junctions, and models where both types can co-exist. The details of the numerical simulations that we shall present are analogous to those in \cite{SIMS1,SIMS2,PRESS}. We then go beyond these particular models, and investigate an
ideal class of models from the point of view 
of obtaining frustrated networks. We study
the properties of such networks and investigate some realizations by simulating two and three-dimensional networks. 

This chapter is organized as 
follows. In Sec. \ref{intr11} we present the Press Ryden Spergel (PRS) algorithm
code and introduce an analytic model for describing the domain wall evolution. 
In Sec. \ref{intr12} we study some models of domain wall network with junctions. In particular, we consider the BBL and Kubotani model and present the results of the simulations carried out in matter dominated era for two-dimensional boxes.
In Sec. \ref{intr122} we investigate the wall lattice properties considering the geometry, energy and topology of the domain wall network generated by random initial conditions in order to understand what the characteristics of an ideal model from the point of view of frustration. We also consider the stability of square and hexagonal networks constructed by hand. In Sec. \ref{intr1222} we introduce a realization of the ideal model and present the snapshots of numerical simulations for some particular cases. Also, we study the scaling behavior of the ideal model and compare it with the BBL model. Finally, in Sec. \ref{intr122222} we summarize 
our results and discuss our no-frustration conjecture. 

\section{\label{intr11}Domain Wall Evolution}
Consider the Lagrangian density given by
\be
\label{lalaga}
\mathcal{L} = \frac12 \partial_\mu \varphi \partial^\mu \varphi - V(\varphi),
\ee
where $V(\varphi)$ is the potential. Variation of the action 
with respect to $\varphi$
in a Friedmann-Robertson-Walker universe leads to the equation of motion 
\begin{equation}
{\frac{{\partial^{2}\varphi}}{\partial 
t^{2}}}+3H{\frac{{\partial\varphi}}{\partial 
t}}-\nabla^{2}\varphi=-{\frac{{\partial 
V}}{\partial\varphi}}\,,\label{dynamics}
\end{equation}
where $\nabla$ is the Laplacian in physical coordinates and 
$H$ is the Hubble parameter.

\subsection{PRS code}
We simulate the domain wall network evolution using the algorithm introduced by Press, Ryden and Spergel in Ref. \cite{PRESS} in
the comoving thickness of the domain wall is fixed. This modification to the equation of motion was introduced in order to be able to resolve the domain walls in a comoving grid.

The equation of motion (\ref{dynamics}) in comoving coordinates can be written as
\be
\frac{\partial^2 \varphi}{\partial \tau^2} + 2 \left(\frac{d\ln a}{d\ln\tau}\right)\,\frac1{\tau}\,\frac{\partial \varphi}{\partial \tau} - \nabla^2 \varphi = -a^2\,\frac{\partial V}{\partial \varphi},
\ee
where $\tau$ is the conformal time. This equation can be generalized to
\be
\label{prsprs}
\frac{\partial^2 \varphi}{\partial \tau^2} + \hat c_1
\left(\frac{d\ln a}{d\ln\tau}\right)\,\frac1{\tau}
\frac{\partial \varphi}{\partial \tau} 
- \nabla^2 \varphi = -a^{\hat c_2} \,\frac{\partial V}{\partial \varphi},
\ee
with constant $\hat c_1$ and $\hat c_2$. The choices $\hat c_1=3$ and  $\hat c_2=0$ 
ensure energy-momentum conservation and
constant comoving thickness, respectively. 

\begin{figure}
\begin{center}
\includegraphics*[width=6cm]{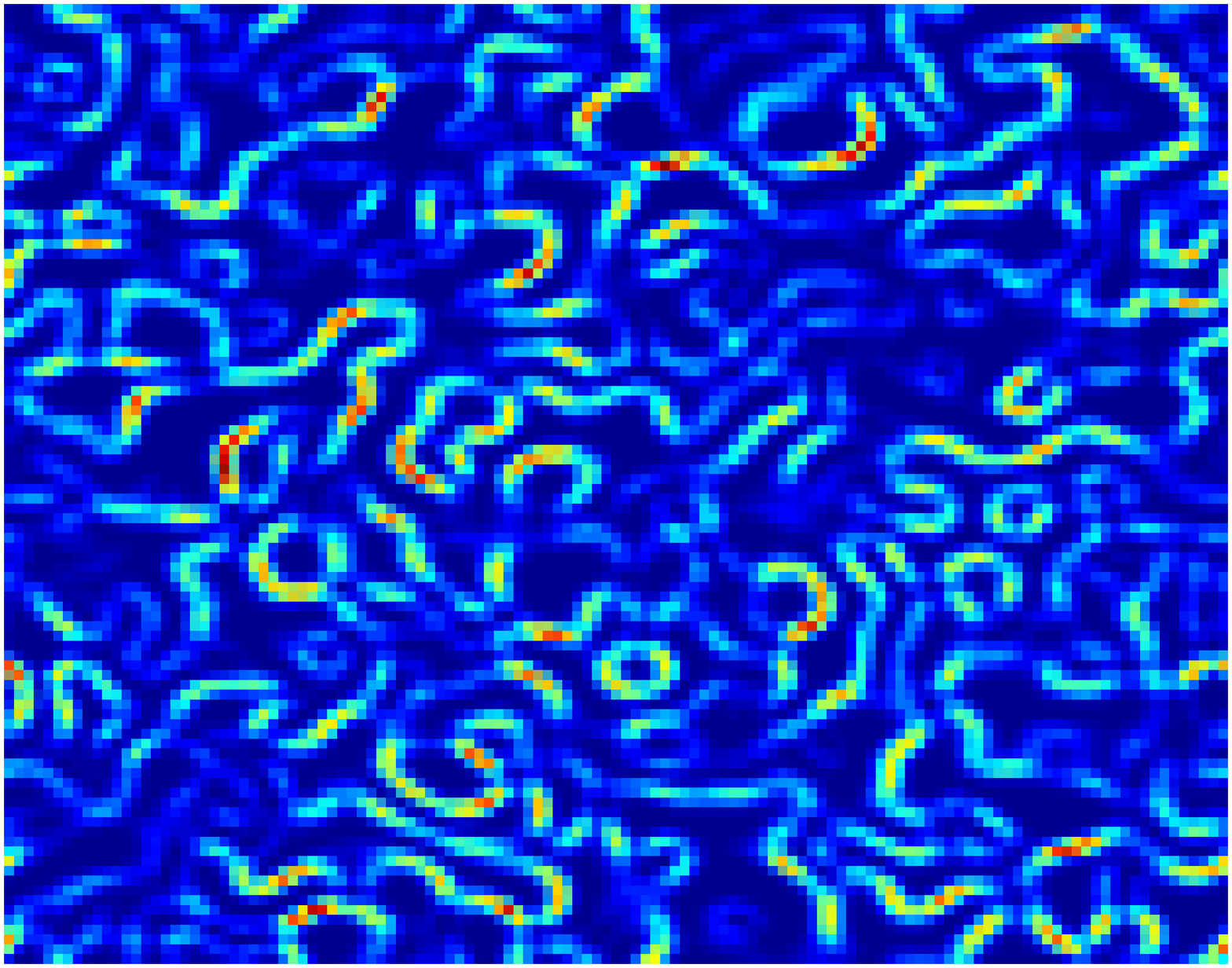}
\includegraphics*[width=6cm]{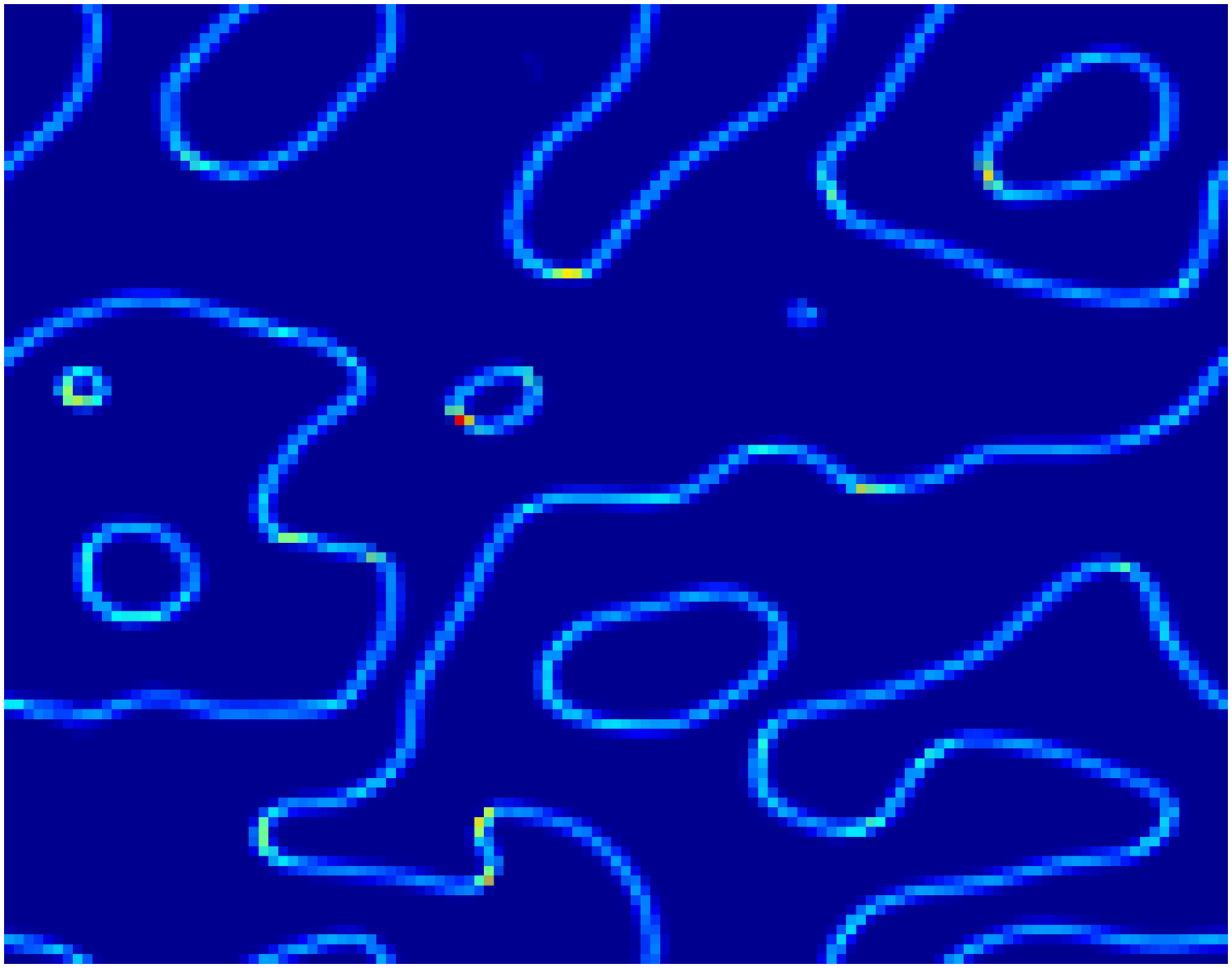}
\includegraphics*[width=6cm]{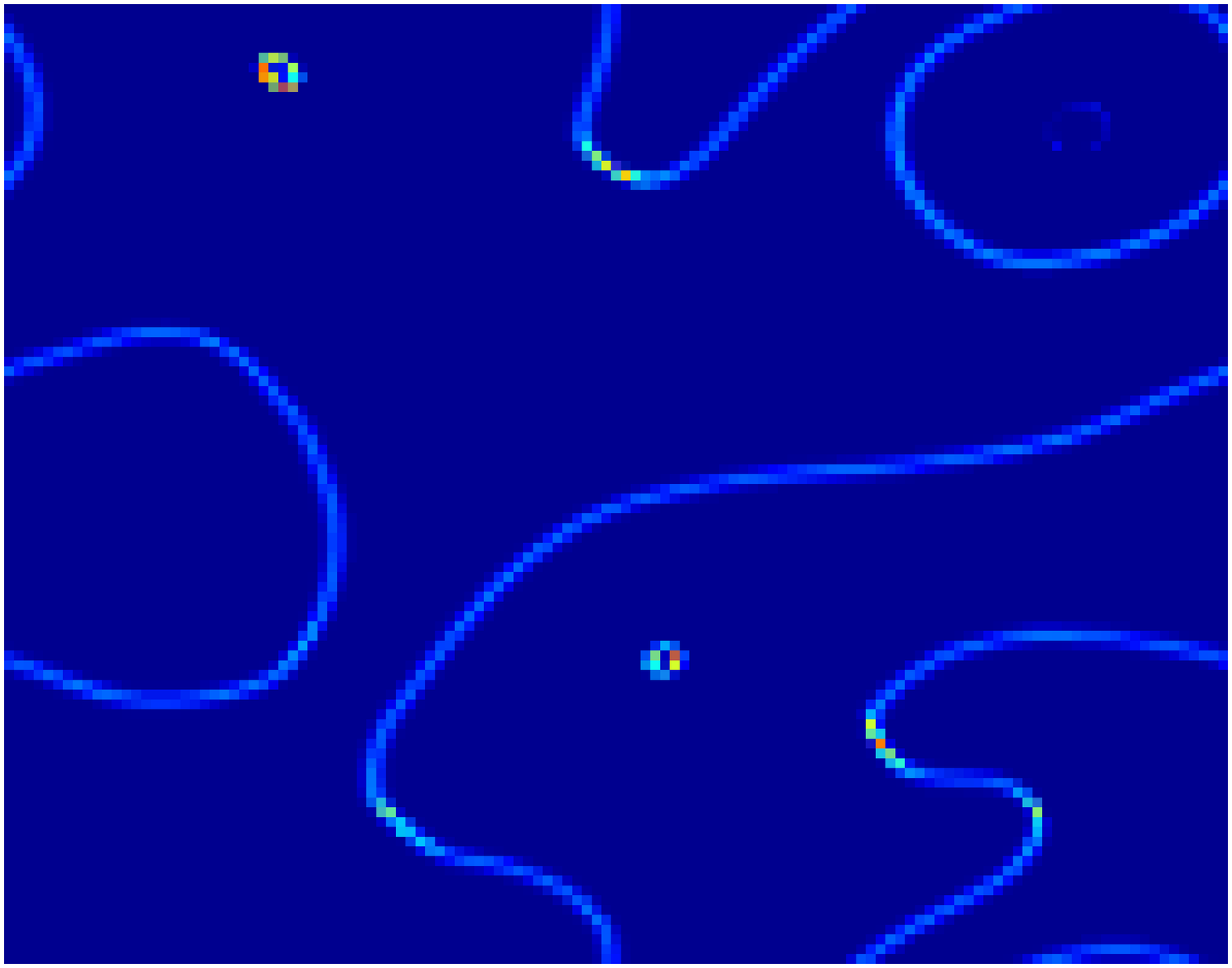}
\includegraphics*[width=6cm]{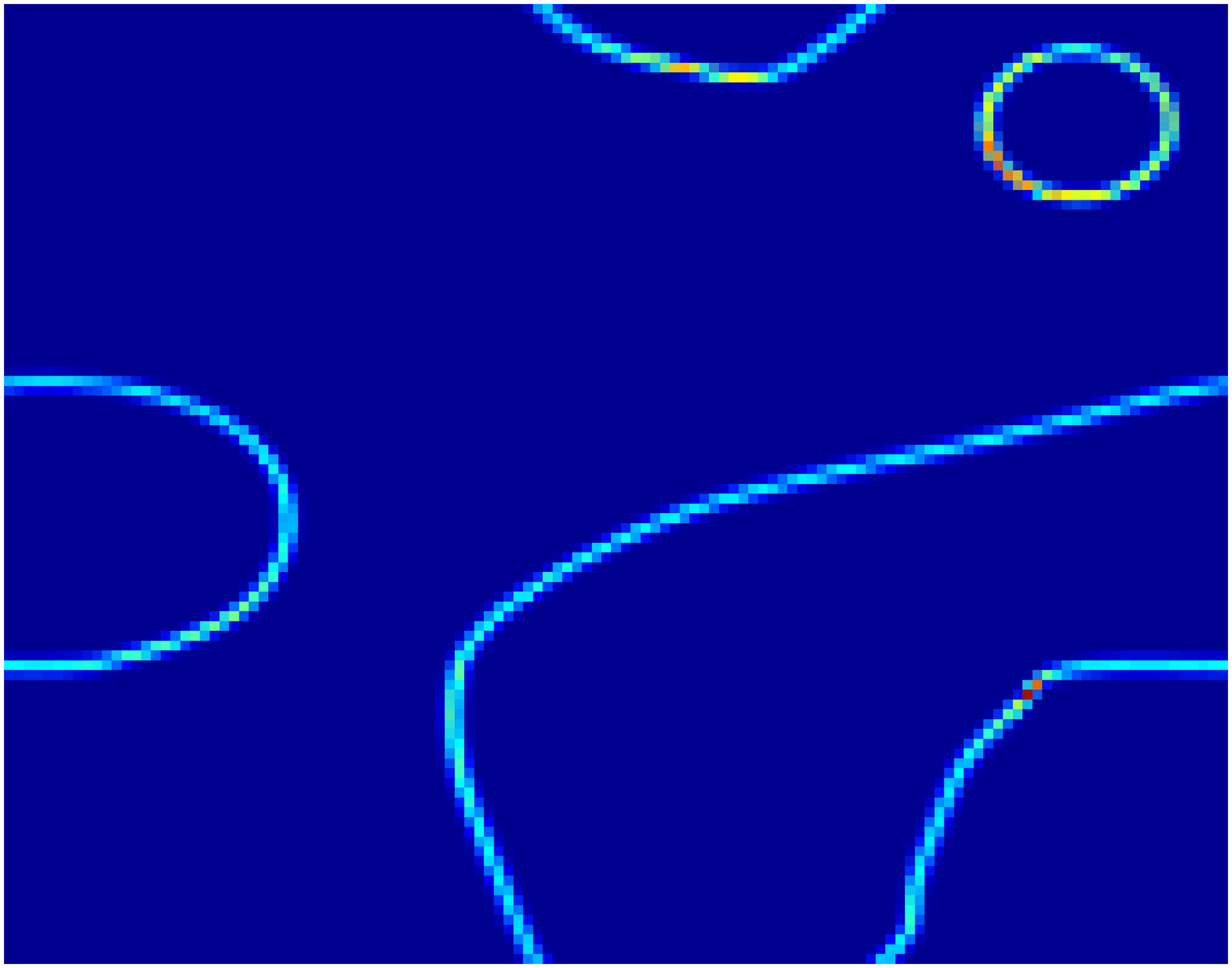}
\end{center}
\caption[Evolution of domain wall network: $\lambda \varphi^4$ potential.]{The matter-era evolution of the domain wall network with of $\lambda \varphi^4$ potential (Eq. \ref{lldw}) and a parameter choice $\lambda=\eta=1$.  The simulation starts with random initial conditions inside the range $(\pm 1) $. 
From left to right and top to bottom, the horizon is approximately 1/16, 1/8, 1/4 and 1/2 of the box size respectively.}
\label{lal}
\end{figure}

We integrate Eq. (\ref{prsprs}) using finite-difference methods on two-dimensional and three-dimens\-ional boxes. 
We assume the initial value of $\varphi$ to be a random variable between $\pm \eta$ and the initial value of $\partial \varphi/\partial \tau$ to be equal to zero everywhere. We normalize the numerical simulations so that $\eta=1$. In addition, we set the conformal time at the start of the simulation and the comoving spacing between the mesh points to be respectively $\tau_i = 1$ and $\Delta x =1$.
Fig. \ref{lal} shows the snapshots of a two-dimensional $512^2$ simulation carried out for the $\lambda \varphi^4$ potential (\ref{lldw}). 
Note that the characteristic scale of the network is roughly proportional to the horizon.

\subsection{An Analytic Model for the Domain Wall Evolution}
Let us consider an analytic \textit{one scale model} to describe the domain wall evolution \cite{SIMS1,SIMS2}.

Firstly, we notice that 
for a planar domain wall the momentum per unit comoving area is proportional to
$a^{-1}$.  Since the physical area in a given comoving volume increases proportionally to $a^{-2}$, one has $\gamma v \propto a^{-3}$. Therefore
\be
\label{am11}
\frac{dv}{dt} = (1-v^2)(-3H\,v).
\ee

Secondly, if we consider a network of planar domain walls and the interactions between the domain walls are neglected the average energy density of such domain wall network
is proportional to $\gamma a^{-1}$. Thus it is straightforward to write
\be
\label{am12}
\frac{d\rho}{dt} = - (1+3\,v^2)(H\rho).
\ee
 
Finally, we define the characteristic length $L$ of the network as $L \equiv \sigma/\rho$. 
Using Eq. \ref{am12} we obtain
\be
\label{am13}
\frac{dL}{dt} = (1+3\,v^2)(H\,L).
\ee
\subsubsection{A More Realistic Scenario}

We have introduced in Refs. \cite{IDEAL1}--\cite{IDEAL3}
a more realistic description of domain wall evolution that takes into account
the curvature of the domain walls, energy loss mechanisms and interactions with other fields.

Firstly, since the wall is not planar there is an acceleration term inversely proportional to the curvature radius of the walls, which we assume to be equal to $L$ in the one scale model. Therefore the equation for the wall velocity (\ref{am11}) can be written as
\be
\frac{dv}{dt}=(1-v^2)\left(\frac{k_w}{L}-3\,H\,v\right),
\label{121}
\ee
where $k_w$ is the \textit{curvature parameter}.

Secondly, we take into account that beyond the damping caused by the Hubble expansion, there is a further damping term coming from 
friction due to particle scattering off the domain walls. Hence we define the \textit{damping length scale} $\ell_d$ as
\be
\frac1{\ell_d} \equiv 3\,H + \ell_f,
\ee
where $\ell_f$ accounts the frictional effects due to the interactions of the walls. 
 
The equation for the wall velocity (\ref{121}) is now
\be
\frac{dv}{dt}=(1-v^2)\left(\frac{k_w}{L}-\frac{v}{\ell_d}\right),
\label{am1111}
\ee
where $\ell_d$ includes both the effects of Hubble damping and particle scattering\footnote{If there are no particles scattering off the walls, $\ell_f \to \infty$, only the damping from the Hubble expansion is computed.}. 
Note that the friction term $\ell_f$ is equal to
\be
\ell_f = \frac{\sigma}{N_w\,T^4} \propto a^4,
\ee
where $T$ is background temperature and $N_w$ is the number of light particles changing their mass across the walls, 
\cite{KIBBLE}.
Therefore $\ell_f$ is expected to dominate at early times, while the Hubble term dominates at late times.

Thirdly, since in a realistic scenario the walls interact there are energy losses to radiation\footnote{We expect the probability
of a wall of size $L$ encountering  another of the same size within a time $dt$ to be proportional to $v\,dt/L$ \cite{KIBBLE}.} 
Eqs. (\ref{am13}) thus become
\ben
\frac{dL}{dt}=HL+ \frac{L}{\ell_d}\,v^2 +c_w\,v\,,\label{vevoldw1}
\een
where the \textit{wall energy loss parameter} $c_w$ parametrizes such energy loss. 

Consider $c_w = 0$ and $\ell_f \to \infty$. If 
$a \propto t^\alpha$ with $\alpha > 1/4$, Eqs. (\ref{am1111}) and (\ref{vevoldw1})
admit a \textit{linear scaling solution}, with
$L \propto t$ and $v = \rm{constant}$. 
For example,
deep into the radiation dominated epoch ($\alpha = 1/2$) one has
\be
L = \frac{2}{\sqrt{3}}\,k_w\,t,\,\,\,\,\,\,\,\,v = \frac1{\sqrt{3}},
\label{uls3}
\ee
whereas
deep into the matter era ($\alpha = 2/3$) one has
\be
L = \frac{\sqrt{6}}{2}\,k_w\,t,\,\,\,\,\,\,\,\,v = \frac1{\sqrt{6}}. 
\ee

Moreover, since no strong signatures on the cosmic microwave background have been observed, the dark energy should be 
approximately homogeneous and isotropic on large scales. The CMB temperature fluctuations generated by domain walls have to be smaller than $10^{-5}$ down to 
scales 
of the order of $10^{-2}\,H^{-1}$. 
This implies that 
$
L \lesssim 10 {\rm kpc}
$
which leads to a very small curvature parameter
$
k_w \lesssim 10^{-6}.
$
This clearly shows that the simplest domain wall scenario without junctions, with $k_w \sim 1$, cannot be the dark energy, regardless of any other considerations. 
Note that
$c_w  \neq 0$ leads to larger values of $L$ and does not 
help to \textit{frustration}, i.e., $L \propto a$ and $v = 0$.
Furthermore, the addition of the friction term also does not help much due to the limited amount of energy with which domain walls can interact conserving energy and momentum \cite{SIMS1,SIMS2}. Hence the only possible candidates for dark energy are non-standard networks, that is those with junctions where walls intersect.

\section{\label{intr12}Domain Wall Networks with Junctions}
In the previous section we have shown that only domain wall networks with junctions can be considered as candidates to describe dark energy.
We have studied several models where different types of junctions are formed. 
In this section, we will investigate two cases: the Bazeia-Brito-Losano and the Kubotani models \cite{BAZEIA,KUBOTANI}. 

\subsection{The BBL Model}
The Bazeia-Brito-Losano (henceforth referred as BBL model) model was introduced as a possible explanation of why
the universe presents only three spatial dimensions \cite{Elphick}--\cite{Durrer}. 
The Lagrangian density in the BBL model is given by \cite{BAZEIA}
\be
\mathcal{L} = \frac12 \sum_{i=1}^n (\partial_\mu \varphi_i \partial^\mu \varphi_i)  + V(\varphi_i),
\ee
where $\varphi_i$ are real scalar fields and $n$ is a integer number. The potential is
\be
\label{bbln}
V = \frac12 \sum_{i=1}^n \left(r - \frac{\varphi_i^2}{r} \right)^2 + \frac{\epsilon}2 \mathcal{C}(\varphi_1, \varphi_2,...,\varphi_n),
\ee
where we define the coupling function $\mathcal{C}$ between the fields as
\be
\mathcal{C}(\varphi_1, \varphi_2,...,\varphi_n) = \mathcal{C}(\varphi_1,\varphi_2) + \mathcal{C}(\varphi_1,\varphi_3) + ... +\mathcal{C}(\varphi_{i-1},\varphi_i)+...\mathcal{C}(\varphi_{n-1},\varphi_n),
\ee
with
\be
\mathcal{C}(\varphi_j,\varphi_k) = \frac12 \left(\varphi_j^2 + \varphi_k^2 - 3 \varphi_j^2 \varphi_k^2 + \frac92 \right)
\ee
\begin{figure}
\begin{center}
\includegraphics*[width=9cm]{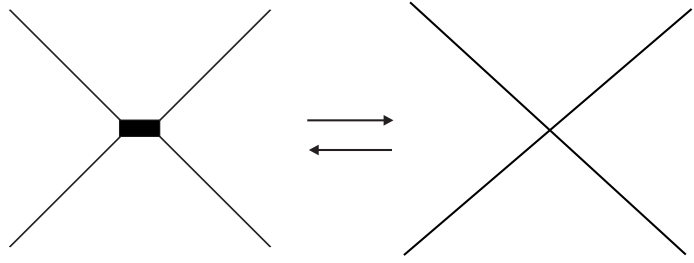}
\end{center}
\caption[Y-type/X-type junction on energetic grounds.]{Depending on whether the thick wall has a tension smaller/larger than twice that of the lower tension ones the 
formation of a Y-type/X-type junction will be favored on energetic grounds.}
\label{threefour}
\end{figure}
where $i,j,k=1,2,...,n$. 
Of course, if the constant coupling $\epsilon$ vanishes, one has a system formed by $n$ decoupled fields.
\subsubsection{BBL Model - ${\bf 1}$ field}
The BBL model
with just one field is given by
\be
\label{bblonepot}
V(\varphi) = \frac12 \left( r - \frac{\varphi^2}r \right)^2
\ee
with minima $\varphi = \pm r$ ($\mathcal{Z}_2$ symmetry). The potential (\ref{bblonepot}) is similar to the $\lambda \varphi^4$ given by Eq. (Eq. \ref{lldw}) with  
$r = \sqrt{2/\lambda} = \sqrt{\lambda/2}\,\eta^2$. (See Fig. \ref{lal}).

\subsubsection{BBL Model - ${\bf 2}$ fields}
Consider the two-field BBL model given by 
\be
\label{potbbl22}
V(\varphi_i) = \frac12 \sum_{i=1}^2 \left(r-\frac{\varphi_i^2}r \right)^2 + \frac{\epsilon}4 \left(\varphi_1^4 + \varphi_2^4 - 6 \varphi_1^2 \varphi_2^2 + 9\right)\,.  
\ee
\begin{figure}
\begin{center}
\includegraphics*[width=9cm]{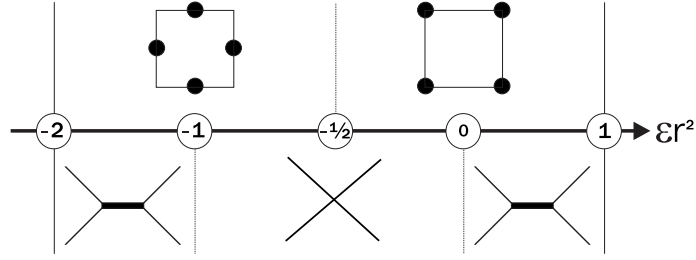}
\end{center}
\caption[Junctions and minima in the two-field BBL model.]{Types of junctions (bottom) and configuration of the minima (top), as a function of the parameter $\epsilon r^2$ in the two-field BBL model \protect\cite{BAZEIA}.}
\label{config2d}
\end{figure}
\begin{figure}[ht]
\begin{center}
\includegraphics*[width=6cm]{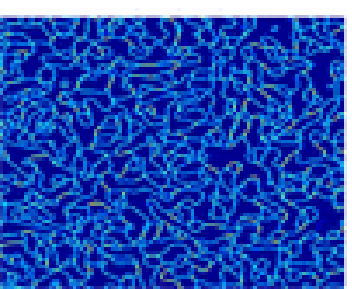}
\includegraphics*[width=6cm]{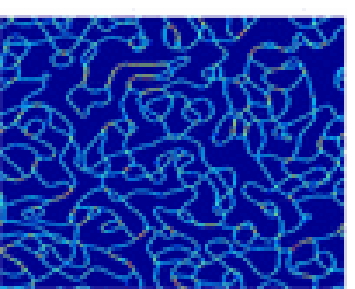}
\includegraphics*[width=6cm]{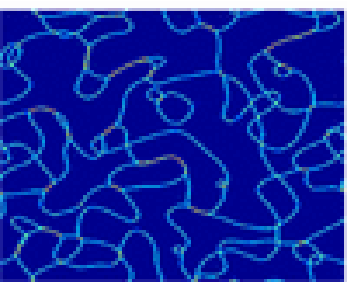}
\includegraphics*[width=6cm]{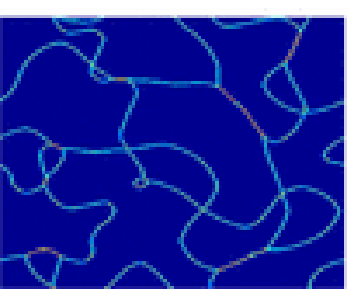}
\end{center}
\caption[Two-field BBL model with $r=\sqrt{3/2}$ and $\epsilon = 0.2$.]{The matter-era evolution of the domain wall network for
the BBL model (\protect\ref{potbbl22}) with $r=\sqrt{3/2}$ and $\epsilon = 0.2$.  The simulation starts with random initial conditions inside the square whose vertexes are the vacua $(\pm \sqrt{15/7},\pm \sqrt{15/7}) $. Note that stable $Y$-type junctions are preferred. From left to right and top to bottom, the horizon is approximately 1/16, 1/8, 1/4 and 1/2 of the box size respectively.}
\label{bazeia2positivo}
\end{figure}
\begin{figure}[ht]
\begin{center}
\includegraphics*[width=6cm]{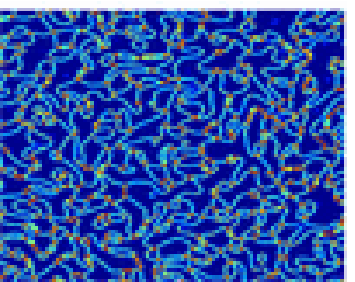}
\includegraphics*[width=6cm]{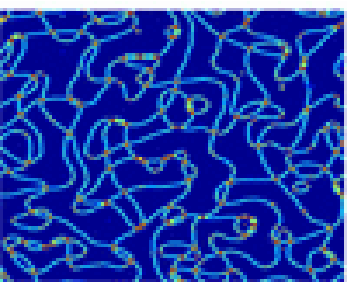}
\includegraphics*[width=6cm]{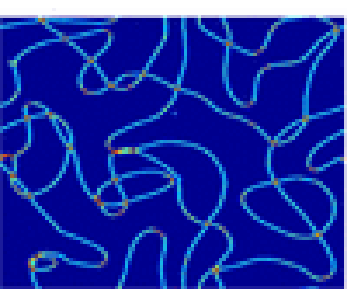}
\includegraphics*[width=6cm]{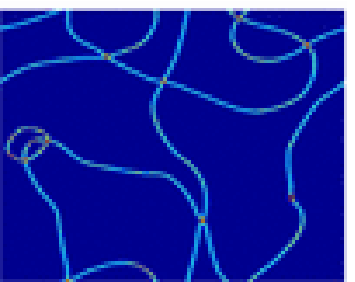}
\end{center}
\caption[Two-field BBL model with $r=\sqrt{3/2}$ and $\epsilon = -0.2$.]{Same as Fig. \protect\ref{bazeia2positivo}, except that now $\epsilon = - 0.2$. Here the simulation starts with initial conditions chosen randomly in the square whose vertexes are the vacua $(\pm \sqrt{15/13},\pm \sqrt{15/13}) $. Note that in this case only stable $X$-type junctions are present in the network. This simulation can be interpreted as the two-field Kubotani model given in Eq. \ref{twokubo} with the parameters $\lambda = 0.3$, $\eta = \sqrt{5/3}$ and $\xi = 5/6$.}
\label{bazeia2negativo1}
\end{figure}
\begin{figure}[ht]
\begin{center}
\includegraphics*[width=6cm]{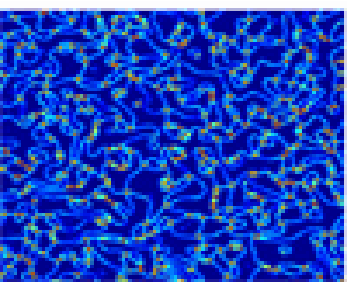}
\includegraphics*[width=6cm]{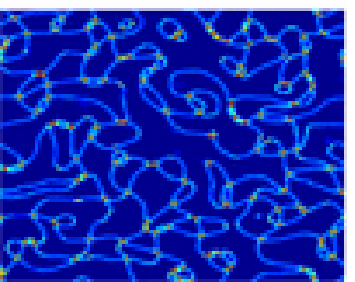}
\includegraphics*[width=6cm]{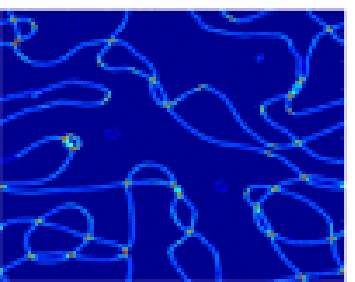}
\includegraphics*[width=6cm]{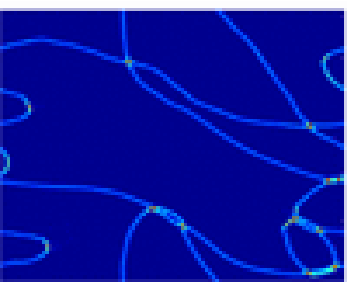}
\end{center}
\caption[Two-field BBL model with $r=\sqrt{3/2}$ and $\epsilon = -0.4$.]{Same as Fig. \protect\ref{bazeia2positivo}, except that now $\epsilon = - 0.4$. In this case we start with initial conditions chosen randomly in the square whose vertexes are the vacua $(\pm \sqrt{15/7},0) $ and $(0,\pm \sqrt{15/7}) $. Here again only stable $X$-type junctions are present in the network.}
\label{bazeia2negativo2}
\end{figure}
The minima of potential (\ref{potbbl22}) are displayed at the vertexes of a square 
in the plane $(\varphi_1, \varphi_2)$ whose orientation is defined by the parameter $\epsilon$. 

Firstly, let us assume $\epsilon = 0$. In this case, the fields are decoupled and the potential (\ref{potbbl22}) is reduced to
\be
\label{potbbl2no}
V(\varphi_i) = \frac12 \sum_{i=1}^2 \left(r-\frac{\varphi_i^2}r \right)^2,
\ee
whose minima $\varphi_1 = \varphi_2 = \pm r$. 
Therefore the vacua are located at
the vertexes of a square in the plane $\varphi_1,\varphi_2$ ($\mathcal{Z}_4$ symmetry). They form six possible topological sectors connecting two different minima: four \textit{edges} (connecting two neighboring minima in field plane) and two \textit{diagonals} (linking two opposite minima).  

The one-dimensional equations of motions are \cite{BAZEIA}
\be
\frac{d \varphi_i}{dx} = r - \frac{\varphi_i^2}{r}
\ee
with the soliton solutions ($i=1,2$) $\varphi_{1a}^2 = r^2$ and $\varphi_{2a}^2 = r^2 \tanh(x)^2$ for the edges sectors,
and $\varphi_{1b}^2 = \varphi_{2_b}$ and 
$\varphi_{2a}^2 = r^2 \tanh(x)^2$ for the diagonal ones,
where $a=1,...,4$ and $b=5,6$.

Secondly, let us now consider $\epsilon \neq 0$. 
The potential is well defined only for $ -2\,<\,\epsilon r^2\,<\,1$, where if
$ -1/2\,<\,\epsilon r^2\,<\,1$  the minima are 
\be
\varphi_i^2=\frac{r^2}{1-\epsilon r^2},\quad i=1,2\,,\label{solplus}
\ee
while for $ - 2\, <\, \epsilon r^2\, <\, - 1/2$ the minima are
\be
\varphi_i^2 = \frac{r^2}{1+\epsilon r^2/2}, \quad \varphi_{j \neq i}^2=0\,.\label{solminus}
\ee

Except if stated otherwise we assume $r=\sqrt{3/2}$ throughout.
The stability of the junctions is determined by the choice of $\epsilon$ because it determines the tension of the walls.
In Ref. \cite{BAZEIA} the authors find that the tension of the one-dimensional soliton solutions in the limit of very small $\epsilon$ are $ E^a = 2 + \frac{21\,\epsilon}4$ and 
$E^b = 4 + 6 \epsilon$,
for the \textit{edges} and \textit{diagonal} walls respectively \cite{Almeida}.  
\begin{figure}
\begin{center}
\includegraphics*[width=6cm]{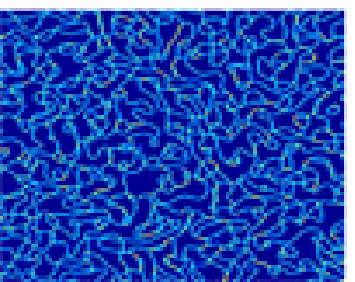}
\includegraphics*[width=6cm]{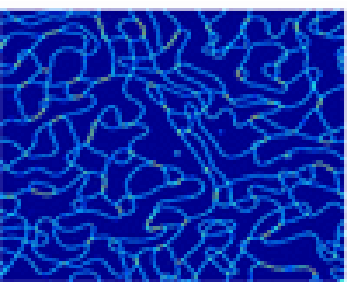}
\includegraphics*[width=6cm]{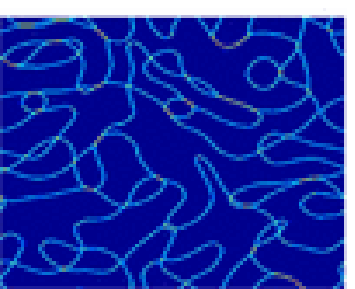}
\includegraphics*[width=6cm]{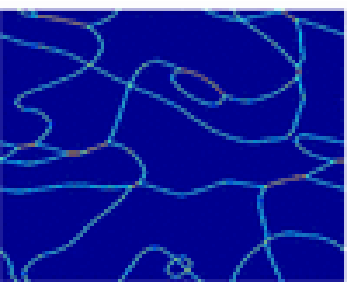}
\end{center}
\caption[Two-field BBL model with $r=\sqrt{3/2}$ and $\epsilon = -0.8$.]{Same as Fig. \protect\ref{bazeia2positivo}, except that now $\epsilon = - 0.8$. Here we start with initial conditions chosen randomly in the square whose vertexes are the vacua $(\pm \sqrt{15/4},0) $ and $(0,\pm \sqrt{15/4}) $. Note that in this case $Y$-type junctions are favored in the network.}
\label{bazeia2negativo3}
\end{figure}

Although they have selected
the specific case where only stable X-type junctions survive
we are interested in all possibles cases: stable Y-type and X-type junctions.

We confirmed their results in the small $\epsilon$ limit, where the ratio of the tension of the walls of \textit{edge} and \textit{diagonal} types, $\sigma_e$ and $\sigma_d$ respectively, is given by 
\be
\frac{\sigma_d}{\sigma_e} =\frac{2+3\epsilon}{1 + 21 \epsilon/8}\,,\label{sigma}
\ee
joining the vacua at $\varphi_i^2=(2/3-\epsilon)^{-1/2}$ \cite{BAZEIA}. 
Depending on whether the diagonal walls have a tension smaller ($\epsilon > 0$) or larger ($\epsilon < 0$) than twice that of the edge ones, the formation of a Y-type or X-type junction is favored on energetic grounds, as it is illustrated in Fig. \ref{threefour}.

Although the analytic expression (\ref{sigma}) is valid only for
$\epsilon \rightarrow 0$, the energetic argument is valid generically. 
In the general case, 
\begin{itemize}
\item
X-type junctions: $\sigma_d > 2\sigma_e$ for  $-1 <\epsilon r^2< 0$.
\item
Y-type junctions: $\sigma_d < 2\sigma_e$ for $\epsilon r^2> 0$ and $\epsilon r^2< -1$.
\end{itemize}
Fig. \ref{config2d} illustrates the complete region allowed for $\epsilon r^2$ and the respective subsets that yield Y-type and X-type junctions.
Nevertheless, if the angle between the walls is close to $\pi/2$ then 
X-type junctions may still be formed for
$\sqrt{2}\,\sigma_e < \sigma_d < 2\,\sigma_e$. 
Hence, there will be an intermediate case  
for which Y-type junctions are favored but X-type junctions will still be present (see Figs. \ref{bazeia2positivo} and \ref{bazeia2negativo3}).

\begin{figure}
\begin{center}
\includegraphics*[width=6cm]{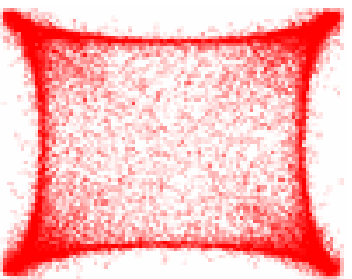}
\includegraphics*[width=6cm]{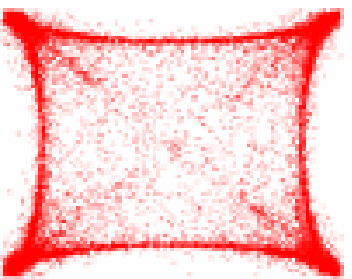}
\includegraphics*[width=6cm]{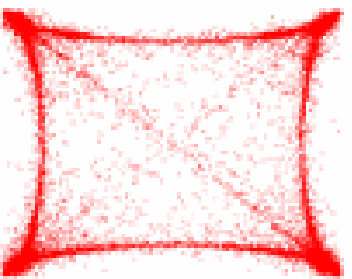}
\includegraphics*[width=6cm]{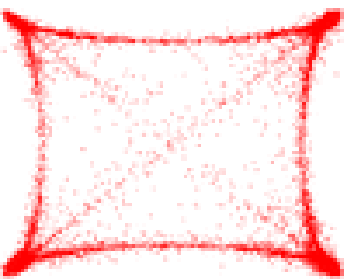}
\end{center}
\caption[Configuration space for Fig. \protect\ref{bazeia2positivo}.]{The configuration space distribution for the four timesteps of the simulation in Fig. \protect\ref{bazeia2positivo}. The vertexes of the square in the plane $(\varphi_1,\varphi_2)$ are the minima of the potential, given by $(\pm \sqrt{15/7}), \pm \sqrt{15/7})$.}
\label{bazeia2positivofase}
\end{figure}

Now, we stress that 
$\epsilon r^2 = -1$ is not a marginal case between the formation of the X-type and Y-type junctions. Actually, the two fields are decoupled as in the case described by the potential in Eq. \ref{potbbl2no} (for $\epsilon r^2 = 0$). As a result 
the fields evolve independently. 
To be clearer we can define a new couple of fields $\psi_1$ and $\psi_2$ given by
\ben
\psi_1=\frac1{\sqrt{2}}\,\left( \varphi_1 + \varphi_2 \right),\\
\psi_2=\frac1{\sqrt{2}}\,\left( \varphi_1 - \varphi_2 \right),
\een
and write the potential in terms of these. One finds that these rotated fields explicitly decouple for $ \epsilon r^2 = -1 $. 
\begin{figure}[ht]
\begin{center}
\includegraphics*[width=6cm]{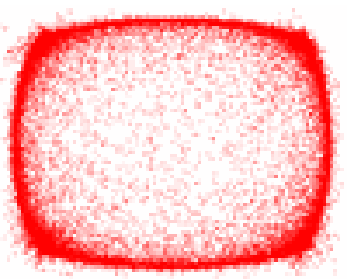}
\includegraphics*[width=6cm]{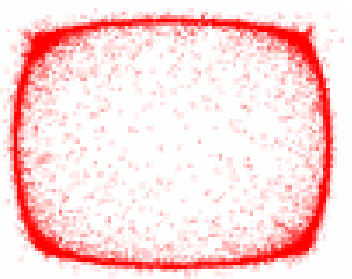}
\includegraphics*[width=6cm]{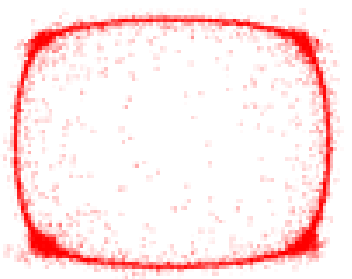}
\includegraphics*[width=6cm]{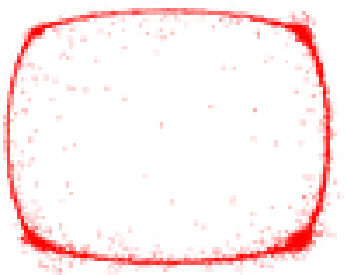}
\end{center}
\caption[Configuration space for Fig. \protect\ref{bazeia2negativo1}.]{The configuration space distribution for the four timesteps of the simulation in Fig. \protect\ref{bazeia2negativo1}. The vertexes of the square in the plane $(\varphi_1,\varphi_2)$ are the minima of the potential, given by $(\pm \sqrt{15/13}), \pm \sqrt{15/13})$.}
\label{bazeia2negativofase1}
\end{figure}
\begin{figure}[ht]
\begin{center}
\includegraphics*[width=6cm]{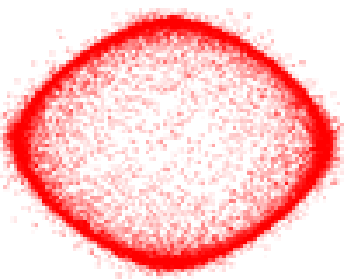}
\includegraphics*[width=6cm]{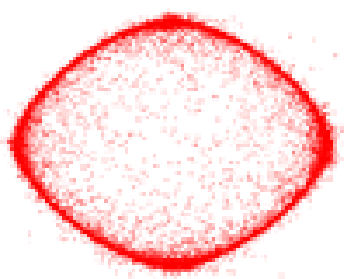}
\includegraphics*[width=6cm]{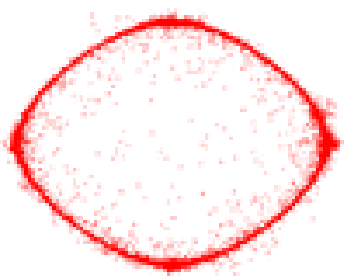}
\includegraphics*[width=6cm]{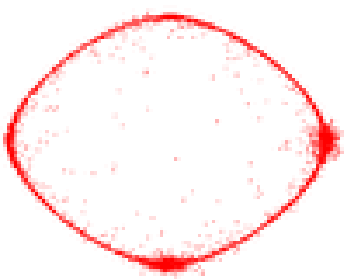}
\end{center}
\caption[Configuration space for Fig. \protect\ref{bazeia2negativo2}.]{The configuration space distribution for the four timesteps of the simulation in Fig. \protect\ref{bazeia2negativo2}. The vertexes of the square in the plane $(\varphi_1,\varphi_2)$ are the minima of the potential, given by $(\pm \sqrt{15/7}), 0)$ and $(0, \pm \sqrt{15/7})$.}
\label{bazeia2negativofase2}
\end{figure}
For an interesting example of a marginal case where both types of junctions are allowed, we can consider the model described by a complex scalar field $\Phi$
\be
\mathcal{L} = \partial_\mu \Phi \partial^\mu \bar{\Phi} - V(\Phi)
\ee
with the potential
\be
V(\Phi)=\tilde\lambda \left|\Phi^N-1\right|^2,\label{pot2}
\ee
where $\tilde\lambda$ is a real parameter. This model has vacua at
\be
\Phi=e^{i\frac{n}{N}},\quad n=0,1,\ldots,N-1\,.
\ee
We can alternatively write this field as 
\be
\Phi=|\Phi|e^{i\phi}
\ee
\begin{figure}
\begin{center}
\includegraphics*[width=6cm]{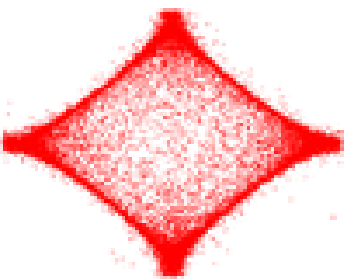}
\includegraphics*[width=6cm]{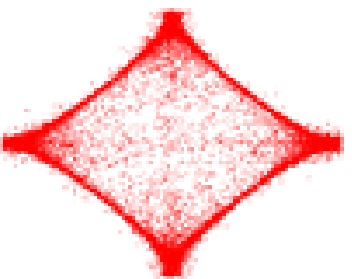}
\includegraphics*[width=6cm]{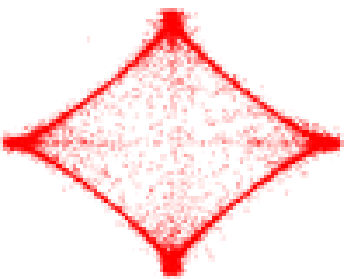}
\includegraphics*[width=6cm]{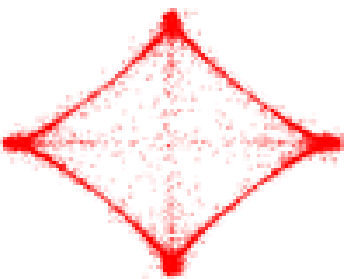}
\end{center}
\caption[Configuration space for Fig. \protect\ref{bazeia2negativo3}.]{The configuration space distribution for the four timesteps of the simulation in Fig. \protect\ref{bazeia2negativo3}. The vertexes of the square in the plane $(\varphi_1,\varphi_2)$ are the minima of the potential, given by $(\pm \sqrt{15/4}), 0)$ and $(0, \pm \sqrt{15/4})$.}
\label{bazeia2negativofase3}
\end{figure}
\begin{figure}
\begin{center}
\includegraphics*[width=5.1cm]{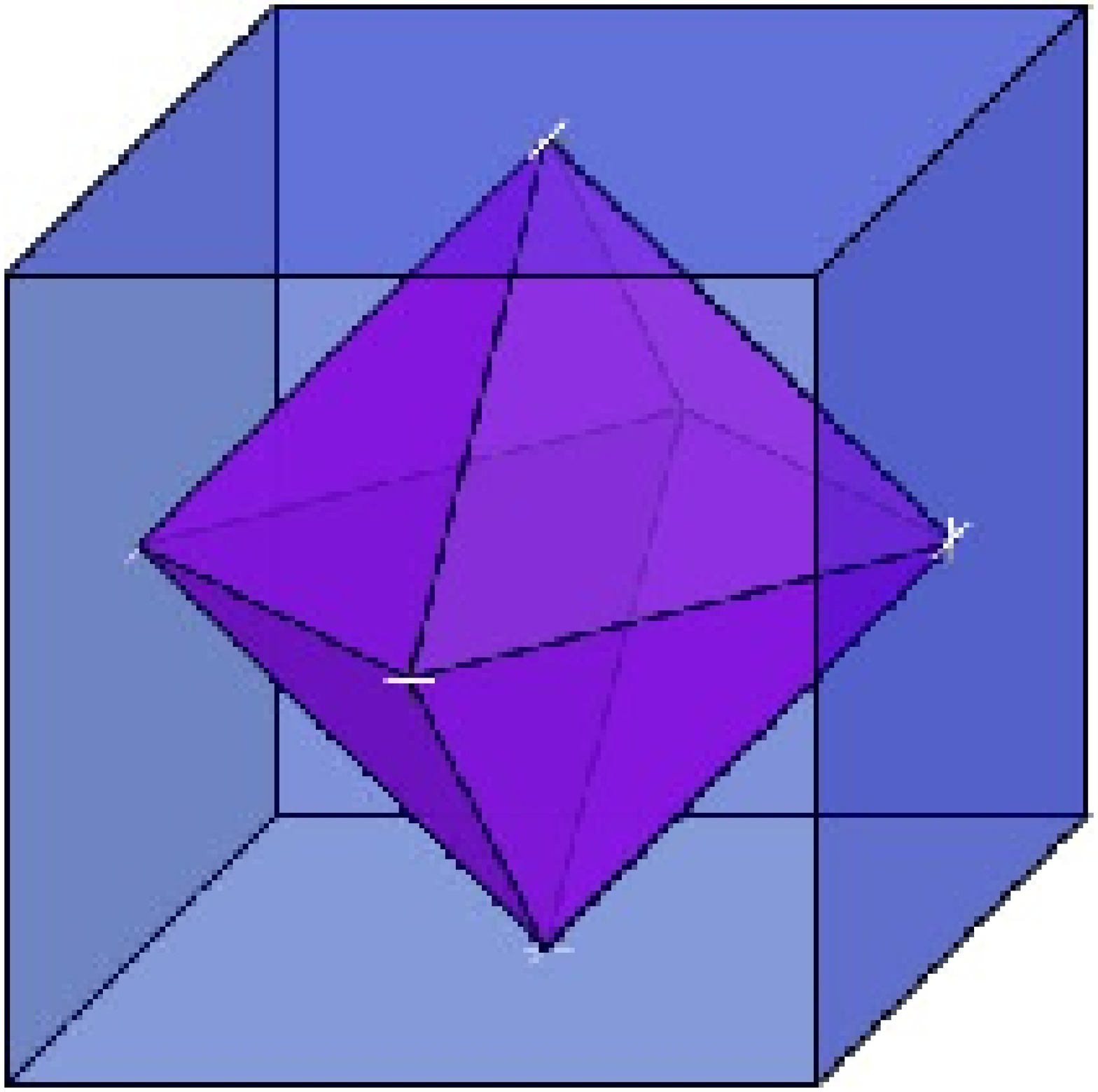}
\end{center}
\caption[The configuration of the vacua in the three field models of BBL.]{
The configuration of the vacua in the three fields BBL model \protect\cite{BAZEIA}, for the branch (\protect\ref{solplus3}) (Kubotani model \protect\cite{KUBOTANI} for $\xi >0$) and for (\protect\ref{solminus3}) (Kubotani model \protect\cite{KUBOTANI} for $\xi <0$). They are displayed at the vertexes of a cube and an octahedron, respectively. Note that in the former case, the formation of Y-type junctions is contingent on the relation between the tensions of the walls in the \textit{diagonal} and \textit{edge} sectors whereas in the latter one, any X-type junctions or Y-type junctions involving the \textit{axis} sector correspond to field space configurations where one of the three fields vanishes.}
\label{cubeocta}
\end{figure}
and it is then obvious that we will have $N$ minima, evenly spaced around the phase $\phi$. The case $N=2$ produces standard domain walls and $N=3$ produces Y-type junctions, but the case $N=4$ is slightly more subtle. The potential (\ref{pot2}) has supersymmetric properties, and hence the energy of a specific solution depends only on the initial and final vacua. In other words, the possible ways of connecting two opposite vacua (directly or through the intermediate vacuum) will have the same energy. Hence this case is an example of the scenario where $\sigma_d = 2\,\sigma_e$, which means that the formation of Y-type junctions is as likely as the formation of X-type ones, and the two will always co-exist. This can also be confirmed numerically, and visually the results can be approximately described as the superposition of those ones shown in the two above cases with $\epsilon > 0$ and $\epsilon < 0$.

We confirmed numerically all the properties of the two-field BBL model by performing $256^2$ simulations with $r=\sqrt{3/2}$. 
In this particular case:
\begin{itemize}
\item
For $-\frac13 < \epsilon < \frac23$ the vacua are at $ \big\{ \pm (\frac23- \epsilon)^{-\frac12},\pm (\frac23- \epsilon)^{-\frac12} \big\}$
\item
For $-\frac43 < \epsilon < -\frac13$ the vacua are at $ \big\{0,\pm (\frac23+\frac{\epsilon}2)^{-1/2}\big\}$ and $\big\{\pm (\frac23+\frac{\epsilon}2)^{-\frac12},0\big\}$. 
\end{itemize}

The outcome of two such simulations is illustrated by the four snapshots displayed in Figs. \ref{bazeia2positivo} and \ref{bazeia2negativo1}, where we have respectively taken $\epsilon= \pm 0.2$. 
Note that in the former case Y-type junctions are preferred while in the latter case only X-type junctions survive.
We also verified that the results 
confirm that for any chosen value of $\epsilon > 0$ (that is in the range $0 < \epsilon < 2/3$, given our choice of $r$), the behavior of the system is similar to that in Fig. \ref{bazeia2positivo}: the formation of Y-type junctions is always preferred. On the other hand, we confirm our expectation that two distinct cases appear for $\epsilon <0$. For the interval $-2/3 < \epsilon < 0$, only X-type junctions are formed, as can be seen in Figs. \ref{bazeia2negativo1} and \ref{bazeia2negativo2} for $\epsilon =-0.2$ and $\epsilon =-0.4$ respectively. However, for $\epsilon <-2/3$ X-type junctions are no longer stable, and instead Y-type junctions are preferred, as illustrated in Fig. \ref{bazeia2negativo3} for the case $\epsilon=-0.8$.

We also consider the behavior of the configuration space formed by the two scalar fields, $\varphi_1$ and $\varphi_2$. 
For this purpose
we plot in  
Figs. \ref{bazeia2positivofase}--\ref{bazeia2negativofase3}  field configurations corresponding to the snapshots of Figs. \ref{bazeia2positivo}--\ref{bazeia2negativo3} respectively. In agreement with our description,
in the cases $\epsilon>0$ and $\epsilon <-2/3$ the field values tend to concentrate along the solutions of both the \textit{edges} and the \textit{diagonal} sectors of the square. In addition, we stress that the orientation of the square is different in the two cases, corresponding to the solutions given by Eq. (\ref{solplus}) and (\ref{solminus}) respectively. 
On the contrary, for the range $-2/3 < \epsilon <0$ the \textit{diagonal} sector of the square gradually becomes depopulated and only the \textit{edges} behave as attractors. 

\subsubsection{\label{threefield}BBL Model - $3$ fields}
We now move on to the case of models with three scalar fields. The discussion turns out to be quite similar to the two-field case, though at some points the effect of the increased dimensionality provides for different phenomenology. 
The three-field BBL model (\ref{bbln}) is given by 
\be
V(\varphi_i) = \frac12 \sum_{i=1}^3 \left[\left(r-\frac{\varphi_i^2}{r}\right)^2 + \epsilon \left(\varphi_i^4 + \frac92\right) \right]
 - 3 \epsilon\left( \varphi_1^2 \varphi_2^2 + \varphi_1^2 \varphi_3^2 +\varphi_2^2 \varphi_3^2 \right)\,.
 \label{bbl3}  
\ee
As in the two-field case, there are two branches for the minima (see Fig. \ref{cubeocta}):

\begin{itemize}
\item
For $ -\frac25 < \epsilon r^2 <  \frac12$ there are $8$ vacua located at the vertexes of a cube in the space $(\varphi_1,\varphi_2,\varphi_3)$ with values
\be
\varphi_i^2=\frac{r^2}{1 - 2 \epsilon r^2},\quad i=1,2\,\label{solplus3}.
\ee
There are twenty-eight topological sectors and three kinds of walls, which for obvious reasons we can refer to as \textit{edges}, 
\textit{external diagonals} and \textit{internal diagonals}. The number of different walls of each type is respectively twelve, twelve and four. 
\item
For $ - 1 < \epsilon r^2 < - \frac25$ there are 6 minima which are located at the vertexes of an octahedron in the space $(\varphi_1,\varphi_2,\varphi_3)$ where
\be
\varphi_i^2 = \frac{r^2}{1+\epsilon r^2}, \quad \varphi_{j \neq i}^2=0\,.\label{solminus3}
\ee
There are fifteen topological sectors and two kinds of walls, which we can refer to as \textit{edges} and \textit{axes}. The number of different walls of each type is respectively twelve and three. 
\end{itemize}

\begin{figure}
\begin{center}
\includegraphics*[width=9cm]{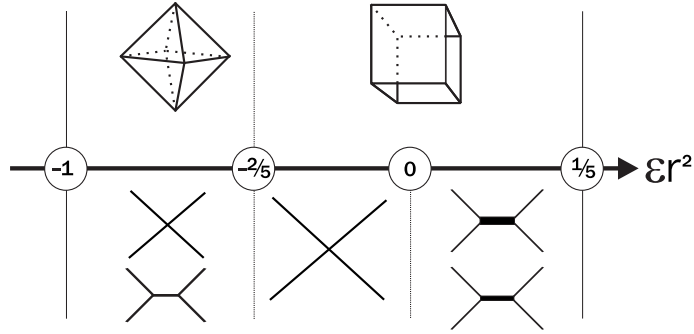}
\end{center}
\caption[Junctions and minima of three-field BBL model.]{Types of junctions (bottom) and configuration of the minima (top), as a function of the parameter $\epsilon r^2$ in the three-field BBL model \protect\cite{BAZEIA}. Notice that in the case $\epsilon r^2>0$ there are effectively two kinds of Y-type junctions, involving either the \protect\textit{internal diagonal} or the \protect\textit{external diagonal} sector.}
\label{config3d}
\end{figure}

Again the choice of the parameter $\epsilon$ determines what type of junctions will be present. For the branch $\epsilon r^2 >-2/5$, corresponding to the cubic solution of Eq. (\ref{solplus3}), X-type junctions survive only if the junctions involving walls from the diagonal sectors are energetically disfavored. 
In other words, they do not necessarily form unless $\sigma_{id} > 3 \sigma_{e}$ (for the internal diagonals) and $\sigma_{ed} > 2 \sigma_{e}$ (for the external ones) which is realized only in the interval $-2/5 < \epsilon r^2< 0$, so in this range we do have stable X-type junctions. Conversely, for $0 < \epsilon r^2 < 1/5$ the Y-type junctions are preferred, and these may involve either of the \textit{diagonal} sectors and so we effectively have two types of Y-junctions. On the other hand, for the branch $\epsilon r^2 <-2/5$, corresponding to the octahedral solution of Eq. (\ref{solminus3}), both Y-type and X-type junctions can survive. Note that in this octahedral branch X-type junctions, as well as any Y-type junctions which involve the \textit{axes} sector (as opposed to only the \textit{edge} sector), correspond to field space configurations where one of the three fields vanishes. The cases $\epsilon = 0$ and $\epsilon = - 2/5$ are again not interesting because they represent the evolution of the three decoupled fields. A numerical study was performed along the same lines as what was done for the two-field case, and its outcome is summarized in Fig. \ref{config3d}.

\subsection{The Kubotani Model}
In this section we investigate the Kubotani model \cite{KUBOTANI,BATTYE} and relate it with the BBL model studied in the previous section. 
The potential is given by  
\be
V(\varphi)=\lambda\left(\sum_{i=1}^N \varphi_i^2-\eta^2\right)^2 + \xi \sum_{i=1}^N\left(\varphi_i^2-\zeta^2\right)^2,\label{onpot}
\ee
where the parameters $\xi$ and $\lambda$ are defined such that
$\xi + \lambda>0$ and $\xi+ N\lambda>0$.

\begin{figure}[ht]
\begin{center}
\includegraphics*[width=6cm]{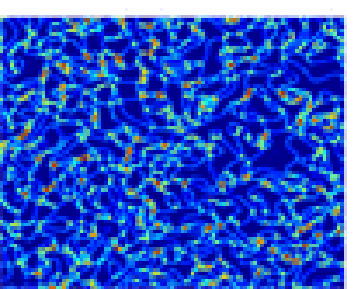}
\includegraphics*[width=6cm]{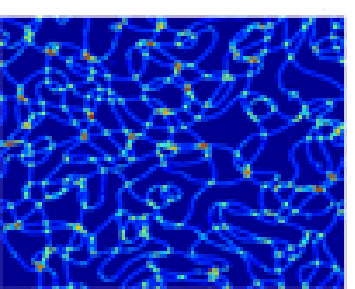}
\includegraphics*[width=6cm]{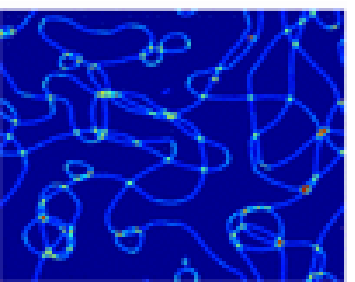}
\includegraphics*[width=6cm]{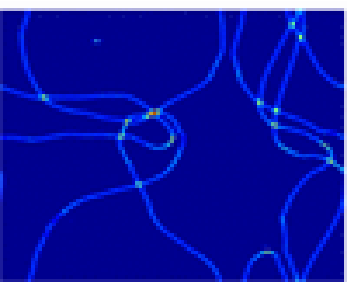}
\end{center}
\caption[Three-field BBL model with $r=\sqrt{3/2}$ and $\epsilon = -0.2$.]{The matter-era evolution of the domain wall network of a Kubotani-type model with a parameter choice $\lambda=3/20$, $\eta^2=10/3$ and $\xi=1/12$. This also mimics a three-field BBL model with $r=\sqrt{3/2}$ and $\epsilon = -0.2$.  Note that only stable X-type junctions survive. From left to right and top to bottom, the horizon is approximately 1/16, 1/8, 1/4 and 1/2 of the box size respectively.}
\label{kuboplus}
\end{figure}
\begin{figure}[ht]
\begin{center}
\includegraphics*[width=6cm]{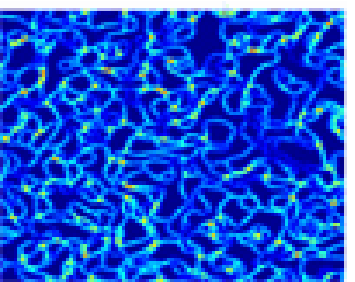}
\includegraphics*[width=6cm]{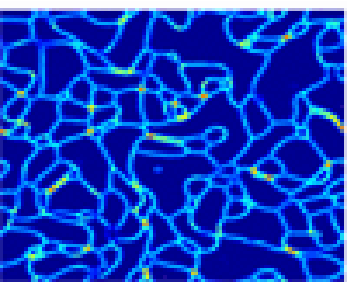}
\includegraphics*[width=6cm]{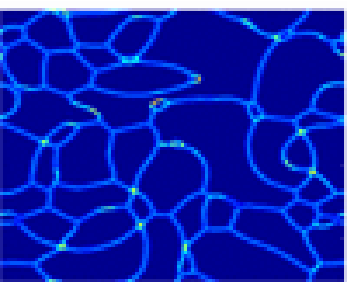}
\includegraphics*[width=6cm]{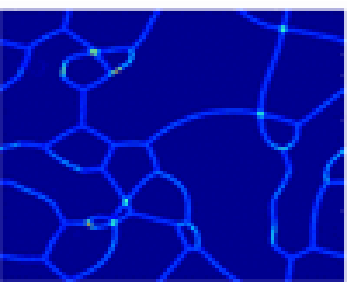}
\end{center}
\caption[Three-field BBL model with $r=\sqrt{3/2}$ and $\epsilon = -0.4$.]{
The matter-era evolution of the domain wall network of a Kubotani-type model with a parameter choice $\lambda=3/10$, $\eta^2=5/3$ and $\xi=-1/6$. This also mimics a three-field BBL model with $r=\sqrt{3/2}$ and $\epsilon = -0.4$.  Note that both Y-type and X-type junctions survive. From left to right and top to bottom, the horizon is approximately 1/16, 1/8, 1/4 and 1/2 of the box size respectively.}
\label{kubominus}
\end{figure}
Let us start by discussing the decoupled case where 
\be
V(\varphi)= \xi \sum_{i=1}^N\left(\varphi_i^2-\zeta^2\right)^2. \label{onpotdec}
\ee
We clearly see
that the fields $\varphi_i$ evolve separately: no junctions are formed.
The minima of the potential (\ref{onpotdec}) are  
$\varphi_i^2=\zeta^2$. 
Note that 
when $\xi=0, \lambda>0$ the model presents $O(N)$ symmetry. 
The N-th case has a total of $2^N$ minima, corresponding to the vertexes of an N-dimensional hypercube with $\varphi_i^2=\eta^2/N$.

In general, the potential (\ref{onpot}) for $N$ fields has the minima 
\be
\varphi^2_i=\frac{\lambda \eta^2+\xi \zeta^2}{N\lambda+\xi},\quad i=1,\ldots, N\,,\label{solkuboplus}
\ee
for  $\xi\ge0$ and 
\be
\varphi^2_i=\frac{\lambda \eta^2+\xi \zeta^2}{\lambda+\xi},\quad \varphi^2_{j\neq i}=0\,,\label{solkubominus}
\ee
for  $\xi <0$. 
Note that the former has $2^N$ minima, whose vertexes form an N-dimensional (hypercube), while the latter has $2N$ minima, that can be thought of as lying in the center of the faces of the said N-dimensional cube.

\subsubsection{Relating Kubotani and BBL Models}
Although the Kubotani model (\ref{onpot}) and the BBL model (\ref{bbln}) have been introduced in different contexts, they can be related by introducing 
the generalized potential
\be
V = A \sum_i \varphi_i^4 + B \sum_i \varphi_i^2 + C \sum_{i \neq j} \varphi_i^2 \varphi_j^2 + D,\label{genericpot}
\ee
where $A$, $B$, $C$ and $D$ are real parameters. It is straightforward to show that for 
\ben
A\,&=&\,\frac{1}{2 r^2}+\frac{\epsilon}{2}, \quad \,B\,=\,-1, \\ 
C\,&=&\,-\frac{3 \epsilon}{2}, \,\,\,\,\,\quad D\,=\,\frac{r^2}{2}+\frac{27 \epsilon}{4},\label{coefbbl}
\een
we recover the  
BBL model while for
\ben
A\,&=&\,\lambda+\xi,\,\,\,\quad B\,=\,-2 \lambda \eta^2,\\
C\,&=&\,2 \lambda,\,\,\,\quad D\,=\,\lambda \eta^4,\label{coefkubo}
\een
leads to the potential for the Kubotani model ($\zeta=0$).
Therefore
assuming 
\be
\label{coefrel2}
\epsilon = - \frac{4\,\lambda}{3} = - \frac{2}{3\,\eta^2}, \,\,\,\quad r = \sqrt{\frac{3}{6\,\xi+10\,\lambda}},
\ee
one relates both models, with the following restrictions 
\begin{itemize}
\item
For $\lambda >0$ and $\eta^2 >0$, only the interval $\epsilon < 0$ of the BBL model can be mapped into the Kubotani model. 
\item
The assumption $\xi >0$ is related to the range $-2/5 < \epsilon r^2 < 0$ while $\xi<0$ is related to $-1/2 < \epsilon r^2 < -2/5$.
\item
The two consistency conditions: $2\eta\lambda^2=1$ on the Kubotani side, and $\epsilon r^2=-27\epsilon^2/2-2/3$ on the BBL side (though the latter affects only the constant term in the potential) indicate that even in this restricted $\epsilon <0$ range the correspondence is not one-to-one.
\end{itemize}
\subsubsection{Kubotani Model - $2$ fields}
The two-field case for the Kubotani model is given by
\be
\label{twokubo}
V = (\lambda + \xi)\,(\varphi_1^4 + \varphi_2^4) - 2\,\eta^2\,\lambda\,(\varphi_1^2 + \varphi_2^2) + 
2\,\varphi_1^2\,\varphi_2^2 + \lambda\,\eta^4,
\ee
that is the same that 
the BBL model with $-1/2 < \epsilon < 0$. 
Since the value of the tensions of the walls always obeys $\sigma_d > 2\, \sigma_e$, only stable X-type junctions are formed here.
Note that
the snapshots of the numerical simulation results shown in Figs. (\ref{bazeia2negativo1},\ref{bazeia2negativofase1}) are the same as for the two-field Kubotani model with $\lambda = 0.3$, $\eta = \sqrt{5/3}$ and $\xi = 5/6$. 

\subsubsection{Kubotani Model - $3$ fields}
Let us now consider $\zeta=0$ and $N=3$.
Again, the key difference from the BBL models is that the tension of the \textit{internal diagonal} walls is always higher than $3$ times the value of the \textit{edge} walls. In other words, unlike the BBL model, Y-type junctions will never be stable in the cubic branch because $\sigma_{id} > 2\,\sigma_{ed} > 3\,\sigma_e$. This is shown in Fig. \ref{kuboplus} for $\lambda=3/20$, $\eta^2=10/3$ and $\xi=1/12$.

\begin{figure}
\begin{center}
\includegraphics*[width=6cm]{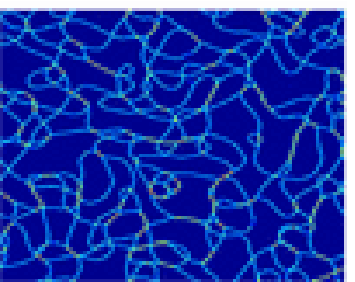}
\includegraphics*[width=6cm]{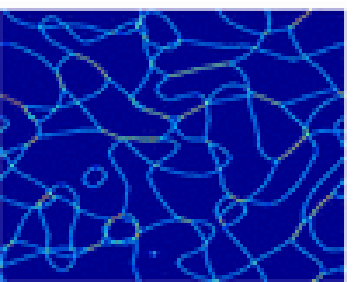}
\includegraphics*[width=6cm]{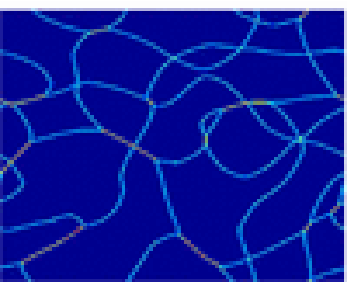}
\includegraphics*[width=6cm]{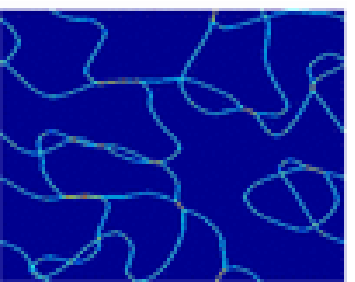}
\end{center}
\caption[Three-field BBL model with $r=\sqrt{3/2}$ and $\epsilon = 0.1$.]{
The matter-era evolution of the domain wall network for a three-field BBL model with $r=\sqrt{3/2}$ and $\epsilon = 0.1$.  Note that only stable Y-type junctions survive. From left to right and top to bottom, the horizon is approximately 1/16, 1/8, 1/4 and 1/2 of the box size respectively.
}
\label{bazeia3cubeY}
\end{figure}

Despite the non-standard definition, it is also possible phenomenologically to assume $\xi <0$. Here
Y-type junctions will form. But just like in the BBL case, one can also envisage having X-type junctions, which can arise from field space configurations where one of the three scalar fields vanishes. In other words, the scenario with only stable X-type junctions would be possible only if the four vacua
lying on a plane are selected. However, no realistic spontaneous symmetric breaking mechanism that selects only a subset of the vacua is expected to occur. Therefore we expect that the existence of both kinds of junctions in any realistic simulation. 
This fact can be confirmed in Fig. \ref{kubominus} for the parameter choices $\lambda=3/10$, $\eta^2=5/3$ and $\xi=-1/6$. We also note that even though the majority of the junctions are of Y-type, the fraction of X-type junctions remains approximately constant during the evolution. 
For completeness we present the case for $0 < \epsilon r^2< 1/2$ in the BBL model, which has no correspondence in the Kubotani case. We have chosen $r=\sqrt{3/2}$ and $\epsilon=0.1$ to illustrate this scenario in Fig. \ref{bazeia3cubeY}. 
\begin{figure}[ht]
\begin{center}
\includegraphics*[width=6cm]{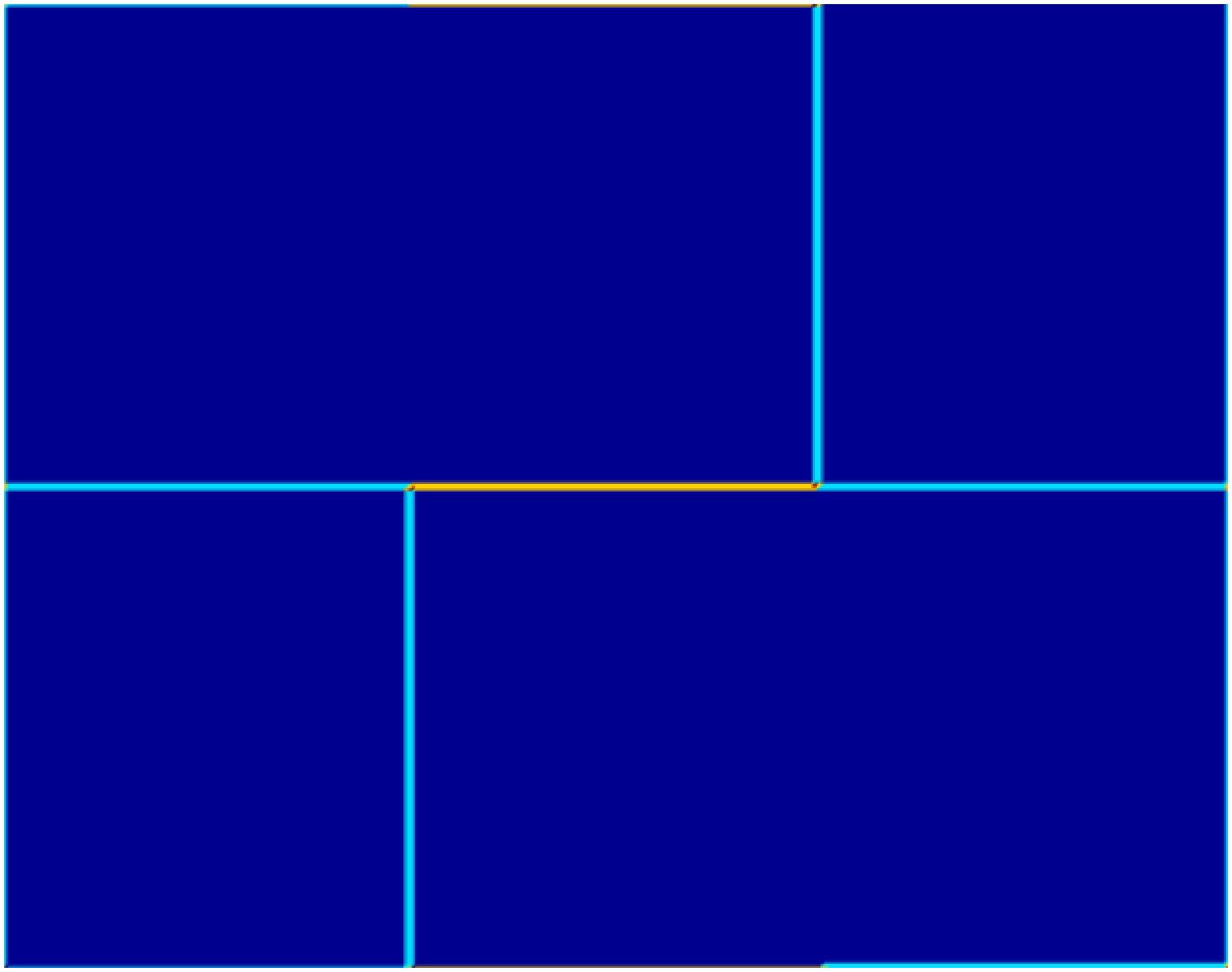}
\includegraphics*[width=6cm]{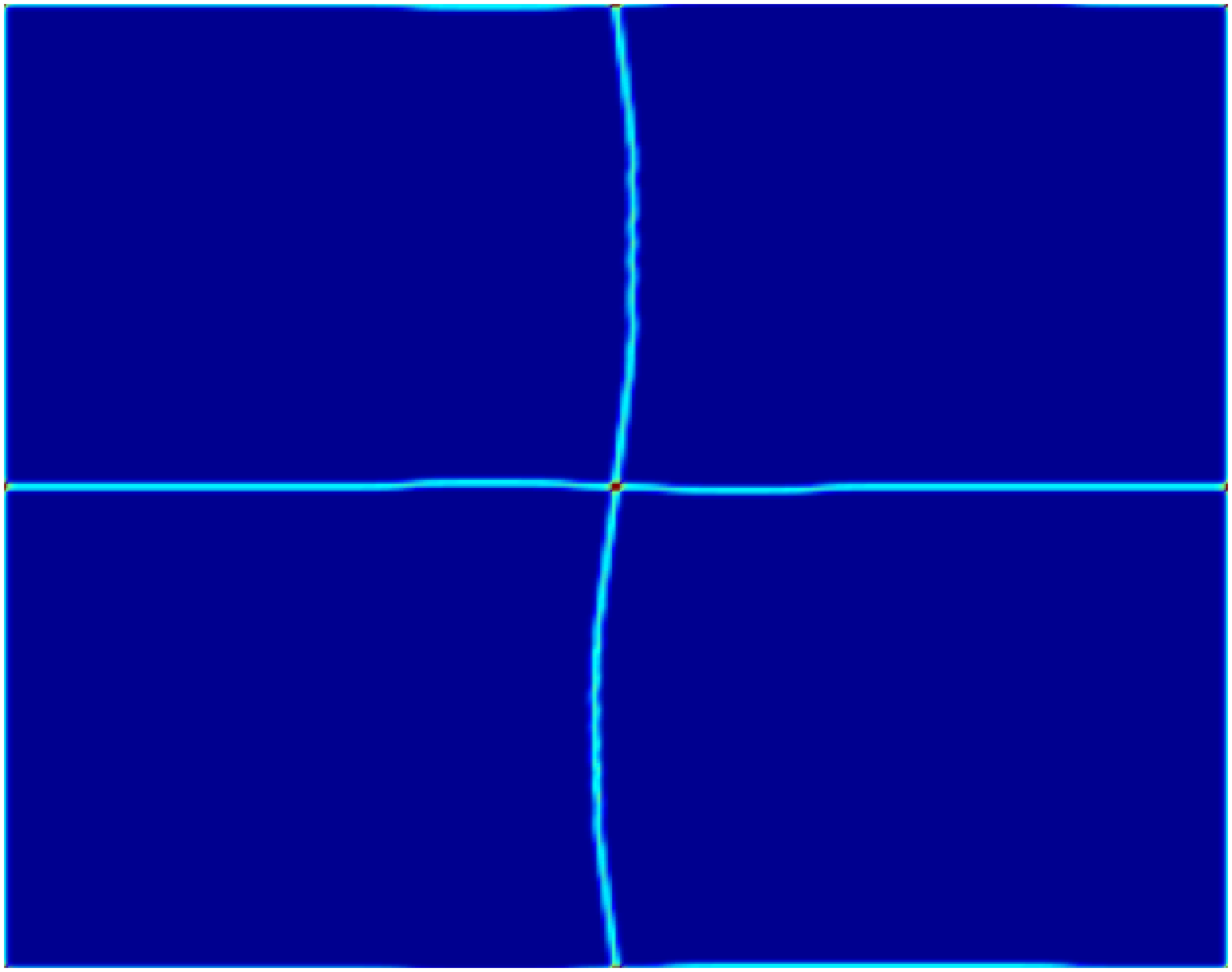}
\end{center}
\caption[Two Y-type junctions decaying into a stable X-type junction.]{Example of two Y-type junctions in a Kubotani-type model which decay into a stable X-type junction; see the text for a detailed discussion.}
\label{kubo3to4}
\end{figure}
\begin{figure}
\begin{center}
\includegraphics*[width=6cm]{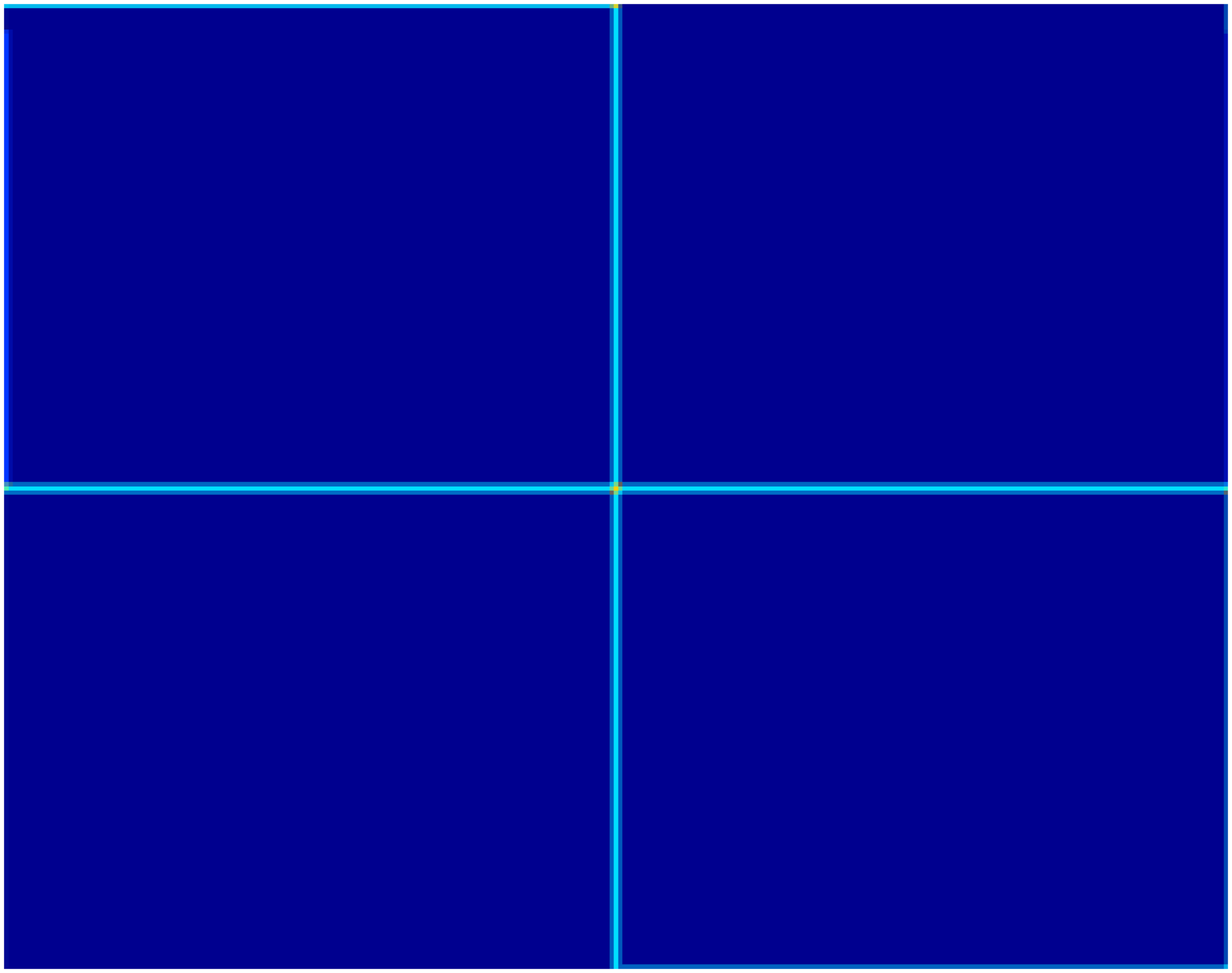}
\includegraphics*[width=6cm]{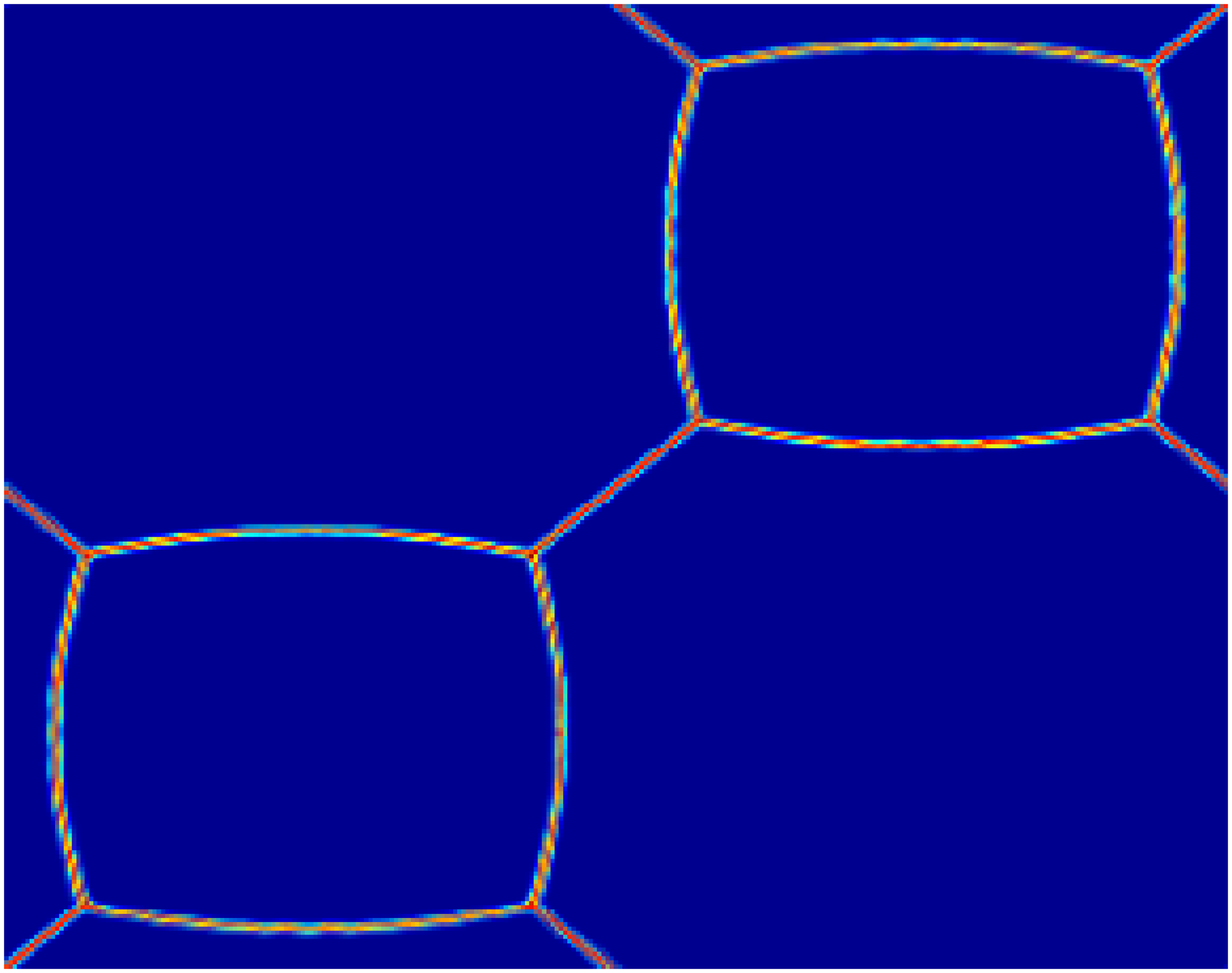}
\end{center}
\caption[A X-type junction  decaying into two Y-type junctions.]{Example of a X-type junction in a Kubotani-type model which decays into two Y-type junctions; see the text for a detailed discussion.}
\label{kubo4to3}
\end{figure}

For further confirmation, we construct by hand\footnote{Construction of a network by hand means selection of
special initial conditions in each point into the grid. Each point is set to be in a vacuum.} a simple square network containing four distinct vacua. First, Fig. \ref{kubo3to4} shows the evolution of $\xi < 0$ Kubotani model starting with four vacua lying on a plane and forming two Y-type junctions: these are destroyed and a new X-type junction is formed.
Although this simulation was performed for the three-field Kubotani model, it is valid in the general $N$-field case. Regardless of the $N$ chosen, if the vacua located in one of the faces of the $N$-hypercube are displayed initially as in Fig. \ref{kubo3to4}, the Y-type junctions will decay into stable X-type ones.
In contrast, in Fig. \ref{kubo4to3} we have a set of four vacua that are not on a plane in field space: in this case any X-type junction is unstable, and decays into a pair of Y-type ones.
Again, this conclusion is valid in the general case with arbitrary $N$.

\section{\label{intr122}Wall Lattice Properties}
In this section we shall look at geometrical properties of 
polyhedra in order to investigate if equilibrium 
flat domain wall configurations can be the natural result of 
domain wall network evolution in two-dimensions. 
We also aim to understand the role played by the model topology and the wall tensions on the evolution of the lattice. 

In the following discussion, we will assume that the energy
associated with the junctions is negligible
which in practice means that they are free to move. This is a reasonable assumption, at least for the purposes of the present discussion. 
If it were not the case, one would have to take into account the 
contribution of the junctions when calculating the equation of state associated with the
domain wall network, and consequently $w = p/\rho$ would necessarily be
greater than $-2/3$ even for a fully static configuration. Such networks would hardly be compatible with observational bounds.

\subsection{Geometrical Considerations}
One polyhedron is a solid figure with four or more faces, all of which are polygons. In other words, a polyhedron is simply
a three-dimension solid which consists of a collection of polygons, usually joined at their edges. 
Its number of vertexes ($\mathcal{V}$),
faces ($F$), edges ($\mathcal{E}$) and genus\footnote{A topological invariant property of a surface defined as the largest number of non-intersecting simple closed curves that can be drawn on the surface without separating it.} $g$ surfaces are related by
\be
\mathcal{V}-\mathcal{E}+F=2-2g,
\label{poincare}
\ee
that is known as Poincar\'e Formula.
We consider 
simulations in a square box with
periodic boundary conditions so that $g=1$. 
However the choice of boundary conditions will not affect our results. 

Let us start 
by assuming that the number of edges of each
polygon, $x$, and the number of edges, $d$, meeting at a vertex 
are fixed. 
Let us denote the number of polyhedron faces by $F=N_x$.
The number of polyhedron vertexes is $\mathcal{V}=N_x x/d$ since each polygon has $N_x$
vertexes but each one of them is shared with $d-1$ other polygons.
Also the number of polyhedron edges is equal to $\mathcal{E}=N_x x/2$ since each
polygon has $N_x$ edges but each one of them is shared with another polygon.
As a result, in this case Eq. (\ref{poincare}) becomes
\begin{equation}
N_x\left(1+\frac{x}{d}-\frac{x}{2}\right)=0\ .
\label{poincare2}
\end{equation}
This equation has the following solutions $(x=6,d=3)$, $(x=4,d=4)$,
$(x=3,d=6)$. These are the well known hexagonal
type lattices with odd Y-type junctions, square lattices
with even X-type junctions and triangular lattices with even 
`$*$'-type junctions in $2$ dimensions.

However, in general we do not expect
that all the polygons have the same
number of edges. 
Therefore we move to the more 
interesting case where any $d$ is fixed and $x$ is kept free. 
In this case it is straightforward to show that
\be
\langle x \rangle \equiv
\frac{\sum_{x=1}^\infty x N_x}{\sum_{x=1}^\infty N_x} =
\frac{2d}{d-2}\,
\label{poincare3}
\ee
with solutions 
$\langle x \rangle = 6$
if $d=3$, $\langle x \rangle = 4$ if $d=4$, $\langle x \rangle = 3$ if $d=6$
and $\langle x \rangle \to 2$ if $d \to \infty$, where $\langle\,\rangle$ represent the average value. If $d$ is not fixed, Eq. (\ref{poincare3})
can be written as
\be
(\langle x \rangle - 2)\,(\langle d \rangle - 2) = 4.
\ee
Note that increasing the average dimensionality of the junctions
leads to a smaller value of $\langle x \rangle$.

\begin{figure}
\begin{center}
\includegraphics*[width=9cm]{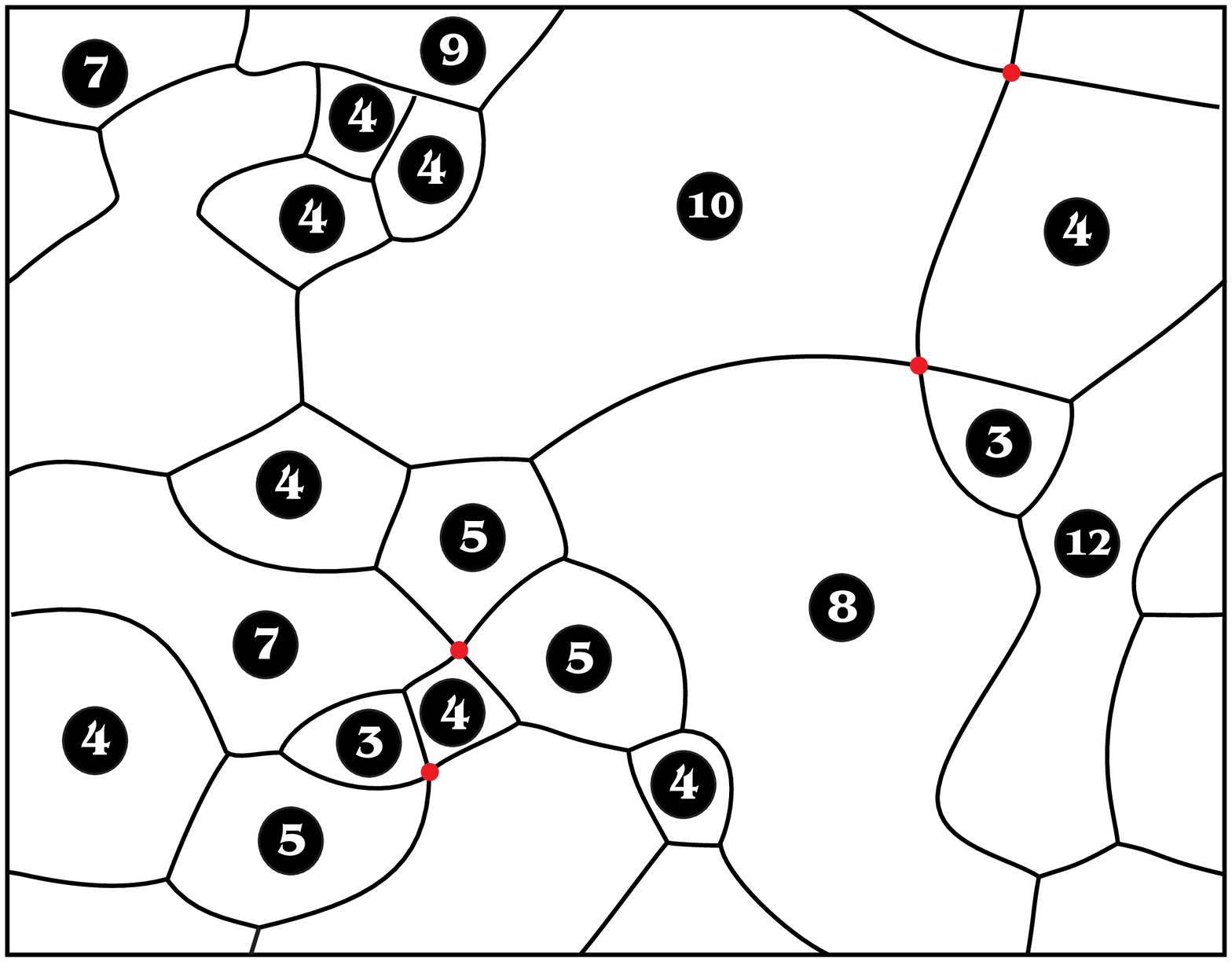}
\end{center}
\caption[Illustration of a random domain distribution in a planar network.]{Illustration of a random domain distribution in a planar network taken from the fourth snapshot of Fig. \ref{kubominus}. Each domain has $x$ edges and each junction is formed by $d$ walls. The average values are $\langle d \rangle = 53/17 $ and $\langle x \rangle = 106/19$.
}
\label{euler}
\end{figure}

We notice that 
the domains in a realistic domain wall network will not in general have straight edges and consequently two edge
domains are possible. However, these domains will be unstable and collapse
due to the domain wall curvature independently of the number of elements
meeting at each junction.

\begin{figure}
\begin{center}
\includegraphics*[width=6cm]{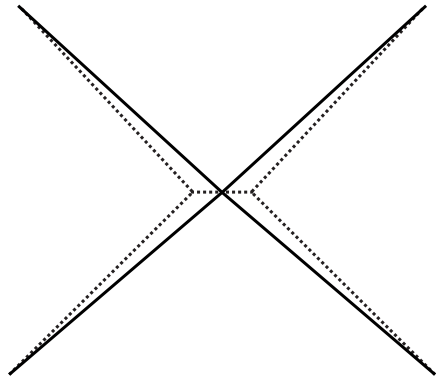}
\end{center}
\caption[An illustration of the decay of an unstable X-type junction.]{An illustration of the decay of an unstable X-type junction 
into stable Y-type junctions. The decay is energetically favorable 
since it leads to a reduction of the total length of the walls (if all have 
the same tension).}
\label{xtoy}
\end{figure}
\begin{figure}[ht]
\begin{center}	
\includegraphics*[width=9cm]{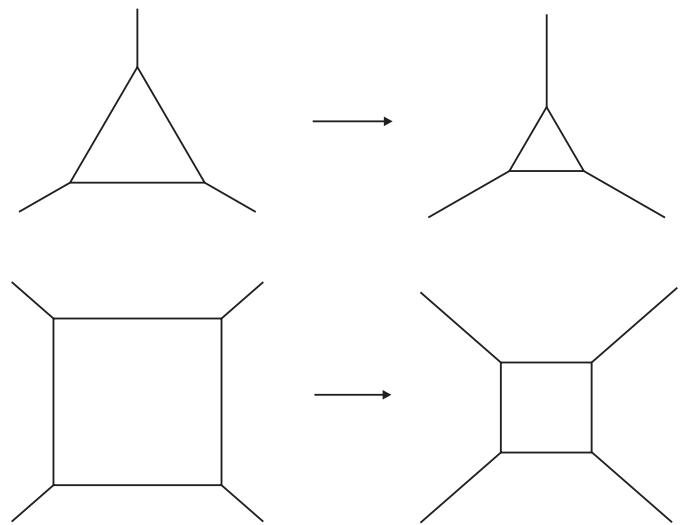}
\end{center}
\caption[An illustration of collapses of 
domains with Y-type junctions.]{An illustration of the collapse of three (top) and four (down) 
edge domains with Y-type junctions. The collapse is energetically favorable 
since it leads to a reduction of the total length of the walls.}
\label{colapse}
\end{figure}
\begin{figure}[ht]
\begin{center}
\includegraphics*[width=9cm]{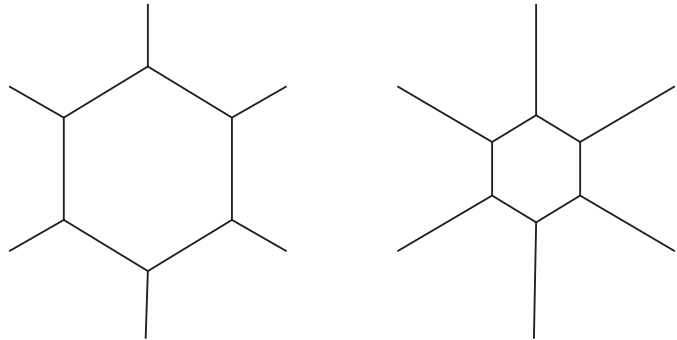}
\end{center}
\caption[An illustration of six-edged polygons with Y-type junctions.]{An illustration of two different six-edged polygons with Y-type junctions of walls with same tension. Both configurations have the same energy.}
\label{noncolapse}
\end{figure}
In a two-dimensional domain wall network generated dynamically by a cosmological phase transition, the number of edges of the domains can assume several values, with a particular distribution. The dimensionality of the junctions in such a network, that is, the number of walls meeting at each junction, depends on the structure of the chosen potential.
We have verified this by using the numerical results in Figs. \ref{kuboplus} (for $\xi >0$) and \ref{kubominus} (for $\xi <0$). While in the former case only X-junctions are present, in the latter one there are both stable Y-type and X-type junctions. 
As a simple illustration, we can take the fourth snapshot in Fig. \ref{kubominus}, which is schematically shown in Fig. \ref{euler}. The average number of faces $x$ of each domain is related to the dimensionality $d$ of the junctions by 
\be
\left( \langle x \rangle -2 \right)\left( \langle d \rangle- 2 \right)\,=\,4.
\label{eulerequation}
\ee
In Fig. \ref{euler}, these values are $\langle d \rangle = 53/17\sim3.11 $ and $\langle x \rangle = 106/19\sim5.58$.
We have also verified that these numbers are approximately constant during the evolution of the network. 

Finally, we emphasize
that no equilibrium
configurations
exist with $d > 6$. This means that if we started with a domain
wall network with $d > 6$, unstable two edge domains would
necessarily be present.
\subsection{Energy Considerations}
\subsubsection{Wall Tensions}
Let us suppose that walls with different tensions are present in the network.
This assumption adds
a further source of instability. In fact, the walls with higher tension
will tend to collapse thus increasing the dimensionality of the junctions which, in general,  
leads to the production of further unstable two edge 
domains. In the previous section we have illustrated this process in Fig. \ref{threefour},  which 
shows the collapse of two Y-type junctions into one X-type junction  
occurring when the thick wall has a tension larger than twice 
that of the lower tension ones. 
We also simulate the BBL Model (see Fig. \ref{kubo3to4} and Fig. \ref{kubo4to3}) for both the cases with the X-type and Y-type being energetically favorable.

In addition, if all the walls have equal tensions, then higher order junctions are unstable and decay to stable Y-type junctions. Fig. \ref{xtoy} illustrated this point, since the total length of the walls (proportional to the total energy of the wall) is smaller for the Y-type junction.
\subsubsection{Stability of Domains}
Let us assume that $d$ is fixed to be equal to $3$. 
This will necessarily be the case if all possible domain walls have the same tension (see Fig. \ref{xtoy}).
In this case it is possible to
show using 
a local stability analysis that regular polygons with $x<6$
are unstable.
This is illustrated in Fig. \ref{colapse}.
Note that in both cases, $x=3$ (top) 
and $x=4$ (bottom), the total length of walls decreases. Consequently, the 
polygons will tend to collapse thus minimizing their potential energy. 
On the 
other hand in Fig. \ref{noncolapse}, for $x=6$, the length remains constant 
and both configurations have the same energy. 
Hence, given that $d=3$ implies $\langle x \rangle = 6$, the
only possible equilibrium configuration with only $Y$-type junctions 
is a hexagonal type lattice. Otherwise unstable two, three, four and five 
edge domains will be formed, as well as polygons with $x>6$.
\begin{figure}
\begin{center}
\includegraphics*[width=9cm]{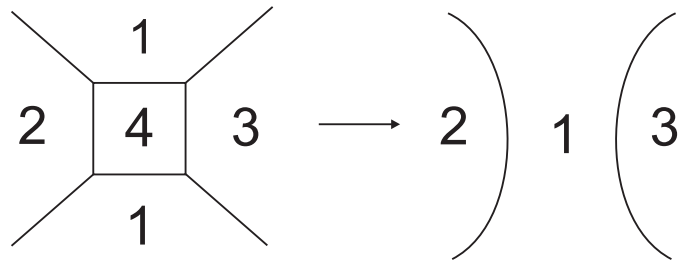}
\end{center}
\caption[First illustration of the collapse of a four-edged polygon.]{An illustration of the collapse of a four-edged polygon in the case where two of the surrounding domains are on the same vacuum state. The collapse leads to a reduction of the number of edges of contiguous domains.}
\label{separate}
\end{figure}
\begin{figure}
\begin{center}
\includegraphics*[width=9cm]{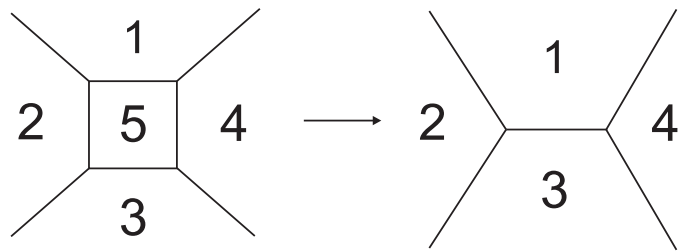}
\end{center}
\caption[Second illustration of the collapse of a four-edged polygon.]{An illustration of the collapse of a four-edged polygon in the case where all the surrounding domains are on different vacuum states and the domain walls all have the same tension. Again, the collapse leads to a reduction of the number of edges of contiguous domains.}
\label{nonseparate}
\end{figure}

\subsection{Topology Considerations}
Finally, we investigate
the collapse of a particular domain on the network.
Consider
a square with only Y-type junctions
with an initial configuration having four contiguous domains with $E_1$, $E_2$, $E_3$ and $E_4$ edges. 
Since the model has only four vacua, 
we can, for example, assume
that the first two will join; the resulting domain will
have $E_1+E_2-4$ edges, while the other two will have $E_3-2$ and $E_4-2$
edges after the collapse of the four edge domain. The production of three
hexagons as a result of the collapse of a four edge domain
would require $E_1+E_2=10$ and $E_3=E_4=8$. 
Hence, we do not expect that a domain wall network in two dimensions will naturally evolve towards a hexagonal lattice from realistic initial conditions.
We illustrate this in Fig. \ref{separate}.  
We see that
the collapse of unstable domains with two, three, four and
five edges will always result in a decrease in the number of edges of some
of the contiguous domains. Of course, if there are at least five vacua in the model then the evolution of the local structures of the lattice can be different. In fact as the number of vacua increases, the probability of the configuration shown in Fig. \ref{nonseparate} 
can be
much layer than the other one in Fig. \ref{separate}. 
\begin{figure}[ht]
\begin{center}
\includegraphics*[width=6cm]{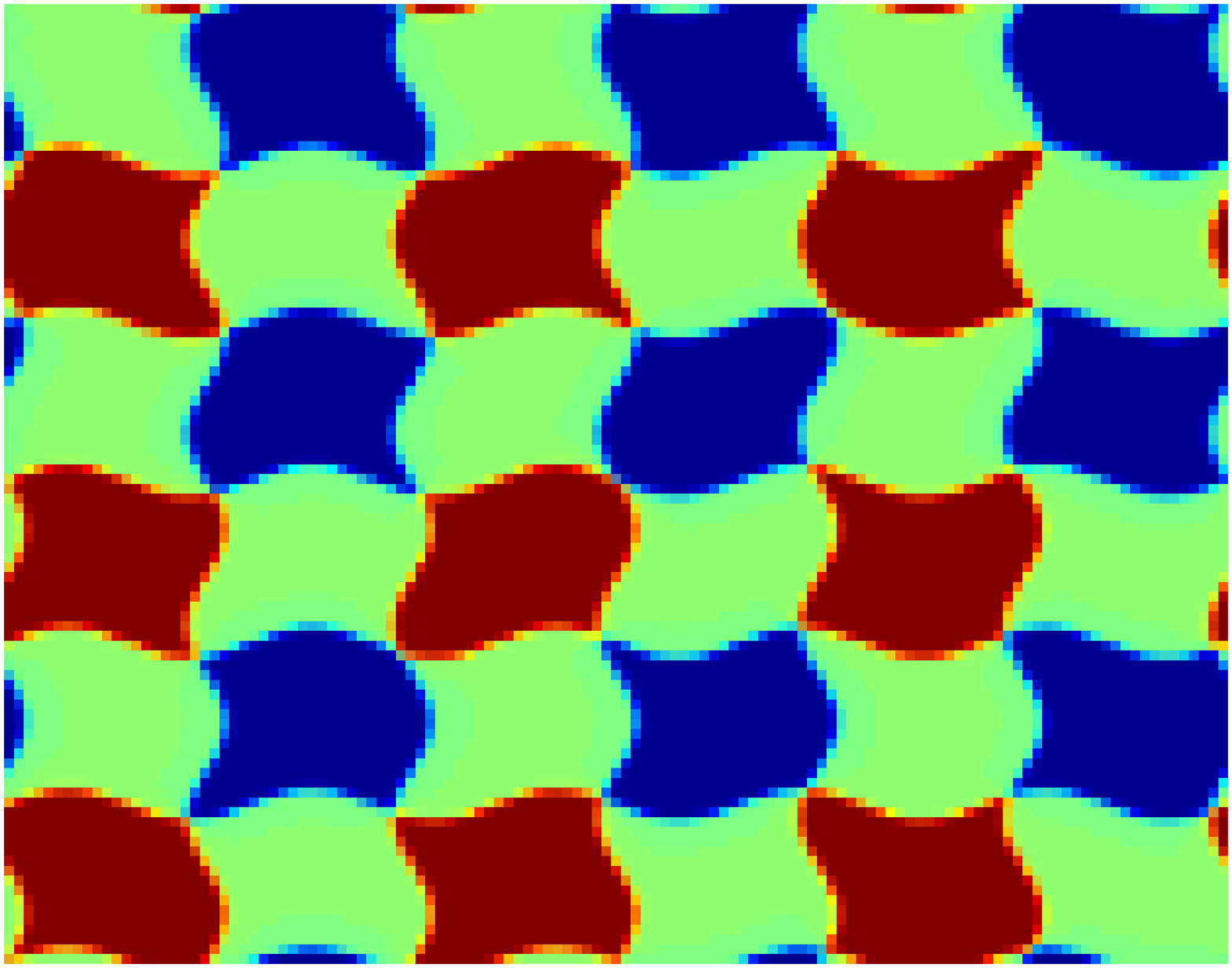}
\includegraphics*[width=6cm]{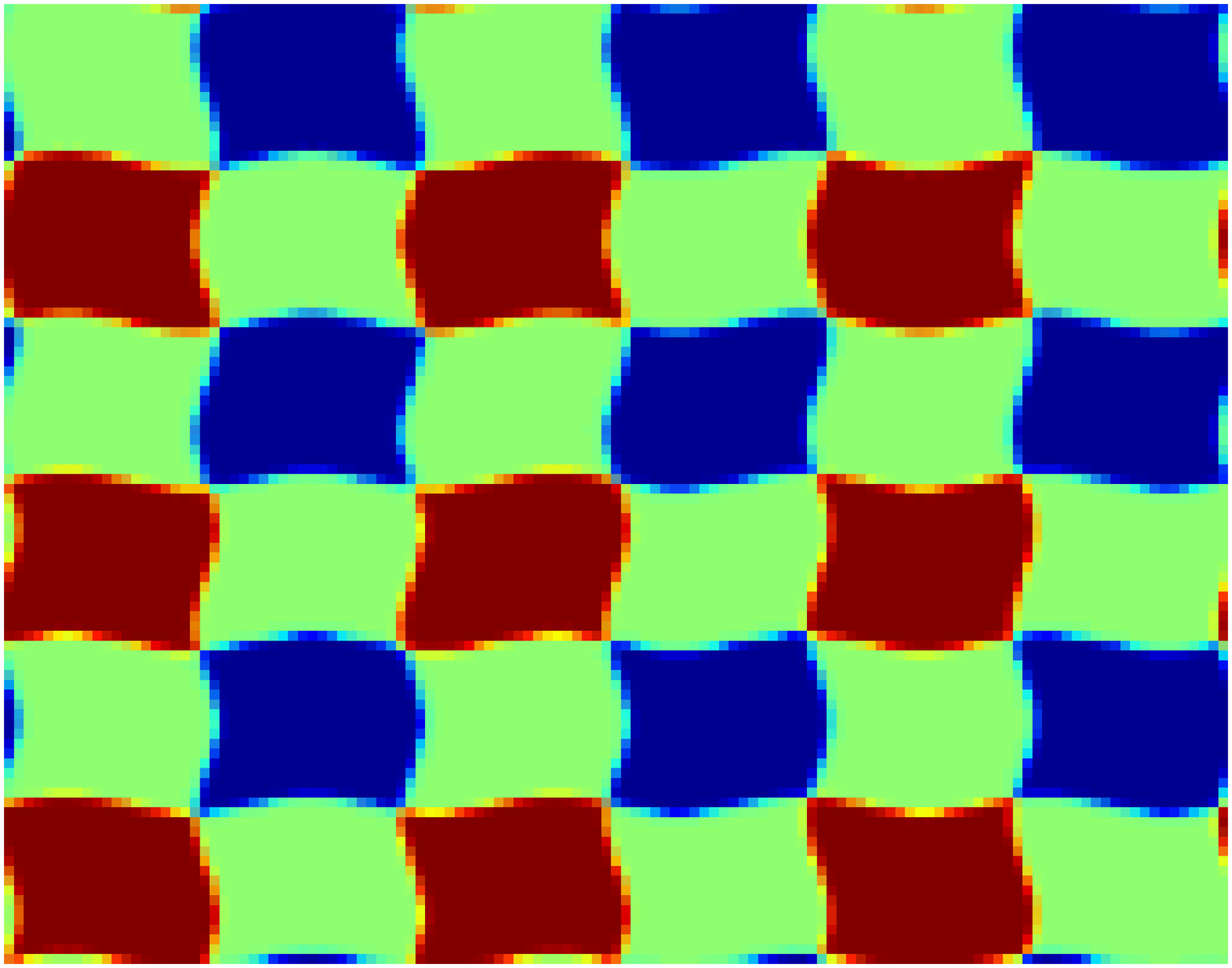}
\includegraphics*[width=6cm]{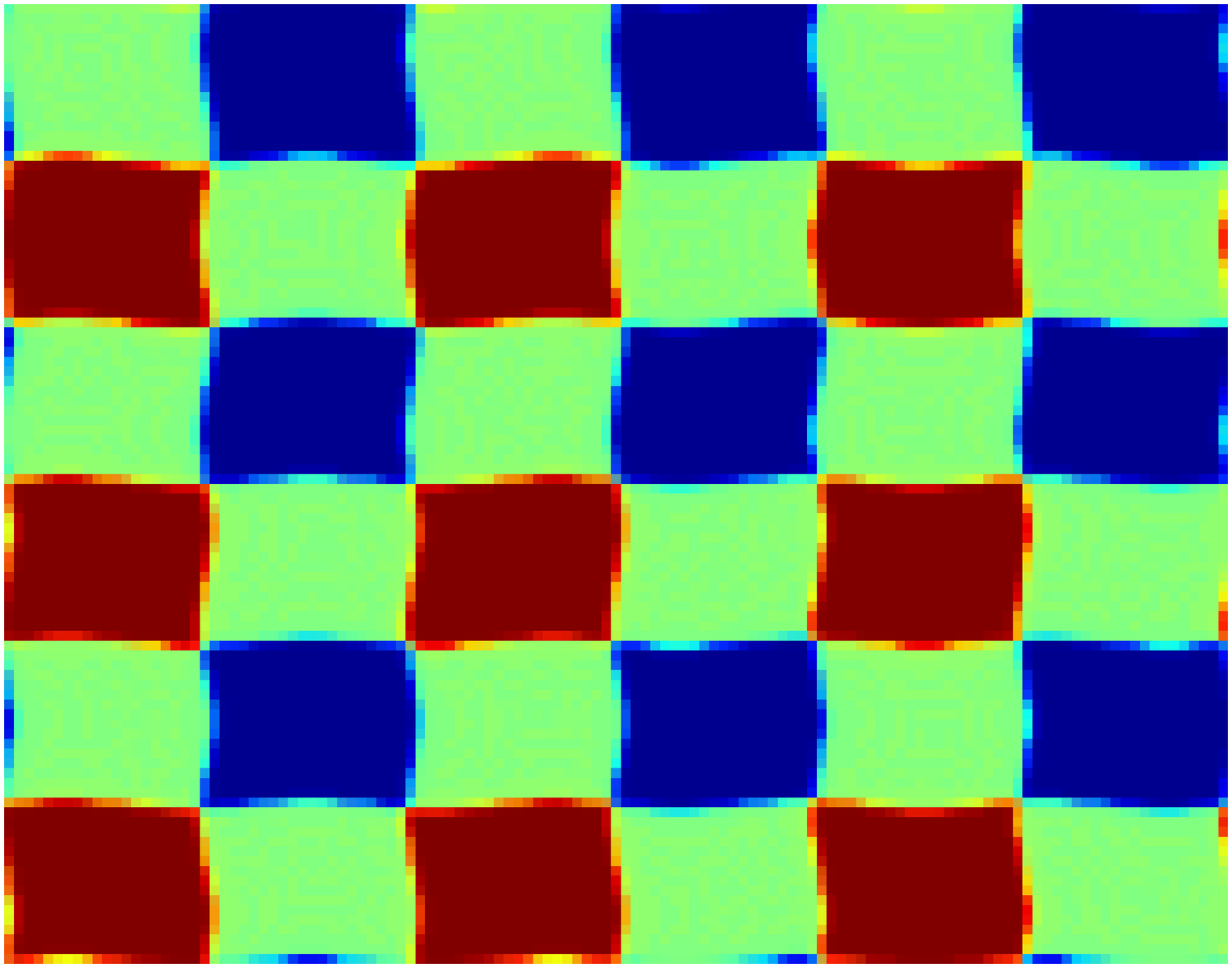}
\includegraphics*[width=6cm]{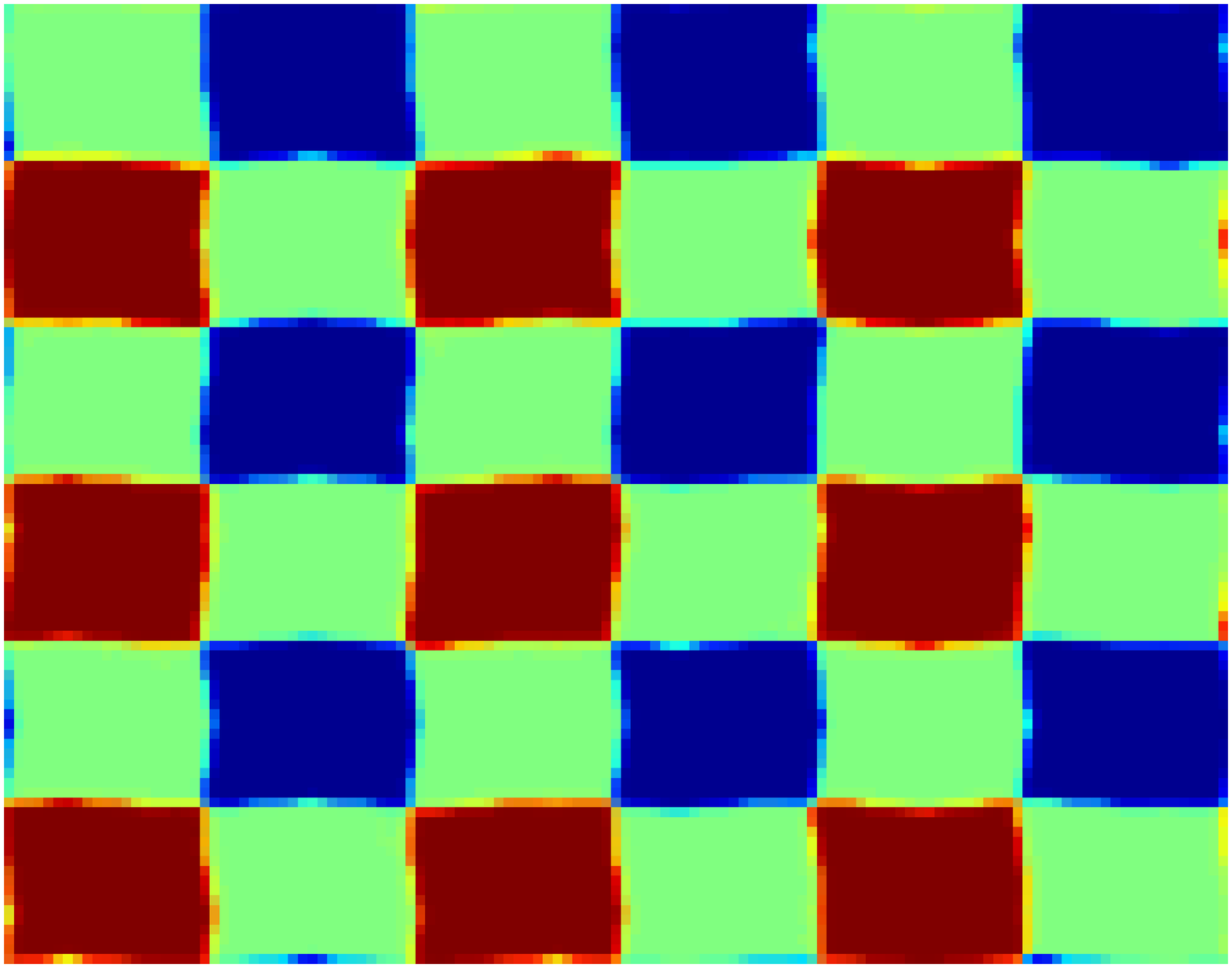}
\end{center}
\caption[Evolution of a perturbed square lattice.]{The evolution of a perturbed square lattice with even X-type junctions in a matter-dominated universe. The top left panel is the initial configuration. From left to right and top to bottom panels the horizon is approximately 1/256, 1/20, 1/10 and 1/5 of the box size respectively. The lattice stabilizes in the right bottom panel configuration.}
\label{square}
\end{figure}
\begin{figure}[ht]
\begin{center}
\includegraphics*[width=6cm]{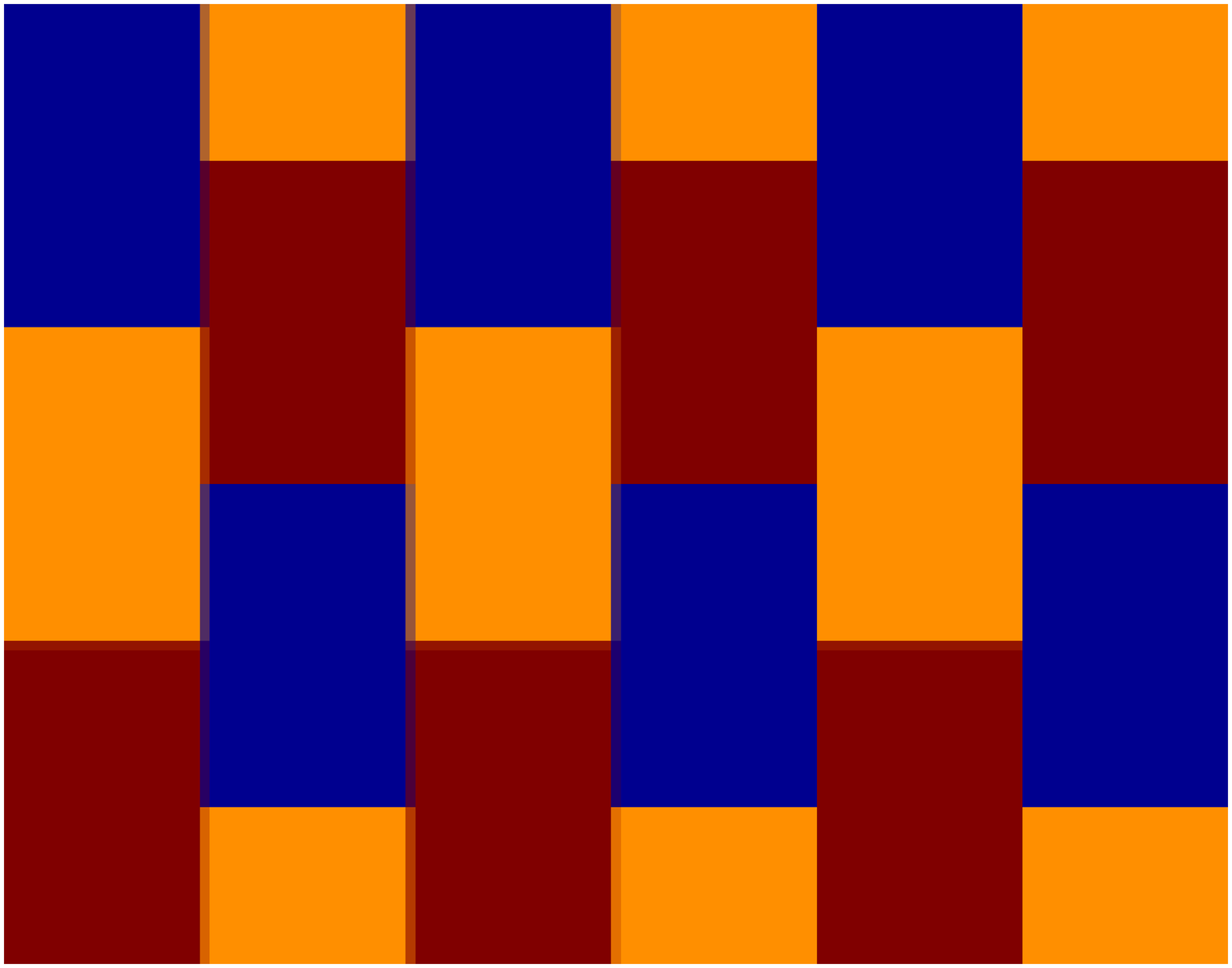}
\includegraphics*[width=6cm]{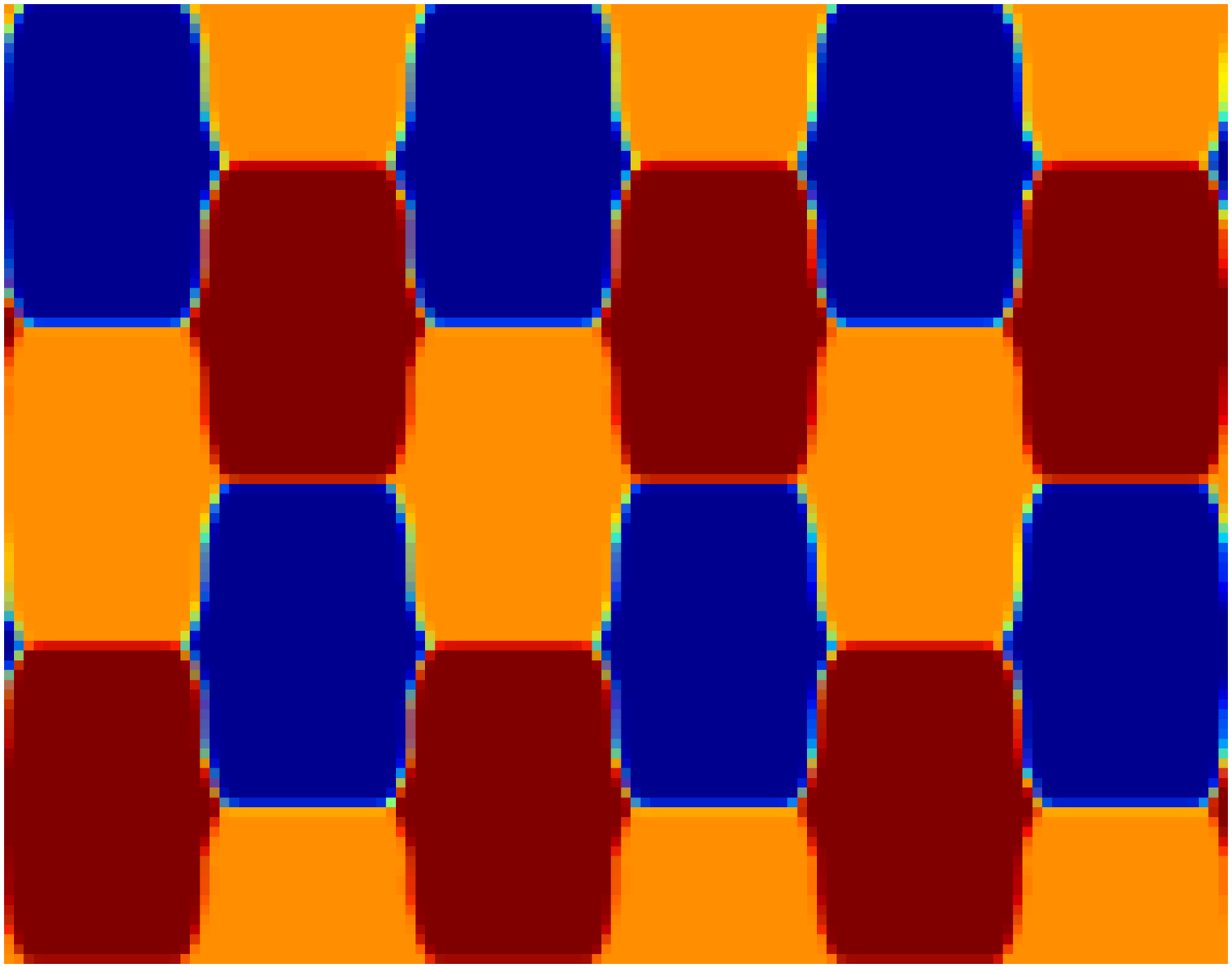}
\includegraphics*[width=6cm]{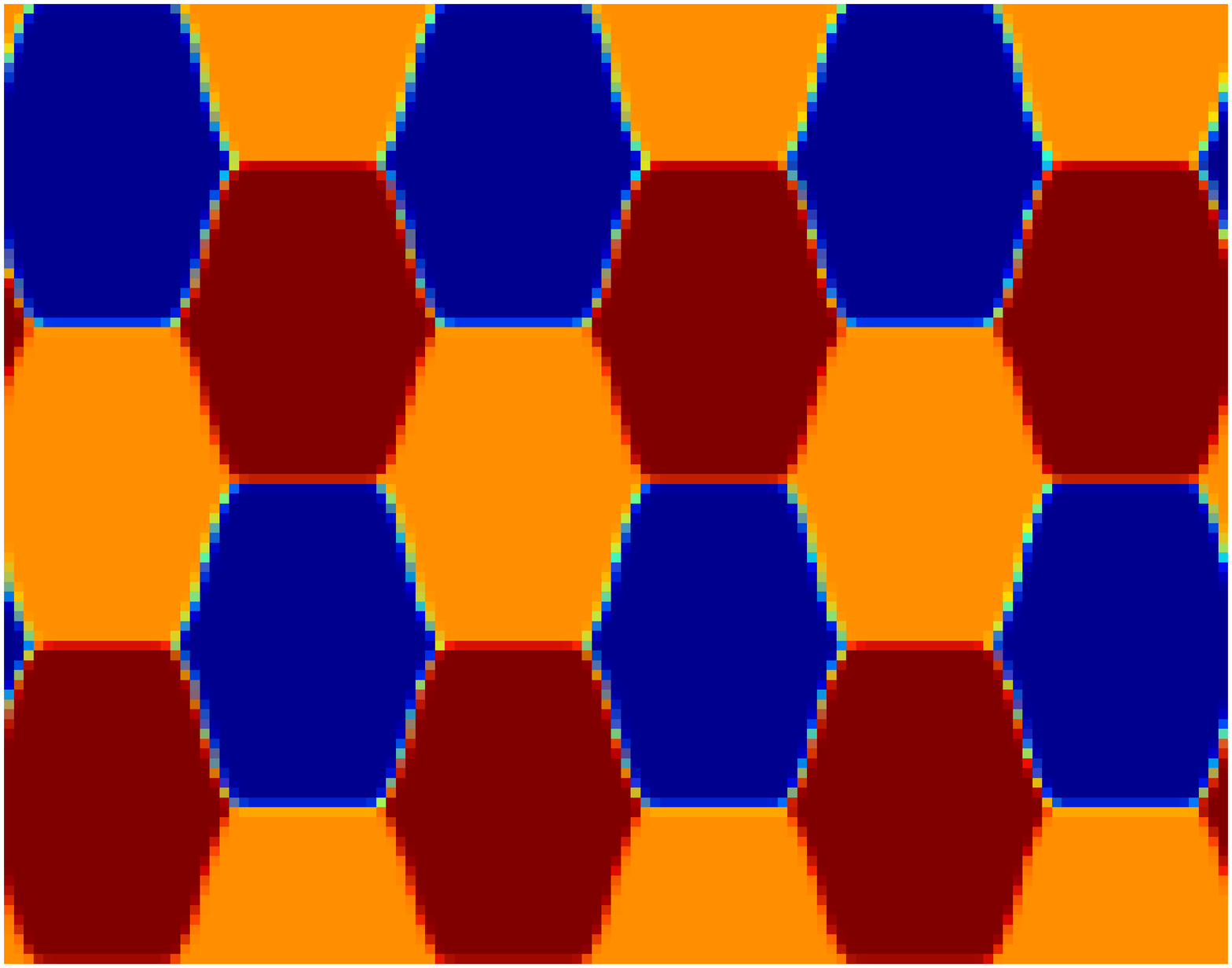}
\includegraphics*[width=6cm]{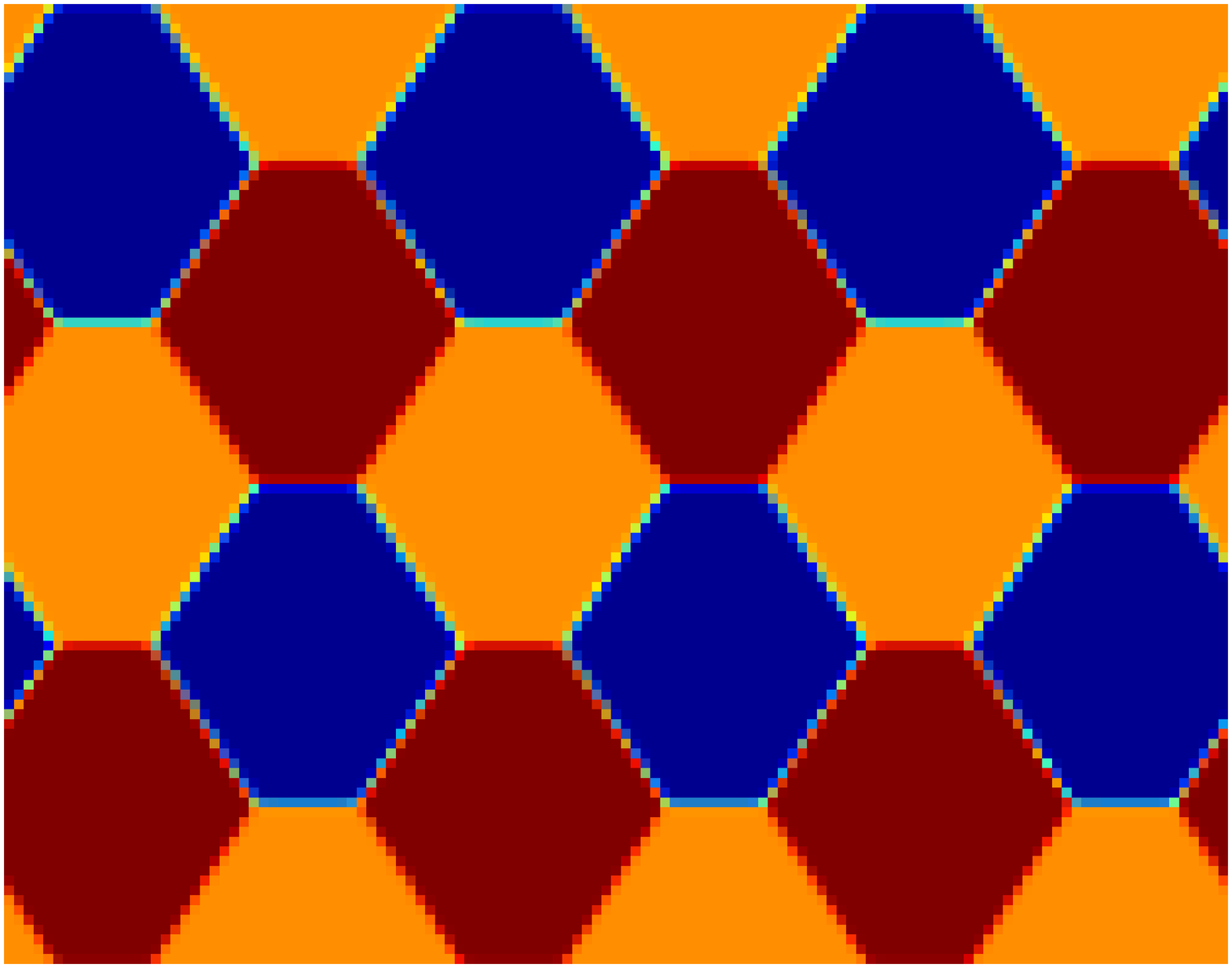}
\end{center}
\caption[Evolution of a perturbed hexagonal lattice.]{The evolution of a perturbed hexagonal lattice with odd Y-type junctions in the matter-dominated epoch. The top left panel is the initial configuration. From left to right and top to bottom panels the horizon is approximately 1/256, 1/10, 1/5 and 2/5 of the box size respectively. The lattice stabilizes in the right bottom panel configuration.}
\label{hex}
\end{figure}
\subsection{Stability of Networks Constructed By Hand}
The stability of simple examples of triangular, hexagonal and square
domain wall lattices in 2D was studied in \cite{BATTYE,CARTER} by looking at their macroscopic
elastic properties. The authors assess the stability of various lattice
configurations and find that some of them are stable. Here, we confirm and generalize their results numerically showing that there are
certain lattice configurations which are stable against large deformations.
Fig. \ref{square} shows a field theory numerical simulation in the matter epoch for a square lattice with even-type X junctions. We see that the network evolves to the minimum energy configuration and stabilizes there. 
The case of the hexagonal lattice with odd Y-type junctions in the matter epoch is shown in Fig. \ref{hex}. As was mentioned this is the only possible equilibrium configuration with only Y-type junctions. 
Although hexagonal lattices with Y-type junctions allow for locally 
confined energy conserving deformations,
Hubble damping may prevent the collapse of such configurations (even if they 
are perturbed) and consequently one should not
completely discard them on this basis. 

Hence, although we agree with the claim that in Minkowski space such a lattice is unstable \cite{BATTYE} we find that in an expanding universe it may be stable since
domain wall velocities are damped by the expansion of the universe. 
However, we do not
expect that hexagonal lattices will be an attractor for the evolution
of domain wall network simulations. 
Also, we recall that we aim to provide an understanding of domain wall networks generated
by spontaneous symmetry breaking, and not only the stability of such special configurations.

These results clearly indicate that the crucial question is not the existence of stable lattice
configurations but whether these can be the natural result of domain wall evolution with
realistic initial conditions.
\section{\label{intr1222}The Ideal Model for Frustrated Networks}

In the previous sections we have studied the dynamics of domain wall networks with junctions. Here we consider an `ideal' class of models 
from point of view of frustration \cite{IDEAL1}. This model, presented in Ref. \cite{IDEAL2}, was constructed by considering the energetic, geometrical and topological considerations of planar domain wall networks discussed in the previous section.

The above energy, geometrical and topological considerations led us to propose a class of models such that satisfy the following conditions \cite{IDEAL1,IDEAL2}:
\begin{itemize}
\item
\textit{There are a very large number of minima}. 
In this case the collapse of a single domain will only very rarely lead to the fusion of two of the surrounding domains. Note that this fusion would lead to a further reduction of the total number of edges of the contiguous domains which as we saw is undesirable since it would increase the probability that some of the contiguous domains would themselves become unstable to collapse. 
\item
\textit{All possible domain walls have equal tensions}.
If that were not the case we would be adding a further source of instability since the walls with higher tension would tend to collapse, thereby increasing the dimensionality of the junctions which, in turn, would lead to a decrease in the average number of edges of individual domains and to the production of further unstable two-edge domains.
\end{itemize}

\subsection{Realizations of the Ideal Model}
Let us consider the model given by
\be
V=\frac{\lambda}{4}\sum_{j=1}^{N+1} r_j^2 \left(r_j^2 - r_0^2\right)^2 \label{ideal}
\ee
with
\be
r_j^2=\sum_{i=1}^N (\varphi_i - p_{{i}_{j}})^2,
\ee
where $p_{{i}_{j}}$ are the $N+1$ coordinates of the vacua of the potential. 
There are $N$ real scalar fields, and the energetic cost for a specific transition between any two of them is the same. 
Geometrically speaking, the $p_{{i}_{j}}$ are located at the vertexes of a ($N+1$)-dimensional regular polygon.
The distance between the vacua is fixed to be equal to $r_0$ and for large $N$ we approach the ideal model
($N \to \infty$ is the ideal case, actually). From the field theory point of view, 
the potential (\ref{ideal})
is the sum of ($N+1$) $\varphi^6$-type potentials. Each of them has one minimum located at the center, and a continuum of minima at a distance $r_0$ from the center. Note that $N$ of these vacua are located exactly at the centers of the other potentials.

\subsubsection{$N=2$: the $Z_3$ Model}

\begin{figure}
\begin{center}
\includegraphics*[width=9cm]{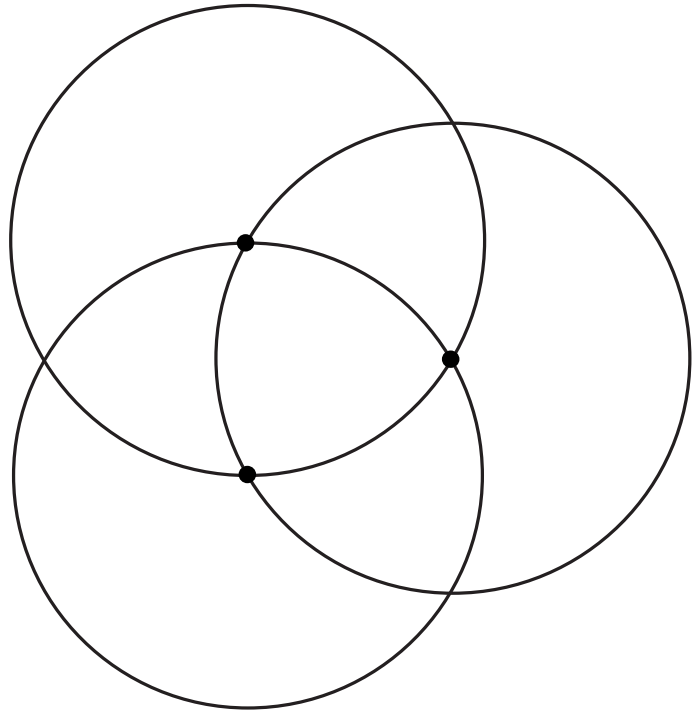}
\end{center}
\caption[Illustration of the vacua displacement for fields.]{The illustration of the vacua displacement for fields. Note that there are only three equidistant points where these minima coincide.}
\label{three11}
\end{figure}

\begin{figure}
\begin{center}
\includegraphics*[width=9cm]{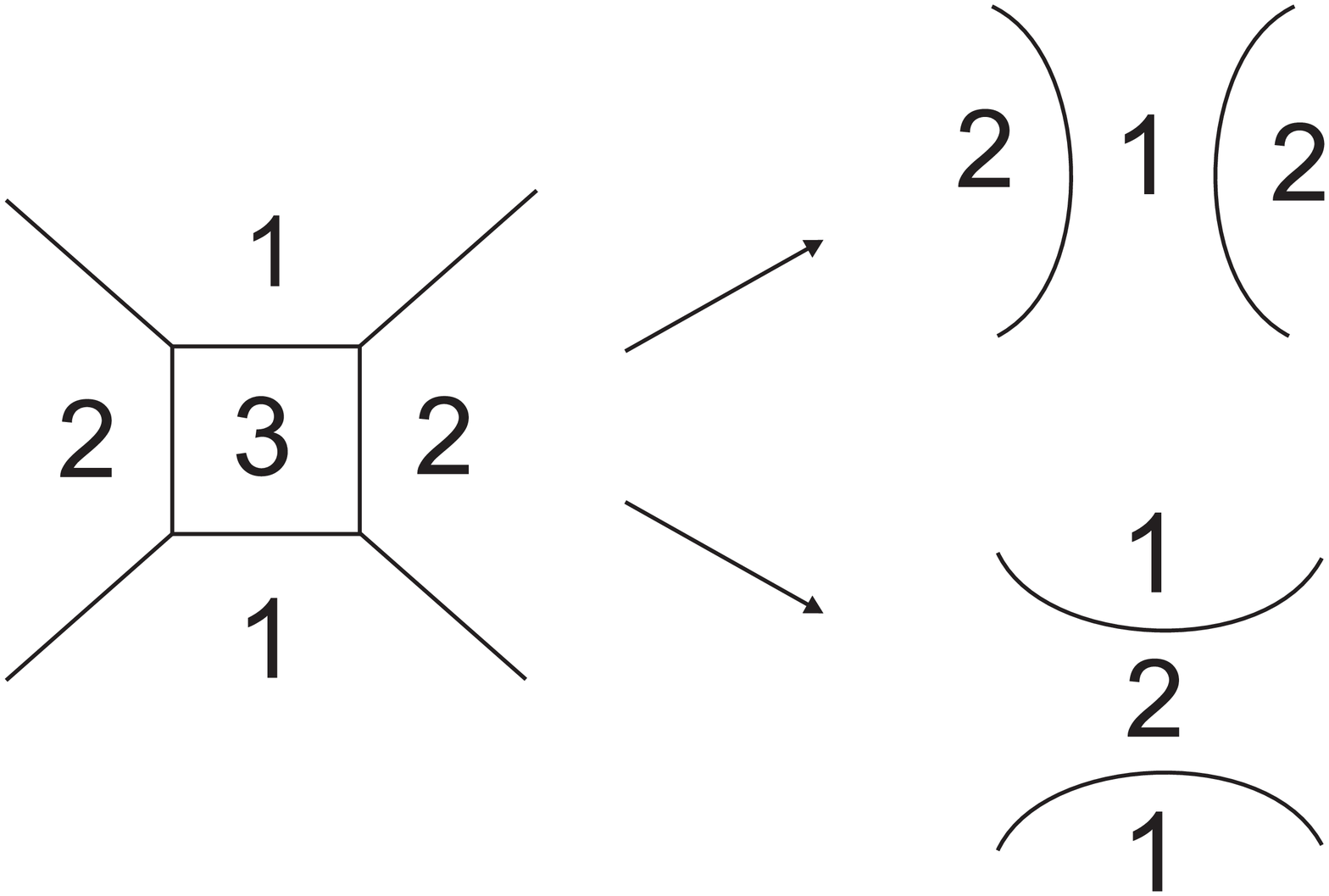}
\end{center}
\caption[Possible collapses of a domain with four edges: three vacua.]{
Possible collapses of a domain with four edges in the model with only three vacua.}
\label{colapso1}
\end{figure}

Let us consider the simplest case $N=2$ whose potential is written as
\be
V=\frac{\lambda}{4}\sum_{j=1}^{3} r_j^2 \left(r_j^2 - r_0^2\right)^2 \label{ideal2case}
\ee
with
\ben
r_1^2 &=& (\varphi_1 - 1)^2 + \varphi_2^2, \\
r_2^2 &=& \left(\varphi_1 + \frac12 \right)^2+ \left(\varphi_2 - \frac{\sqrt{3}}2 \right)^2, \\
r_3^2 &=& \left(\varphi_1 + \frac12 \right)^2+ \left(\varphi_2 + \frac{\sqrt{3}}2 \right)^2 .
\een
There are three equidistant vacua at the vertexes of an equilateral triangle with coordinates in the $(\varphi_1, \varphi_2)$ plane given by $(1,0)$, $(-1/2, \sqrt{3}/2)$ and $(-1/2, -\sqrt{3}/2)$.
(See Fig. \ref{three11}). This particular case is in fact analogous to the model given by Eq. (\ref{pot2}) for $N=3$, i.e., the $Z_3$ model. 
Fig. \ref{colapso1} shows possible collapses of a domain with four edges in the model with only three vacua.

\subsubsection{$N=3$: The Tetrahedral Model}
Let us now consider $N=3$.
The potential is given by
\be
V=\frac{\lambda}{4}\sum_{j=1}^{4} r_j^2 \left(r_j^2 - r_0^2\right)^2 \label{ideal3case}
\ee
with
\ben
r_1^2 &=& (\varphi_1 - 1)^2 + \varphi_2^2 + \varphi_3^2, \\
r_2^2 &=& \left(\varphi_1 + \frac12 \right)^2+ \left(\varphi_2 - \frac{\sqrt{3}}2 \right)^2 + \varphi_3^2,\\
r_3^2 &=& \left(\varphi_1 + \frac12 \right)^2+ \left(\varphi_2 + \frac{\sqrt{3}}2 \right)^2 + \varphi_3^2,\\
r_4^2 &=& \varphi_1^2+ \varphi_2^2 + (\varphi_3 - \sqrt{2})^2.
\een
\begin{figure}[ht]
\begin{center}
\includegraphics*[width=7.5cm]{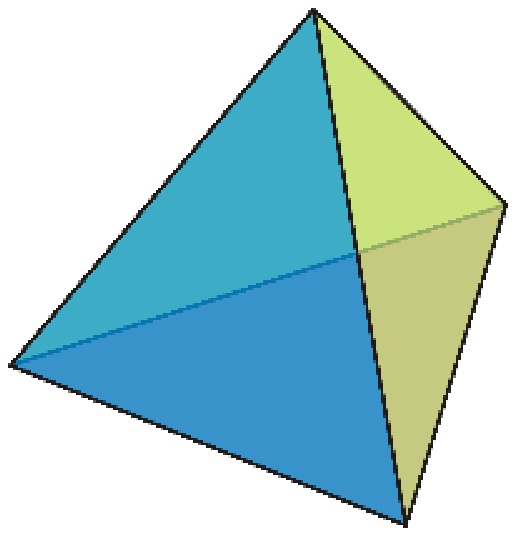}
\end{center}
\caption[Configuration of the vacua of the ideal model with $N=3$.]{Configuration of the vacua of the model of Eq. (\protect\ref{ideal}) with $N=3$.}
\label{tetra}
\end{figure}
There are four vacua located at the vertexes of a tetrahedron as illustrated in Fig. \ref{tetra}. Their coordinates in the field space ($\varphi_1,\varphi_2,\varphi_3$) are ($1,0,0$), ($-1/2,\sqrt{3}/2,0$), ($-1/2,-\sqrt{3}/2,0$) and ($0,0,\sqrt{2}$).
The presence of a fourth vacuum allows for the configuration illustrated in Fig. \ref{separate}. However, the probability that such situation occurs is small and the potential is not enough to evade the collapses of domains cutting the network. Another aspect to be noted is that although the two-field BBL model also has four minima, the two models are completely different. 
While here only stable Y-type junctions are formed, in the BBL case stable X-type junctions may be formed depending on the parameter $\epsilon$. Indeed, if Y-type junctions are formed in the BBL model, different kinds of walls necessarily will join the junction.
On the contrary, here only Y-type junctions are formed and all of them have the same tension.

\subsubsection{$N=4$: The Pentahedral Model}

Let us assume $N=4$. In this case the five vacua are displayed at the vertexes of a pentahedron with coordinates in the field space ($\varphi_1,\varphi_2,\varphi_3,\varphi_4$) given by ($1,0,0,0$), ($-1/2,\sqrt{3}/2,0,0$), ($-1/2,-\sqrt{3}/2,0,0$), ($0,0,\sqrt{2},0$) and ($0,0,\sqrt{2}/4,\sqrt{30}/4$). Here the potential is written as
\be
V=\frac{\lambda}{4}\sum_{j=1}^{5} r_j^2 \left(r_j^2 - r_0^2\right)^2 \label{ideal4case}
\ee
with
\ben
r_1^2 &=& (\varphi_1 - 1)^2 + \varphi_2^2 + \varphi_3^2 +\varphi_4^2,\\
r_2^2 &=& \left(\varphi_1 + \frac12 \right)^2+ \left(\varphi_2 - \frac{\sqrt{3}}2 \right)^2 + \varphi_3^2 + \varphi_4^2,\\
r_3^2 &=& \left(\varphi_1 + \frac12 \right)^2+ \left(\varphi_2 + \frac{\sqrt{3}}2 \right)^2 + \varphi_3^2 + \varphi_4^2,\\
r_4^2 &=& \varphi_1^2+ \varphi_2^2 + (\varphi_3 - \sqrt{2})^2 + \varphi_4^2,\\
r_5^2 &=& \varphi_1^2+ \varphi_2^2 + \left(\varphi_3 - \frac{\sqrt{2}}{4} \right)^2 + \left(\varphi_4 - \frac{\sqrt{30}}{4} \right)^2.\\
\een
Now, we have 
the further possibility of the configuration described in Fig. \ref{nonseparate}. Meanwhile the chance of Fig. \ref{separate} occurring grows. 

In Ref. \cite{CARTER}, the author introduced a very similar model with five evenly spaced minima (all the walls having same tension).
Despite the same geometrical structure in the field space,  
the Carter potential is constructed by summing $\varphi^4$-type potentials, while the ideal model introduced here is of $\varphi^6$-type. 
Of course, both models allow for the configuration in Fig. \ref{nonseparate}. 
Carter claims that this model will form stable X-type junctions, 
but our results clearly show that
only Y-type junctions survive.

\subsection{Two-dimensional simulations}
Firstly, two-dimensional simulations were carried out to be contrasted with the BBL and Kubotani model results shown in the previous section. 
A large number of realizations of the ideal model in Eq. (\ref{ideal}) for several values of $N$ up to $20$ were run in the matter dominated epoch.
In all these simulations we have confirmed that only stable Y-type junctions are formed. In addition, 
if one imposes X-type junctions in the network by hand they are destroyed quickly. 
However, the stable \textit{honeycomb} network is not formed by starting with
random
initial conditions.
As an illustration of
the simulation results, we have chosen $N=4$ and $N=20$. 

\begin{figure}[ht]
\begin{center}
\includegraphics*[width=6cm]{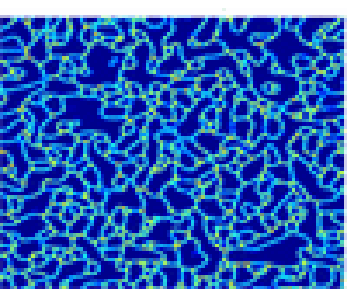}
\includegraphics*[width=6cm]{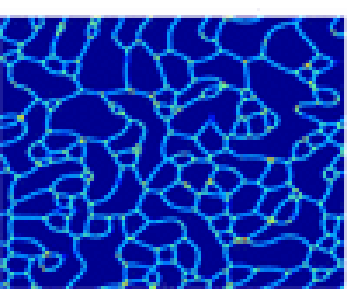}
\includegraphics*[width=6cm]{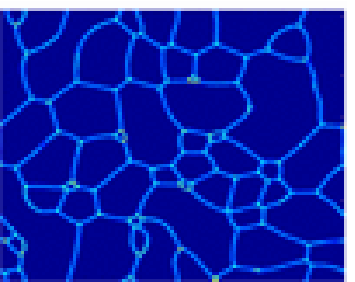}
\includegraphics*[width=6cm]{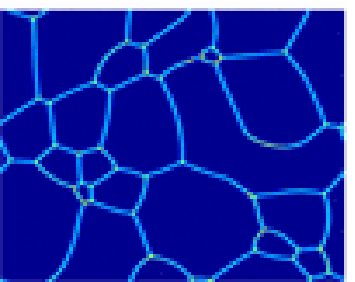}
\end{center}
\caption[The ideal class of model: $N=4$.]{
The matter-era evolution of the domain wall network for the ideal model with $N=4$. Note that this model is effectively analogous to that of \protect\cite{CARTER}, and that only stable Y-type junctions survive. From left to right and top to bottom, the horizon is approximately 1/16, 1/8, 1/4 and 1/2 of the box size respectively.
}
\label{ideal4}
\end{figure}
\begin{figure}[ht]
\begin{center}
\includegraphics*[width=6cm]{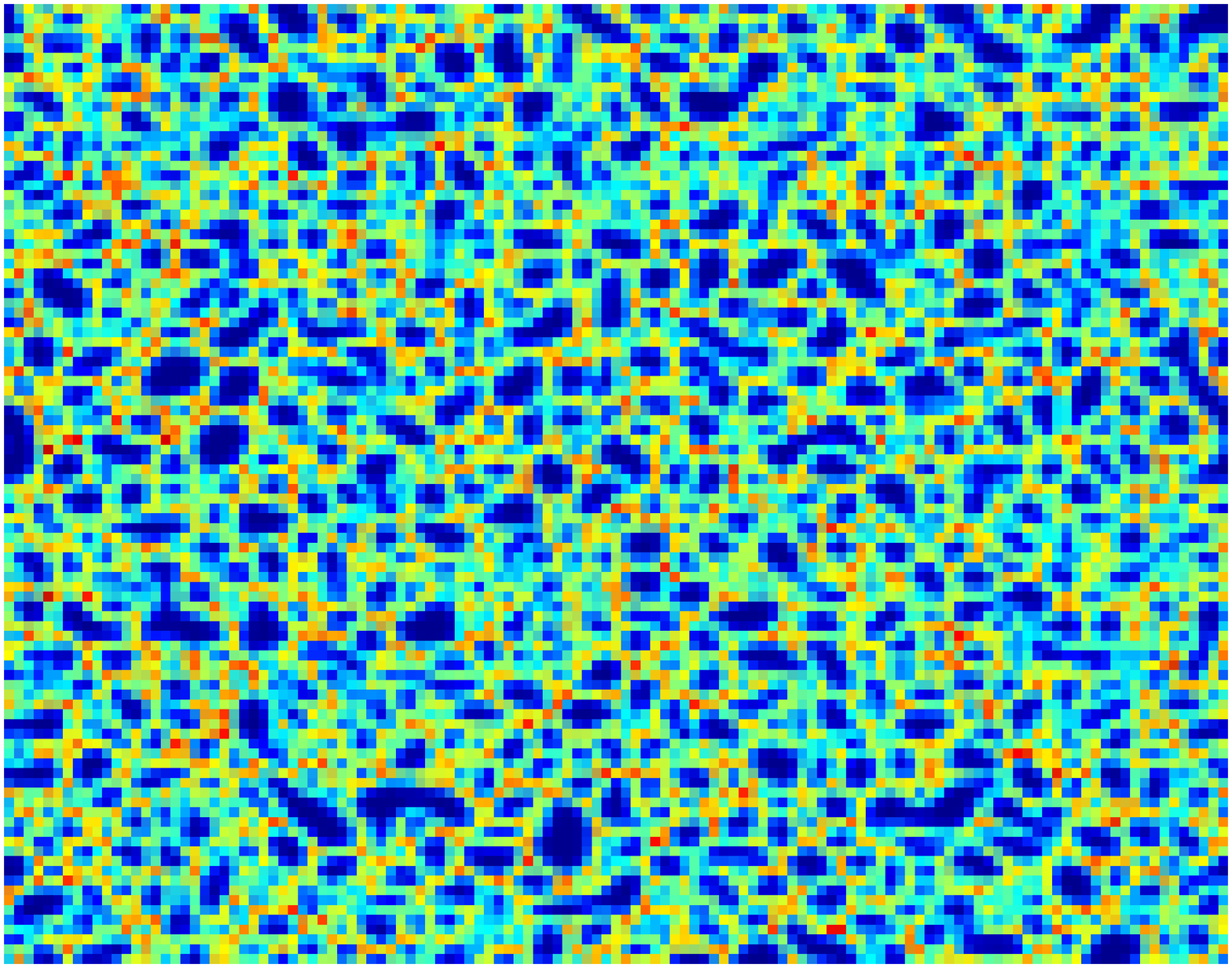}
\includegraphics*[width=6cm]{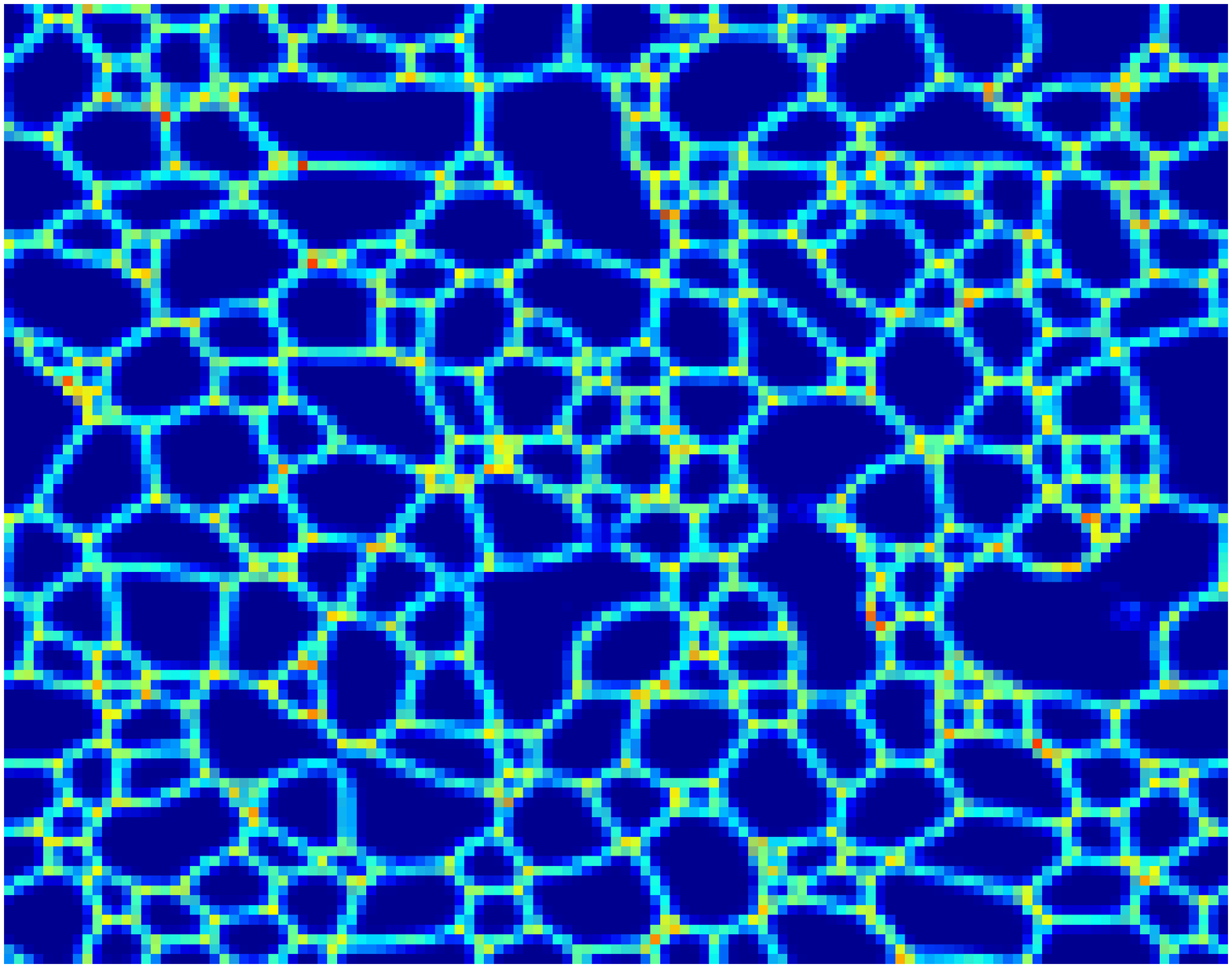}
\includegraphics*[width=6cm]{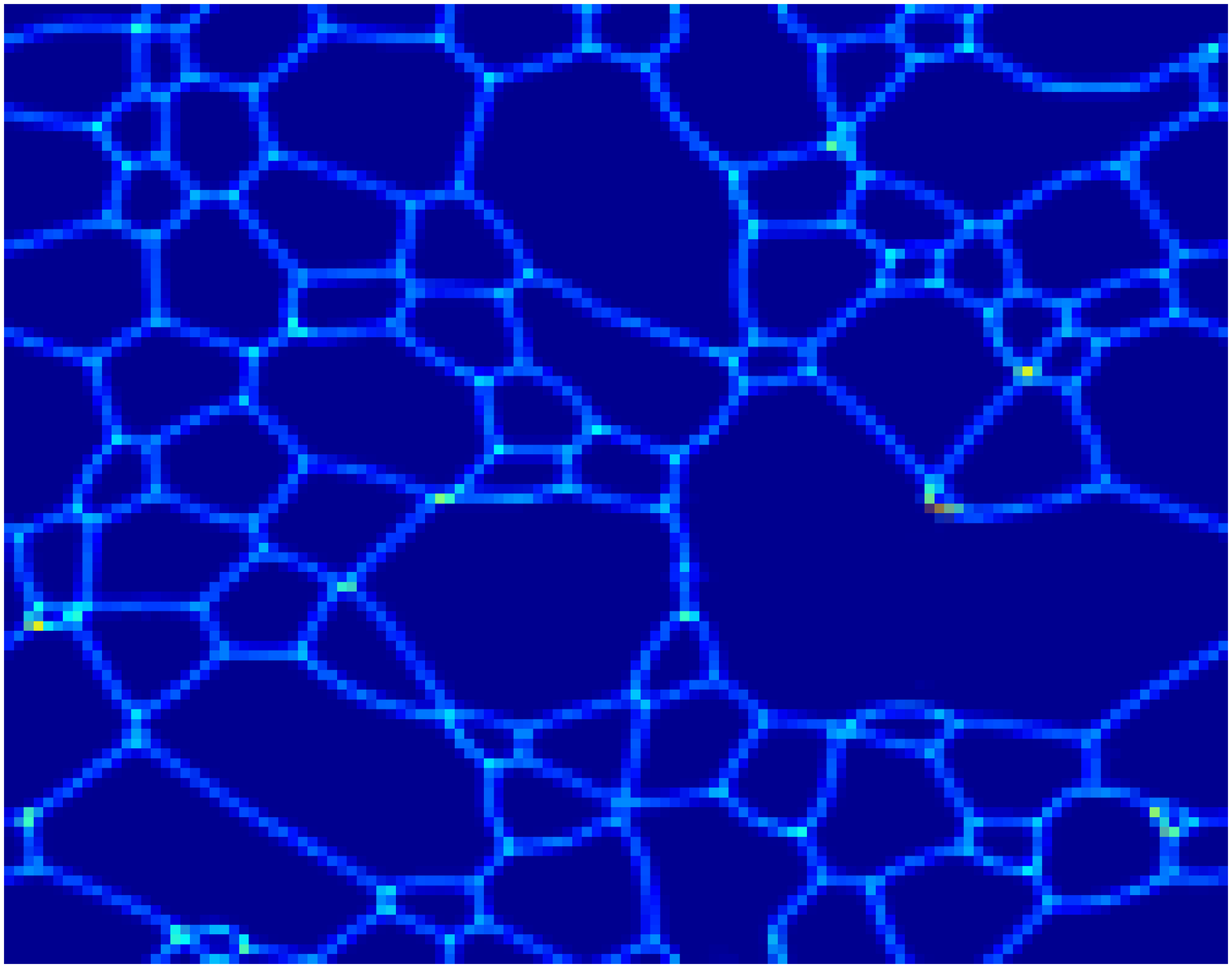}
\includegraphics*[width=6cm]{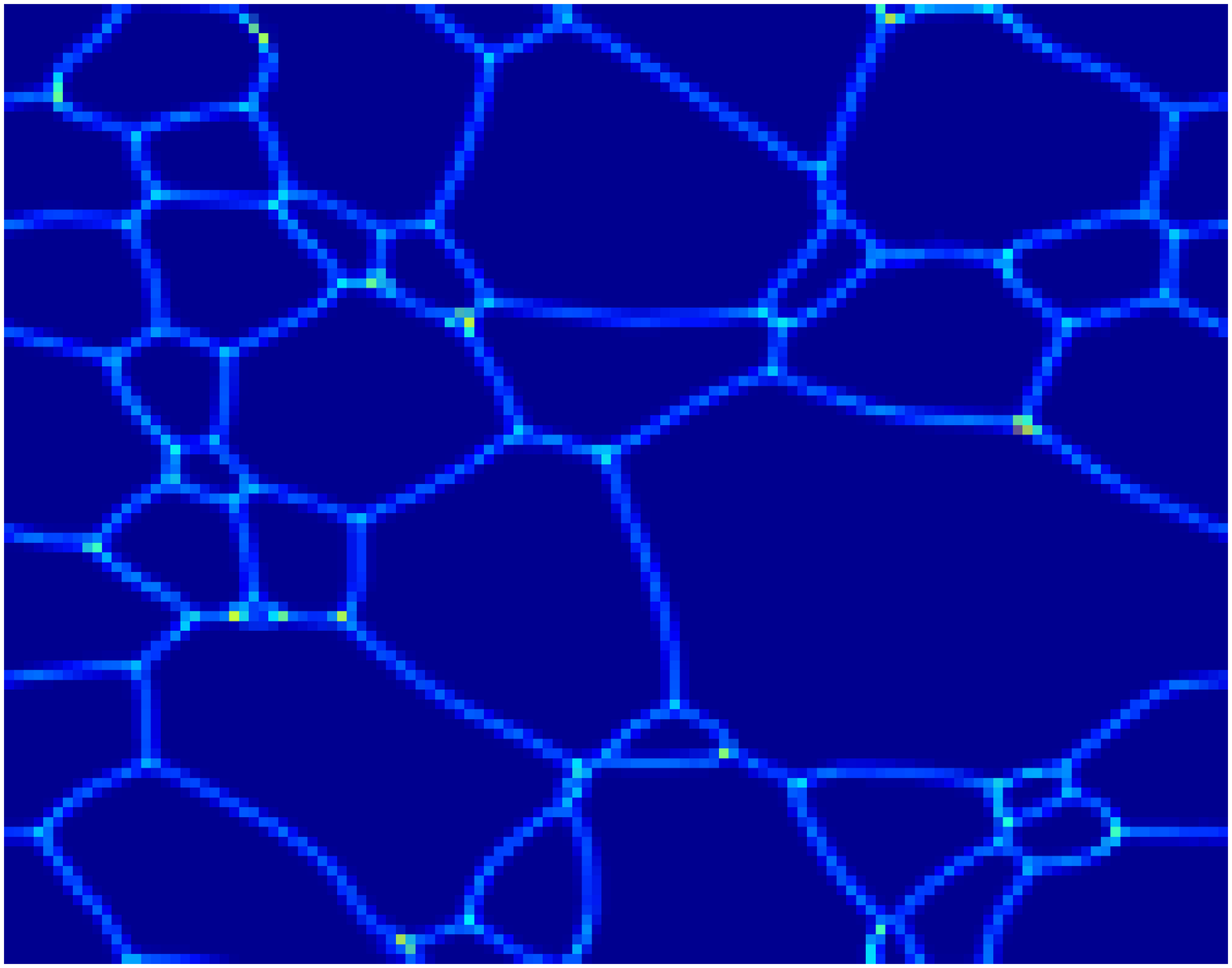}

\end{center}
\caption[The ideal class of model: $N=20$.]{
The matter-era evolution of the domain wall network for the ideal model with $N=20$. Note that only stable Y-type junctions survive. From left to right and top to bottom, the horizon is approximately 1/16, 1/8, 1/4 and 1/2 of the box size respectively.
}
\label{ideal7}
\end{figure}
We plot in Fig. \ref{ideal4} the snapshots for the evolution of the Pentahedral model ($N=4$). This simulation shows the matter era evolution of the domain wall network. Since this case is effectively analogous to that the Carter's potential in \cite{CARTER}, it confirms our claim that only stable Y-type junctions are formed instead of X-type one as was suggested by Carter. Fig. \ref{ideal7} shows the corresponding results for $N=20$.

\subsection{\label{sfield}Three-dimensional parallel simulations}
The second part of our results were obtained on the UK Computational Cosmology Consortium's COSMOS supercomputer\footnote{See \textit{http://www.damtp.cam.ac.uk/cosmos} for additional information.}. They constitute the most accurate three-dimensional field theory simulations in the literature of domain wall networks with junctions and were presented in Ref. \cite{IDEAL3}. 
We have used the same modified version of the algorithm of Press, Ryden and Spergel \cite{PRESS,RYDEN} described in the previous sections and
the code was parallelized with OpenMP directives. 
We have also optimized the code for the shared memory architecture of COSMOS. 
We will discuss the results from series of different eras simulations of $128^3$, $256^3$ and $512^3$ boxes, for the ideal class of models (\ref{ideal}) performed for all values of $N$ between 2 and 20. In addition, we ran 10 different runs for each box size and number of fields. 

These parallel simulations are memory-limited. We need
need approximately $DIM^3*N*4.5*8/1024^2$ MB, where $DIM$ is the box size and $N$ is the number of fields.
For $N=20$ and $512^3$ boxes, not less than 90 GB were required\footnote{A $512^3$ simulation with $N=3$ (requiring about 14.5 Gb of memory) takes about 14 seconds per step on 16 processors and only 5 seconds per step on 32 processors (as the memory ratio becomes favorable) without box output, and a complete run takes just 85 minutes. For larger runs the scalability is good if one keeps the memory smaller than 1 GB per processor. The largest simulation we have performed, a $512^3$ box with 20 scalar fields and box outputs at every timestep, took about 2 hours and 15 minutes on 128 processors.}. The complete set of numerical results will appear soon in Ref. \cite{IDEAL4}.
An output box binary file can also be produced at specified timesteps which can then be used to generate animations, an example of which is available at 

\textit{http://www.damtp.cam.ac.uk/cosmos/viz/movies/evo2\_25620\_msmpeg.avi}.

\subsubsection{Wall Observables}

We measure the domain wall velocities using an algorithm analogous to that described in \cite{AWALL} which removes the radiated energy from the walls which otherwise would contaminate the estimate of the velocities. This is clearly an important advantage over previous velocity estimations (see for example \cite{PRESS}). 

We defined a domain wall as the region where $V(\phi) > \varrho \, V_{\rm max}$ ($V_{\rm max}$ being the maximum of the potential) and estimated the velocities as
\be
v_*^2 \equiv \langle v^2 \gamma^2 \rangle \sim \sum_{V(\phi_i) > 0.2 V_{\rm max}}  \frac{{\dot \phi_i}^2}{2 V(\phi_i)}\, ,
\ee
where a dot represents a derivative with respect to conformal time and $\gamma = (1-v^2)^{-1/2}$. 
We verified that, for $0.2 \leq \varrho \leq 0.6$, our
results are almost independent of $\varrho$,
as long as the domain wall is sufficiently resolved.
The chosen threshold value $\varrho$ will thus define two properties of the network: the corresponding thickness of the static domain wall $\delta$ and the volume fraction of the box with domain walls $f$. They can be used to estimate the \textit{comoving characteristic scale} of the network
\be
L_c \equiv \frac{L}a \sim \frac{\delta}f.
\ee

In order to check whether the domain wall network is evolving according to a linear scaling solution we
define the \textit{scaling exponent} $\lambda$ so that $
L_c \propto \tau^{1-\lambda}$,
where $\tau$ is the conformal time. If $\lambda = 0$ we have linear scaling. 
On the contrary, the solution frustrates for $\lambda=1$.
The scaling exponent can be calculated at two different values of the conformal time $\tau_1$ and $\tau_2$ by using the following relation \cite{SIMS1}
\be
\lambda(\tau_1,\tau_2) = \frac
{\ln{(R_1/R_2)}}
{\ln{(\tau_1/\tau_2)}},
\ee
where $R \equiv \tau\,L_c^{-1}$. 
In general, since the simulations are carried out until $\tau$ becomes equal to one half of the box comoving size, the values of $\tau_1$ and $\tau_2$ are chosen in order to consider the second half of the dynamical range of the simulations.
Note that the earlier part is not included to avoid contamination of the results by our particular choice of initial conditions, while beyond one half of the comoving size, the effect of the finite box size will contaminate the results.
\begin{figure}
\begin{center}
\includegraphics[width=9cm]{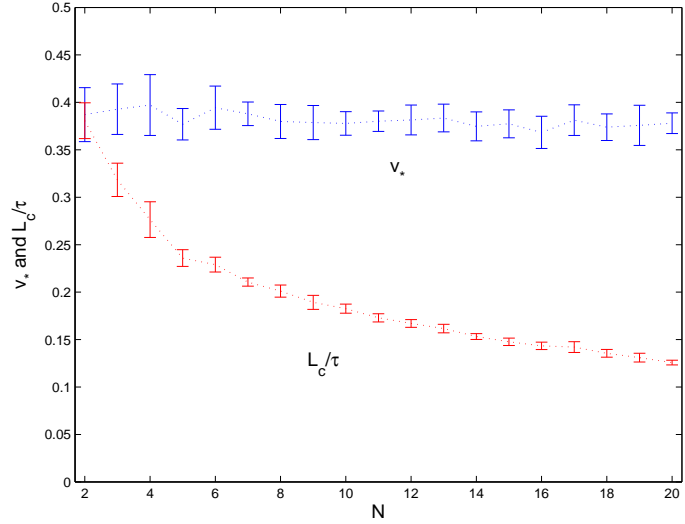}
\caption[The ideal class of model: $v_*$ and $L_c/\tau$ in matter era for $128^3$ boxes.]{\label{scalmat} The asymptotic values of $v_*$ and $L_c/\tau$ for 
the ideal class of models with $N$ ranging from 
$2$ to $20$. The error bars represent 
the standard deviation in an ensemble of 10 simulations. The simulations were performed in the matter era for $128^3$ boxes.}
\end{center}
\end{figure}
\begin{figure}
\begin{center}
\includegraphics[width=9cm]{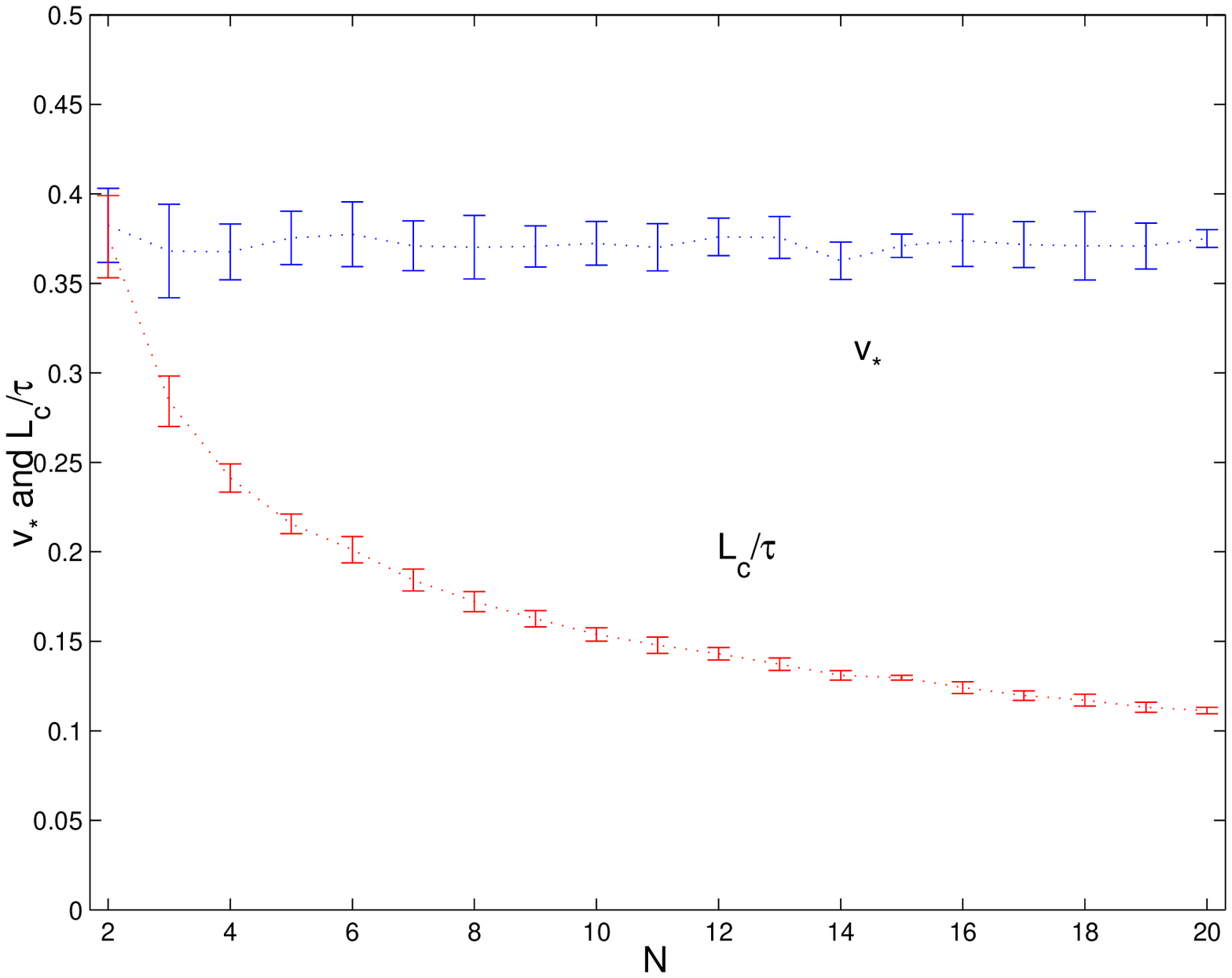}
\caption[The ideal class of model: $v_*$ and $L_c/\tau$ in matter era for $512^3$ boxes.]{\label{vlmatter} The asymptotic values of $v_*$ and $L_c/\tau$ for 
the ideal class of models with $N$ ranging from 
$2$ to $20$. The error bars represent 
the standard deviation in an ensemble of 10 simulations. The simulations were performed in the matter era for $512^3$ boxes.}
\end{center}
\end{figure}
\begin{figure}
\begin{center}
\includegraphics[width=9cm]{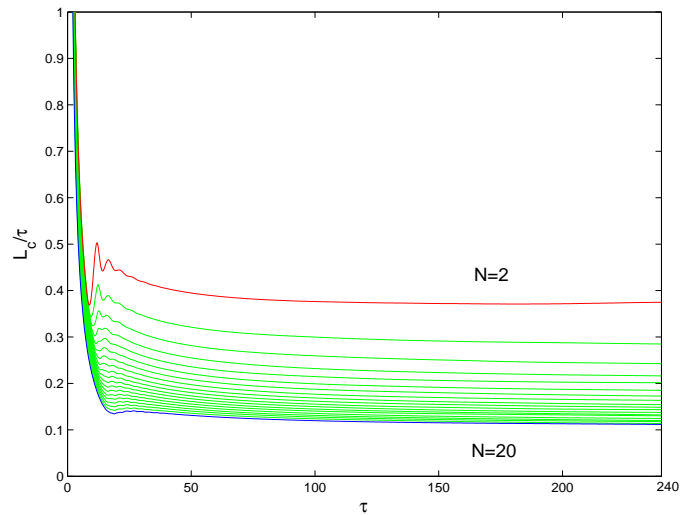}
\caption[The ideal class of model: $L_c/\tau$ in matter era for $512^3$ boxes.]{\label{lmatter} $L_c$ as a function of the conformal time $\tau$ for the class of ideal models ranging from $2$ to $20$. The simulations were carried out in the matter dominated era. Note that as $N$ increases the asymptotic value of $L_c$ becomes smaller.}
\end{center}
\end{figure}
\begin{figure}
\begin{center}
\includegraphics[width=9cm]{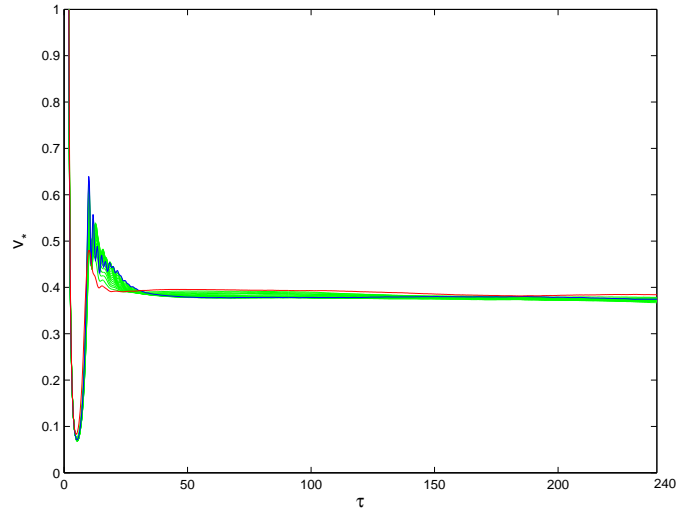}
\caption[The ideal class of model: $v_*$ in matter era for $512^3$ boxes.]{\label{vmatter} $v_*$ as a function of the conformal time $\tau$ for the class of ideal models ranging from $2$ to $20$. The simulations were performed in the matter dominated era. Note that 
there is no significant dependence of the velocities with $N$.}
\end{center}
\end{figure}
\begin{figure}
\begin{center}
\includegraphics[width=9cm]{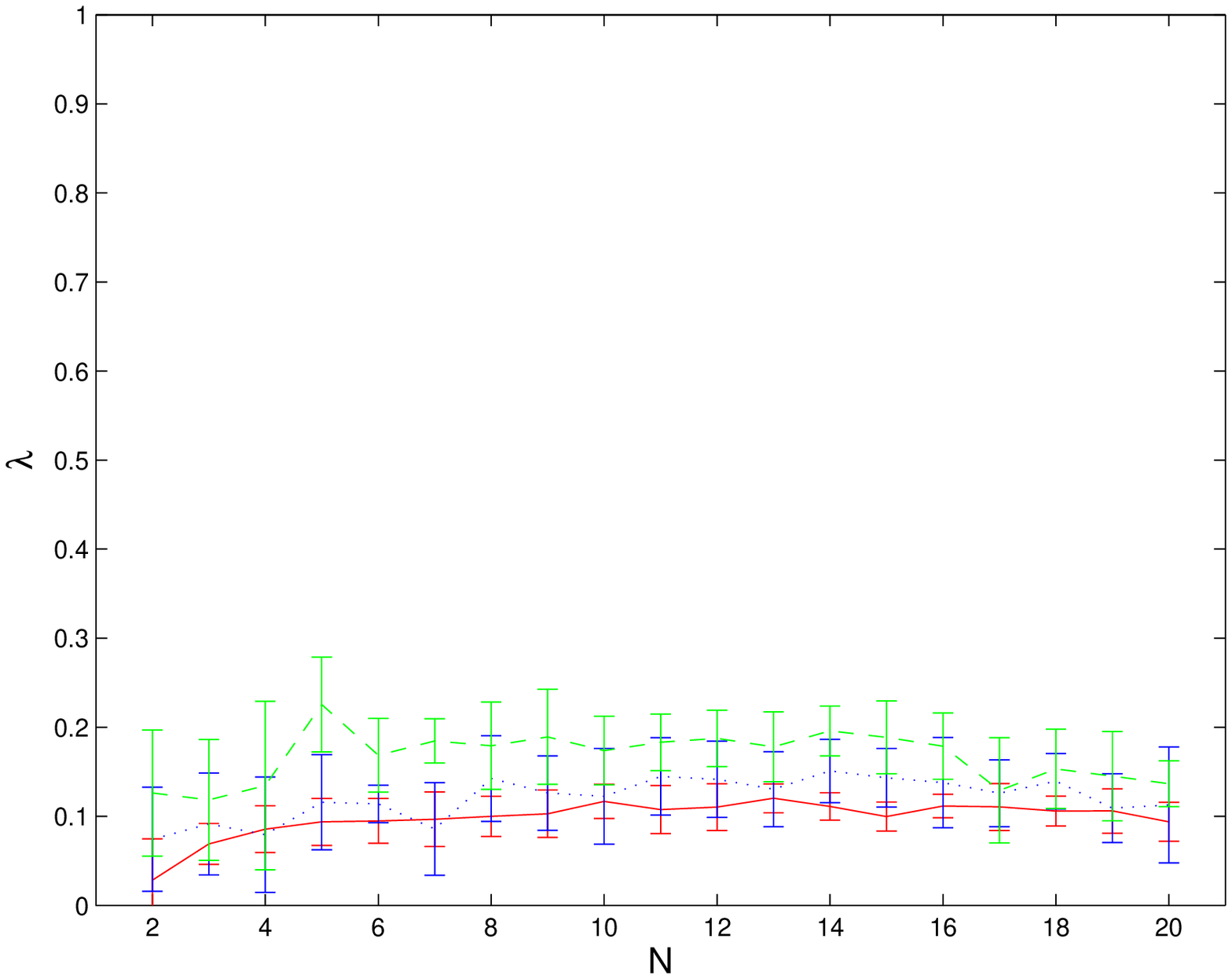}
\caption[The ideal class of models: $\lambda$ in matter era for $128^3$, $256^3$ and $512^3$ boxes.]{\label{exponents512}  The scaling exponents, $\lambda$, for all $N$'s between 2 and 20 in matter era for $128^3$ (dashed line), $256^3$ (dotted line) and $512^3$ (solid line) boxes.
Note that frustration would correspond to $\lambda=1$ and linear scaling to $\lambda =0$. Note also that as the box size simulation increases, $\lambda$ gets closer to zero.}
\end{center}
\end{figure}
\subsubsection{Numerical Results}
We performed series of simulations of $128^3$, $256^3$ and $512^3$ boxes, for $2 \leq N \leq 20$ in matter era. We carried out $10$ different simulations for each box size and number of fields. 

In Figs. \ref{scalmat} and \ref{vlmatter} we plot the asymptotic values of $v_*$ and $L_c/\tau$ for the ideal class of models with $2 \leq N \leq 20$ run in $128	^3$ and $512^3$ boxes in the matter dominated era, respectively. 
We have performed $10$ simulations for each value of $N$, and shown the standard deviation with appropriate error bars \cite{AWALL}. 
The differences between the results obtained for boxes of different sizes are visible but quite small---in general the results are consistent with one another, given the error bars. The way the differences go is also what one expects. Mindful of these differences, we can still say that for many of our purposes the $256^3$ or even $128^3$ boxes (each of which we can usually run in about an hour or less) will produce adequate results.
A larger $N$ yields a smaller asymptotic value of $L_c/\tau$. This aspect is highlighted in Fig. \ref{lmatter}, where the differences between the 
successive $N$ results for $L_c/\tau$ become increasingly smaller. Consequently, 
the results obtained up to $N=20$ can be interpreted to be already close 
to the $N \to \infty$ limit results. 
On the other hand, there is no significant 
dependence of the velocities, $v_*$, with $N$, which is confirmed in Fig. \ref{vmatter}.
Overall we find
\be
v_{mat}=0.36\pm0.02\, \label{velocjmat}
\ee
where $v_{mat}$ is the asymptotic value of the velocity in the matter era simulations. We compare (\ref{velocjmat}) with the value obtained in
Ref. \cite{AWALL} for $3D$ $512^3$ simulations of standard domain walls $v_{st,mat}$:
\be
v_{st,mat}=0.39\pm0.02. \label{veloc0mat}
\ee
Therefore the velocities are slightly smaller in the case of networks with junctions. This is of course what we expect, though given the various numerical uncertainties the difference between the two is not very significant.  
A more important comparison can be made with the value we would expect if the network had no energy losses. In this case we would expect $v\sim0.41$, and the difference means that energy losses are still noteworthy (at about the ten percent level), despite the existence of the junctions. This is ultimately the physical reason why the frustration mechanism can never be realized.

This difference between the behavior of the two averaged quantities is crucial---although changing the number of fields $N$ will change the network's characteristic length (and therefore its density), the fact that the velocities do not change is an indication that in some sense the `local' dynamics of each individual wall will effectively be the same. Note that for this to be the case it is obviously necessary that all the walls have the same tension, so we do not expect this result to hold for non-ideal models. A second necessary condition is that the junctions themselves must be dynamically unimportant.

Fig. \ref{exponents512} shows the scaling exponents, $\lambda$, for $2 \leq N \leq 20$ for $128^3$ (dashed line), $256^3$ (dotted line) $512^3$ (solid line) boxes. Again, the error bars represent the standard deviation in a set of $10$ simulations carried out in the matter dominated era.
We see that $\lambda$ is slightly greater than zero which indicates that there are small departures a the scaling solution (keep in mind that exact linear scaling obviously corresponds to $\lambda=0$). Indeed we find
\begin{equation}
\lambda_{mat}=0.11\pm0.05\,, \label{scalingjmat}
\end{equation}
where $\lambda_{mat}$ is the asymptotic value of the scaling exponent in the matter era simulations of domain walls with junctions.
Comparing (\ref{scalingjmat}) with the value obtained in Ref. \cite{AWALL} for analogous simulations of standard domain walls
\begin{equation}
\lambda_{st,mat}=0.04\pm0.02\,,
\label{scaling0mat}
\end{equation}
one has that the two values are in fact relatively similar; the larger exponents in the case with junctions should be attributable (at least in part) to the longer time needed for the relaxation to scaling when there are several coupled fields. Actually, 
$\lambda$ is slightly 
greater than zero but
as we increase the box size simulations,
$\lambda$ gets closer and closer to zero.
Larger simulations have a larger dynamical range which means that the network evolves for a longer period of time. Hence, it is not surprising that those networks are able to approach closer a linear scaling solution. On the other hand, frustration (which would correspond to $\lambda=1$) is clearly ruled out.

\begin{figure}
\begin{center}
\includegraphics[width=9cm]{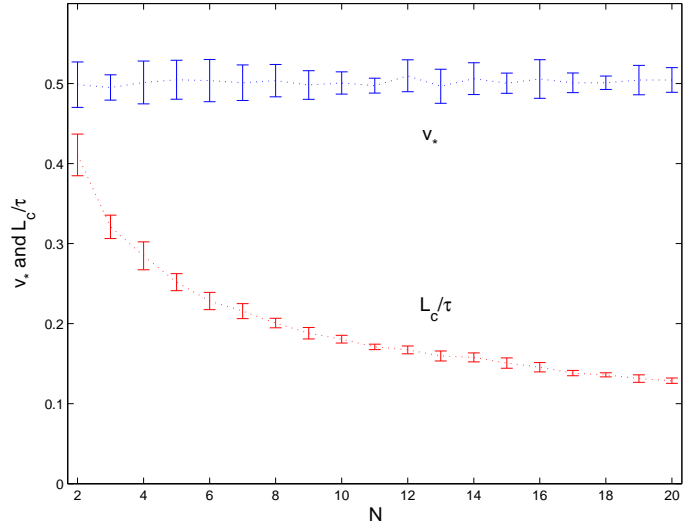}
\caption[The ideal class of models:  $v_*$ and $L_c/\tau$ in the radiation era for $256^3$ boxes.]{\label{vlrad} The asymptotic values of $v_*$ and $L_c/\tau$ for 
the ideal class of models with $N$ ranging from 
$2$ to $20$. The simulations were performed in the radiation era for $256^3$ boxes.}
\end{center}
\end{figure}
\begin{figure}
\begin{center}
\includegraphics[width=9cm]{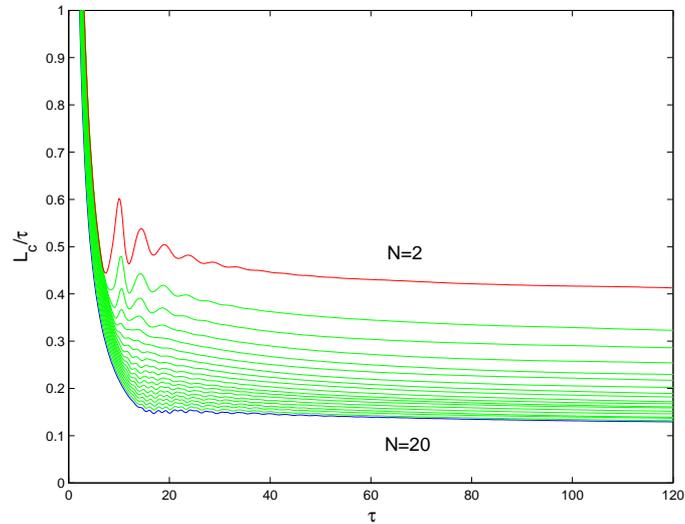}
\caption[The ideal class of model: $L_c/\tau$ in radiation era for $256^3$ boxes.]{\label{lrad} $L_c$ as a function of the conformal time $\tau$ for the class of ideal models ranging from $2$ to $20$. The simulations were performed in radiation dominated era. Note that as $N$ increases the asymptotic value of $L_c$ becomes smaller.}
\end{center}
\end{figure}
\begin{figure}
\begin{center}
\includegraphics[width=9cm]{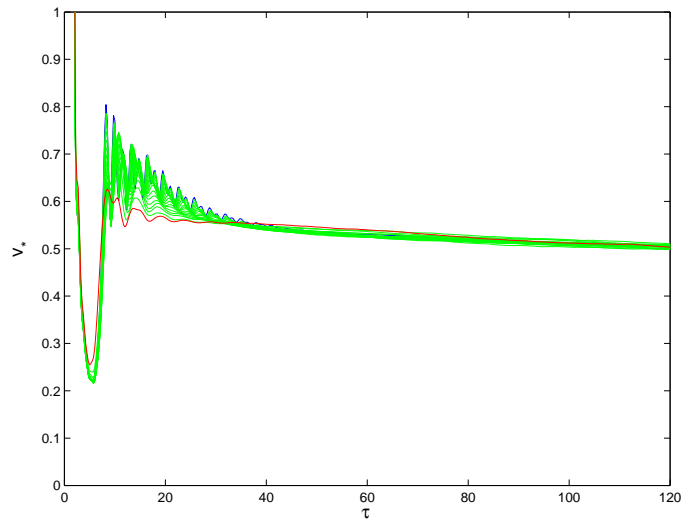}
\caption[The ideal class of model: $v_*$ in radiation era for $256^3$ boxes.]{\label{vrad} $v_*$ as a function of the conformal time $\tau$ for the class of ideal models ranging from $2$ to $20$. The simulations run in the radiation dominated era. Note that 
there is no significant dependence of the velocities with $N$.}
\end{center}
\end{figure}

\begin{figure}
\begin{center}
\includegraphics[width=9cm]{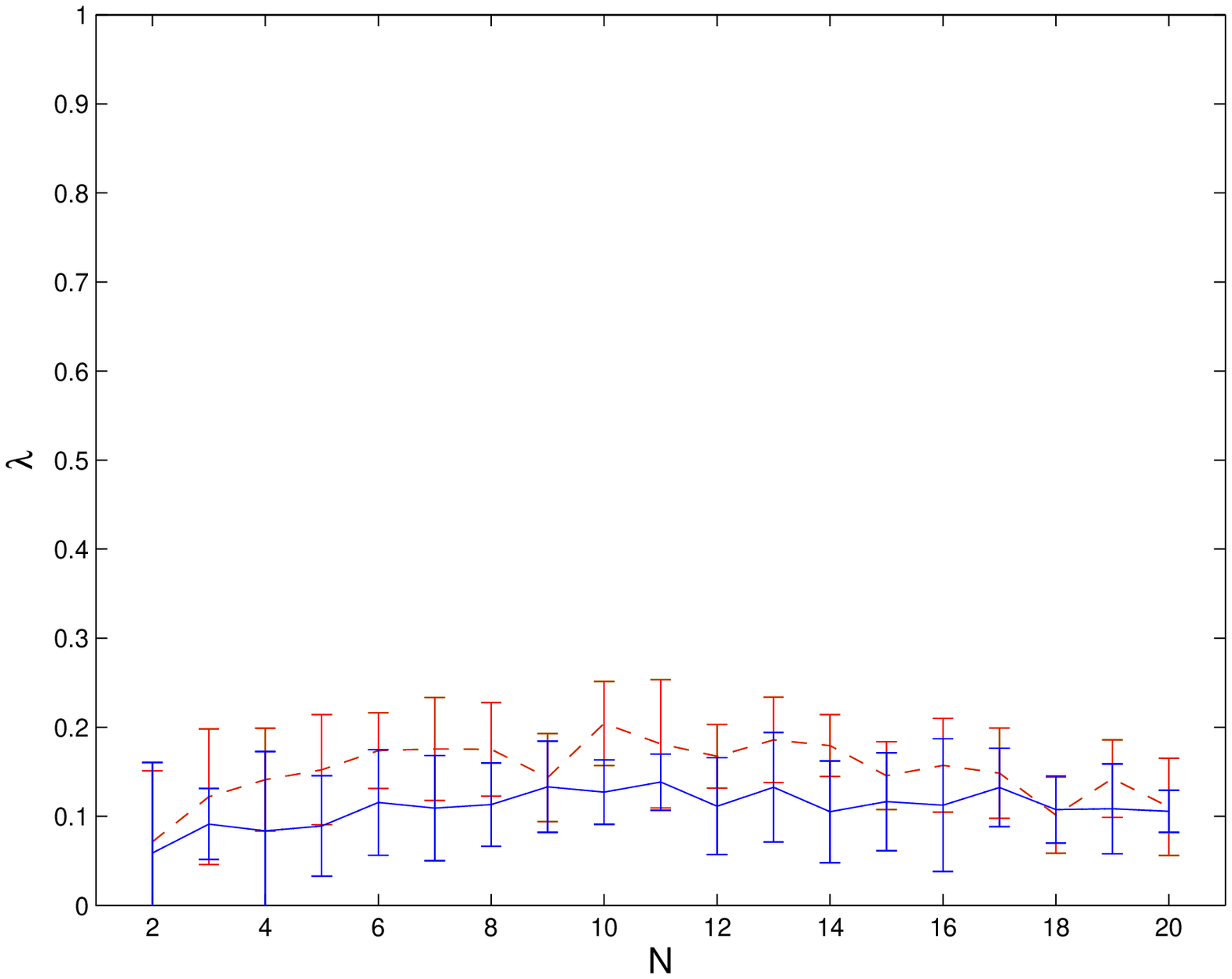}
\caption[The ideal class of models: $\lambda$ in the radiation era for $128^3$ and $256^3$ boxes.]{\label{exponentsrad}The scaling exponents, $\lambda$, for all $N$'s between 2 and 20 in the radiation era, for $128^3$ (dashed line) and $256^3$ (solid line) boxes.
Note that frustration would correspond to $\lambda=1$ and linear scaling to $\lambda =0$. Note also that as the box size simulation increases, $\lambda$ gets closer to zero.}
\end{center}
\end{figure}

Even though simulations of the matter era (that is, with the scale factor evolving as $a\propto t^{2/3}$) are the most relevant from the point of view of cosmological scenarios involving domain walls, it is also interesting to study their evolution in other cosmological epochs. Results of analogous simulations of the ideal model for the radiation era ($a\propto t^{1/2}$) are shown in Figs. Figs. \ref{vlrad}-- \ref{exponentsrad}. The scaling exponents and properties are defined and measured as before, and the simulation parameters are also similar, the only difference being that in this case we only have carried out $128^3$ and $256^3$ simulations. 

The results are qualitatively identical to the ones in the matter era, although as expected there are some quantitative differences in the scaling parameters. We have computed the scaling exponent $\lambda_{rad}$:  
\begin{equation}
\lambda_{rad}=0.10\pm0.05\,, \label{scalingjrad}
\end{equation}
which is again comparable to the one obtained for analogous simulations of standard domain walls \cite{AWALL}
\begin{equation}
\lambda_{st,rad}=0.04\pm0.02\,. \label{scaling0rad}
\end{equation}
Likewise, we do not find any significant dependence the velocities $v_{rad}$ with $N$. We now obtain
\begin{equation}
v_{rad}=0.45\pm0.03\, \label{velocjrad}
\end{equation}
whereas for standard domain walls in the radiation era
\begin{equation}
v_{st,rad}=0.48\pm0.02\,, \label{veloc0rad}
\end{equation}
and the value we would expect if the network had no energy losses is now $v\sim0.57$ (see Eq. \ref{uls3}). The only noteworthy difference is that energy losses are comparatively more important in this case---at about the twenty percent level, as opposed to ten percent in the matter era. There is also an enhancement of the scaling density that is similar to the one in the matter era.

\begin{figure}
\begin{center}
\includegraphics[width=9cm]{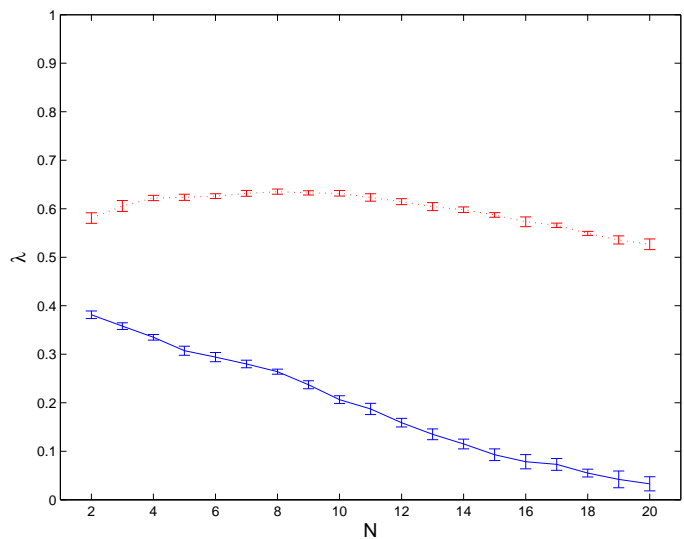}
\caption[The ideal class of models: $\lambda$ in the slow-expansion era for $128^3$ and $256^3$ boxes.]{\label{exponslw}The scaling exponents, $\lambda$, for all $N$'s between 2 and 20, for the $128^3$, $256^3$ boxes (dotted and solid lines respectively) in a slow-expansion era, $a\propto t^{1/5}$. The error bars represent the standard deviation in an ensemble of 10 simulations. Note that frustration would correspond to $\lambda=1$.}
\end{center}
\end{figure}

\begin{figure}
\begin{center}
\includegraphics[width=9cm]{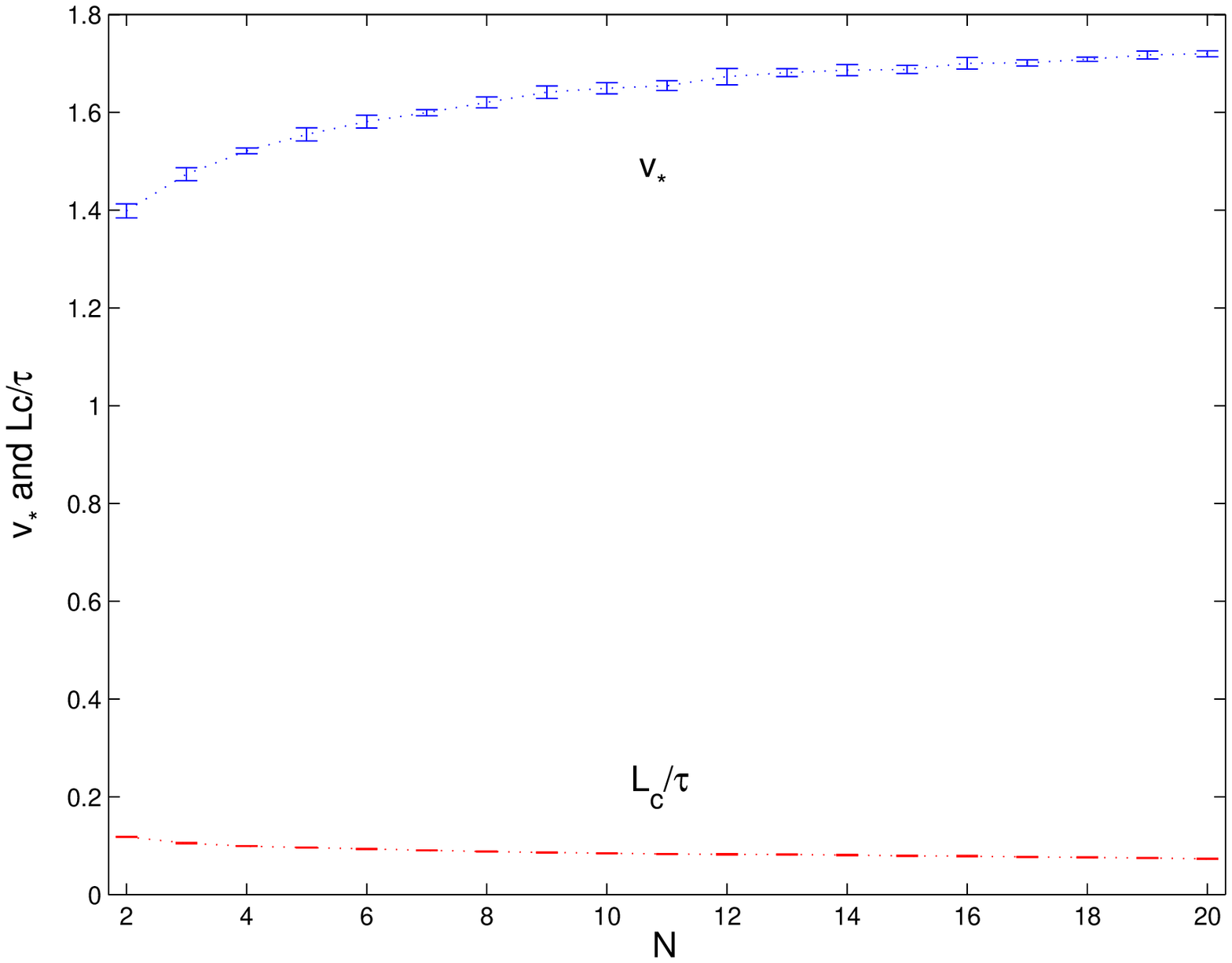}
\caption[The ideal class of models:  $v_*$ and $L_c/\tau$ in the slow-expansion era for $128^3$ boxes.]{\label{scalslw1}The asymptotic values of $v_*$ and $L_c/\tau$ for slow-expansion era simulations of the ideal class of models with $N$ ranging from $2$ to $20$ (top and bottom lines respectively). These come from simulations of $128^3$, and the error bars represent the standard deviation in an ensemble of 10 simulations.}
\end{center}
\end{figure}

\begin{figure}
\begin{center}
\includegraphics[width=9cm]{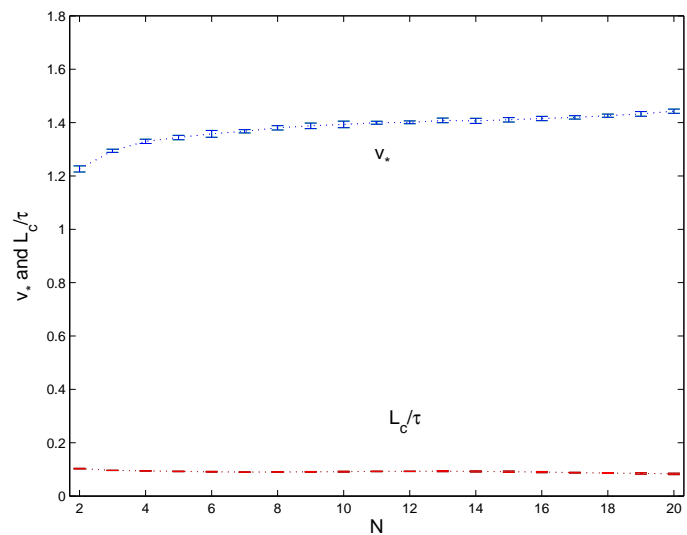}
\caption[The ideal class of models:  $v_*$ and $L_c/\tau$ in the slow-expansion era for $256^3$ boxes.]{\label{scalslw2}The asymptotic values of $v_*$ and $L_c/\tau$ for slow-expansion era simulations of the ideal class of models with $N$ ranging from $2$ to $20$ (top and bottom lines respectively). These come from simulations of $256^3$, and the error bars represent the standard deviation in an ensemble of 10 simulations.}
\end{center}
\end{figure}

Figs. \ref{exponslw}---\ref{scalslw2} display results of analogous simulations for a cosmological epoch with an expansion rate even slower than the radiation era, namely $a\propto t^{1/5}$. This case might be cosmologically relevant in some toy models, for a transient epoch in the very early universe, but in any case it is an important numerical test because our analytic modeling leads us to expect that for $a\propto t^\kappa$ and $\kappa<1/4$ a linear scaling solution can only exist if there are energy losses.


We emphasize that an accurate numerical evolution of this case is technically difficult. The slower expansion on this case means little dissipation, which in turn implies that there is a large thermal bath in the box. Nevertheless, we do find evidence for an approach to a linear scaling solution in this case, although the evidence is not as strong and unambiguous as that in the matter and radiation eras. The main reason for this is that the large thermal bath present will tend to make the relaxation to scaling be significantly slower. This explains why the $128^3$ boxes have much larger $\lambda$ than the $256^3$ ones-- the latter simulations have a dynamic range (in conformal time) that is twice that of the former. This analysis therefore confirms the existence of energy losses in these domain wall networks, as well as their dynamical importance.

Note that the exponents shown in Fig. \ref{exponslw} are effective parameters, calculated in the final part of each simulation. Dividing this dynamic range into smaller bins would lead to exponents becoming progressively closer to $\lambda=0$ as the simulations evolve. The scaling parameters $v_*$ and $L_c/\tau$ shown in Figs. \ref{scalslw1} and  \ref{scalslw2}, on the other hand, are not averaged throughout each simulations but are measured at the end of each of them. They can therefore still be compared with the values we obtained in the radiation and matter eras. As one would expect, in this case we find that the wall networks have much higher densities and velocities ($v_{slw}=0.81$). The dependence of the scaling density on the number of fields is still seen but is now quite weaker than in the matter or radiation cases. Moreover, the scaling velocity now also has a mild dependence on $N$, which as we saw is not present in the other cases. This different behavior is, to a large extent, due to the presence of the larger thermal bath, whose effects on a particular field will depend on $N$. The reason for this is the way we set up the simulations. Our choice of initial conditions is such that there is always a similar amount of energy in the box, which implies that the average energy per field decreases as we increase $N$. On the other hand, it is also conceivable that our algorithm for measuring velocities becomes relatively less precise for ultra-relativistic velocities.

\subsubsection{Scaling of the Non-Ideal Models}
\begin{figure}
\begin{center}
\includegraphics[width=9cm]{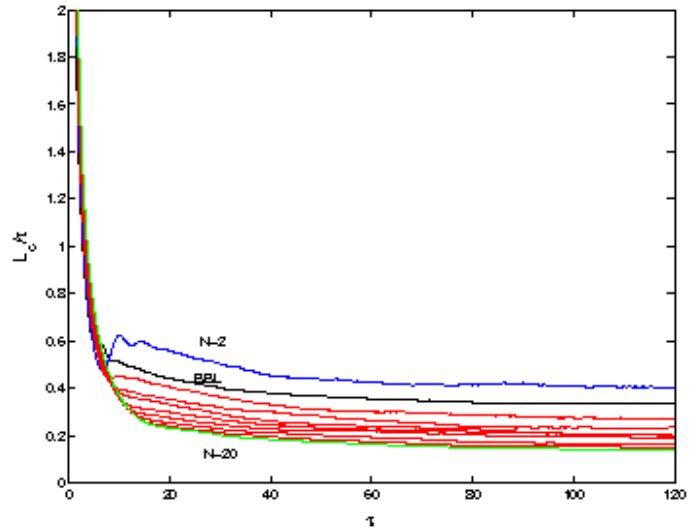}
\caption[$L_c/\tau$ for the BBL model and the ideal class of models.]{\label{non1}The plot of $L_c/\tau$ as a function of the conformal time $\tau$ for the three-field version of the BBL model (dashed line) with only X-type junctions and for the ideal model (solid line) for even $N$ between 2 and 20. The simulations were carried out in $256^3$ boxes. Although the ideal model with largest $N$ has a smaller $L_c/\tau$ than the BBL model, the result shows that the frustration scenario is not realized.}
\end{center}
\end{figure}

In order to compare the scaling behavior of the ideal model with the non-ideal ones, we have performed some $3D$ simulations of the three-field BBL Model (\ref{bbl3}) in the parameter range where only X-type junctions can be formed. We recall that this scenario is realized when the parameters $\epsilon r^2$ are chosen between $-2/5$ and $1/2$ as in Eq. (\ref{solplus3}) (this case is analogous to the Kubotani model with $\xi >0$).
Fig. \ref{non1} shows the results for $L_c/\tau$ as a function of the conformal time $\tau$ for the BBL model with X-type junctions (dashed line) and the ideal class of models for even $ 2 \leq N \leq 20$ . Note that the model with $N=2$ does 
not perform better 
than the BBL model from the point of view of frustration, this only happen for $N > 3$. However, it is clear that 
$L_c/\tau \to {\rm constant}$ even for large $N$.
as $N \to \infty$.
So even the ideal model fails to produce a frustrated domain wall network.


\section{\label{intr122222}Conclusions}
We have studied the possibility that a domain wall network could account for the
the dark energy or a significant part of it thus 
explaining the recent acceleration of the universe. For this purpose, we considered necessary conditions for 
frustration.
We presented a one scale velocity dependent analytic model which describes 
some of the main properties of 
the domain wall network evolution. We derived a strong bound on the curvature of the walls, which shows that viable candidate networks must be fine-tuned and non-standard. We also discussed various requirements that any stable lattice of frustrated walls would need to obey. 

We have studied several domain wall models in two spatial dimensions, and discussed the conditions under which various types of defect junctions can exist. In particular, we focused on the BBL \cite{BAZEIA} and Kubotani \cite{KUBOTANI} models.
We have highlighted the situations where only stable
Y-type junctions or X-type ones are formed. In addition we showed the case where both types co-exist. 

Using geometrical, topological and energy arguments we
isolated model 
features 
which we expected to be crucial from of the point of view of frustration.
However, our
results strongly lead us to a
no frustration conjecture. In all our numerical simulations, including those of the ideal model, we find a dynamical behavior consistent with a convergence towards a scaling solution \cite{SIMS1,SIMS2,AWALL} and no evidence of any behavior that could eventually lead to frustration. Indeed,
 
the results of an 3D simulations clearly indicate that no frustrated network can be formed dynamically out of random initial conditions that would mimic a realistic phase transition.
Notice that we are making no claims about the possibility of designing (by hand) a frozen lattice-type configuration. 
We have presented the most compelling evidence to date that domain wall networks can not be the dark energy.

\chapter{\label{cap3}{\sf Varying-$\alpha$ Cosmic Strings}}

\section{Overview}

The space-time variations of the so-called $``$constants$"$ of Nature  have attracted much attention motivated by models with extra spatial 
dimensions \cite{Polchinski}.
For example, in the context of varying-$\alpha$ models such as the one proposed 
by Bekenstein \cite{Bekenstein}, the variation  
is sourced by a scalar field  
coupling minimally to the metric and to the electromagnetic term.

We recall that the interest in this type of models increased due to the results 
coming from quasar absorption systems 
\cite{Webb2}--\cite{Chand}
suggesting a cosmological 
variation of $\alpha$ at low redshifts \cite{Srianand2}. Other constraints at low redshift 
include atomic clocks \cite{Marion} and meteorites \cite{Olive}. At high 
redshifts there are also upper limits to the allowed variations of $\alpha$ 
coming from either the Cosmic Microwave Background 
\cite{Avelino2}--\cite{sigurdson}
or 
Big Bang Nucleossynthesis 
\cite{Campbell,Cyburt}.

In this chapter we will consider
a second class of topological defects, the cosmic strings, in the context of the varying-$\alpha$ theories. 
In the model studied here, the electromagnetic energy 
localized along a stable string-like core acts as a source for 
spatial variations of $\alpha$ in the vicinity of the string. 
Although a change in the value of $\alpha$ can be accompanied by a variation of other fundamental constants, we will consider 
a phenomenological approach and neglect possible variations of other 
fundamental constants. 

We generalize the results presented initially in Ref.
\cite{Magueijo} for the case of  
a generic gauge kinetic function.
We find numerical solutions for the fields and compare them with the standard solution with no variation of $\alpha$ and
compute the spatial variations of the fine-structure constant in their vicinity
(taking into account 
the limits imposed by tests to the Weak Equivalence Principle \cite{Will}).

This chapter is organized as 
follows. In Sec. \ref{cap2cs} we briefly introduce Bekenstein-type models and obtain the 
equations describing a static string solution. We describe the numerical 
results obtained for a number of possible choices of the gauge kinetic 
function in Sec. \ref{csnum} discussing the possible cosmological implications of 
string networks of this type in the light of equivalence principle 
constraints in Sec. \ref{cscc}. Finally, in Sec. \ref{csdis} 
we summarize 
our results. 

\section{\label{cap2cs}Varying-$\alpha$ in Abelian Field Theories}
In this section we describe Bekenstein-type models with a charged
complex scalar field $\phi$ undergoing spontaneous symmetry breaking.

Let us consider a complex scalar field $\phi$ with a gauged $U(1)$ symmetry. 
The Lagrangian density is given by
\be
\label{laga21}
\mathcal{L} = D_\mu \phi^* D^\mu \phi  - \frac14 F_{\mu\nu}\,F^{\mu\nu} - V(\phi),
\ee  
where $V(\phi)$ is the potential and
the covariant derivatives are defined as
\be
\label{deriva21}
D_\mu \phi = \partial_\mu \phi - i\,e\,A_\mu \phi,
\ee
where $e$ is the electric charge and $A_\mu$ is the vector gauge field. The electromagnetic field strength is defined as
\be
\label{primeirafmn}
F_{\mu\nu} = \partial_\mu A_\nu - \partial_\nu A_\mu.
\ee
In fact, the Lagrangian in Eq. (\ref{laga21}) is invariant under the local transformations
\ben
\label{cg1}
&&\phi(x) \rightarrow \phi'(x) = e^{i\Lambda(x)}\,\phi(x),\\
\label{cg2}
&&A_\mu(x) \rightarrow A'_\mu(x) = A_\mu(x) + \frac1e \partial_\mu \Lambda(x),
\een
where $\Lambda(x)$ is a scalar function of space-time. This set of transformations obeys the
$U(1)$ symmetry group.

Varying the action 
\be
\label{1acig}
\mathcal{S} = \int d^4 x \left[D_\mu \phi^* D^\mu \phi  - \frac14 F_{\mu\nu}\,F^{\mu\nu} - V(\phi) \right]
\ee
with respect to the fields $\phi^*$ and $A_\mu$, one obtains the equations of motion
\ben
\label{eqphi0}
&&D_\mu D^\mu \phi + \frac{\partial V}{\partial \phi} = 0,\\
\label{fmn0}
&&\partial_\mu F^{\mu\nu} =j^\nu =-i\,e [\phi^* D^\nu \phi -(D^\nu \phi)^* \phi],
\een
where $j^\nu$ is a conserved current. 

In this chapter we
will assume the Mexican hat potential $V(\phi)$ given by
\be
\label{amhp}
V(\phi) = \frac{\lambda}4 \left(\phi^* \phi - \eta^2 \right)^2,
\ee
where $\eta$ represents the symmetry breaking scale.
\subsection{Bekenstein's theory}
Let us now assume that the electric charge $e$ is a function
of space and time coordinates,
\be
e=e_0 \epsilon(x^\mu),
\ee 
where $\epsilon(x^\mu)$ is a real scalar field and $e_0$ is an arbitrary constant charge.

The Lagrangian density (\ref{laga21}) is then generalized to include the field $\epsilon$
\be
\label{laga22}
\mathcal{L} = D_\mu \phi^* D^\mu \phi  - \frac14 F_{\mu\nu}\,F^{\mu\nu} - V(\phi) - \frac{\omega}{2\,\epsilon^2}\,\partial_\mu \epsilon \partial^\mu \epsilon,
\ee 
where $\omega$ is a real parameter and the 
the gauge invariant electromagnetic field tensor (\ref{primeirafmn}) is now 
\be
\label{segundafmn}
F_{\mu\nu} = \frac1{\epsilon}\, \left[\partial_\mu (\epsilon A_\nu) - \partial_\nu (\epsilon A_\mu) \right].
\ee
Introducing an auxiliary gauge field given by $a_\mu \equiv \epsilon A_\mu$, 
the covariant derivatives (\ref{deriva21}) can be written as
\be
\label{dcawithout}
D_\mu\,\phi=\left(\partial_\mu-i e_0 a_\mu\right)\,\phi.
\ee
In the same way, defining the tensor
\be
f_{\mu\nu}\, \equiv \,\epsilon\,F_{\mu\nu} = \,\partial_\mu  a_\nu -\partial_\nu a_\mu,
\ee 
the Lagrangian density can be rewritten as
\be
\label{laga23}
\mathcal{L} = D_\mu \phi^* D^\mu \phi  - \frac1{4\,\epsilon^2} f_{\mu\nu}\,f^{\mu\nu} - V(\phi) - \frac{\omega}{2\,\epsilon^2}\,\partial_\mu \epsilon \partial^\mu \epsilon
\ee 
which simplifies the variational problem enormously. 
\subsection{Generalizing the Bekenstein model}
Now, we generalize the Bekenstein model by assuming that the coupling to the 
Maxwell term is not limited to be an exponential function of the real scalar field.

Let us consider another real scalar field $\varphi$ defined as
\be
\label{varphi21}
\varphi \equiv \sqrt{\omega}\,\ln{\epsilon}.
\ee
Therefore the Lagrangian density (\ref{laga23}) becomes
\be
{\cal L}\, = \left(D_\mu\,\phi\right)^{*}\left(D^\mu\,\phi\right)
-\frac{1}{4}\,B_F(\varphi)\,f_{\mu\nu}\,f^{\mu\nu} -V(\phi)\,
+\frac{1}{2}\,\partial_\mu\,\varphi\,\partial^\mu
\,\varphi \, , \label{lll}
\ee
where we have also introduced the gauge kinetic function 
$B_F(\varphi) \equiv \epsilon^{-2}$, and allowed it to be a generic function of the field $\varphi$.  
In the Bekenstein model, $B_F(\varphi) = \exp(-2\,\varphi/\sqrt{\omega})$.

\subsection{Equations of Motion}
By varying the action with respect to the complex conjugate of
$\phi$, i.e., $\phi^\star$, one gets
\be\label{eqphi11}
D_\mu\,D^\mu\,\phi\,=\,-\frac{\partial\, V}{\partial\,\phi^\star}\,.
\ee
Variation with respect to $a_\mu$ leads to:
\be\label{eqfmnw}
\partial_\nu\left[B_F(\varphi)\,f^{\mu\nu}\right]\,=\,j^\mu\,
\ee
with the current $j^\mu$ defined as
\be
j^\mu = i\,e_0\,\left[\phi\,\left(D^\mu\,\phi\right)^\star
-\phi^\star\,\left(D^\mu\,\phi\right)\right].
\ee
Note that for $B_F$ equal to an arbitrary constant, Eq. (\ref{eqfmnw}) reduces to Eq. (\ref{fmn0}) with an effective charge $e = e_0/\sqrt{B_F}$.

Finally, variation with respect to $\varphi$ gives
\be\label{eqbox11}
\partial_\mu\,\partial^\mu\,\varphi\,=\,-\frac{1}{4}\,
\frac{\partial B_F(\varphi)}{\partial\, \varphi}\, f^2\,.
\ee
\subsection{The Ansatz}
Searching for soliton solutions in this theory, we assume the \textit{ansatz} proposed by Nielsen and Olesen \cite{Nielsen},
with $\phi$ and $a_\mu$ written as 
\ben
\phi &=& \chi(r)\,e^{i n \theta}\, , \label{n}\\
a_\theta  &=& a(r)\, , \label{bansatz}
\een
where $\chi(r)$ and $a(r)$ are real functions of $r$ and
all other components of $a_\mu$ are set to zero. 

Substituting the ansatz given in (\ref{n}-\ref{bansatz}) into 
Eqs. (\ref{eqphi11}--\ref{eqbox11}) one gets the equations of motion
\ben
&&\frac{1}{r}\frac{d}{dr}\left(r\frac{d\chi}{dr}\right)-
\left[\left(\frac{n}{r}-e_0\,a\right)^2
-\frac{\eta^2\,\lambda}{2}+\frac{\lambda}2 \chi^2\right]\chi=0, \label{pri}\\
&&\frac{d}{dr}\left(B_F\frac{1}{r}\frac{d}{dr} \left(r a\right)\right)
+2e_0\left(\frac{n}{r}-e_0a\right)\chi^2=0,\label{seg}\\
\label{eqvarphi}
&&-\frac{1}{r}\frac{d}{dr}\left(r\frac{d\varphi}{dr}\right)+\frac{1}{2}
\frac{dB_F(\varphi)}{d\varphi}
\left(\frac{1}{r}\frac{d}{dr} \left(ra\right)\right)^2=0,\label{ter}
\een
where we have taken into account that 
\be
f^2 = 2 f^{r\theta} \, f_{r\theta} = 2 \left[ \frac1{r} \frac{d}{dr}(ra) \right]^2.
\label{error}
\ee
\subsection{The Energy Density}
We can also investigate
the dependence of the energy density on the radial coordinate $r$.
For static strings the stress-energy tensor takes a diagonal
form with the energy density of the vortex being
$\rho = T^0_0 = g^{00} T_{00}=T_{00} $,
with
\be
\rho =  \left( \frac{d \chi}{dr}  \right)^2 + \left( \frac{d \varphi}{dr}  \right)^2 + \frac{1}{2\,r^2} B_F \left( \frac{dv}{dr}  \right)^2 +  \left( \frac{n - e_0 v}{r} \right)^2 \chi^2  + \frac{\lambda}4 
\left( \chi^2 - \eta^2 \right)^2,
\ee
where  $v \equiv a\,r$, while the spatial components of the stress-energy tensor are given
by $T^{i}_{j} = diag(p_r, p_\theta, p_z)$ with $p_z = - \rho$. 
Therefore, the energy density of the
vortex, $\rho$, is everywhere positive, while the longitudinal pressure $p_z$
is negative. In fact, this is also one of the defining 
features of standard cosmic strings.
\section{\label{csnum}Numerical Implementation of the Equations of Motion}
We look for solutions to the coupled non-linear equations for a static straight string.
We first reduce
equations (\ref{pri}-\ref{ter}) to a set of first order
differential equations for numerical implementation.

Let us introduce three new variables
\ben
\frac{d\,\chi}{dr}&=&  \xi\, , \label{Ia}\\
\frac{d\,v}{dr}&=& b\,r \, ,\label{IIa}\\
\frac{d\varphi}{dr}&=&s \, .\label{IIIa}
\een
Then by substituting (\ref{Ia}-\ref{IIIa}) into
equations (\ref{pri}-\ref{ter}), one gets
\ben
\label{IV}
\frac{d\xi}{dr} &=&-\frac{\xi}{r}+\left[\left(\frac{n-e_0\,v}{r}\right)^2
-\frac{\eta^2\lambda}2+\frac{\lambda}2\,\chi^2\right]\chi\,, \\
\label{V}
\frac{d\,b}{dr}\, &=&\,\frac{1}{B_F}\,\left[-\frac{dB_F}{d\varphi}\,s\,b
-2e_0\left(\frac{n-e_0\,v}{r}\right)\,\chi^2\right]\,, \\
\label{VIaa}
\frac{d\,s}{dr}\,&=&\,-\frac{s
}{r}+\frac{1}{2}\,\frac{dB_F}{d\varphi}\,b^2\,.
\een
One has a set of six ordinary first order differential
equations, which requires, at least, six boundary conditions.

\subsection{Boundary Conditions}

Far away from the core the
field $\chi$ must be equal to $\eta$ (we take $\eta=1$). 
Consequently, substituting this condition in Eq. (\ref{pri}) one 
immediately sees that $v(r)=n/e_0$ far away from the core ($r \to + \infty$). 
In addition, there are four boundary conditions at the string core. As the phase $\theta$ is undefined for $r=0$, we impose that $\chi$ must vanish at that point. Also, we set $v=0$ at the center of the string. Otherwise, the magnetic energy density would diverge at $r=0$.
Normalizing the electric charge such that $e = e_0$ at the origin, we impose that the gauge kinetic function is equal to unity at $r=0$. Finally, by using the Gauss law 
to solve equation 
(\ref{ter}) assuming that there are no sources of $\alpha$ variation 
other than the string, one has that $s=0$ at the string core.

In summary, the appropriate boundary conditions are
\be
\label{bca1}
\lim_{r \rightarrow 0} \chi(r) = 0,\,\,\,\,\,\,\,\,\,\,
\lim_{r \rightarrow \infty} \chi (r) =  1\, ,
\ee
\be
\label{bca2}
\lim_{r \rightarrow 0} v (r) = 0,\,\,\,\,\,\,\,\,\,\,\,
\lim_{r \rightarrow \infty} v(r) =  \frac{n}{e_0} , 
\ee
\be
\label{bca3}
\lim_{r \rightarrow 0} B_F (r) = 1,\,\,\,\,\,\,\,\,\,\,\,
\lim_{r \rightarrow 0} s(r) =  0\, . 
\ee
Therefore we have a two point boundary value problem with four conditions
at the origin and two conditions far from the core. 
\subsection{Numerical Technique}
In order to solve this problem numerically we used the relaxation
method which replaces the set of six ordinary differential equations by
finite-difference equations on a mesh of points covering the range
of the integration. This numerical
method is very efficient if a good initial
guess is supplied. In our case, the solutions of the standard Nielsen-Olesen
vortex were used to generate a good initial guess.
We checked that our code reproduces the results for the standard 
Nielsen-Olesen vortex if $B_F = 1$. 

The results that we will describe here are obtained by assuming in the numerical implementation that $\lambda = 2$ and $n = 1$ for definiteness and use units in 
which $\eta^2=1$. We also emphasize that although the Principle Equivalence tests constrain $dB_F/d\varphi$
to be very small, we will often use larger values in order to better investigate the effects associated with $\alpha$ variability.

\subsection{Exponential coupling}

The general prescription detailed above can be particularized to specific
choices of gauge kinetic function. 
First, let us consider the particular case  
\be
B_F(\varphi)\,=\,e^{-\frac{2\varphi}{\sqrt{\omega}}}\,, \label{expoa}
\ee
where $\omega$ 
is a coupling constant. Actually, this case corresponds to the original Bekenstein model described by Eq. (\ref{laga23}).
\begin{figure}
\begin{center}
\includegraphics*[width=9cm]{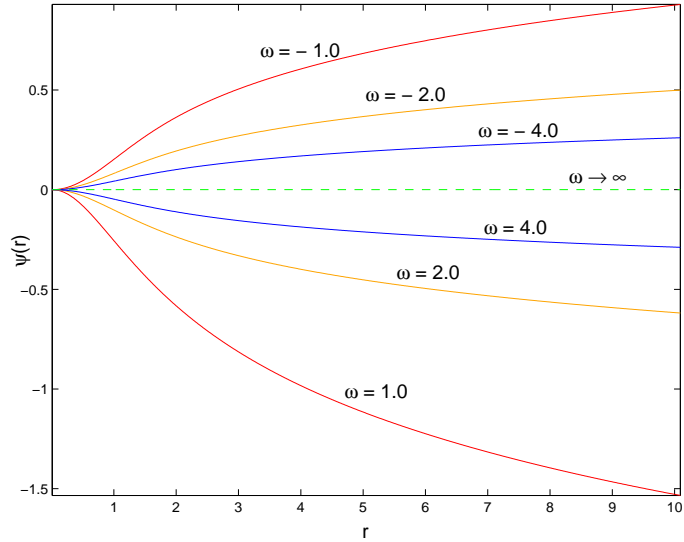}
\end{center}
\caption[Varying-$\alpha$ strings: $\psi(r)$ for exponential $B_F$.]{The numerical solution of the scalar field 
$\psi \equiv \ln{\epsilon}$ as a function of distance, $r$, 
from the core of string, for the original Bekenstein model. 
If $\omega < 0$ then $\epsilon \to \infty$ when $r \to \infty$.
On the other hand, if $\omega > 0$ then $\epsilon \to 0$ when $r \to \infty$.
The dashed line represents the constant-$\alpha$ theory, which corresponds 
to the limit $ \omega \rightarrow \infty$.} \label{variavel}
\end{figure}

It is easy to show that in the limit $\omega\,\rightarrow\,\infty$ one 
recovers the Nielsen-Olesen vortex with constant $\epsilon$. Although 
the gauge kinetic function in Eq. (\ref{expoa}) is only well defined for 
$\omega > 0$ the model described by the Lagrangian density in 
Eq. (\ref{laga23}) allows for both negative and positive values 
of $\omega$. 
However, note that if $\omega <0$ the energy density is no longer positive 
definite. 

In  Fig. \ref{variavel} we plot the numerical solution of the scalar field 
$\psi \equiv \ln{\epsilon}$ as a function of distance, $r$, 
from the core of string, in the context of the original Bekenstein model. 
Note that if $\omega < 0$ then $\epsilon$ 
diverges asymptotically away from the string core. On the other 
hand if $\omega > 0$ then $\epsilon$ goes to zero when $r  \to \infty$. 
In the large $\omega$ limit the curves for positive and negative 
$\omega$ are nearly symmetric approaching the dashed line representing 
the constant-$\alpha$ model when $\omega \rightarrow \infty$.

\subsection{Polynomial coupling}
Now let us move to a more general case where the gauge kinetic function takes a polynomial
form
\be
\label{polipoli}
B_F(\varphi)=1.0+\sum_{i=1}^{N}\,\beta_i\,\varphi^i \, ,
\ee
where $\beta_i$ are dimensionless coupling constants and $N$ is an integer.
We have solved the equations of motion assuming the gauge kinetic function given in Eq. (\ref{polipoli}), and 
have verified some important properties.

First, we note that if $\beta_1=0$ the classical Nielsen-Olesen
vortex solution with constant $\alpha$ is still a valid solution.  
This means that there is a class of gauge kinetic functions for which the
classical static solution is maintained despite the modifications to the model. It can be easily verified by substituting the gauge kinetic function (\ref{polipoli}) in Eqs. (\ref{pri}--\ref{ter}) and taking $\beta_1=0$, that the standard Nielsen-Olesen vortex solution is recovered.

Substituting the gauge kinetic function in Eq. (\ref{ter})
one gets
\be\label{oddeven}
\frac{1}{r}\frac{d}{dr}\left(r\frac{d\varphi}{dr}\right)=
\frac{b^2}{4}\left(\sum_{k=1}^{N}(2k-1)\beta_{2k-1}\varphi^{2k-2}\right)\,
+ \frac{b^2}{4}\left(\sum_{k=1}^{N}(2k)\beta_{2k}\varphi^{2k-1}\right),
\ee
which shows that 
the transformation $\beta_i \rightarrow -\beta_i$ for odd $i$
modifies the sign of the solution of $\varphi(r)$ without changing 
$\chi$ or $b$ since $B_F$ is kept invariant.
\begin{figure}
\begin{center}
\includegraphics*[width=9cm]{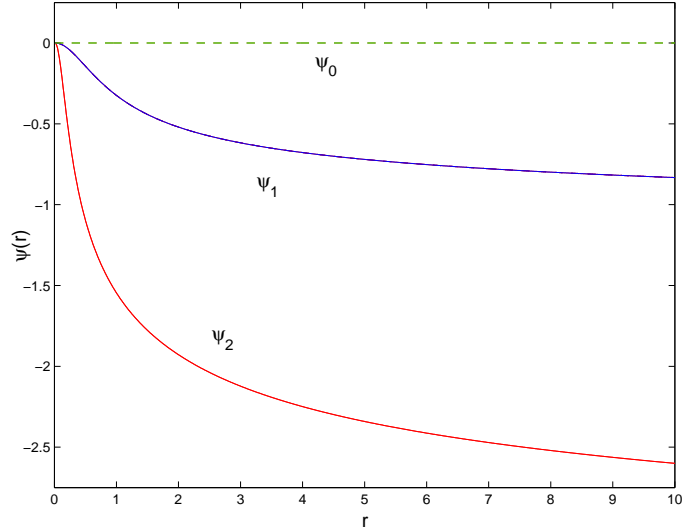}
\end{center}
\caption[Varying-$\alpha$ strings: $\psi(r)$ for polynomial $B_F$.]{The numerical solution of the scalar field 
$\psi \equiv \ln \epsilon$ as a function of distance, $r$, 
from the core of string, for a polynomial gauge kinetic function. 
Models 0, 1 and 2 are defined by $\beta_1=0$ ($\beta_2$ arbitrary),  
$\beta_1=-3,\,\beta_2=0$ (linear coupling) and 
$\beta_1=-5,\,\beta_2=10$ respectively.  
}
\label{bfs}
\end{figure}
We will see that both for $\beta_1 \, >\,0$ and $\beta_1 \, <\,0$ the 
behavior 
of $\psi \equiv \ln \epsilon$  is similar to that of the original Bekenstein 
model described by the Lagrangian density in Eq. (\ref{laga23}), 
with $\omega\,>\,0$, in particular in the 
limit of small $|\beta_1|$/large $\omega$.
In fact a polynomial expansion of the exponential gauge kinetic function of 
the original Bekenstein model has 
$\beta_1=-2/{\sqrt \omega}$. 
This relation between the
models arises from the fact that $B_F$ given in Eq. (\ref{expoa})
can be expanded in a series of powers of $\varphi$ according to 
\be
\beta_i = \frac{(-2)^i}{w^{i/2} i!}\,.
\ee
On the other hand, as mentioned before, if $\beta_1\,=\,0$ one recovers 
the standard result for the Nielsen-Olesen vortex with constant-$\alpha$, 
irrespective of the chosen values of $\beta_i$ for $i > 1$.

We have studied the behavior of the solutions of Eqs. (\ref{Ia}--\ref{VIaa}) 
for various values of $N$ but for simplicity we shall only consider 
$N \leq 2$. In particular, we consider 
\be
B_F = 1.0 + \beta_1 \varphi + \beta_2 \varphi^2, \label{nosso}
\ee
with two free parameters.
\begin{figure}
\begin{center}
\includegraphics*[width=9cm]{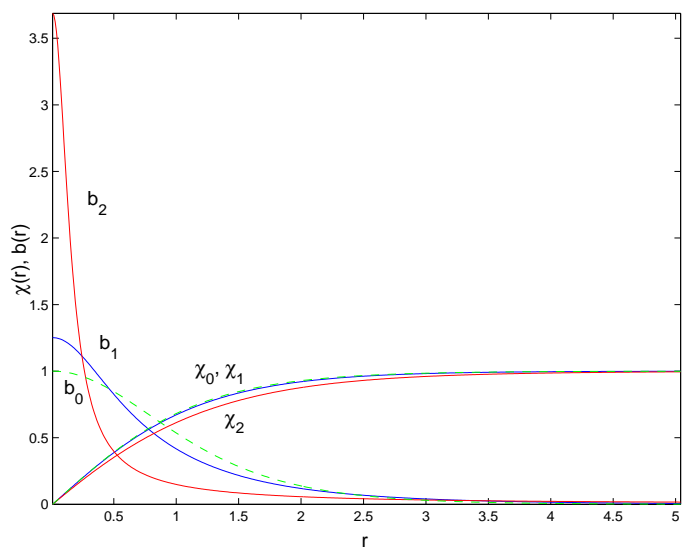}
\end{center}
\caption[Varying-$\alpha$ strings: $\chi(r)$ and $b(r)$ for polynomial $B_F$.]{
The numerical solution of the fields $\chi(r)$ and $b(r)$ as a function of 
distance, $r$, from the core of string, for models $0$, $1$ and $2$. Note 
that the change in $b(r)$ with respect to the standard 
constant-$\alpha$ result is much more dramatic than the change in $\chi(r)$.}
\label{final1}
\end{figure}
In Fig. \ref{bfs} we plot the numerical solution of the scalar field 
$\psi \equiv \ln \epsilon$ as a function of distance, $r$, 
from the core of string, for a polynomial gauge kinetic function. 
Models $1$ and $2$ are defined by $\beta_1=-3,\,\beta_2=0$ (linear coupling) and 
$\beta_1=-5,\,\beta_2=10$ respectively. 
Model 0 (dashed line) represents any model with 
$\beta_1 = 0$ and has $\alpha= {\rm constant}$. Note that the replacement 
$\beta_1 \to -\beta_1$ does not modify the solution for $\psi$.
\begin{figure}
\begin{center}
\includegraphics*[width=9cm]{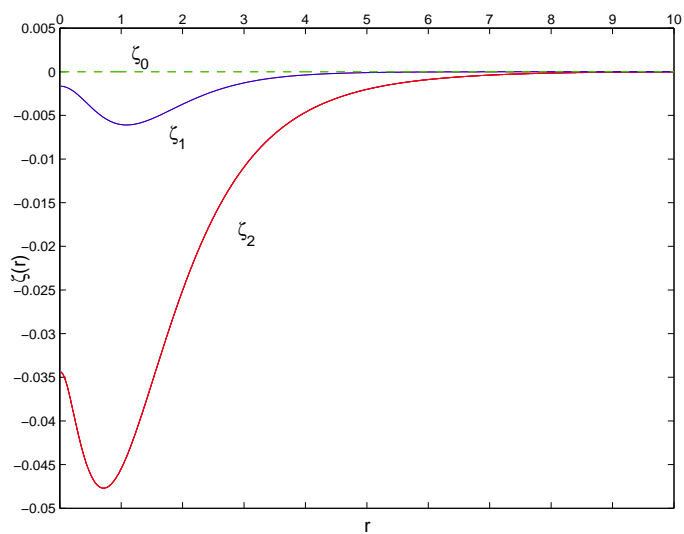}
\end{center}
\caption[Varying-$\alpha$ strings: $\zeta(r)$ for polynomial $B_F$.]{Plot of $\zeta_i(r)=\log(\chi_i/\chi_0)$ for the different 
polynomial gauge kinetic functions. Note that although this was not very 
visible in Fig. \ref{final1}, even a small value of 
$\beta_1$ leads to a change of the vortex solution.}
\label{final4}
\end{figure}
In Fig. \ref{final1} we plot the numerical solution of the fields 
$\chi(r)$ and $b(r)$ as a function of distance, $r$, from the core of string, 
for models $0$, $1$ and $2$. We see that the change in $b(r)$ with respect 
to the standard constant-$\alpha$ result is much more dramatic than the 
change in $\chi(r)$. In order to verify the modification to $\chi(r)$ in more 
detail we define the function
\be
\zeta_i(r) = \log\left( \frac{\chi_i}{\chi_0}\right)\label{zeta}
\ee
and plot in Fig. \ref{final4} the results for the different models. 
We clearly 
see that even a small value of $\beta_1$ leads to a modification of the vortex 
solution with respect to the standard Nielsen-Olesen solution.
\begin{figure}
\begin{center}
\includegraphics*[width=9cm]{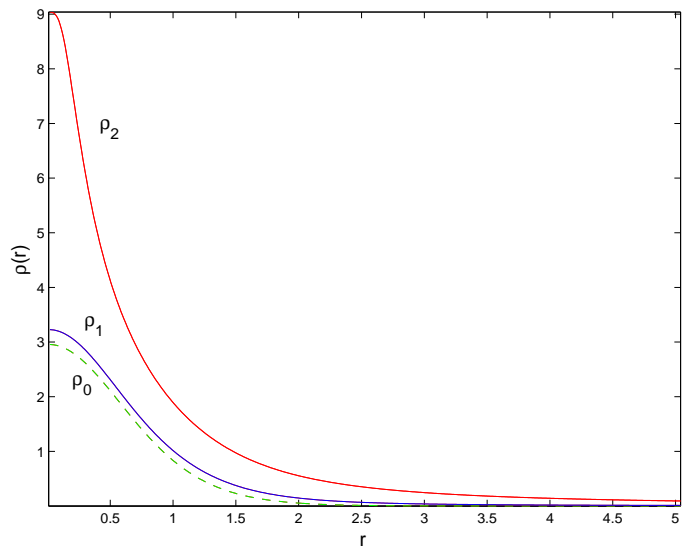}
\end{center}
\caption[Varying-$\alpha$ strings: $\rho(r)$ for polynomial $B_F$.]{The energy density as a function of the 
distance, $r$ from the string core for models 0, 1 and 2. 
The dashed line represents the constant-$\alpha$ model
These results clearly show an increase of the energy density due to the 
contribution of the extra field $\varphi$.}
\label{energy}
\end{figure}
Finally, we have also studied the behavior of the energy density in this
model. In fact, since the fine structure constant varies we have a new 
contribution due to the field $\varphi$, to the total energy of the 
topological 
defect. As has been previously discussed in ref. \cite{Magueijo} the 
contributions to the energy density of the string can be divided into two 
components. One component is localized around the string 
core (the local string component), the other is related to the 
contribution of the kinetic term associated 
with the spatial variations of the fine structure constant and is not 
localized in the core of the string. The energy profile of this last 
contribution is analogous to that of a global string, 
whose energy per unit length diverges asymptotically far from the core. 
In Fig. \ref{energy} we plot the 
string energy density 
as a function of the distance, $r$ from the string core for models $0$, $1$ 
and $2$. 
The dashed line represents the constant-$\alpha$ model. We clearly see an 
increase of the energy density due to the contribution of the extra field 
$\varphi$.

\section{\label{cscc}Cosmological Consequences of Varying-$\alpha$ Strings}
In the previous section we studied static solutions of varying-$\alpha$ cosmic strings.
Now, let us
investigate whether such cosmic string networks can induce measurable
space-time variations of $\alpha$ in a cosmological setting. In order
to answer this question we recall that $\varphi$
satisfies the Poisson equation given by
\be
\nabla^2 \varphi = \frac{1}{4}\, 
\beta\, f^2\,.\label{poisson111}
\ee
In Eq. (\ref{poisson111}) we have assumed for simplicity that the gauge 
kinetic function is a linear function of $\varphi$, i.e.,
\be
\label{lineline}
B_F(\varphi) = 1 + \beta \varphi.
\ee

The variation of the fine structure constant away from the
string core is then 
\be
2 \pi r \frac{d \varphi}{dr}=  \beta I(r)  \mu (r_{\rm max})\,,
\ee
where 
\be
\mu(r)=2 \pi \int^r_0 \rho(r') r' dr'\,,
\ee
and 
\be
I(r)=\frac{\pi}{2\mu (r_{\rm max})} \int_0^r f^2(r') r' dr'\,,
\ee
is a function of $r$ smaller than unity. Here $r_{\rm max}$ represents a 
cut-off scale which is in a cosmological context of the order of the string 
correlation length. Far away from the string core $I(r)$ is a slowly 
varying function of $r$ which is always smaller than unity. An approximate 
solution for the behavior of the field $\varphi$ may be obtained by taking 
$I(r) \sim {\rm const}$
\be
\varphi \sim  \frac{\beta I  \mu(r_{\rm max})}{2\pi} \ln \left(\frac{r}{r_0}\right)\,,
\ee
where $r_0$ is an integration constant. Since $\epsilon=B_F(\varphi)^{-1/2}$ 
we have 
\be
\epsilon \sim 1-\frac{\beta^2 I  \mu(r_{\rm max})}{4\pi} \ln \left(\frac{r}{r_0}\right)\,.
\label{epsilonrrrr}
\ee
We see that the variation of the fine structure constant away from the
string core is proportional to the gravitational potential induced by
the strings. 
\subsection{The upper limit to the variation of $\alpha$}
In order to
compute an upper limit on the variation of the fine-structure constant in the vicinity of a cosmic string, we
first recall that 
the value of $G \mu$ is constrained to be small
($\lesssim 10^{-6}-10^{-7}$) in order to avoid conflict with CMB and LSS
results \cite{Durrer,mu1,Wu1,mu2} (or even smaller depending on the 
decaying channels available to the cosmic string network \cite{mu3}). 
We stress that these constraints are for standard local strings described by the 
Nambu-Goto action. Even though the cosmic strings studied here 
are non-standard, having a local and a global component, we expect the 
limits on $G\mu(r_{\rm max})$ to be similar as those on $G \mu$ for 
standard local strings.

On the other hand, the factor $\beta$ 
appearing in
Eq. (\ref{epsilonrrrr}) is constrained by equivalence principle tests to be 
$|\beta| < 10^{-3} \, G^{1/2}$ \cite{Will,OlivePos}.  

Hence, 
taking into 
account that we cannot observe scales larger than the horizon 
($\sim \, 10^4 \, {\rm Mpc}$) and are unlikely to 
probe variations of $\alpha$ at a distance much smaller than $1 \, {\rm pc}$ 
from a cosmic string, Eq. (\ref{epsilonrrrr}) implies that a conservative 
overall limit on observable variations of $\alpha$ seeded by cosmic strings 
is 
\be
\frac{\Delta \alpha}{\alpha} \lesssim 10^{-12}
\ee
which is too small to have any 
significant cosmological impact. We thus see that, even 
allowing for a large contribution coming from the logarithmic factor in 
Eq. (\ref{epsilonrrrr}), the spatial variations of $\alpha$ induced by such 
strings are too small to be detectable.

\section{\label{csdis}Conclusions}

We have investigated cosmic strings in the context of varying-$\alpha$ models 
considering a generic gauge kinetic 
function. 
We showed that there
is a class of models of this type for 
which the classical Nielsen-Olesen vortex is still a valid solution. We also found that
the spatial variations of $\alpha$ (sourced by 
the electromagnetic energy concentrated along the string core)  
are roughly proportional to the gravitational potential induced by the strings
which is constrained to be small. However, we showed that
Equivalence Principle constraints impose tight limits on the allowed
variations of $\alpha$ on cosmological scales induced by
cosmic string networks of this type. 

\chapter{\label{cap4}{\sf Varying-$\alpha$ Magnetic Monopoles}}

\section{Overview}

In the previous chapter we have studied 
a model where spatial variations of 
the fine-structure constant were induced by cosmic strings.
In this chapter we study 
another class of topological defects, local magnetic monopoles, which  
can arise in the context of non-Abelian field theories with $SU(2)$ spontaneous symmetry breaking 
\cite{thooft}--\cite{forgacs}. 
In the context of varying-$\alpha$ models, the  
electromagnetic energy in their core may also induce spatial variations of $\alpha$ \cite{Nosso2}.

We recall that the \textit{monopole problem} of the standard cosmological model can be solved by considering, for instance, the inflationary universe \cite{Guth}. Inflation introduces a period of exponential expansion of the universe, and all of the present observed universe arises from a tiny region which was initially smaller than the causal horizon. As a result, not only monopoles, but all defects formed before the inflationary period, are diluted by an enormous factor. After inflation the universe thermalizes at some temperature and some monopole-antimonopole pairs may be produced by thermal fluctuations, but their density is supressed by a very large value \cite{Lazarides,Turner}. However, in some models the predicted monopole abundance produced at thermalization is acceptably small, but still detectable \cite{Preskill}.
A number of other theories were elaborated in order to solve the monopole problem. Among them, there is the possibility that
monopoles and anti-monopoles are linked by strings \cite{Langacker} which can also lead to the annihilation of magnetic monopoles.

In this chapter
we start by solving the standard problem for a fixed $\alpha$ and 
recover the classical solution presented by 't Hooft and Polyakov 
\cite{thooft,polyakov} (see for instance Refs. \cite{Vilenki,Manton,forgacs,kirkman}). 
Then we proceed to investigate other solutions where $\alpha$ varies in space, comparing them with the 
standard one.
Finally, 
we find the 
upper limits to the variation of $\alpha$ in their vicinity.

This chapter is organized as 
follows. In Sec. \ref{bmna} we introduce Bekenstein-type models in Yang-Mills theories 
and obtain the equations of motion for a static magnetic monopole. We write the 
energy density of the monopole and show that the standard 
electromagnetism is recovered outside the core if $\alpha$ is a constant. In. Sec. \ref{nimm} 
we present the 
numerical technique applied to solve the equations of motion and give 
results for several choices of the gauge kinetic 
function. In Sec. \ref{cvmm} we study the limits imposed by the Equivalence Principle 
on the allowed variations of $\alpha$ in the vicinity of the monopole, as a 
function of the symmetry breaking scale. Finally, in Sec. \ref{cccmm}, we summarize 
and discuss the results.

\section{\label{bmna}Varying-$\alpha$ in Non-Abelian Field Theories}

In this section we describe Bekenstein-type models in the context of non-Abelian Yang-Mills theories \cite{Magueijona}.
We shall consider the non-Abelian gauge field theory described by the Lagrangian density
\be
\label{laga31}
\mathcal{L} = D_\mu \Phi^a D^\mu \Phi^a - \frac14 F^a_{\mu\nu} F^{a\mu\nu} + V(\Phi^a),
\ee
where $\Phi^a$ with $a=1,2,3$ is a triplet whose elements are scalar fields and $V(\Phi^a)$ is the potential.
The covariant derivatives are 
\be
D_\mu \Phi^a = \partial_\mu \Phi^a - e \epsilon^{abc} A^b_\mu \Phi^c
\ee
and the electromagnetic strength tensor has the form
\be
\label{fmnaa}
F^a_{\mu\nu}=\partial_\mu A^a_\nu - \partial_\nu A^a_\mu + e \epsilon^{abc} A^b_\mu A^c_\nu,
\ee
where $A_\mu^a$ is an isovector, $\epsilon^{abc}$ is the Levi-Civita tensor and $e$ is the electric charge. 

Analogously
to the Mexican hat potential $V(\phi)$ taken for the string case in Eq. (\ref{amhp}), we assume the potential
\be
V(\Phi^a) = \frac{\lambda}{4}\,\left(\Phi^a\,\Phi^a-\eta^2\right)^2,
\ee
where
$\lambda > 0$ is the coupling of the scalar self-interaction, and $\eta$ is the vacuum
expectation value of the Higgs field.
\subsection{Bekenstein-type models}
Now we proceed as in Chapter \ref{cap3} and assume that the electric charge is a function of the space-time coordinates, i.e., $e = \epsilon\,e_0$, where $\epsilon$ is a real scalar field. 
In this case, 
the  
Lagrangian density can be written as
\be
\mathcal{L}\,=\, \frac12\,\left(D_\mu\,\Phi^a \right)\left(D^\mu\,\Phi^a\right) - V(\Phi^a) 
-\frac{B_F(\varphi)}{4}\,f_{\mu\nu}^{a}\,f^{a\mu\nu}
+\frac12\,\partial_\mu\,\varphi\,\partial^\mu\,\varphi, \label{laga}
\ee
where $a_\mu^a=\epsilon\,A_\mu^a$ is an auxiliary gauge field and $\varphi$ is a real scalar field.

The covariant derivatives are now
\be
D_\mu \Phi^a = \partial_\mu \Phi^a + e_0\,\epsilon^{a\,b\,c} a_\mu^b\,\Phi^c \label{cd}
\ee
and the non-Abelian gauge field strength is given by
\be
f_{\mu\nu}^a \equiv \,\epsilon\,F_{\mu\nu}^a=\partial_\mu a_\nu^a - \partial_\nu a_\mu^a + e_0 \epsilon^{a\,b\,c}\,a_{\mu}^b a_{\nu}^c, \label{gfs}
\ee
where $F_{\mu\nu}^a$ is defined in Eq. (\ref{fmnaa}).
We recall that
$B_F(\varphi) \equiv \epsilon^{-2}$ is the gauge kinetic function which acts as the effective dielectric 
permittivity and can phenomenologically be taken as an arbitrary function of $\varphi$.
 
Note that the Lagrangian density in Eq. (\ref{laga}) is invariant 
under SU(2) gauge transformations of the form 
\ben
\delta\, \Phi^a &=& \epsilon^{a\,b\,c} \Phi^b \Lambda^c,\\
\delta\, a_\mu^a&=& \epsilon^{a\,b\,c} a_{\mu}^b \, \Lambda^c + e_0 \partial_\mu\,\Lambda^a,
\een
where $\Lambda^a$ is a generic isovector. 
This symmetry is broken down to $U(1)$ because there is a non vanishing expectation value of the Higgs field. Thus the two components of the vector field develop a mass $M_W = \eta\,e_0$, while
the mass of the Higgs field is $M_H = \eta\,\sqrt{2\,\lambda}$.

It is convenient to define the dimensionless ratio
\be
\label{dra}
\zeta\,=\,\frac{M_H}{M_W}\,=\frac{\sqrt{2\,\lambda}}{e_0},
\ee
and to rescale the radial coordinate by $M_W$, so that distance is expressed in units of $M_W^{-1}$.
\subsection{Equations of Motion}
Varying the action with respect to the adjoint Higgs field $\Phi^{a\dag}$ one gets
\be\label{eqphi}
D_\mu\,D^\mu\,\Phi^a \,=\,-2\,\lambda\,\Phi^a\,\left(\Phi^b\,\Phi^b - \eta^2 \right).
\ee
Variation with respect to $a_\mu^a$ leads to
\be\label{eqfmn}
D_\nu\,\left[ B_F(\varphi) f^{a\,\mu\nu} \right] = j^{a\nu}
\ee
with the current $j^{a\mu}$ defined as
\be
j^{a\mu} =  e_0\,\epsilon^{a\,b\,c} \,\Phi^b\, (D^\mu\,\Phi^c).
\ee
We stress that if $B_F$ is a constant the standard theory is recovered
with an effective electric charge $e = e_0/\sqrt{B_F}$.
 
Finally, variation with respect to $\varphi$ gives
\be\label{eqbox}
\partial_\mu\,\partial^\mu\,\varphi\,=\,-\frac{1}{4}\,
\frac{\partial B_F(\varphi)}{\partial\, \varphi}\, f^2\\
\ee
with $f^2\,=\,f^a_{\mu\nu}\,f^{a\,\mu\nu}$.

\subsection{The ansatz}
We are now interested in static, spherically symmetric, magnetic monopole solutions. 
Hence we take the $``$hedgehog$"$ ansatz 
\ben
\Phi^a(r)&=&X(r)\,\frac{x^a}{r},\label{Hig}\\
a_0^a(r)&=&0\label{a0},\\
a_i^a(r)&=&\epsilon_{iak}\,\frac{x_k}{e_0\,r^2}\,[W(r)-1],\label{ai}
\een
where $x^a$ are the Cartesian coordinates and $r^2=x^k\,x_k$. $X(r)$ and $W(r)$ are dimensionless radial functions which
minimize the self-energy, i.e., the mass of the monopole
\ben
E&=&\frac{4\pi\eta}{e_0}\int_{0}^{\infty}dr \Big\{\frac{r^2}{2}\left(\frac{dX}{dr}\right)^2 + X^2 W^2 + \frac{\zeta^2 r^2}{8}(1-X^2)^2 \nn\\
&+&B_F\left[\left(\frac{dW}{dr}\right)^2 + \frac{(1-W^2)^2}{2r^2}\right] + \frac{r^2}{2}\left(\frac{d\varphi}{dr}\right)^2 \Big\}, \label{act}
\een
where the coordinate $r$ and the functions $X$ and $\varphi$ have been rescaled as
\be
r\,\rightarrow\,\frac{r}{e_0\,v},\,\,\,\,\,\,\,\,\,\,X\,\rightarrow\,v\,X,\,\,\,\,\,\,\,\,\,\,\varphi\,\rightarrow\,v\,\varphi.
\ee
So that $r$, $X$ and $\varphi$ are now dimensionless.

It will prove useful to compute the energy density which is given by
\ben
\rho &=& \frac{\eta}{e_0} \Big\{ B_F\left[\frac1{r^2}\,\left(\frac{dW}{dr} \right)^2 + \frac12 \left(\frac{1-W^2}{r^2}\right)^2 \right]\nn \\
&+& \frac12 \left(\frac{dX}{dr}\right)^2  + \left(\frac{WX}{r} \right)^2  + \frac{\zeta^2}{8}(1-X^2)^2 + \frac12\,\left(\frac{d\varphi}{dr}\right)^{2} \Big\}. \label{ro}
\een
Substituting the ansatz given in (\ref{Hig}--\ref{ai}) into the equations of motion
(\ref{eqphi}--\ref{eqbox}) one obtains
\ben
&&\frac1{r^2}\,\frac{d}{dr}\left(r^2\,\frac{dX}{dr}\right)- \left[\frac{\zeta^2}{2}(1-X^2)+\frac{2W^2}{r^2}\right]\,X=0, \label{one}\\
&&\frac{d}{dr}\left(B_F\, \frac{dW}{dr}\right)-\left[\frac{B_F}{r^2}(1-W^2) + X^2 \right]W=0, \label{two}\\
&&\frac1{r^2} \frac{d}{dr}\left(r^2 \frac{d\varphi}{dr} \right)-\frac{d\,B_F}{d\varphi}\left[\left(\frac{dW}{dr}\right)^2+ \frac12\left(\frac{1-W^2}{r^2}\right)^2\right]=0.\label{three}
\een
\subsection{The electromagnetic tensor outside the monopole}
Defining
\be
f_{\mu\,\nu}=\frac{\Phi^a}{|\Phi|}\,f_{\mu\,\nu}^{a}+\frac1{e_0\,|\Phi|^3}\,\epsilon^{a\,b\,c}\,\Phi^a\,(D_\mu\Phi^b)(D_\nu\Phi^c) \label{fmn}
\ee
and choosing the gauge where $\Phi^a=\delta^{a\,3}\,|\Phi|$, i.e., one gets
\be
f_{\mu\nu}=\partial_\mu\,a_\nu^3-\partial_\nu\,a_\mu^3. \label{f3}
\ee
By writing $a_\mu^3 = a_\mu$, one identifies (\ref{f3}) with the usual electromagnetic tensor which for 
constant $\varphi$ satisfies the ordinary Maxwell equations everywhere except 
in the region where $X \sim 0$.

\section{\label{nimm}Numerical Implementation of the Equations}
In order to solve the equations of motion numerically, we reduce them to a set of first order equations. For that purpose we define the variables
\ben
&&\frac{dX}{dr}=V, \label{varr1}\\
&&\frac{dW}{dr}=U, \label{varr2}\\
&&\frac{d\varphi}{dr}=s. \label{varr3}
\een			
Equations (\ref{one}-\ref{three}) then can be written as
\ben
\label{IV}
\frac{dV}{dr}&=&-\frac{2V}{r}+\frac{\zeta^2}{2}\,X(X^2-1)+\frac{2\,W^2\,X}{r^2},\\
\label{V}
\frac{dU}{dr}&=&\frac{1}{B_F}\left[\frac{dB_F}{d\varphi}\,s\,U
+\frac{B_F}{r^2}(W^2-1)W+ X^2\,W \right],\\
\label{VI}
\frac{ds}{dr}&=&\,-\,\frac{2\,s}{r}+\frac{dB_F}{d\varphi}\left[U^2+\left(\frac{1-W^2}{r^2}\right)^2\right],
\een
which require at least six boundary conditions.

At far distances from the core ($r \rightarrow \infty$), the Higgs field $X(r)$ falls off to its vacuum value, $X=1$. Using this boundary condition in Eq. (\ref{one}), one gets that $W(r)$ must vanish far from the core. On the other hand, 
since the symmetry at the core is not broken, then the Higgs field vanishes at $r=0$ and from the regularity of the energy-momentum tensor, $W=1$ at $r=0$.
The other two boundary conditions come from the normalization of the electric charge at the origin. At the core $e=e_0$, which means that the gauge kinetic function is equal to unity. Finally, by using the Gauss law to solve the Eq. (\ref{three}) without sources of $\alpha$ variation other than the monopole, one gets that $s$ must vanish at $r=0$. 
Therefore, we have at all six boundary conditions, four of them at origin and the other two far away from the core:
\ben
&&\lim_{r \rightarrow 0} X(r) = 0,\,\,\,\,\,\,\,\,\,\,
\lim_{r \rightarrow 0} W(r) = 1,  \label{condi1}\\
&&\lim_{r \rightarrow 0} B_F (r) = 1,\,\,\,\,\,\,\,\,\,\,\,
\lim_{r \rightarrow 0} s(r) =  0, \label{cond2}\\
&&\lim_{r \rightarrow \infty} X(r) = 1,\,\,\,\,\,\,\,\,\,\,\,
\lim_{r \rightarrow \infty} W(r) = 0.  \label{cond3}
\een
As the boundary conditions are at different points of the domain of the functions to be found, we use the relaxation numerical method replacing the set of differential equations by finite-difference equations on a grid of points that covers the whole range of the integration, as in the previous chapter.

\subsection{'t Hooft-Polyakov Standard Solution}
In this section we take $B_F=1$. This recovers the standard 't Hooft-Polyakov monopole solution \cite{thooft} with no variations of the fine-structure constant, i.e., for which $\varphi(r) = {\rm{constant}}$.

It proves to be useful to write the equations of motion which are
\ben
\frac{dX}{dr}&=&\,V,\label{carm0}\\
\frac{dW}{dr}&=&\,U, \label{carm1} \\
\frac{dV}{dr}&=&-\frac{2V}{r}+\frac{\zeta^2}{2}\,X(X^2-1)+\frac{2\,W^2\,X}{r^2}, \label{carm2} \\
\frac{dU}{dr}&=&\frac1{r^2}(W^2-1)W+ X^2W.  \label{carm3}
\een		

Let us first consider $\zeta=0$, i.e., a massless Higgs field. In this case the equations of motion are 
\ben
\frac{dX}{dr}&=&\,V,\label{carm34}\\
\frac{dW}{dr}&=&\,U, \label{carm4} \\
\frac{dV}{dr}&=&-\frac{2V}{r}+\frac{2\,W^2\,X}{r^2}, \label{carm5} \\
\frac{dU}{dr}&=&\frac1{r^2}(W^2-1)W+ X^2W. \label{carm6}
\een
Substituting Eqs. (\ref{carm34}--\ref{carm6}) in Eq. (\ref{act}), the energy of the monopole can be written as
\be
\label{ener}
E = \frac{4\pi\eta}{e_0}\Big\{\int_0^{\infty} dr\left[ \frac12 \left(r\frac{dX}{dr}-\frac{1-W^2}{r}\right)^2 + \left(\frac{dW}{dr}+WX \right)^2 
\right]+ \int_0^{\infty}dP \Big\},
\ee 
where we have introduced a new scalar field $P(r)$ as
\be
P(r)=X\,(1-W^2).
\ee
We notice that the equations of motion (\ref{carm34}-\ref{carm6}) can be completely factorized into the set of
two first order equations 
\ben
\frac{dX}{dr} &=& \frac{1-W^2}{r^2}, \label{monof}\\
\frac{dW}{dr} &=& -W\,X. \label{monog}
\een
These are 
precisely equivalent to the equations of motion (\ref{carm34}-\ref{carm6}). This is the Bogomoln'yi method \cite{bogomolnyi}--\cite{menezes2}. Using Eq. (\ref{ener}) to calculate the energy one gets 
\be
E = \frac{4\,\pi\,\eta}{e_0}\Big[P(r \rightarrow \infty) - P(r=0) \Big],
\ee
which from (\ref{carm0}-\ref{carm3}) gives $E = 4 \pi\,\eta/e_0$, that is the Bogomoln'yi energy.
\begin{figure}
\begin{center}
\includegraphics*[width=9cm]{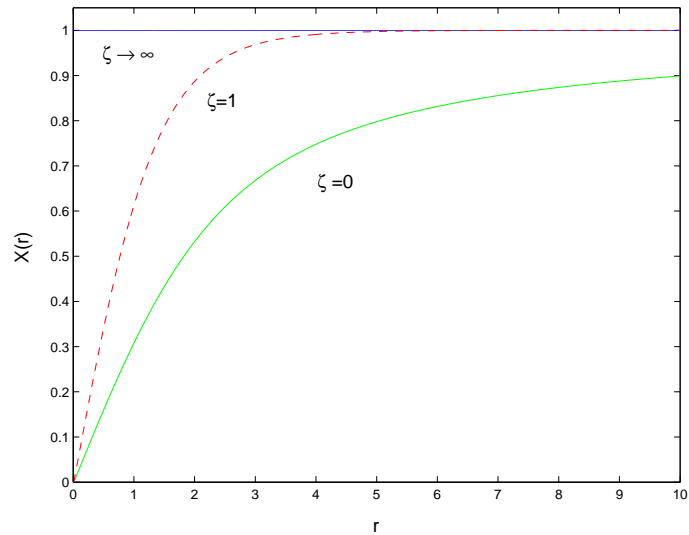}
\end{center}
\caption['t Hooft-Polyakov monopole: $X(r)$.]{
The numerical solution of the field $X(r)$  as a function of 
distance, $r$, from the core of the 't Hooft-Polyakov monopole for several values of $\zeta$. For $\zeta=0$ the Higgs field is massless and for $\zeta \rightarrow \infty$ the Higgs field is frozen at its vacuum value except at the origin.}
\label{kovH}
\end{figure}
\begin{figure}
\begin{center}
\includegraphics*[width=9cm]{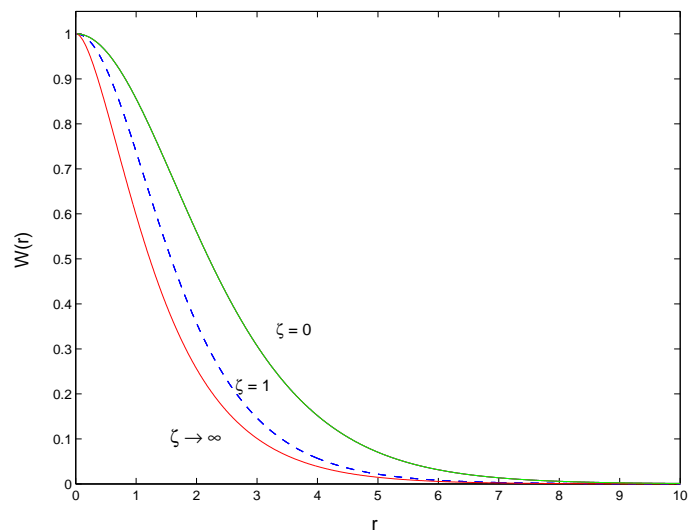}
\end{center}
\caption['t Hooft-Polyakov monopole: $W(r)$.]{
The numerical solution of the field $W(r)$  as a function of 
distance, $r$, from the core of the 't Hooft-Polyakov monopole for three values of $\zeta$.}
\label{kovW}
\end{figure}
\begin{figure}[ht]
\begin{center}
\includegraphics*[width=9cm]{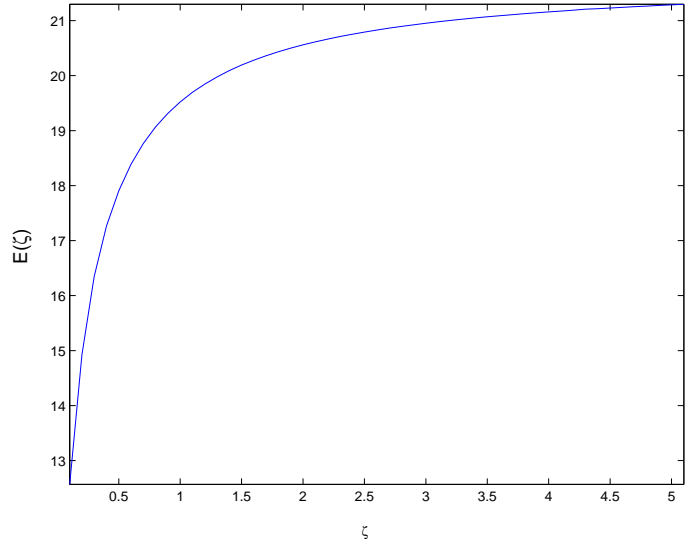}
\end{center}
\caption['t Hooft-Polyakov monopole: $E(\zeta)$.]{
The numerical solution of the energy of the 't Hooft-Polyakov monopole, $E(\zeta)$,  as a function of 
the ratio of the Higgs mass to the vector field one, $\zeta$. Note that the monopole energy remains finite when $\zeta \rightarrow \infty$.}
\label{enerkov}
\end{figure}
Note that the solutions of equations 
(\ref{monof}-\ref{monog}) are
\ben
X(r) &=& \frac{r}{\sinh{(r)}}, \label{bps1}\\
W(r) &=& \frac1{\tanh{(r)}} - \frac1{r}, \label{bps2}
\een
which are the Bogomol'nyi, Prasad, Sommerfield (BPS) solutions \cite{bogomolnyi,prasad}. 
In Figs. \ref{kovH} and \ref{kovW} we plot $X(r)$ and $W(r)$ given by Eqs. (\ref{bps1}) and (\ref{bps2}).

Let us now consider the opposite limit for $\zeta$, i.e., $\zeta \rightarrow \infty$ which 
for fixed $e_0$ implies 
$\lambda \rightarrow \infty$, i.e., the Higgs potential is much larger than its kinetic term.
As a result, the Higgs field
is forced to be frozen at its vacuum value everywhere except at the origin. 
The equations of motion (\ref{carm0}--\ref{carm3}) are reduced to 
\be
\frac{d^2W}{dr^2}\,=W\left(1-\frac{1-W^2}{r^2}\right)
\ee
and $X =0$.

In Figs. \ref{kovH} and \ref{kovW}, we plot $X(r)$ and $W(r)$ for $\zeta \rightarrow \infty$.
As expected for $\zeta = 0$ the Higgs field is massless while it is  frozen at its vacuum value (except the origin) for $\zeta \to \infty$.

The previous values for $\zeta$ simplify very much the equations of motion. However, we have solved them numerically with a very good precision
for $\zeta$ in the interval $10^{-4} \leq \zeta \leq 10^{3}$.
We also show (see Fig. \ref{enerkov}) that the energy of the monopole increases with $\zeta$ asymptoting to a finite constant
for $\zeta \rightarrow \infty$ .  

In the next subsections we will find magnetic monopole solutions with varying fine-structure constant. We will set the parameter $\zeta=1$.

\subsection{Bekenstein Model}
We already know from Chapter \ref{cap3} that for  
\be
B_F(\varphi)\,=\,e^{-\frac{2\varphi}{\sqrt{\omega}}}, \label{expo}
\ee
with $\omega > 0$, one recovers the original Bekenstein model \cite{Bekenstein}. 
Now defining
\be
\psi=\frac{\varphi}{\sqrt{\omega}}
\ee
and substituting in (\ref{expo}) one gets that Eq.(\ref{laga})
can be written as
\be
\mathcal{L}= \frac12\left(D_\mu\,\Phi^a \right)\left(D^\mu\,\Phi^a\right) 
- \frac{e^{-2\,\psi}}{4}f_{\mu\nu}^{a}\,f^{a\mu\nu}- V(\Phi^a) 
+\frac{\omega}2\partial_\mu\,\psi\,\partial^\mu\,\psi \label{lagabek},
\ee 
where now $\omega$ can take either positive or positive values.

\begin{figure}
\begin{center}
\includegraphics*[width=9cm]{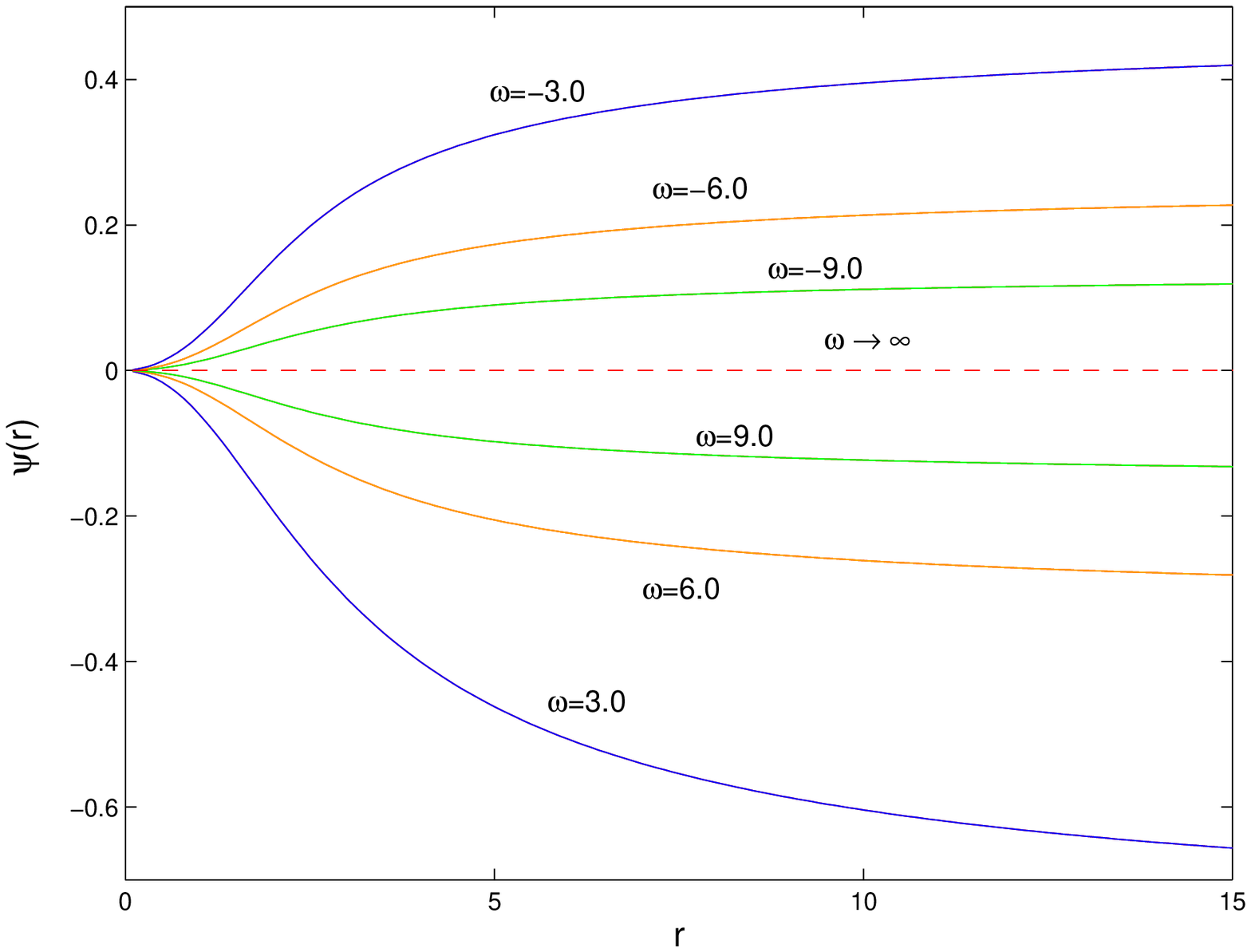}
\end{center}
\caption[Varying-$\alpha$ monopoles: $\psi(r)$ for exponential $B_F$.]{The numerical solution of the scalar field 
$\psi \equiv \ln{\epsilon}$ as a function of distance, $r$, 
from the core of the monopole, in the Bekenstein model. 
Note that if $\omega < 0$,  $\epsilon \to \infty$ when $r \to \infty$.
On the other hand, if $\omega > 0$ then $\epsilon \to 0$ when $r \to \infty$.
The dashed line represents the constant-$\alpha$ theory, which corresponds 
to the limit $ \omega \rightarrow \infty$.} 
\label{psibek}
\end{figure}
\begin{figure}
\begin{center}
\includegraphics*[width=9cm]{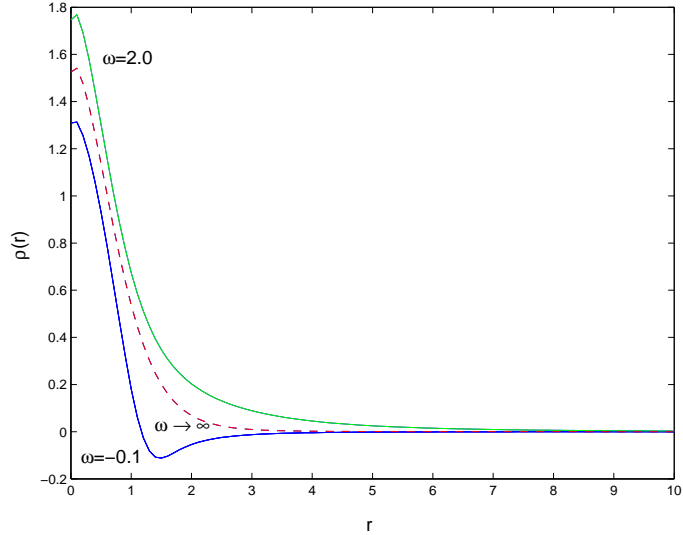}
\end{center}
\caption[Varying-$\alpha$ monopoles: $\rho(r)$ for exponential $B_F$.]{The numerical solution of the energy density $\rho$ as a function $r$, 
the distance from the core of the monopole. The dashed-line is for $\omega \rightarrow \infty$. Note that for $\omega=2.0$, the energy density is non negative, while for $\omega=-0.1$ it becomes negative at some distance from the core.} 
\label{rhobek}
\end{figure}
First we note that in the limit $\omega\,\rightarrow\,\infty$ one recovers 
the 't Hooft-Polyakov monopole described in the previous subsection.  
In  Fig. \ref{psibek} we plot the numerical solution of the scalar field 
$\psi(r) = \ln \epsilon$ for several values of $\omega$. 
One concludes that if $\omega < 0$ then $\epsilon$ 
diverges asymptotically away from the core of the monopole, and the energy density 
\ben
\rho &=& \frac{\eta}{e_0}\Big\{ e^{-2\,\psi}\left[\frac1{r^2}\,\left(\frac{dW}{dr} \right)^2 + \frac12 \left(\frac{1-W^2}{r^2}\right)^2 \right]  \nn \\
&+& \frac12 \left(\frac{dX}{dr}\right)^2 + \left(\frac{WX}{r} \right)^2 + \frac18 \zeta^2(1-X^2)^2\, + \frac{\omega}2\,\left(\frac{d\psi}{dr}\right)^2 \Big\}
\een
is no longer positive definite. 
However if $\omega > 0$ then $\epsilon$ vanishes asymptotically when $r  \to \infty$, and the energy density is, in this case, positive definite (See Fig. \ref{rhobek}). We also note from Fig. \ref{psibek} that in the large $\omega$ limit the curves for positive and negative 
$\omega$ are approximately symmetric approaching the dashed line which represents 
the constant-$\alpha$ model when $\omega \rightarrow \infty$.

\begin{figure}
\begin{center}
\includegraphics*[width=9cm]{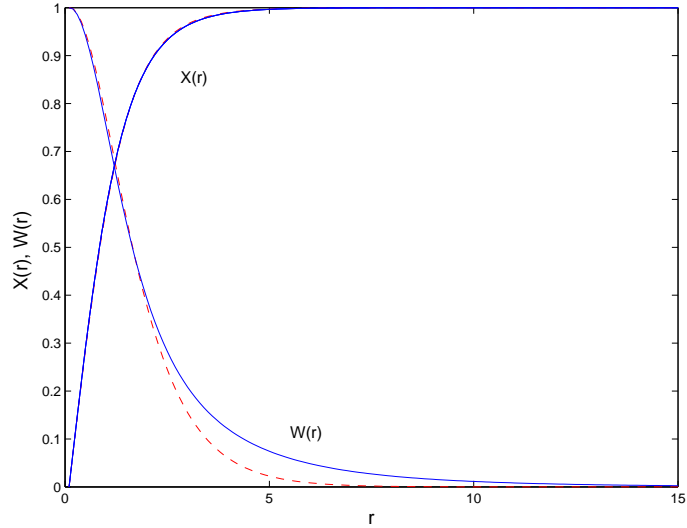}
\end{center}
\caption[Varying-$\alpha$ monopoles: $X(r)$ and $W(r)$ for exponential $B_F$.]{The numerical solution of the Higgs field $X(x)$ and the gauge field $W(x)$ as 
functions of distance $r$, from the core of the monopole. Note that the change in $W(r)$ with
respect to the 't Hooft-Polyakov solution(dashed-line) is more relevant than the change in $X(r)$.} 
\label{hwbek}
\end{figure}

Finally, in Fig. \ref{hwbek} we plot the numerical solution of the scalar field $X(r)$ and the gauge field $W(r)$ as a function of distance, $r$, from the core of the monopole. The dashed-line represents the constant-$\alpha$ solution and the solid line represents the Bekenstein one for $\omega=2.0$. 
We verified that even in the $\omega \rightarrow 0$ limit, the change in $X(r)$ with respect to 't Hooft-Polyakov solution is still negligible.


\subsection{Polynomial Gauge Kinetic Function}
Let us now consider a class of gauge kinetic functions:
\be
B_F(\varphi)=1.0+\sum_{i=1}^{N}\,\beta_i\,\varphi^i \, , \label{poli}
\ee
where $N$ is an integer. As $\varphi$ is dimensionless then $\beta_i$ is dimensionless as well.
We have shown in Chapter \ref{cap3} that
these models are equivalent to the Bekenstein one (\ref{expo}) since
\be
\beta_i = \frac{(-2)^i}{w^{i/2} i!}\,
\ee 
This relationship between the coupling constants $\beta_i$ and $\omega$ has 
interesting consequences for the model given by Eq. (\ref{poli}). 

First we verified that the behavior of $\psi$ both for $\beta_1 \, >\,0$ and $\beta_1 \, <\,0$  is similar to that of the Bekenstein model with $\omega\,>\,0$ which is recovered in the 
limit of small $|\beta_1|$/large $\omega$. 
We also verified that if one takes $\beta_1=0$ one gets the 't Hooft-Polyakov limit, for any $\beta_i$ with $i>1$. This means that there is a class of gauge kinetic functions for which the
classical static solution is maintained despite the modifications to the model. 
This property also is present in the case of the cosmic strings as we have mentioned in the previous chapter.
 
Another property can be noticed when one substitutes the gauge kinetic function in the equation of motion for $\varphi$ (\ref{three}). One gets
\be\label{oddeven}
\frac{1}{r^2}\,\frac{d}{dr}\left(r^2\frac{d\varphi}{dr}\right)=
C^2\left[\sum_{k=1}^{N}(2k-1)\beta_{2k-1}\varphi^{2k-2}
+ \sum_{k=1}^{N}(2k)\beta_{2k}\varphi^{2k-1}\right]\,
\ee
with
\be
C^2\,=\,\left(\frac{dW}{dr} \right)^2+\frac12\left(\frac{1-W^2}{r}\right)^2. \label{synn}
\ee
Since that $C^2\,>\,0$, when $\beta_i \rightarrow -\beta_i$ for odd $i$ one sees by Eq. (\ref{oddeven}) that $\varphi(r)\,\rightarrow\,-\varphi(r)$. However, as $B_F$ in Eq. (\ref{poli}) is kept invariant, $X(r)$ or $W(r)$ do not vary.

\begin{figure}
\begin{center}
\includegraphics*[width=9cm]{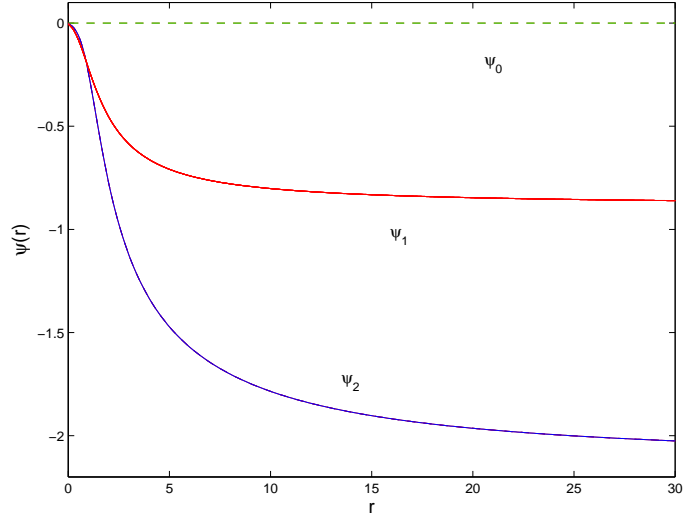}
\end{center}
\caption[Varying-$\alpha$ monopoles: $\psi(r)$ for polynomial $B_F$.]{The numerical solution of the scalar field 
$\psi \equiv \ln \epsilon$ as a function of distance, $r$, 
from the core of monopole, for a polynomial gauge kinetic function. 
Models 0, 1 and 2 are defined by $\beta_1=0$ ($\beta_2$ arbitrary),  
$\beta_1=-3,\,\beta_2=0$ (linear coupling) and 
$\beta_1=-2,\,\beta_2=5$ respectively.}
\label{psipoli}
\end{figure}

We have found the set of solutions for several values of $N$. Without loss of generality
we took $N=2$ in Eq. (\ref{poli}) to write 
\be
B_F = 1.0 + \beta_1 \varphi + \beta_2 \varphi^2 \label{nosso}
\ee
and define the model $0$ with $\beta_i=0$ (standard monopole solution), model $1$ with $\beta_1=-3,\,\beta_2=0$ (linear coupling) and model $2$ with
$\beta_1=-2,\,\beta_2=5$ (quadratic coupling), respectively. Fig. \ref{psipoli} shows the numerical solution of the scalar field $\psi(r)$. As a result we have verified that the replacement 
$\beta_1 \to -\beta_1$ does not modify the solution for $\psi$ as it was expected.

\begin{figure}
\begin{center}
\includegraphics*[width=9cm]{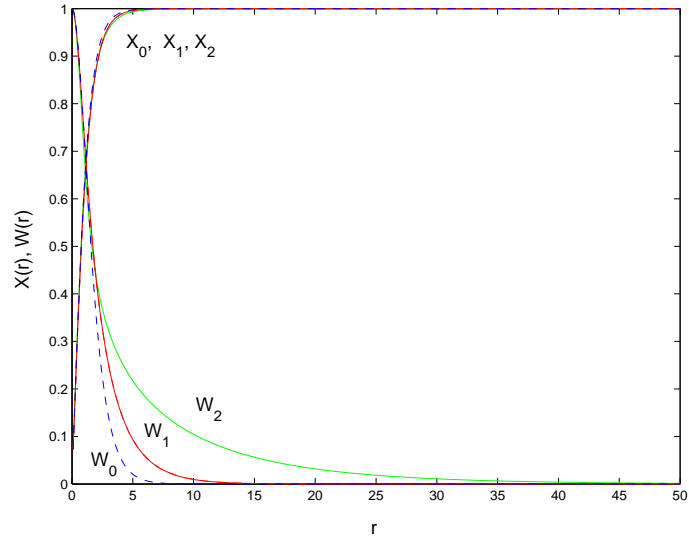}
\end{center}
\caption[Varying-$\alpha$ monopoles: $X(r)$ and $W(r)$ for polynomial $B_F$.]{
The numerical solution of the fields $X(r)$ and $W(r)$ as functions of 
distance, $r$, from the core of monopole, for models $0$, $1$ and $2$. Note 
that the change in $W(r)$ with respect to the standard 
constant-$\alpha$ result is much more dramatic than the change in $H(r)$.}
\label{hwpoli}
\end{figure}
\begin{figure}
\begin{center}
\includegraphics*[width=9cm]{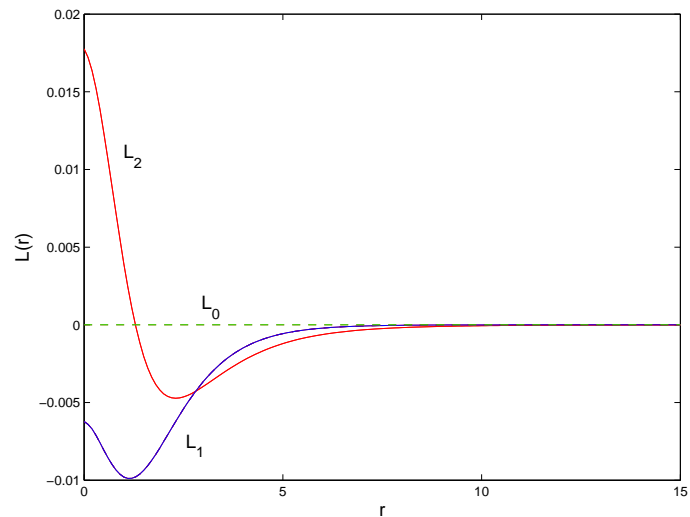}
\end{center}
\caption[Varying-$\alpha$ monopoles: $L(r)$ for polynomial $B_F$.]{Plot of $L_i(r)=\log(X_i/X_0)$ for the models $0$, $1$ and $2$. One clearly sees that even a small value of 
$\beta_1$ leads to a different vortex solution from the standard 't Hooft-Polyakov one.}
\label{lpoli}
\end{figure}

From the equations of motion (\ref{one}-\ref{three}) a change in $B_F$ is directly manifest in a change of $U$, i.e., of $W$ and indirectly results in a change of $V$, i.e., of $X$. This can be checked through Fig. \ref{hwpoli} where one can see clearly that the change in $W(r)$ with respect 
to the standard constant-$\alpha$ result is much more dramatic than the 
change in $X(r)$. This is more explicit in Fig. \ref{lpoli} where we define the function
\be
L_i(r) = \log\left( \frac{X_i}{X_0}\right) \label{zeta}
\ee
with $i=0,1,2$.
We note that even a small value of $\beta_1$ leads to a modification of the magnetic monopole 
solution with respect to the standard 't Hooft-Polyakov solution.

\begin{figure}
\begin{center}
\includegraphics*[width=9cm]{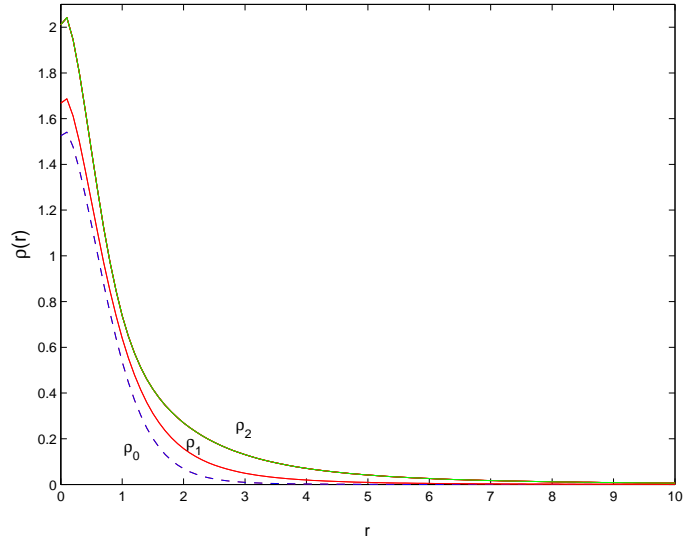}
\end{center}
\caption[Varying-$\alpha$ monopoles: $\rho(r)$ for polynomial $B_F$.]{The energy density as a function of the 
distance, $r$ from the monopole core for models 0, 1 and 2. 
The dashed line represents the constant-$\alpha$ model. 
Note that there is an increase of the energy density due to the 
contribution of the extra field $\varphi$ in the case models $1$ and $2$.}
\label{rhopoli}
\end{figure}

We have also studied the behavior of the energy density in these
models which are plotted in Fig. \ref{rhopoli}.
The increase of the energy density for models $1$ and $2$ when compared to model $0$ is due to the contribution of
the spatial variations of the structure constant.
In fact, as in the case of cosmic strings, the energy density of the monopole can be divided into two 
components: one that is localized inside the core of the monopole 
and another related to the 
contribution of the kinetic term associated 
with the spatial variations of the fine structure constant. 

\section{\label{cvmm}Constraints on Variations of $\alpha$}
$ $\par
We now look for an overall limit on the spatial variations of the fine-structure constant seeded by magnetic monopoles.
We consider a monopole network with the same core size, $r_c$. We impose a cut-off by considering that there is a separation of $2\,r_{\rm max}$ to its first neighbor. In other words, we set $r_{\rm max}/r_c \gg 1$ so that we can neglect the interactions of each monopole with its neighbor and therefore approximate it as an isolated monopole, such as the one in the previous sections, with the difference that the space outside the core now extends up to a finite distance: the correlation length of the system $r_{\rm max}$.

Let us now calculate the spatial variations of the fine-structure constant outside the core.
For simplicity we assume
\be
B_F(\varphi)\,=\,1\,+\,\beta\varphi, \label{bfl}
\ee 
i.e., the gauge kinetic function is a linear function in $\varphi$ that satisfies the spherically symmetric Poisson equation given by
\be
\nabla^2 \varphi = \frac{1}{4}\, 
\beta \, f^2\,.\label{poisson}
\ee 
Note that $\beta$ is constrained by Equivalence Principle tests to be such that $|\beta| < 10^{-3}\,G^{1/2}$ (see \cite{Will,OlivePos}).

Integrating Eq.(\ref{bfl}) from  the core up to $r_{max}$, the cosmological cut-off scale, one gets
\be
4 \pi r^2 \frac{d \varphi}{dr}=  \beta I(r)  M(r_{\rm max})\,. \label{poi}
\ee
In Eq.(\ref{poi}) we used the mass of the monopole which is given by
\be
M(r)=4 \pi \int^r_0 \rho(r') r'^2 dr'\,
\ee
and
\be
I(r)=\frac{\pi}{M(r_{\rm max})} \int_0^r f^2(r') r'^2 dr'\,,
\ee
which is a slowly varying function of $r$ outside the core always smaller than unity. Thus we can take $I(r) \sim {\rm const}$ and integrate Eq. (\ref{poi}) to get
\be
\varphi \sim  \frac{\beta I\,M(r_{\rm max})}{4\pi} \left(\frac1{r}-\frac1{r_0}\right)\,,
\ee
where $r_0$ is a integration constant which could be identified as the core radius. Using $B_F(\varphi)=\epsilon^{-2}$ one gets
\be
\epsilon \sim 1-\frac{\beta^2 I\,M(r_{\rm max})}{8\pi} \left(\frac1{r}-\frac1{r_0}\right)\,,
\label{epsilonr}
\ee
which means that the variation of the fine structure constant away from the
monopole core is proportional to the gravitational potential induced by
the monopoles.

\subsection{GUT Monopoles }

Let us estimate an overall limit for the spatial variation of $\alpha$ outside the core of GUT monopoles. In this context the mass of the monopole is of order of
\be
M(r_{max}) \sim 10^{16}\,GeV.
\ee
The variation of $\alpha$ from the core up to infinity is
\be
\frac{\Delta \alpha}{\alpha} = \frac{\alpha(r \rightarrow \infty)-\alpha(r_0)}{\alpha(r_0)} = \epsilon^2 -1 \lesssim 10^{-13},
\ee
where we have used $\alpha=\alpha_0/B_F(\varphi)$, with $\alpha=\alpha(r \rightarrow \infty)$ and $\alpha_0=\alpha(r_0)$.

\subsection{Planck Monopoles}
Proceeding as above for the Planck scale symmetry with 
\be
M(r_{max}) \sim 10^{19} {\rm Gev}
\ee
one obtains an overall limit for the spatial variation of the fine-structure seeded by magnetic 
monopoles
\be
\frac{\Delta \alpha}{\alpha} \lesssim 10^{-7},
\ee
which is still very small even for Planck scale monopoles.

\section{\label{cccmm}Conclusions}
We introduced Bekenstein-type models in the Non-Abelian Yang-Mills theory in order to investigate the variations of the fine-structure constant in the vicinity of a static local monopole.
We studied various models with constant $\alpha$ and reviewed the standard static 't Hooft-Polyakov magnetic monopole 
solution. 
Then we considered models with varying-$\alpha$ and confirmed that despite the existence of a class of models for which the 't Hooft-Polyakov standard solution is still valid. This property had been verified in Chapter \ref{cap3} in the context of the cosmic strings in varying-$\alpha$ theories. However, in general, the solutions of the local defects depart from the standard case.
We also showed that Equivalence Principle 
constraints impose tight limits on the variations of $\alpha$ induced by
magnetic monopoles. This confirms 
the difficulty to generate significant large-scale spatial variation of 
the fine structure constant found in the previous section, even in the most favorable case 
where these variations are seeded by magnetic monopoles.
\chapter{\label{cap5}{\sf Evolution of $\alpha$ in the Non-Linear Regime} \label{nev}}
\section{Overview} 

We have seen in this thesis that a 
real scalar field coupled to the matter sector could be responsible for the variation of $\alpha$.
In fact, given that
in the course of the cosmological evolution, inhomogeneities grow, become non-linear and decouple from the background evolution, it has been argued that the same could happen to the local variations of $\alpha$. 
A number of authors have studied this problem and have shown using linear theory that
the spatial variations of the fine 
structure constant induced by fluctuations in the matter fields are 
proportional to the gravitational potential. Although such spatial variations of $\alpha$ are typically 
very small to be detected directly with present day technology, 
some authors have argued that 
large variations could occur in the vicinity of compact objects with strong gravitational 
fields \cite{mota1}--\cite{mota3}.

In previous chapters, we have considered two examples of compact objects, cosmic strings and magnetic monopoles. We have shown that both kinds of topological defects do not seed large variations of $\alpha$ in their vicinity despite the electromagnetic energy concentrated in their core.
In this chapter, we will 
study the variations of $\alpha$ induced by the collapse of large scale structures.
In fact, this has been previously studied in \cite{mota1}--\cite{mota3} using a simple spherical infall model for the evolution of infinite wavelength density perturbations and a particular generalization of the Bekenstein model \cite{Sandvik} for the evolution of $\alpha$. 
It was found that in the linear regime and in the matter era the variation of $\alpha$ would follow the density contrast. Moreover, it was also claimed that this approximation was valid in the non-linear regime (meaning turnaround and collapse). Here we revisit and question these results. 
Our work was divided into two parts. Firstly, we confirm the results obtained by Barrow
and Mota by using the infinite wavelength approximation to study the growth of a uniform 
spherical matter inhomogeneity with a final size $r_0$ \cite{mota1}--\cite{mota3} . Secondly, we argue that a local approximation for 
determining the variation of the fine-structure constant outside virialized regions is the correct approach \cite{Magueijonl}.

An alternative approximation was followed  
in Ref. \cite{Shaw}. This provides a more detailed and mathematically-inclined analysis of local variations in physical `constants', but does confirm our results. Our approximation, while much simpler, has the advantage of making explicit the reasons why spatial variations have to be small, and why the use of the spherical collapse model of a infinite wavelength perturbation is inadequate. Other papers also have later confirmed our results  \cite{Barrow3}--\cite{Barrow2}.

This chapter is organized as follows. 
In Sec. \ref{lbm}, we overview the linearized Bekenstein model.
In Sec. \ref{nonlin} we discuss the non-linear evolution of the fine structure constant using two different approximations. Our results are described in Sect. \ref{res} and further discussion will be found in Sect. \ref{disc}. 
\section{\label{lbm}The Model}
Let us consider the linearized Bekenstein model (see for example Ref. \cite{linear}).
In this model a neutral scalar field is non-minimally coupled to electromagnetism by means of
the gauge kinetic function, $B_F(\varphi)$. The Lagrangian density is given by
\be
{\cal L} = {\cal L}_\varphi + {\cal L}_{\varphi F} + {\cal L}_{\rm other}\, , 
\ee
where
\ben
{\cal L}_\varphi &=& \frac{1}{2}\partial_\mu \varphi\, \partial^\mu \varphi - V(\varphi)\, ,\\
{\cal L}_{\varphi F} &=& -\frac{1}{4}B_F(\varphi) F_{\mu \nu} F^{\mu \nu}\,, 
\een
and ${\cal L}_{\rm other}$ is the Lagrangian density of the other fields. 
We assume that $V(\varphi)$ is a linear function of the field $\varphi$, namely,
\be
V(\varphi) = V(\varphi_0) + \frac{dV}{d\varphi} \left(\varphi-\varphi_0\right)\, , 
\ee
where $dV/d\varphi$ is constant. 
The subscript `0' refers to
the 
present time.
Also, we consider that $B_F$ is a linear function of $\varphi$, 
\be
B_F (\varphi) = 1-\beta (\varphi-\varphi_0)\,.
\ee 
Equivalence Principle tests impose constraints on the value of $\beta$, namely $|\beta| < 10^{-3}\,G^{1/2}$.
We recall that $B_F(\varphi)$ acts as the effective dielectric 
permittivity which can be phenomenologically taken to be an arbitrary function of $\varphi$.

Using the definition of the gauge kinetic function, $B_F=\alpha_0/\alpha$, we note that
\be
\frac{\alpha}{\alpha_0}= 1+\beta \left(\varphi-\varphi_0\right)\,  
\ee
up to linear order. Assuming that 
\be
\frac{d\,V(\varphi)}{d\,\varphi} < 0,\,\,\,\,\,\,\,\,
\frac{d\,\alpha}{d\,\varphi} > 0,
\ee
the results lead to a fine-structure constant that had a smaller value in the past than at the present time. 
The assumptions of linear functions of $\varphi$ for $V(\varphi)$ and $\alpha(\varphi)$ are necessarily verified for some period
of time around the present day.  

The equation of motion for the field $\varphi$ is given by  
\be 
\label{phieq222} 
\Box \varphi= - \frac{d\,V(\varphi)}{d\,\varphi} + \frac{\beta}{4} F_{\mu \nu} F^{\mu \nu}\, 
\ee 
and we define
\be
\frac{F_{\mu \nu} F^{\mu \nu}}{4} = - \gamma_F \rho_m,
\ee 
where $\gamma_F$ represents the matter density fraction which contributes 
to the right hand side of Eq. (\ref{phieq222}).  
For baryons 
$\gamma_F$ is constrained to be very small since the electromagnetic 
corrections to the mass of protons and neutrons are of the order of 1 part 
in $10^4$ \cite{Damour}. Also, since the variations of $\alpha$ are very small, $\gamma_F$ 
is expected to be a nearly constant parameter.
Bearing in mind that
\be
F_{\mu \nu} F^{\mu \nu} = 2\,(B^2-E^2) < 0,
\ee 
we note that the last two terms in Eq. (\ref{phieq222}) have opposite signs.

In this chapter we will consider the linearized Bekenstein model with Eq. (\ref{phieq222}) written as
\be
\label{varphil}
\Box \varphi = - \beta \gamma_F \rho_m \,,
\ee
where we have set $d V/d \varphi = 0$ for simplicity.
Of course a generalization to models with $dV/d\varphi \neq 0$ would modify the 
background dynamics of $\alpha$  but would not 
otherwise change our conclusions. 

First,
by using Eq. (\ref{varphil}) we can compute the evolution of the background value of $\alpha$ with physical 
time, that is given approximately by
\be 
\label{lphieq3} 
{\ddot {\bar \alpha}}+3H{\dot {\bar \alpha}} \sim - \xi \rho_m {\bar \alpha}\,,
\ee
where 
\be
\xi=\gamma_F \beta^2 < 6 \times 10^{-6} \gamma_F.
\ee 
Note that we have used the definition of the field $B_F$ in terms of $\alpha$ to write Eq. (\ref{lphieq3}).
On the other hand, 
since we aim in this chapter to provide a description for the evolution of $\alpha$ in the non-linear regime
by using a local static approximation, we linearize Eq. (\ref{varphil}) to obtain a local static 
solution in a slightly perturbed Minkowski space 
\be 
\label{phieq2} 
\frac{\nabla^2 \alpha}{\bar \alpha} \sim \xi \delta \rho_m\, . 
\ee
Here, we neglected the $\ddot \varphi$ term in Eq. (\ref{varphil}) in order to obtain equation Eq. (\ref{phieq2}). This is a valid approximation if $|\ddot \varphi| \ll |\nabla^2 \varphi|$. 

Note that Eq. (\ref{lphieq3}) implies that the time variation of $\alpha$ 
induced in one Hubble time is 
\be
\frac{\Delta \alpha}{\alpha}\, \sim\, \xi\, \delta \rho_m\, 
\alpha\, H^{-2},
\ee
while Eq. (\ref{phieq2}) tell us that the spatial 
variation 
of $\alpha$ seeded by a perturbation of size $L \ll H^{-1}$ is given 
by 
\be
\frac{\Delta \alpha}{\alpha}\, \sim\, \xi\, \delta \rho_m\, \alpha\, L^2.
\ee
Hence, we 
expect that the spatial variations of $\alpha$ generated 
on small cosmological scales (where non-linear gravitational clustering is 
currently taking place) will be small.

\section{\label{nonlin}Non-Linear Evolution of $\alpha$}
We will consider the evolution of $\alpha$ due to the growth of a uniform spherical matter inhomogeneity with a final size $r_0$ using two different approaches.
\subsection{Infinite Wavelength}
We will start by considering 
the evolution of two homogeneous and isotropic universes. 
This approach was used in \cite{mota1,mota2} to study the evolution of a uniform (that is, \textit{infinite wavelength}) perturbation in  $\alpha$, with 
the initial conditions set at some initial time $t_i$ deep in the matter era.
Thus we name this the \textbf{infinite wavelength approximation}.

\subsubsection{The background universe}
The background is a spatially flat Friedman-Robertson-Walker 
universe containing matter, radiation and a cosmological constant. The
Friedmann equation is
\be
H^2=H_i^2\left[\Omega_{mi} \left(\frac{a}{a_i}\right)^{-4} 
+\Omega_{ri} \left(\frac{a}{a_i}\right)^{-4} +\Omega_{\Lambda i} \right]\, , 
\ee
where $\Omega_{mi}$, $\Omega_{ri}$ and $\Omega_{\Lambda i}$ 
are respectively the density parameter
for matter, radiation and cosmological constant at some initial time $t_i$.
In order to ensure the flatness for the background universe,
we assume that 
\be
1 - \Omega_{mi} - \Omega_{ri} - \Omega_{\Lambda i} =0.
\ee
The acceleration of the universe is given by 
\be
\label{Fried}
\frac{\ddot a}{a}=-H_i^2\left[\frac{\Omega_{mi}}{2} \left(\frac{a}{a_i}\right)^{-3} 
+\Omega_{ri} \left(\frac{a}{a_i}\right)^{-4} -\Omega_{\Lambda i} \right]\, . 
\ee
Here for simplicity we include the contribution of the energy density of the field $\varphi$ in the value of $\Omega_\Lambda$, thus neglecting the contribution of the kinetic contribution to energy density of the field $\varphi$, which is nevertheless constrained to be small. Also, since any variation of the fine-structure constant from the epoch of nucleosynthesis onwards is expected to be very small 
\cite{Avelino2,Martins2} we neglect its minor contribution to the evolution of the baryon density, included
in $\Omega_{m}$.
\subsubsection{The perturbed universe}
We also consider a second universe, a spatially closed Friedmann-Robertson-Walker universe with matter, radiation and a cosmological constant\footnote{We have neglect the contribution of the scalar field $\varphi$ for the background as well as for the closed universe but it can be included. This does not modify our conclusions.} with
\be
\Omega_{ki}=1-\Omega_{mi}-\Omega_{ri}-\Omega_{\Lambda i} = - \Delta \Omega_{mi} \neq 0.
\ee 
This closed universe is obtained by perturbing the matter density of the background universe with a small infinite wavelenght perturbation. 
We take the initial time $t_i$ to be deep into the radiation epoch in such way that
\ben
&&H_{Pi} \sim H_i,\\ 
&&\Omega_{Pri}=\Omega_{ri},\\
&&\Omega_{P\Lambda i}=\Omega_{\Lambda i},\\ 
&&\Omega_{Pmi}=\Omega_{mi}+\Delta \Omega_{mi},\\
&&\Omega_{Pki}=-\Delta \Omega_{mi}.
\een
where the subscript $P$ indicates that the quantity is related to the perturbed (closed) universe.

In this approximation the 
values of $\alpha$ in the background and perturbed universes 
are computed using Eq. (\ref{lphieq3}) so that 
\be
\label{phieq5} 
\left( \ddot{\frac{\Delta \alpha}{\alpha}} \right)_{b,\infty}+3H\left(\dot{\frac{\Delta \alpha}{\alpha}}\right)_{b,\infty} \sim - \xi 
\frac{3 \Omega_m H^2}{8\pi}\,,
\ee
where
\be
\frac{\Delta \alpha}{\alpha} = 
\frac{{\bar \alpha}-{\bar \alpha}_i}{{\bar \alpha}_i}
\ee
and the subscripts $b$ and $\infty$ indicate that ${\bar \alpha}$ is 
calculated for the background and perturbed universes respectively.

\subsection{\label{lolocal}Local Approximation}
We have introduced another approach in Ref. \cite{Nosso3} which we have referred to as the {\bf{local approximation}}. 
In this approach, we consider a slightly perturbed Minkowski space and assume that the 
background value of the
fine structure constant
is fixed.
Of course, 
this is a good approximation only on length scales much smaller than the 
Hubble radius.
Although we follow the evolution of $\alpha$ into the non-linear regime, where the density perturbations become much larger than unity, the fractional variations of $\alpha$ are always constrained to be small.
Thus we name this the \textbf{local approximation} \cite{linear}. 

We solve the Poisson equation (\ref{phieq2}) outside the spherical distribution of mass, $M$, with radius $r$, to determine the spatial
variation of $\alpha$ as
\be 
\label{phieq1} 
\frac{\delta \alpha}{\alpha}\equiv \frac{\alpha(r)-{\bar \alpha}}{\bar \alpha} 
\sim - \xi \frac{M}{4\pi r} \,
\ee
identifying ${\bar \alpha}$ with $\alpha(r=\infty)$.
Given that 
\be
M=\frac{4\pi \delta \rho_m r^3}3,
\ee
we have
\be 
\label{model1} 
\frac{\delta \alpha}{\alpha}=- \frac{1}{3} \xi \delta \rho_m r^2.
\ee

This calculation assumes that the spherical inhomogeneity has a 
radius significantly smaller than the horizon in order for Eq. (\ref{phieq2}) 
to be a valid approximation locally. 
It is straightforward to show that, in this limit, the condition 
\be
\Big|\frac{\partial^2 (\delta \alpha/\alpha)}{\partial t^2}\Big| \sim \Big|\frac{\delta \alpha}{\alpha} \Big|\,H^2 \ll \Big|\frac{\partial^2 (\delta \alpha/\alpha)}{\partial r^2}\Big|
\ee
is required by self-consistency.

Therefore $\Delta\alpha/\alpha$ can be computed as
we have
\be
\label{diff}
\left(\frac{\Delta \alpha}{\alpha}\right)_\ell \sim 
\left(\frac{\Delta \alpha}{\alpha}\right)_b +\frac{\delta \alpha}{\alpha}.
\ee
If one considers the matter density computed for the background and perturbed universes in
the previous subsection, the local variation of $\alpha$ can be written as
\be
\label{phieq6} 
\frac{\delta \alpha}{\alpha} \sim - \xi 
\frac{\left(H^2 \Omega_m\right)_P-\left(H^2 \Omega_m\right)}{8\pi H^2} 
\left(a_P r_0 H\right)^2\,,
\ee
where the $P$ represents the closed perturbed universe. We can already anticipate that these two different approaches for estimating spatial variations of $\alpha$ will produce very different results since in the first approach there is no reference to the size of the fluctuation (an infinite approximation is considered).

\section{\label{res}Numerical Results}
\begin{figure} 
\begin{center}
\includegraphics*[width=9cm]{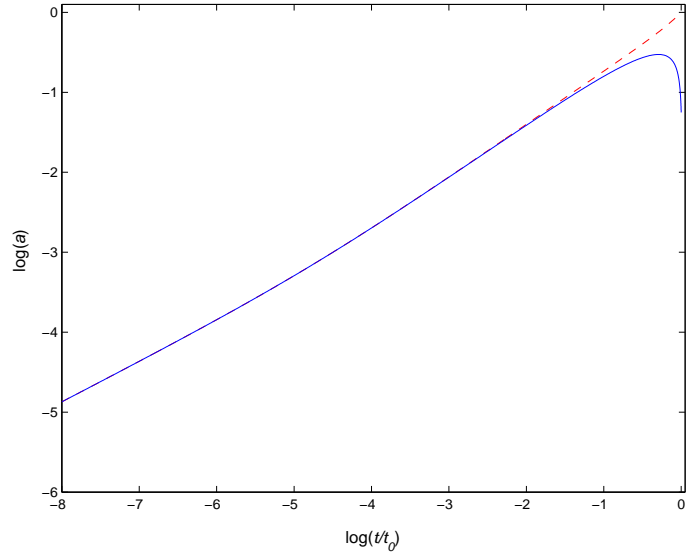}  
\end{center}
\caption[Evolution of $\alpha$ in the non-linear regime: $a(t)$.]{\label{fig1}Evolution of the scale factor, $a$, as a function of 
physical time, $t$, for the background and perturbed universes (solid and 
dashed lines respectively)} 
\end{figure} 
\begin{figure} 
\begin{center}
\includegraphics*[width=9cm]{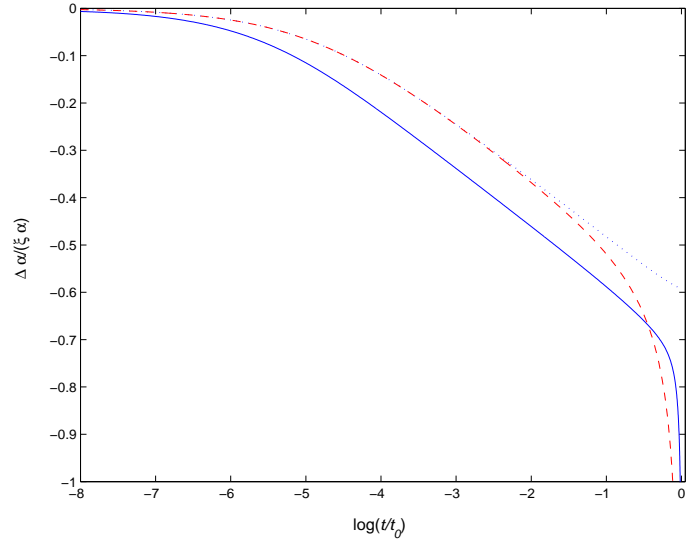} 
\end{center}
\caption[Evolution of $\alpha$ in the non-linear regime: $\alpha(t)$.]{\label{fig2}Evolution of the fine structure constant, $\alpha$, 
as a function of physical time, $t$, in the vicinity of spherical 
distribution of mass including the spatial variations of $\alpha$ calculated 
using the local approximation with $r_0=2H_0^{-1}$ (solid line) and the infinite wavelength approximation (dashed line). The dotted line represents the background evolution of $\alpha$. Note that $\xi < 6 \times 10^{-6} \gamma_F$ is a 
very small number.}
\end{figure} 
\begin{figure} 
\begin{center}
\includegraphics*[width=9cm]{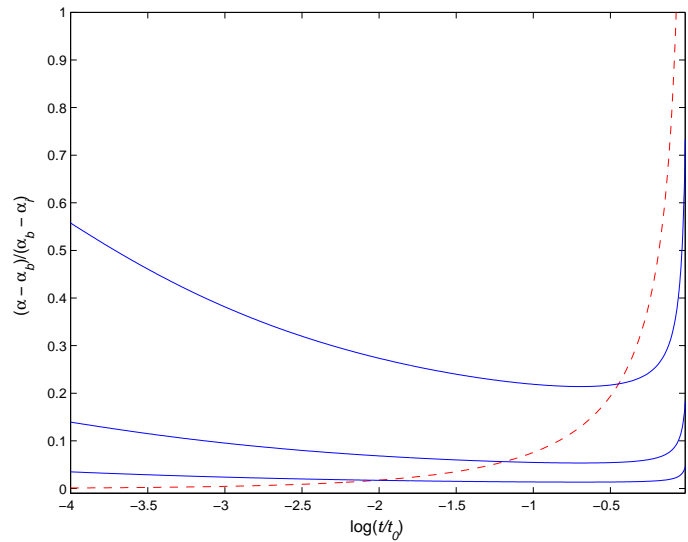} 
\end{center}
\caption[Evolution of $\alpha$ in the non-linear regime: $(\alpha-\alpha_b)/(\alpha_b-\alpha_i)$.]
{\label{fig3} Evolution of $(\alpha-\alpha_b)/(\alpha_b-\alpha_i)$, in various 
models as a function of physical time, $t$. The dashed line represents the 
infinite wavelength approximation while the solid lines use the local 
approximation with $r_0=2H_0^{-1}$, $r_0=H_0^{-1}$ and $r_0=H_0^{-1}/2$ 
(top to bottom respectively).}
\end{figure} 
The initial values of the various cosmological parameters were chosen in such way that at present time we have 
$\Omega_m^0=0.29$, $\Omega_{\Lambda_0}=0.71$ and $\Omega_{r0}=8.4 \times 10^{-5}$
for the background universe. 
Also, the value of $\Delta \Omega_{mi}$ is such that the collapse of the perturbed 
universe occurs near the present epoch. This is clearly seen in Fig. \ref{fig1}
which plots the evolution of the scale factor $a$ as 
as a function 
of the physical time, $t$, for the background and perturbed universes 
(solid and dashed lines respectively). 
Note that, according to the spherical 
collapse model, a uniform density perturbation will become infinite in a 
finite time. Clearly this is not realistic (at least for local fluctuations) 
since in practice the collapse is stopped by virialization.

We plot in Fig. \ref{fig2} 
the b evolution of $\Delta \alpha/\alpha$ 
obtained using the two different models. 
The background evolution is shown by the dotted line, while $\Delta \alpha/\alpha$ for \textit{infinite wavelength} and
\textit{local approximation} models are plotted with dashed and solid lines, respectively.
The crucial difference among the two approximations is seen immediately by solving the equations of evolution: while the  
evolution of $\Delta \alpha/\alpha$ in the local approximation depends on the final size of the perturbation, $r_0$, no reference to the perturbation size is made 
in the second case. Actually, for the latter model a uniform infinite wavelength is assumed. Hence, the latter model will produce a single result 
while the first method will produce results which depend explicitly 
on the final perturbation size, $r_0$. 
For example, the results shown in Fig. \ref{fig2}  for 
the local approximation are obtained by assuming $r_0=2H_0^{-1}$. If one works in the non-linear regime, this is clearly an unrealistic value 
of $r_0$. However,  we have used it in order to get a sizable depart from the background 
$\Delta \alpha/\alpha$ (note that in order to be self-consistent the model 
requires that $r \ll H^{-1}$) and the result can easily be rescaled 
for realistic values of $r_0$ using (\ref{phieq6}).
A more realistic value of $r_0$ is the radius of a typical cluster of galaxies $(r_0 \sim 10 {\rm Mpc})$. In this case the spatial variations of $\alpha$ calculated using the local approximation need to be rescaled down by roughly six orders of magnitude with respect to the unrealistic $r_0 \sim 2\,H_0^{-1}$ case shown in Figs. \ref{fig2} and \ref{fig3}.

For observing explicitly the behavior of the evolution of  the fine-structure constant when
smaller values of $r_0$ are considered, that is, when it is assumed mass distributions into the non-linear
regime we compute the results for $(\alpha-\alpha_b)/(\alpha_b-\alpha_i)$  using the different models as a function of physical time, $t$. In Fig. \ref{fig3}, the dashed line represents the one result obtained by using the 
infinite wavelength approximation and the solid lines represent the results obtained by the local 
approximation with $r_0=2H_0^{-1}$, $r_0=H_0^{-1}$ and $r_0=H_0^{-1}/2$ 
(top to bottom respectively). Although we have started by non-realistic values of the radius of the 
spherical distribution, it is clear that as we move towards 
smaller values of $r_0$ the local approximation will 
give negligible variations of $\alpha$, that is indeed the correct conclusion for realistic cases. 

\section{\label{disc} Discussion}
It is interesting to discuss the evolution of $\delta \alpha/\alpha$ in the linear regime during the matter and radiation eras.
In 
the linear regime the physical radius of the spherical overdensity grows as
$r \propto a$,
and consequently $M$ varies as
\begin{equation}
M \propto \delta \rho_m a^3 = \delta_m {\bar \rho}_m a^3\,.
\end{equation}
We conclude that the spherical mass distribution is approximately constant during the radiation era and grows proportionally to $a$ during the matter era. 

Using the local approximation one has that 
according to the solution of the Poisson equation outside the overdensity, deep into the radiation era 
\be
\frac{\delta \alpha}{\alpha} \propto a^{-1}\,,
\ee
while deep into the matter era
\be
\frac{\delta \alpha}{\alpha} \sim  {\rm constant}\,.
\ee
It is also important to point out 
that the amplitude of the $\alpha$ variation (for a given $\delta \rho_m$) 
depends quadratically on the radius of the spherical overdensity 
($\Delta \alpha/\alpha \propto r^2$). Consequently, for $r_0 \ll H_0^{-1}$ 
the spatial variations predicted by the local approximation will be very small. The background evolution of $\Delta \alpha/\alpha$ in linearized Bekenstein 
models has been previously studied in \cite{linear} and it is well known that 
$\Delta \alpha/\alpha \to 0$ deep into the 
radiation era.

This is confirmed by the results plotted in Fig. \ref{fig2} which clearly 
confirm the above discussion. In both approximations
\begin{equation}
\frac{\Delta \alpha}{\alpha} \to \left(\frac{\Delta \alpha}{\alpha}\right)_b 
\to 0\,
\end{equation}
deep into the radiation era. On the other hand  
\begin{equation}
\left(\frac{\Delta \alpha}{\alpha}\right)_\ell- 
\left(\frac{\Delta \alpha}{\alpha}\right)_b \sim {\rm constant}\,
\end{equation}
deep into the matter era. Also, as expected the evolution of 
$\Delta \alpha/\alpha$ in the infinite wavelength approximation only departs significantly from the background one near the present time (when the collapse of the perturbed universe occurs).

\section{\label{capconclusions}{\sf Conclusions}}
We have presented and contrasted two different approximations for the non-linear evolution of $\alpha$. 
While in the local approximation 
we assume that the size of the perturbation which gives rise to spatial 
variation of $\alpha$ is much smaller than the Hubble radius, $H^{-1}$, 
in the infinite wavelength approximation the opposite is required for self-consistency. Hence, it is not surprising that the results obtained using the two models are so different. The crucial question is which one provides the right answer when applied to cosmology.
Clearly, the infinite wavelength approximation is not 
a good approximation in general because we are not interested in an homogeneous 
perturbation with a size larger than the Hubble radius.  Note that in this 
approximation not only the gradients are neglected but it is also assumed 
that the Hubble parameter, $H$, is the local 
one (not the global $H$). In addition, the fact that the matter density is homogeneously 
distributed inside a given spherical region does not imply that $\alpha$ will 
also be homogeneous within that region.

Therefore we are convinced that local approximation provides the right answer since the non-linear effects are only expected to be important 
in this context on scales much smaller than $H_0^{-1}$ (scales smaller than the typical galaxy cluster size). On such small scales, we therefore predict that the spatial variations of $\alpha$ generated in 
the simplest models should be too small to be of cosmological interest.

These results confirm that it is difficult to generate significant 
large-scale spatial variations of $\alpha$ as it is shown in previous chapter for the special cases of cosmic strings and magnetic monopoles \cite{Nosso,Nosso2} 
even when we account for the evolution of the fine structure constant in the 
non-linear regime, when density perturbations become much larger than unity. This results were also confirmed by other authors in Ref. \cite{Barrow3}--\cite{Barrow2}.

\chapter{\label{capconclusions}{\sf Outlook}}
In this thesis we have studied the cosmological consequences of topological defects, in particular for the dark energy scenario and varying fundamental constants theories. In the following, we summarize our main results and conclusions and address future works.

In Chapter \ref{cap2}, we have considered the possibility of a domain wall network as a source of dark energy to explain the recent acceleration of the universe. 
\begin{itemize} 
\item
We have presented an analytic model that describes very well the domain wall evolution. We have taken into account the curvature and the interactions of the walls with radiation and other particles and derived a strong bound on the curvature of the walls, which shows that viable candidate networks must be fine-tuned and non-standard. 
\item
We have studied several typical domain wall models in two spatial dimensions, and discussed the conditions under which various types of defect junctions can exist.We have highlighted the situations where only stable
Y-type junctions or X-type ones are formed. In addition we have shown the case where both types co-exist. 
\item
Using geometrical, topological and energy arguments we 
distinguished which features of a particular model are crucial to construct 
the best candidate potential for frustration. 
However, our
results strongly lead us to a
no frustration conjecture. In all our numerical simulations, including those of the ideal model, we have found a dynamical behavior consistent with a convergence towards a scaling solution and no evidence of any behavior that could eventually lead to frustration. 
\item
The results of 3D simulations clearly indicate that no frustrated network can be formed dynamically out of random initial conditions that would mimic a realistic phase transition.
Therefore we have presented the most compelling evidence to date that domain wall networks can not be the dark energy.
\end{itemize}

In this context, it is interesting to study the dependence of our results on the number of space-time dimensions.
Another relevant problem is the role of the 
junctions in the dynamics of the network. 
We recall that the type of junctions formed will depend both on energetic considerations and on the topology of the minima of the potential in field space.
In the future we intend to consider further examples of particular models and study this issue with analytic and numerical tools.

In Chapter \ref{cap3} 
we have investigated cosmic strings in the context of varying-$\alpha$ theories.
\begin{itemize}
\item
We have
considered a generic gauge kinetic 
function and
shown that there
is a class of models of this type for 
which the classical Nielsen-Olesen vortex is still a valid solution. 
However, in general,
the solutions of the local cosmic strings depart from the standard case.
\item
We have shown that
the 
Equivalence Principle constraints impose tight limits on the allowed
variations of $\alpha$ on cosmological scales induced by
cosmic string networks of this type. 
\end{itemize}
In Chapter \ref{cap4} we have investigated 
the variations of the fine-structure constant in the vicinity of a static local magnetic monopole. 
\begin{itemize}
\item
We have introduced
the Bekenstein-type models in the Non-Abelian Yang-Mills theory 
and studied various models with varying-$\alpha$ and confirmed that despite the existence of a class of models for which the 't Hooft-Polyakov standard solution is still valid. 
\item
We have verified that, in general,
the solutions of the local magnetic monopoles depart from the standard case.
\item
We have shown that Equivalence Principle 
constraints impose tight limits on the variations of $\alpha$ induced by
magnetic monopoles which confirms 
the difficulty to generate significant large-scale spatial variation of 
the fine structure constant found in the previous section, even in the most favorable case 
where these variations are seeded by magnetic monopoles.
\end{itemize}
In Chapter \ref{cap5} we
have investigated the evolution of the fine-structure constant in the non-linear regime.
\begin{itemize}
\item
We have presented and contrasted two different approximations for the non-linear evolution of $\alpha$. 
While in the local approximation 
we have assumed that the size of the perturbation which gives rise to spatial 
variation of $\alpha$ is much smaller than the Hubble radius, $H^{-1}$, 
in the infinite wavelength approximation the opposite is required for self-consistency. 
\item
We have shown that
the infinite wavelength approximation is not 
a good approximation in general because we are not interested in an homogeneous 
perturbation with a size larger than the Hubble radius.  Note that in this 
approximation not only the gradients are neglected but it is also assumed 
that the Hubble parameter, $H$, is the local 
one (not the global $H$). In addition, the fact that the matter density is homogeneously 
distributed inside a given spherical region does not imply that $\alpha$ will 
also be homogeneous within that region.
\item
We have concluded that local approximation provides the right answer since the non-linear effects are only expected to be important 
in this context on scales much smaller than $H_0^{-1}$ (scales smaller than the typical galaxy cluster size). On such small scales, we therefore predict that the spatial variations of $\alpha$ generated in 
the simplest models should be too small to be of any cosmological interest.
\item
We have presented results which confirm that it is difficult to generate significant 
large-scale spatial variations of $\alpha$ as it is shown in previous chapter for the special cases of cosmic strings and magnetic monopoles even when we account for the evolution of the fine structure constant in the 
non-linear regime, when density perturbations become much larger than unity.

\end{itemize}

Overall, we have assumed that the spatial variations of $\alpha$ in the vicinity of cosmic strings, magnetic monopoles and compact objects are
sourced only by the gauge kinetic 
function $B_F$ coupled to the Maxwell term. 
A generalization of this model,
we intend to investigate in the near future,
includes different source of variation of the fine-structure constant due to a discrete spontaneous 
symmetry breaking which produces domain walls dividing the universe in patches with different values of $\alpha$. 
Of course, 
the symmetry breaking scale must be sufficiently low in order for the domain walls 
to be cosmologically benign.
This is an interesting model, in which domain walls may be responsible for cosmologically relevant spatial variations of the fine-structure constant as well as other fundamental parameters.

\chapter*{{\sf Appendix - Numerical Codes - Domain Wall Network Evolution}}
\small
\section*{The Main Code}
In this section we present the C code used for performing the field theory numerical simulation of domain wall networks evolution. This version carries out the evolution of ideal class of models for $N=20$ and was optimized for the shared memory architeture of the COSMOS supercomputer were the simulations were performed. For a $512^3$ box, this simulation took about 2 hours and 15 minutes on 128 processors.
/*------------------------------------------------------*/\\
/* VT 26/10/06 v10                                       */\\
/* Compile with -DRAW for binary data output            */\\
/* Compile with -DTEST for timings \& fixed seed=0       */\\
/*------------------------------------------------------*/\\
\\
\#include <stdio.h>\\
\#include <math.h>\\
\#include <string.h>\\
\#include <stdlib.h>\\
\#include "walls.h"\\
\#ifdef \_OPENMP\_\\
\#include <omp.h>\\
\#endif\\
\\
int main(int argc, char **argv)\\
\{\\
 static  double S0[dim1][dim2][dim3][dim4][inicampos];\\
 static  double S1[dim1][dim2][dim3][dim4][inicampos];\\
 static  double DS0[dim1][dim2][dim3][dim4][inicampos];\\
 static  double DS1[dim1][dim2][dim3][dim4][inicampos];\\
 static  double GS[dim1][dim2][dim3][dim4][inicampos];\\
  double GT;\\
  static  float GTf[dim1][dim2][dim3];\\
  double VS[inicampos+1];\\
  double K[23][23];\\
  double R[inicampos+1],cons;\\
  int m,wq,qw;\\
  /*integration variables inicialization*/\\
  int i,ip1,im1;\\
  int j,jp1,jm1;\\
  int k,kp1,km1;\\
  int l,lp1,lm1,NQ;\\
  int ncounter;\\
  /*evolution variables inicialization*/\\
  double ct,dct,dx,dx2;\\
  double ctfinal;\\
  double at;\\
  double delta;\\
  double D1,D2;\\
  /*area and energy calculation inicialization*/\\
  double V0t4;\\
  double miu;\\
  double A,Pi;\\
  double Lshi[inicampos];\\
  double ek;\\
  double AREA, area;\\
  double dif1,dif2,dif3,dif4;\\
  double V0, V; \\
  char doutfilme[MAXBUF], doutscaling[MAXBUF];\\
  /* files*/\\
  FILE *dados, *foutfilme, *scaling, *fcp, *fstep;\\
  /*simulation parameters inicialization*/\\
  int N1=dim1;\\
  int N2=dim2;\\
  int N3=dim3;\\
  int N4=dim4;\\
  double lamda;\\
  long idum;\\
  int skipcp = 0;\\
  char sstep[MAXBUF];\\
  char scp[MAXBUF];\\
  double nfields1;\\
  double epsilon=-0.1;\\
  double omega0=5.0; \\
  int num,teste;\\
  double alfa=3.0;\\
  double beta0=0.0;\\
  double W0=10.0;\\
  double a,b,e;\\
  /*dynamical range inicialization*/\\
  double zi;\\
  double t0,ti;\\
  int counter1, contar,D,c,s, cont, cont2;\\
  double expoentec,alpha;\\
  double eklim,ekaux,ekauxQ;\\
  int npontos,npontosQ;\\
  double assim=1.0;\\
  double r=0.5;\\
  int arg = (argc < 2 ? 1 : atoi(argv[1]));\\
  char nomeficheiro[20];\\
  /* OpenMP */\\
  double my\_eklim, my\_AREA, my\_ek;\\
  int my\_npontos;\\
  /* timing */\\
  double t1,t2,t3,t4,t5, ta,tb,tc,td;\\
  double csecond();\\
  t1 = csecond();\\
\#ifdef TEST\\
  idum = 0;\\
\#else\\
  idum = time (NULL);\\
\#endif\\
  srand48(idum);\\
  /* get parameters from cmd line or default ini file */\\
  if ( argv[1] != NULL \&\& ((dados=fopen(argv[1],"r")) != NULL) ) \{\\
    fscanf(dados,"\%lf \%lf \%lf \%lf \%d \%d \%d \%d \ n",		    \\  
	   \& lim, \& expoente,\& ctinitial,\& tampasso,\& npassos,\& nfields,\& noutput,\& nckpt);\\
  \} else if ( (dados=fopen(inifile,"r")) != NULL )\\
  \{\\
    fscanf(dados,"\%lf \%lf \%lf \%lf \%d \%d \%d \%d \ n",		      \\
	   \& lim, \& expoente,\& ctinitial,\& tampasso,\& npassos,\& nfields,\& noutput,\& nckpt);\\
  \}\\
    print\_parameters();\\
    contar=0.0;\\
    dct=tampasso*ctinitial;\\
    ctfinal=ctinitial+npassos*dct;\\
    expoentec=expoente/(1-expoente);\\
    Pi=3.1415927;\\
    dx=ctinitial;\\
    dx2=dx*dx;\\
    ct=ctinitial;\\
    nfields1=nfields+1.0;\\
    ncounter=0;	\\
    counter1=0;\\
    lamda=0.3/(nfields+5.0);\\
    V0=0.145; \\
    c=0;\\
    /*******************Minima matrix****************************/\\
    K[1][1]=1.0;	K[1][2]=0.0;           
    K[1][3]=0.0;	K[1][4]=0.0;           
    K[1][5]=0.0;\\	K[1][6]=0.0;             
    K[1][7]=0.0;	K[1][8]=0.0;  
    K[1][9]=0.0;	K[1][10]=0.0; \\
    K[1][11]=0.0;	K[1][12]=0.0;    
    K[1][13]=0.0;	K[1][14]=0.0; 
    K[1][15]=0.0;\\	K[1][16]=0.0; 
    K[1][17]=0.0;	K[1][18]=0.0;
    K[1][19]=0.0;	K[1][20]=0.0; \\
    /***/\\
    K[2][1]=-1.0/2.0;	K[2][2]=sqrt(3.0)/2.0; 
    K[2][3]=0.0;		K[2][4]=0.0;           
    K[2][5]=0.0;\\		K[2][6]=0.0;             
    K[2][7]=0.0;		K[2][8]=0.0;  
    K[2][9]=0.0;		K[2][10]=0.0; \\
    K[2][11]=0.0;		K[2][12]=0.0;  
    K[2][13]=0.0;		K[2][14]=0.0; 
    K[2][15]=0.0;\\		K[2][16]=0.0; 
    K[2][17]=0.0;		K[2][18]=0.0; 
    K[2][19]=0.0;		K[2][20]=0.0; \\
    /***/\\
    K[3][1]=-1.0/2.0;	K[3][2]=-sqrt(3.0)/2.0; 
    K[3][3]=0.0;		K[3][4]=0.0;           
    K[3][5]=0.0;	\\	K[3][6]=0.0;             
    K[3][7]=0.0;		K[3][8]=0.0;  
    K[3][9]=0.0;		K[3][10]=0.0; \\
    K[3][11]=0.0;		K[3][12]=0.0;  
    K[3][13]=0.0;		K[3][14]=0.0; 
    K[3][15]=0.0;\\		K[3][16]=0.0; 
    K[3][17]=0.0;		K[3][18]=0.0; 
    K[3][19]=0.0;		K[3][20]=0.0; \\
    /***/\\
    K[4][1]=0.0;		K[4][2]=0.0;           
    K[4][3]=sqrt(2.0);	K[4][4]=0.0;           
    K[4][5]=0.0;	\\	K[4][6]=0.0;             
    K[4][7]=0.0;		K[4][8]=0.0;  
    K[4][9]=0.0;		K[4][10]=0.0; \\
    K[4][11]=0.0;		K[4][12]=0.0;  
    K[4][13]=0.0;		K[4][14]=0.0; 
    K[4][15]=0.0;	\\	K[4][16]=0.0; 
    K[4][17]=0.0;		K[4][18]=0.0; 
    K[4][19]=0.0; 	K[4][20]=0.0; \\
    /***/\\
    K[5][1]=0.0;    		K[5][2]=0.0;           
    K[5][3]=sqrt(2.0)/4.0;	K[5][4]=sqrt(30.0)/4.0;
    K[5][5]=0.0;	\\		K[5][6]=0.0;             
    K[5][7]=0.0;			K[5][8]=0.0;  
    K[5][9]=0.0;			K[5][10]=0.0; \\
    K[5][11]=0.0;			K[5][12]=0.0;  
    K[5][13]=0.0;			K[5][14]=0.0; 
    K[5][15]=0.0;	\\		K[5][16]=0.0; 
    K[5][17]=0.0;			K[5][18]=0.0; 
    K[5][19]=0.0; 		K[5][20]=0.0; \\
    /***/\\
    K[6][1]=0.0;			K[6][2]=0.0;           
    K[6][3]=sqrt(2.0)/4.0;	K[6][4]=sqrt(30.0)/20.0;
    K[6][5]=3.0*sqrt(5.0)/5.0;	\\ K[6][6]=0.0;             
    K[6][7]=0.0;			K[6][8]=0.0;  
    K[6][9]=0.0;			K[6][10]=0.0; \\
    K[6][11]=0.0;			K[6][12]=0.0;  
    K[6][13]=0.0;			K[6][14]=0.0; 
    K[6][15]=0.0;	\\		K[6][16]=0.0; 
    K[6][17]=0.0;			K[6][18]=0.0; 
    K[6][19]=0.0; 		K[6][20]=0.0; \\
    /***/\\
    K[7][1]=0.0;			K[7][2]=0.0;           
    K[7][3]=sqrt(2.0)/4.0;	K[7][4]=sqrt(30.0)/20.0;
    K[7][5]=sqrt(5.0)/10.0;\\	K[7][6]=sqrt(7.0)/2.0;   
    K[7][7]=0.0;			K[7][8]=0.0;  
    K[7][9]=0.0;			K[7][10]=0.0; \\
    K[7][11]=0.0;			K[7][12]=0.0;  
    K[7][13]=0.0;			K[7][14]=0.0; 
    K[7][15]=0.0;\\			K[7][16]=0.0; 
    K[7][17]=0.0;			K[7][18]=0.0; 
    K[7][19]=0.0; 		K[7][20]=0.0; \\
    /***/\\
    K[8][1]=0.0;			K[8][2]=0.0;           
    K[8][3]=sqrt(2.0)/4.0;	K[8][4]=sqrt(30.0)/20.0;
    K[8][5]=sqrt(5.0)/10.0;\\	K[8][6]=sqrt(7.0)/14.0;  
    K[8][7]=2.0*sqrt(21.0)/7.0;	K[8][8]=0.0;  
    K[8][9]=0.0;			K[8][10]=0.0; \\
    K[8][11]=0.0;			K[8][12]=0.0;  
    K[8][13]=0.0;			K[8][14]=0.0; 
    K[8][15]=0.0;\\			K[8][16]=0.0; 
    K[8][17]=0.0;			K[8][18]=0.0; 
    K[8][19]=0.0;			K[8][20]=0.0; 	 	  \\
    /***/\\
    K[9][1]=0.0;			K[9][2]=0.0;           
    K[9][3]=sqrt(2.0)/4.0;	K[9][4]=sqrt(30.0)/20.0;
    K[9][5]=sqrt(5.0)/10.0;\\	K[9][6]=sqrt(7.0)/14.0;  
    K[9][7]=sqrt(21.0)/28.0;	K[9][8]=3.0*sqrt(3.0)/4.0;  
    K[9][9]=0.0;			K[9][10]=0.0; \\
    K[9][11]=0.0;			K[9][12]=0.0;  
    K[9][13]=0.0;			K[9][14]=0.0; 
    K[9][15]=0.0;		\\	K[9][16]=0.0; 
    K[9][17]=0.0;			K[9][18]=0.0; 
    K[9][19]=0.0;			K[9][20]=0.0; \\
    /***/	 \\
    K[10][1]=0.0;			K[10][2]=0.0;           
    K[10][3]=sqrt(2.0)/4.0;	K[10][4]=sqrt(30.0)/20.0; 
    K[10][5]=sqrt(5.0)/10.0;\\	K[10][6]=sqrt(7.0)/14.0;  
    K[10][7]=sqrt(21.0)/28.0;	K[10][8]=sqrt(3.0)/12.0;  \\
    K[10][9]=sqrt(15.0)/3.0;	K[10][10]=0.0; \\
    K[10][11]=0.0;		K[10][12]=0.0;   
    K[10][13]=0.0;		K[10][14]=0.0; 
    K[10][15]=0.0;	\\	K[10][16]=0.0; 
    K[10][17]=0.0;		K[10][18]=0.0; 
    K[10][19]=0.0;		K[10][20]=0.0; \\
    /***/\\
    K[11][1]=0.0;			K[11][2]=0.0;           
    K[11][3]=sqrt(2.0)/4.0;	K[11][4]=sqrt(30.0)/20.0; 
    K[11][5]=sqrt(5.0)/10.0;\\	K[11][6]=sqrt(7.0)/14.0;  
    K[11][7]=sqrt(21.0)/28.0;	K[11][8]=sqrt(3.0)/12.0;  
    K[11][9]=sqrt(15.0)/30.0;	K[11][10]=sqrt(165.0)/10.0; \\
    K[11][11]=0.0;		K[11][12]=0.0;
    K[11][13]=0.0;		K[11][14]=0.0; 
    K[11][15]=0.0;	\\	K[11][16]=0.0; 
    K[11][17]=0.0;		K[11][18]=0.0; 
    K[11][19]=0.0;		K[11][20]=0.0; \\
    /***/\\
    K[12][1]=0.0;			K[12][2]=0.0;           
    K[12][3]=sqrt(2.0)/4.0;	K[12][4]=sqrt(30.0)/20.0; 
    K[12][5]=sqrt(5.0)/10.0;\\	K[12][6]=sqrt(7.0)/14.0;  
    K[12][7]=sqrt(21.0)/28.0;	K[12][8]=sqrt(3.0)/12.0;  
    K[12][9]=sqrt(15.0)/30.0;	K[12][10]=sqrt(165.0)/110.0; \\
    K[12][11]=3.0*sqrt(22.0)/11.0;K[12][12]=0.0;
    K[12][13]=0.0;		K[12][14]=0.0; 
    K[12][15]=0.0;	\\	K[12][16]=0.0; 
    K[12][17]=0.0;		K[12][18]=0.0; 
    K[12][19]=0.0;		K[12][20]=0.0; \\
    /***/\\
    K[13][1]=0.0;			K[13][2]=0.0;           
    K[13][3]=sqrt(2.0)/4.0;	K[13][4]=sqrt(30.0)/20.0; 
    K[13][5]=sqrt(5.0)/10.0;\\	K[13][6]=sqrt(7.0)/14.0;  
    K[13][7]=sqrt(21.0)/28.0;	K[13][8]=sqrt(3.0)/12.0;  
    K[13][9]=sqrt(15.0)/30.0;	K[13][10]=sqrt(165.0)/110.0; \\
    K[13][11]=sqrt(22.0)/44.0;	K[13][12]=sqrt(26.0)/4.0;
    K[13][13]=0.0;		K[13][14]=0.0; 
    K[13][15]=0.0;\\		K[13][16]=0.0; 
    K[13][17]=0.0;		K[13][18]=0.0; 
    K[13][19]=0.0;		K[13][20]=0.0;  \\
    /***/\\
    K[14][1]=0.0;			K[14][2]=0.0;           
    K[14][3]=sqrt(2.0)/4.0;	K[14][4]=sqrt(30.0)/20.0;           
    K[14][5]=sqrt(5.0)/10.0;\\	K[14][6]=sqrt(7.0)/14.0;            
    K[14][7]=sqrt(21.0)/28.0;	K[14][8]=sqrt(3.0)/12.0; 
    K[14][9]=sqrt(15.0)/30.0;	K[14][10]=sqrt(165.0)/110.0; \\
    K[14][11]=sqrt(22.0)/44.0;	K[14][12]=sqrt(26.0)/52.0;
    K[14][13]=sqrt(273.0)/13.0;	K[14][14]=0.0;\\
    K[14][15]=0.0;	\\	K[14][16]=0.0; 
    K[14][17]=0.0;		K[14][18]=0.0; 
    K[14][19]=0.0;		K[14][20]=0.0;  \\
    /***/\\
    K[15][1]=0.0;			K[15][2]=0.0;           
    K[15][3]=sqrt(2.0)/4.0;	K[15][4]=sqrt(30.0)/20.0;           
    K[15][5]=sqrt(5.0)/10.0;\\	K[15][6]=sqrt(7.0)/14.0;             
    K[15][7]=sqrt(21.0)/28.0;	K[15][8]=sqrt(3.0)/12.0; 
    K[15][9]=sqrt(15.0)/30.0;	K[15][10]=sqrt(165.0)/110.0; \\
    K[15][11]=sqrt(22.0)/44.0;	K[15][12]=sqrt(26.0)/52.0;
    K[15][13]=sqrt(273.0)/182.0;\\	K[15][14]=3.0*sqrt(35.0)/14.0; 
    K[15][15]=0.0; \\		K[15][16]=0.0; 
    K[15][17]=0.0;		K[15][18]=0.0; 
    K[15][19]=0.0;		K[15][20]=0.0;  \\
    /***/\\
    K[16][1]=0.0;			K[16][2]=0.0;           
    K[16][3]=sqrt(2.0)/4.0;	K[16][4]=sqrt(30.0)/20.0;           
    K[16][5]=sqrt(5.0)/10.0;\\	K[16][6]=sqrt(7.0)/14.0;             
    K[16][7]=sqrt(21.0)/28.0;	K[16][8]=sqrt(3.0)/12.0; 
    K[16][9]=sqrt(15.0)/30.0;	K[16][10]=sqrt(165.0)/110.0; \\
    K[16][11]=sqrt(22.0)/44.0;	K[16][12]=sqrt(26.0)/52.0;
    K[16][13]=sqrt(273.0)/182.0;\\	K[16][14]=sqrt(35.0)/70.0; 
    K[16][15]=2.0*sqrt(10.0)/5.0;\\	K[16][16]=0.0; 
    K[16][17]=0.0;		K[16][18]=0.0; 
    K[16][19]=0.0;		K[16][20]=0.0;  \\
    /***/\\
    K[17][1]=0.0;			K[17][2]=0.0;           
    K[17][3]=sqrt(2.0)/4.0;	K[17][4]=sqrt(30.0)/20.0;           
    K[17][5]=sqrt(5.0)/10.0;\\	K[17][6]=sqrt(7.0)/14.0;             
    K[17][7]=sqrt(21.0)/28.0;	K[17][8]=sqrt(3.0)/12.0; 
    K[17][9]=sqrt(15.0)/30.0;	K[17][10]=sqrt(165.0)/110.0; \\
    K[17][11]=sqrt(22.0)/44.0;	K[17][12]=sqrt(26.0)/52.0;
    K[17][13]=sqrt(273.0)/182.0;\\	K[17][14]=sqrt(35.0)/70.0; 
    K[17][15]=sqrt(10.0)/40.0;\\	K[17][16]=sqrt(102.0)/8.0; 
    K[17][17]=0.0;		K[17][18]=0.0; 
    K[17][19]=0.0;		K[17][20]=0.0;  \\
    /***/\\
    K[18][1]=0.0;			K[18][2]=0.0;           
    K[18][3]=sqrt(2.0)/4.0;	K[18][4]=sqrt(30.0)/20.0;          
    K[18][5]=sqrt(5.0)/10.0;\\	K[18][6]=sqrt(7.0)/14.0;            
    K[18][7]=sqrt(21.0)/28.0;	K[18][8]=sqrt(3.0)/12.0; 
    K[18][9]=sqrt(15.0)/30.0;	K[18][10]=sqrt(165.0)/110.0; \\
    K[18][11]=sqrt(22.0)/44.0;	K[18][12]=sqrt(26.0)/52.0;
    K[18][13]=sqrt(273.0)/182.0;\\	K[18][14]=sqrt(35.0)/70.0;
    K[18][15]=sqrt(10.0)/40.0;\\	K[18][16]=sqrt(102.0)/136.0; 
    K[18][17]=3.0*sqrt(51.0)/17.0;K[17][18]=0.0; 
    K[18][19]=0.0;\\		K[17][20]=0.0;  \\
    /***/\\
    K[19][1]=0.0;			K[19][2]=0.0;           
    K[19][3]=sqrt(2.0)/4.0;	K[19][4]=sqrt(30.0)/20.0;         
    K[19][5]=sqrt(5.0)/10.0;\\	K[19][6]=sqrt(7.0)/14.0;             
    K[19][7]=sqrt(21.0)/28.0;	K[19][8]=sqrt(3.0)/12.0; 
    K[19][9]=sqrt(15.0)/30.0;	K[19][10]=sqrt(165.0)/110.0; \\
    K[19][11]=sqrt(22.0)/44.0;	K[19][12]=sqrt(26.0)/52.0;
    K[19][13]=sqrt(273.0)/182.0;\\	K[19][14]=sqrt(35.0)/70.0;
    K[19][15]=sqrt(10.0)/40.0;\\	K[19][16]=sqrt(102.0)/136.0; 
    K[19][17]=sqrt(51.0)/102.0;	K[19][18]=sqrt(57.0)/6.0; 
    K[19][19]=0.0;\\		K[19][20]=0.0;   \\
    /***/\\
    K[20][1]=0.0;			K[20][2]=0.0;           
    K[20][3]=sqrt(2.0)/4.0;	K[20][4]=sqrt(30.0)/20.0;          
    K[20][5]=sqrt(5.0)/10.0;\\	K[20][6]=sqrt(7.0)/14.0;          
    K[20][7]=sqrt(21.0)/28.0;	K[20][8]=sqrt(3.0)/12.0; 
    K[20][9]=sqrt(15.0)/30.0;	K[20][10]=sqrt(165.0)/110.0; \\
    K[20][11]=sqrt(22.0)/44.0;	K[20][12]=sqrt(26.0)/52.0;
    K[20][13]=sqrt(273.0)/182.0;\\	K[20][14]=sqrt(35.0)/70.0;
    K[20][15]=sqrt(10.0)/40.0;\\	K[20][16]=sqrt(102.0)/136.0; 
    K[20][17]=sqrt(51.0)/102.0;	K[20][18]=sqrt(57.0)/114.0; \\
    K[20][19]=sqrt(570)/19.0;	K[20][20]=0.0;   \\
    /***/\\
    K[21][1]=0.0;			K[21][2]=0.0;           
    K[21][3]=sqrt(2.0)/4.0;	K[21][4]=sqrt(30.0)/20.0;          
    K[21][5]=sqrt(5.0)/10.0;\\	K[21][6]=sqrt(7.0)/14.0;            
    K[21][7]=sqrt(21.0)/28.0;	K[21][8]=sqrt(3.0)/12.0; 
    K[21][9]=sqrt(15.0)/30.0;	K[21][10]=sqrt(165.0)/110.0; \\
    K[21][11]=sqrt(22.0)/44.0;	K[21][12]=sqrt(26.0)/52.0;
    K[21][13]=sqrt(273.0)/182.0;\\	K[21][14]=sqrt(35.0)/70.0;
    K[21][15]=sqrt(10.0)/40.0;\\	K[21][16]=sqrt(102.0)/136.0; 
    K[21][17]=sqrt(51.0)/102.0;	K[21][18]=sqrt(57.0)/114.0; \\
    K[21][19]=sqrt(570)/380.0;	K[21][20]=3.0*sqrt(70.0)/20.0;   \\
    /* initialize in parallel to get first touch right - vt */\\
\#pragma omp parallel for default (none) \\
            private(i,j,k,l,cont) shared(N1,N2,N3,N4,nfields,S0,S1,DS0,DS1,GS, GTf)\\
    for(i=0;i<N1;i++)\{\\
      for(j=0;j<N2;j++)\{\\
        for(k=0;k<N3;k++)\{\\
	    GTf[i][j][k] = 0.0;\\
	    for(l=0;l<N4;l++)\{\\
	      for (cont=0;cont<nfields;cont++) \{\\
		S0[i][j][k][l][cont]=0.0;\\
		DS0[i][j][k][l][cont]=0.0;\\
		S1[i][j][k][l][cont]=0.0;\\
		DS1[i][j][k][l][cont]=0.0;\\
		GS[i][j][k][l][cont]=0.0;\\
	      \}
	    \}
	\}
      \}
    \}
    \\
    /*******************************initial conditions*******************************/\\
    sprintf( sstep, "\%s-\%d-\%d", ckpt\_prefix, MDIM, nfields);\\
    if ( (fstep = fopen(sstep, "r")) == NULL )\\
      \{\\
	/* Init from random distribution */\\
	for(i=0;i<N1;i++)\{\\
	  for(j=0;j<N2;j++)\{\\
	    for(k=0;k<N3;k++)\{\\
	      for(l=0;l<N4;l++)\{\\
		\\
		cons= drand48();	 \\
		for (cont2=0;cont2<=(nfields);cont2++)\{\\
		  if ((cons>(cont2/(nfields1))) \&\& (cons<((cont2+1)/nfields1)))\{\\
		    for (cont=0;cont<nfields;cont++)\{                  \\
		      DS0[i][j][k][l][cont]=0.0;\\
		      S0[i][j][k][l][cont]=K[cont2+1][cont+1];\\
		    \}
		  \}
		\}
	      \}
	    \}
	  \}
	\}\\
      \} else \{\\
	/* printf("Have checkpoint \ n"); */ \\
      fscanf( fstep, "\%d \%lf", \&ncounter, \&ct );\\
      sprintf( scp, "\%s-\%d-\%d-\%.2f.bin", ckpt\_prefix, MDIM, nfields, ct);\\
      if ( (fcp = fopen( scp, "rb")) != NULL) \\
	  \{\\
	    fread(\&S0[0][0][0][0][0],  sizeof (double), nfields*N1*N2*N3*N4, fcp);\\
	    fread(\&DS0[0][0][0][0][0], sizeof (double), nfields*N1*N2*N3*N4, fcp);\\
	    fclose(fcp);\\
	  \} else \{\\
	    printf("Error: Cannot open checkpoint data file \%s \ n", scp);\\
	    exit(1);\\
	  \}\\
	/* adjust contar so output file numbering is correct */ \\
	if ( ncounter > 0 )\\
	  contar=ncounter/noutput;\\
	fclose(fstep);\\
	printf("Restarting with ct = \%g , ncounter = \%d, contar = \%d \ n",\\ 
	       ct, ncounter, contar);\\
	skipcp = 1;\\
      \}\\
    filename(doutscaling, doutscaling\_prefix,nfields);\\
    at=pow(ctinitial,expoentec);\\
    /* timing */ \\
    t2 = csecond();\\
    ta = t2 -t1;\\
    printf("Init: time= \%5.2f sec \ n", ta);\\
    /* MAIN LOOP */ \\
\#pragma omp parallel default(none) \\
            private(i,j,k,l,cont,m,ip1,im1,jp1,jm1,kp1,km1,lp1,lm1, \ \\
		    GT,Lshi,R,VS,my\_AREA,my\_ek,my\_eklim,V,ekaux)		\    \\
            shared(N1,N2,N3,N4,nfields,nfields1,GS,S0,dx2,delta,dct,lamda,  \ \\
		   S1,DS1,K,lim,area,DS0,ek,eklim,ct,ctfinal,t1,t2,t3,t4, \ \\
                   ta,tb,tc,td,ncounter,alfa,expoentec,contar,doutfilme, \ \\
                   foutfilme,at,scaling,omega0,V0, noutput, nckpt, fstep, \ \\
		   sstep, scp, fcp, skipcp, doutscaling, GTf, doutfilme\_prefix, \ \\
		   ckpt\_prefix)\\
 \{\\
    while (ct < ctfinal)\{\\
\#pragma omp master\\
      \{\\
\#ifdef TEST\\
	ta=t2-t1;\\
	tb=t4-t2;\\
	tc=0; \\
	td=t4-t1;\\
	printf ("Step:\%3d   ta=\%5.2f   tb=\%5.2f   tot=\%5.2f ", \\
		ncounter, ta,tb,td);\\
	t1 = csecond();\\
\#endif	\\
	delta=0.5*alfa*dct/ct*expoentec;  
	ek=0.0;\\
	area=0.0;\\
        eklim=0.0; \\
      \}\\
 /* end master */\\
\#pragma omp barrier\\
      my\_ek = 0;\\
      my\_eklim = 0.0;\\
      my\_AREA=0;\\
\#pragma omp for\\
      for(i=0;i<N1;i++)\{\\
	for(j=0;j<N2;j++)\{\\
	  for(k=0;k<N3;k++)\{	\\
	    for(l=0;l<N4;l++)\{\\
	      ip1=(i+1) \% N1;\\
	      im1=(i-1+N1) \% N1;\\
	      jp1=(j+1) \% N2;\\
	      jm1=(j-1+N2) \% N2;\\
	      kp1=(k+1) \% N3;\\
	      km1=(k-1+N3) \% N3;\\
	      lp1=(l+1) \% N4;\\
	      lm1=(l-1+N4) \% N4;\\
	      V=0.0;\\
	      GT=0.0;\\
	      /* GS from S0; GT is sum of GS (over cont) */ \\
	      for (cont=0;cont<nfields;cont++)\{          \\
		GS[i][j][k][l][cont]=0.5*(pow((0.5*(S0[ip1][j][k][l][cont]-S0[im1][j][k][l][cont])),2.0)\\
					  +pow((0.5*(S0[i][jp1][k][l][cont]-S0[i][jm1][k][l][cont])),2.0)\\
					  +pow((0.5*(S0[i][j][kp1][l][cont]-S0[i][j][km1][l][cont])),2.0)\\
					  +pow((0.5*(S0[i][j][k][lp1][cont]-S0[i][j][k][lm1][cont])),2.0));\\         
		GT+=GS[i][j][k][l][cont]; \\
	      \} \\
	      GTf[i][j][k] = (float) GT;\\
	      /* Lshi from S0 (over cont) */\\
	      for (cont=0;cont<nfields;cont++)\{\\
		Lshi[cont]=S0[ip1][j][k][l][cont]+S0[im1][j][k][l][cont]\\
		           -2.0*S0[i][j][k][l][cont]+S0[i][jp1][k][l][cont]\\
		           +S0[i][jm1][k][l][cont]-2.0*S0[i][j][k][l][cont]\\
		           +S0[i][j][kp1][l][cont]+S0[i][j][km1][l][cont]\\
		           -2.0*S0[i][j][k][l][cont]+S0[i][j][k][lp1][cont]\\
                           +S0[i][j][k][lm1][cont]-2.0*S0[i][j][k][l][cont]; \\
		Lshi[cont]/=dx2; \\
	      \}	  		          \\
	      /* R  from S0 (over m over cont) */\\
	      for(m=0;m<nfields1;m++)\{\\
		R[m]=0;\\
		for (cont=0;cont<nfields;cont++)\{\\
		  R[m]+=pow(S0[i][j][k][l][cont] - K[m+1][cont+1],2.0);\\
		  VS[cont]=0;\\
		\}	  
	      \}\\
	      /* VS from R (over m over cont) */\\
	      for(m=0;m<nfields1;m++)\{\\
		for (cont=0;cont<nfields;cont++)\{\\
		  VS[cont]+=lamda*3.0*(R[m]-3.0)*(R[m]-1.0)*(S0[i][j][k][l][cont]-K[m+1][cont+1]);\\
		\}
	      \}\\
	      ekaux = 0.0; \\
	      /* DS1 from DS0 \& VS \& Lshi,  S1 from S0 \& DS1 (over cont)*/ \\
	      for (cont=0;cont<nfields;cont++)\{\\
		DS1[i][j][k][l][cont]=((1.0-delta)*DS0[i][j][k][l][cont]\\
				       +dct*(Lshi[cont]-VS[cont]))/(1.0+delta);\\
		S1[i][j][k][l][cont]=S0[i][j][k][l][cont]+dct*DS1[i][j][k][l][cont];\\
		
		ekaux+=DS1[i][j][k][l][cont]*DS1[i][j][k][l][cont];\\
	      \}\\
	      for(m=0;m<nfields1;m++)\{\\
		V+=lamda*0.5*R[m]*(R[m]-3.0)*(R[m]-3.0);\\
	      \}\\
	      if (V >= lim)\{\\
		my\_eklim += ekaux/V;\\
		my\_AREA+=1.0;\\
	      \}
	    \}
	  \} 
	\}
      \}\\
\#pragma omp for\\
      for(i=0;i<N1;i++)\{\\
	for(j=0;j<N2;j++)\{\\
	  for(k=0;k<N3;k++)\{\\
	    for(l=0;l<N4;l++)\{\\
	      for (cont=0;cont<nfields;cont++)\{\\
		S0[i][j][k][l][cont] =S1[i][j][k][l][cont];\\
		DS0[i][j][k][l][cont]=DS1[i][j][k][l][cont];\\
		my\_ek+=DS1[i][j][k][l][cont]*DS1[i][j][k][l][cont];\\
	      \}	  
	    \}
	  \}
	\}   
      \}\\
\#pragma omp atomic\\
   area += my\_AREA;\\
\#pragma omp atomic\\
   ek += my\_ek;\\
\#pragma omp atomic\\
   eklim += my\_eklim;\\
\#pragma omp barrier\\
\#pragma omp master\\
   \{\\
\#ifdef TEST\\
    t2 = csecond();\\
\#endif\\
      ct+=dct;\\
      at=pow(ct,expoentec);\\
      ncounter+=1;\\
      if ( (scaling=fopen(doutscaling,"a")) != NULL  )\\
	\{\\
	   fprintf(scaling,"\%e \%e \%e \%e \ n",ct,area/(dim1*dim2*dim3*dim4),eklim/area,ek/((area)*omega0*V0));\\
	   fclose(scaling);\\
	\}\\
	/****************************output*****************************/\\
	if (((ncounter-1) \% noutput )==0.0) \{\\
	  printf("(out)");\\
	  contar+=1;\\
	  filename(doutfilme,doutfilme\_prefix,nfields*100+contar);\\     
	  if ( (foutfilme=fopen(doutfilme,"w")) != NULL ) \{\\
	  \#ifdef RAW\\
		  fwrite(GTf, sizeof(float), N1*N2*N3, foutfilme);\\
\#else	\\    
	    for(i=0;i<N1;i++)\{\\
	      for(j=0;j<N2;j++)\{\\
		for(k=0;k<N3;k++)\{\\
		  fprintf(foutfilme,"\%f \ n",GTf[i][j][k]);\\
		\}
	      \}
	    \}\\
\#endif\\
	    fclose(foutfilme);\\
	  \}
	\}
\\
	/* checkpoint every nckpt - except if just restored */\\
	if ( skipcp == 1 )\\
	  skipcp = 0;\\
	else \\
	  \{\\
	    if ( ncounter>0 \&\& nckpt>0 \&\& (ncounter \% nckpt == 0 ) ) \{\\	      
	      /* printf("Checkpointing at ct=\%g, ncounter=\%d ", ct, ncounter);*/ \\
	      printf("(ckpt)");\\
	      sprintf( scp, "\%s-\%d-\%d-\%.2f.bin", ckpt\_prefix, MDIM, nfields, ct);\\
	      if ( (fcp = fopen( scp, "w")) != NULL)\\ 
		\{\\
		  fwrite(\&S0[0][0][0][0][0],  sizeof (double), nfields*N1*N2*N3*N4, fcp);\\
		  fwrite(\&DS0[0][0][0][0][0], sizeof (double), nfields*N1*N2*N3*N4, fcp);\\
		  fclose(fcp);\\
		\} else \{\\
		  printf("Error: Cannot open file \%s \ n", scp);\\
		  exit(1);\\
		\}\\
	      sprintf( sstep, "\%s-\%d-\%d", ckpt\_prefix, MDIM, nfields);\\       
	      if ( (fstep = fopen(sstep, "w")) != NULL )\\
		\{\\
		  fprintf( fstep, "\%d \%lf", ncounter, ct );\\
		  fclose(fstep);\\
		\} else \{\\
		  printf("Error: Cannot open file \%s \ n", sstep);\\
		  exit(1);\\
		\}
	    \}
	  \}\\
	  printf("\ n");\\
\#ifdef TEST \\
      t4 = csecond();\\
\#endif\\
\}\\
\#pragma omp barrier\\
    \}\\
 /* end loop over ct */\\
 \}\\
 */ end parallel*/\\
\}
\section*{Simulation Parameters \& default values}
\#include "walls.h"\\
/* Filenames */ \\
char inifile[MAXBUF]           = "dados.txt" ;\\
char doutfilme\_prefix[MAXBUF]  = "box3D"    ;\\
char doutscaling\_prefix[MAXBUF]= "scaling3D"    ;\\ 
char ckpt\_prefix[MAXBUF]       = "ckpt3D"   ;\\
/* Default values */\\
double lim       = 0.0290 ;\\
double expoente  = 0.6666666667;\\
double ctinitial = 1.0  ;\\
double tampasso  = 0.25 ;\\
int npassos      = 1020   ;   \\
/* number of steps */\\
int nfields      = inicampos    ;\\
/* needs to be <= inicampos*/\\
int nckpt        = 255  ;    \\
/* frequency to checkpoint */\\
int noutput      = 1   ;\\
/* frequency to output field array */\\  
void print\_parameters(void )\\
\{\\
  printf("\# \# \# \# \# \# \# \# \# \# \# \# \# \# \# \# \# \# \# \# \# \# \# \# \# \# \# \# \# \# \# \# \# \# \# \# \# \# \# \# \ n \ n ");\\
  printf("Simulation parameters: \ n \ n");\\
  printf(" dim      = \%d  \ n ", MDIM);\\
  printf(" lim      = \%lf \ n", lim);\\
  printf(" exp      = \%lf \ n", expoente);\\
  printf(" ct\_ini   = \%lf \ n", ctinitial);\\
  printf(" tampasso = \%lf \ n", tampasso);\\
  printf(" steps    = \%d \ n ", npassos);\\
  printf(" fields   = \%d \ n", nfields);\\
  printf(" n output = \%d \ n", noutput);\\
  printf(" n ckpt   = \%d \ n", nckpt);\\
  printf("\ n");\\
  printf("\# \# \# \# \# \# \# \# \# \# \# \# \# \# \# \# \# \# \# \# \# \# \# \# \# \# \# \# \# \# \# \# \# \# \# \# \# \# \# \# \ n \ n ");\\
\}

\section*{Filename Input}
void filename(outputname, inputname, outnum)\\
        char outputname[];\\
        char inputname[];\\
        int     outnum;\\
\{\\
        int     dig4, dig3, dig2, dig1;\\
        int     inull;\\
        register int jout;\\
        dig4 = outnum / 1000;\\
        dig3 = (outnum - 1000 * dig4) / 100;\\
        dig2 = (outnum - 1000 * dig4 - 100 * dig3) / 10;\\
        dig1 = outnum - 1000 * dig4 - 100 * dig3 - 10 * dig2;\\
        inull = 1;\\
        while (((inputname[inull - 1] != ' ') \& \& (inputname[inull - 1] != '\ 0') \& \& (inull < 26)))\\
                inull = inull + 1;\\
        for (jout = 1; jout <= inull - 1; jout++)\\
                outputname[jout - 1] = inputname[jout - 1];\\
        outputname[inull - 1] = '.';\\
        outputname[inull + 1 - 1] = 48 + dig4;\\
        outputname[inull + 2 - 1] = 48 + dig3;\\
        outputname[inull + 3 - 1] = 48 + dig2;\\
        outputname[inull + 4 - 1] = 48 + dig1;\\
        outputname[inull + 5 - 1] ='\ 0';\\
\}

\section*{Dimensions and Number of Fields}
\# define MDIM 512\\
\# define dim1 MDIM\\
\# define dim2 MDIM\\
\# define dim3 MDIM\\
\# define dim4 1\\
/* preprocessor */\\
\# define MAXFUB 100\\
/* function definitions */\\
extern void filename ( char [], char [], int);\\
extern double csecond ( void);\\
extern void print\_parameter( void);\\
/* filename */\\
extern char inifile[MAXFUB];\\
extern char doutfilme\_prefix[MAXFUB];\\
extern char doutscaling\_prefix[MAXFUB];\\
extern char ckpt\_prefix[MAXFUB];\\
/*default values*/\\
extern double lim;\\
extern double expoente;\\
extern double ctinitial;\\
extern double tampasso;\\
extern double npassos;\\
extern double nfields;\\
extern double nckpt;\\
extern double noutput;\\

\renewcommand{\baselinestretch}{1.0}
\bibliography{biblio}

\end{document}